\documentclass{bredele}

\usepackage{thesis}

\title{Solvatation de systèmes d’intérêt pharmaceutique : apports de la théorie de la fonctionnelle de la densité moléculaire}

\author{Cédric Gageat}

\institute{l'\'Ecole Normale Supérieure}

\supervisor[Maximilien Levesque]{Daniel Borgis}
\doctoralschool{Chimie physique et chimie analytique de Paris Centre}{388}
\date{24 Novembre 2017}

\jury{
  Mme Francesca Ingrosso\\
  Université de Lorraine, Rapportrice

  M. Thomas Simonson\\
  \'Ecole polytechnique, Rapporteur

  Mme Liliane Mouawad\\
  Institue Curie, Membre du jury

  M. Jean-Philip Piquemal\\
  UPMC, Membre du jury

  M. Ivan Duchemin\\
  CEA/INAC, Membre du jury

  M. Daniel Borgis\\
  \'Ecole Normale Supérieure, Directeur

  M. Maximilien Levesque\\
  \'Ecole Normale Supérieure, Encadrant
}

\frabstract{Le développement d'un nouveau médicament est un processus long et co\^uteux. Entre la détermination d'une cible thérapeutique et la mise sur le marché d'un nouveau médicament, plus de dix ans de recherche sont nécessaires pour un coût supérieur à un milliard d'euros.
L'accélération de ce processus et la réduction de son coût restent un enjeu majeur. Pour y parvenir, les simulations numériques, peu co\^uteuses et rapides, sont massivement utilisées. Malgré cela, elles restent limitées, en partie à cause de la quantité très importante de molécules de solvant à considérer.
La théorie de la fonctionnelle de la densité moléculaire permet d'étudier la solvatation de composés de n'importe quelle taille et de n'importe quelle forme. Elle prédit en quelques secondes seulement à la fois l'énergie libre de solvatation et une carte détaillée de la densité d'équilibre autour de ce soluté.
Ces grandeurs étant à la base de nombreux autres calculs utilisés par l'industrie pharmaceutique, la MDFT ouvre donc une autre voie d'optimisation de ces process. Cette thèse consiste à effectuer le premier pas vers l'ensemble de ces applications. 
Pour cela, nous avons adapté la théorie ainsi que le code associé avant de l'appliquer à des systèmes biologiques.}

\enabstract{Drug development is time and cost-consuming: It takes in average 10 years and 1 billion euros to move from a therapeutic target to a new drug. To speedup this process and reduce its cost, numerical simulation are massively used.
Nevertheless, they remain limited, one reason of which is the huge amount of solvent molecules to consider.
The molecular density functional theory is a liquid state theory that allows the study of the solvation thermodynamics of solutes of arbitrary shape.
MDFT predicts, in few seconds only, the free energy of solvation and the solvent profils.
These parameters are at the heart of many others calculation used by the pharmaceutical industry. 
This thesis is the first step towards these applications. For that purpose, we adapted the theory as well as the associated code to this new target, then applied them to system of biological interest.
}

\frkeywords{solvatation biomolécules théorie de la fonctionnelle de la densité}
\enkeywords{solvation biomolecules Molecular density functional theory watermap}

\begin{document}

\frontmatter

\tikzexternaldisable
\maketitle{}
\tikzexternalenable

\cleardoublepage

\chapter*{Remerciements}

Je souhaite tout d'abord remercier Ludovic Jullien, directeur de l'UMR 8640\\P.A.S.T.E.U.R., pour son accueil au sein de son laboratoire.
\medbreak
Mes remerciements vont ensuite naturellement vers Daniel Borgis et Maximilien Levesque qui ont dirigé et encadré ma thèse, pour leur disponibilité, leur encadrement et leurs conseils.
\medbreak
Je remercie également les membres du jury d’avoir accepté d'évaluer et d’assister à la présentation de ce travail.
\medbreak
Je remercie chaleureusement tous les membres du pôle théorie du département de chimie de l'\'Ecole Normale Supérieur ainsi que ceux de la Maison De La Simulation qui m'ont acceuillis et avec qui j'ai beaucoup appris. Merci Nicolas C., Matthieu et Yacine pour les nombreuses discussions que nous avons eues et vos contributions à la réussite de ce projet. Merci Matthieu et Yacine de m'avoir fait découvrir ce merveilleux monde qu'est le HPC. Merci Nicolas L. pour tes invitations à Montrouge. Je remercie également très chaleureusement pour leur soutien logistique: Victoria Terziyan, Stéphanie Benabria et Valérie Belle.
\medbreak
Un grand merci à ceux qui ont contribué à rendre ces 3 ans beaucoup plus agréables: Elsa, Beno\^it, Geoffrey, Sébastien. Merci à tous les quatre pour votre bonne humeur au quotidien. Merci Sébastien, merci Geoffrey d'avoir toujours été présents même dans les moments les plus critiques. Merci Elsa, sans toi je serais sans doute mort de faim.
\medbreak
J'en profite également pour adresser mes plus sincères remerciements à toute l'équipe du master ISDD. Merci Anne-Claude de m'avoir soutenu et supporté pendant tout ce temps. Merci Leslie d'avoir su rester professionnelle et juste malgré nos différents.
\medbreak
Je tiens également à remercier tous ceux qui ont su m'orienter au bon moment et qui ont finalement contribué à cette réussite: David Lagorce, Anne-Claude Camproux, Ludovic Jullien, Carole Jourdan, Matthieu Haefele et Yacine Ould-Rouis.
\medbreak
Je souhaite enfin te remercier, Marine. Sans toi je ne serais pas là o\`u j'en suis aujourd'hui. Merci.

\clearemptydoublepage

\renewcommand\contentsname{Sommaire}
\tableofcontents
 
 
\renewcommand{\cftdotsep}{\cftnodots}
\cftpagenumbersoff{figure}
\cftpagenumbersoff{table}
\cleardoublepage
\listoffigures
\cleardoublepage
\listoftables

\clearemptydoublepage
\section*{Notations}
\begin{tabular}{l l}
$\boldsymbol{r}$ & Position, en 3D, de la molécule d'eau étudiée \\
$\Omega$ & Orientation de la molécule d'eau étudiée \\
$\rho\left(\boldsymbol{r},\Omega \right)$ & Densité en solvant à la position $\boldsymbol{r}$ et pour l'orientation $\Omega$ [\AA$^{-3}$]  \\  
$\rho_0$ & Densité bulk de référence (1 kg.L$^{-1}$ soit 0.033 \AA$^{-3}$ pour l'eau) \\
$\mathcal{F}[\rho\left(\boldsymbol{r},\Omega \right)]$ & Fonctionnelle de la densité moléculaire $\rho$ \\
$\mathcal{F}_{id}[\rho\left(\boldsymbol{r},\Omega \right)]$ & Partie idéale de la fonctionnelle de la densité moléculaire [kJ.mol$^{-1}$]\\
$\mathcal{F}_{ext}[\rho\left(\boldsymbol{r},\Omega \right)]$ & Partie extérieure de la fonctionnelle de la densité moléculaire  [kJ.mol$^{-1}$]\\
$\mathcal{F}_{exc}[\rho\left(\boldsymbol{r},\Omega \right)]$ & Partie d'excès de la fonctionnelle de la densité moléculaire [kJ.mol$^{-1}$]\\
$\mathcal{F}_{b}[\rho\left(\boldsymbol{r},\Omega \right)]$ & Fonctionnelle de bridge [kJ.mol$^{-1}$]\\
$\phi\left(\boldsymbol{r},\Omega \right)$ & Potentiel d'interaction entre le soluté et le solvant à la position $\boldsymbol{r}$ et pour \\
 & l'orientation $\Omega$ [kJ.mol$^{-1}$]\\
$\mathrm{k_B}$ & Constante de Boltzmann. $\mathrm{k_B}$=8.3144598.10$^{-3}$ [kJ.mol$^{-1}$.K$^{-1}$]\\
$c\left(\boldsymbol{r}-\boldsymbol{r}^\prime,\Omega,\Omega^\prime \right)$ & Fonction de corrélation directe entre la densité à la position $\boldsymbol{r}$ et pour \\
 & l'orientation $\Omega$ et la densité à la position $\boldsymbol{r}^\prime$ et pour l'orientation $\Omega^\prime$\\
$\gamma(\boldsymbol{r},\Omega)$ & Résultat de la convolution entre la  fonction de corrélation directe et la\\
& fonction $\Delta\rho$\\
$\gamma$ & Tension de surface [mJ.m$^{-2}$]\\
$\Delta G_{solv}$ & \'Energie libre de solvatation [kJ.mol$^{-1}$]\\
f$\ast$g & Convolution entre les fonctions f et g\\
$\hat{f}$ & Transformée de Fourier de la fonction f\\
$\boldsymbol{k}$ & Vecteur réciproque\\
$\bar{\rho(\boldsymbol{r})}$ & Densité gros grain à la position $\boldsymbol{r}$ [\AA$^{-3}$] \\
$\beta$ & Inverse du produit de la constante de Boltzmann et de la température \\
 & $(\mathrm{k_B}T)^{-1}$ [mol.kJ$^{-1}$]\\
\end{tabular}

\clearpage
\section*{Acronymes}
\begin{tabular}{ll}
MDFT & Théorie de la fonctionnelle de la densité moléculaire\\
HNC & Hyper-Netted Chain approximation\\
DM & Dynamique moléculaire \\
MC & Monte-Carlo \\
PDB & Protein data bank \\
FT & Transformée de Fourier \\
FFT & Transformée de Fourier rapide \\
HT & Transformée de Hankel \\
FGSHT & Transformée des harmoniques sphériques généralisées rapide \\
RDF & Fonction de distribution radiale\\
\end{tabular}

\clearemptydoublepage
\mainmatter

\part{Introduction et théorie}

\clearemptydoublepage
\chapter{Introduction}
\label{chap:introduction}

\begin{figure}[ht]
  \center
  \begin{tikzpicture}

    \draw[fill=white!30!yellow] (0,1,0)--(3,3,0)--(3,9,0)--(0,11,0)--cycle;
    \draw[fill=white!30!orange] (3,3,0)--(6,4.5,0)--(6,7.5,0)--(3,9,0)--cycle;
    \draw[fill=white!30!red] (6,4.5,0)--(12,5.5,0)--(12,6.5,0)--(6,7.5,0)--cycle;

    \draw (-1.5,6) node[above]{\textasciitilde M} ;
    \draw (-1.5,6) node[below]{composés} ;
    \draw (1.5,6) node[above]{\textasciitilde 10K} ;
    \draw (1.5,6) node[below]{composés} ;
    \draw (4.5,6) node[above]{\textasciitilde 250} ;
    \draw (4.5,6) node[below]{composés} ;
    \draw (9,6) node[above]{\textasciitilde 5} ;
    \draw (9,6) node[below]{composés} ;
    
    \draw (1.5,11.5) node[above]{recherche et} ;
    \draw (1.5,11.5) node[below]{découverte} ;
    \draw (4.5,9.5) node[above]{développement} ;
    \draw (4.5,9.5) node[below]{pré clinique} ;
    \draw (9,8) node[above]{essais} ;
    \draw (9,8) node[below]{cliniques} ;

     \draw[fill=red]  (14.5,6,0) circle (0.3);
     \draw[red, fill=red] (14.5,6.3,0)--(14,6.3,0)--(14,5.7,0)--(14.5,5.7,0)--cycle;
  	 \draw[fill=red] (14.5,6.3,0)--(14,6.3,0)--(14,5.7,0)--(14.5,5.7,0);
     \draw[fill=white] (13.5,6,0) circle (0.3);
     \draw[white, fill=white] (13.5,6.3,0)--(14,6.3,0)--(14,5.7,0)--(13.5,5.7,0)--cycle;
     \draw[fill=white] (13.5,6.3,0)--(14,6.3,0)--(14,5.7,0)--(13.5,5.7,0);

     \draw[line width=2pt, latex-latex] (0,0) -- (3,0);
     \draw[line width=2pt, latex-latex] (3,0) -- (6,0);
     \draw[line width=2pt, latex-latex] (6,0) -- (12,0);

    \draw (1.5,0) node[below]{3 années} ;
    \draw (4.5,0) node[below]{1-2 années} ;
    \draw (9,0) node[below]{6-7 années} ;

  \end{tikzpicture}
    \caption{Schéma simplifié du développement d'un nouveau médicament.}
    \label{fig:entonnoir_drug_design}
\end{figure}
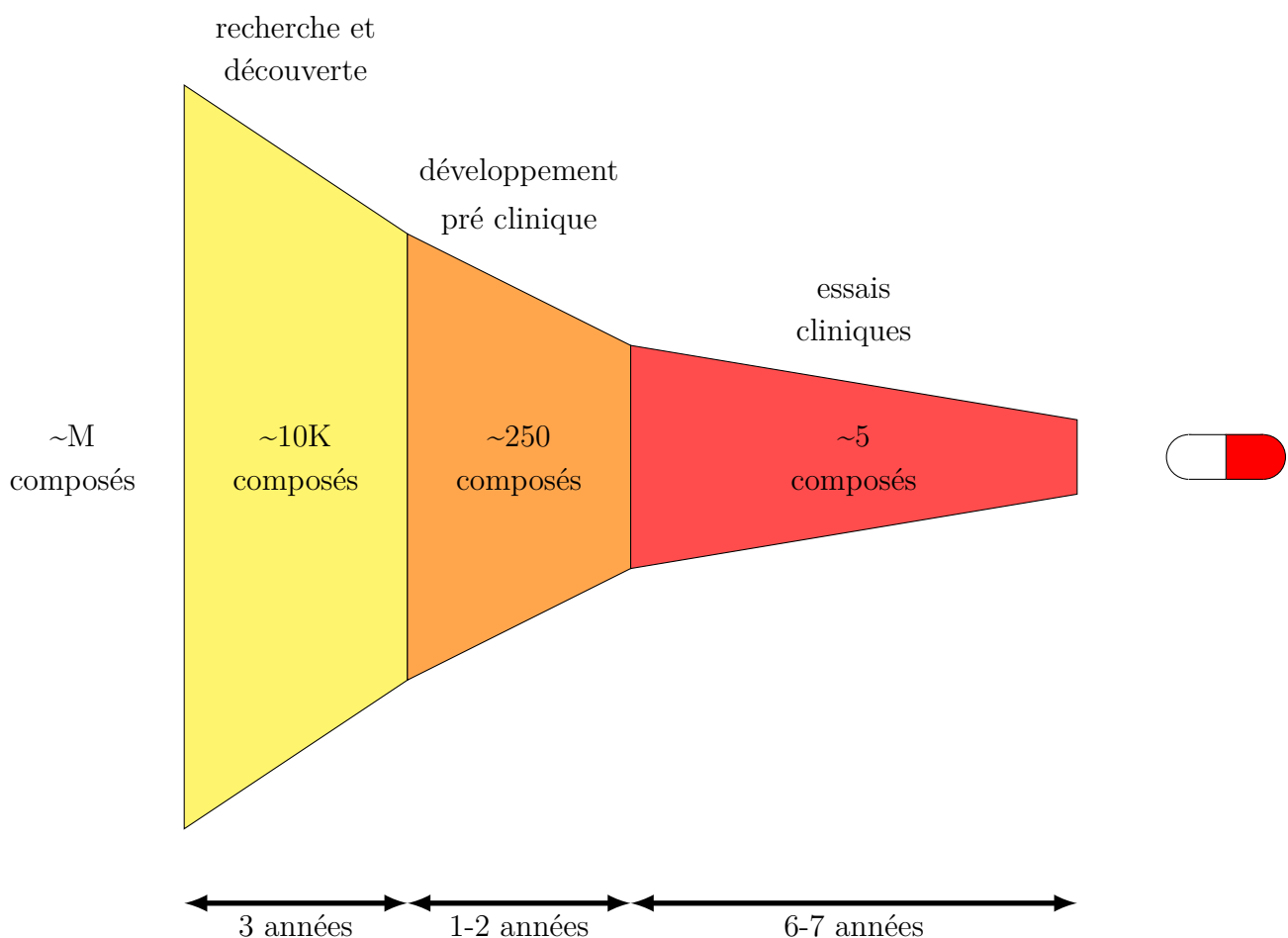

Entre l'identification d'une cible thérapeutique et la mise sur la marché d'un nouveau médicament, une dizaine d'années de recherche et plus d'un milliard d’euros sont nécessaires\cite{DiMasi_price_2003, Siddiqui_high_2012} (voir figure \ref{fig:entonnoir_drug_design}). Afin d'accélérer ce processus et ainsi d'en diminuer le coût, les simulations informatiques sont massivement utilisées. Pour s'approcher au maximum des conditions réelles, et donc de ce qu'il se passe dans le corps humain, ces simulations doivent avoir lieux dans l'eau, c'est à dire en solution. Malgré la puissance de calcul des ordinateurs actuels, ces simulations restent limitées à cause du nombre important de molécules d'eau nécessaires. Afin de s'adapter au mieux aux besoins des différentes études, il existe plusieurs représentations du solvant qui permettent de choisir entre vitesse et précision. Dans ce manuscrit, nous allons présenter la théorie de la fonctionnelle de la densité moléculaire\cite{jeanmairet_molecular_2013-1,jeanmairet_classical_2015,Jeanmairet_introduction_2014,jeanmairet_hydration_2014,levesque_solvation_2012} (MDFT) qui allie vitesse et précision. Mon projet de thèse consiste à effectuer le premier pas vers toutes ces applications en adaptant la théorie ainsi que son implémentation aux systèmes biologiques.

\section{Contexte}
Avant d'envisager le développement d'une nouvelle solution thérapeutique il est nécessaire de comprendre les phénomènes à l'origine de la maladie que l'on souhaite guérir. La première étape consiste donc à comprendre la cascade biologique à l'origine de cette maladie. Une fois le phénomène identifié, les chercheurs vont sélectionner une protéine impliquée dans cette cascade. Cette molécule sera appelée "cible". À partir de cet instant, le développement d'un médicament va consister à trouver  parmi plusieurs millions de petits composés le meilleur candidat avant de l'optimiser. Ce candidat doit répondre à plusieurs critères: il doit (i) pouvoir accéder en quantité suffisante à la protéine cible, (ii) se lier à elle et l'inhiber afin de bloquer la cascade biologique à l'origine de la maladie visée et (iii) se lier le moins possible à d'autres protéines afin de minimiser les effets secondaires. Pour des raisons de sécurité, de temps et de coût, il est bien sûr impossible de tester l'ensemble de ces millions de molécules en laboratoire ou en essai clinique. Les simulations informatiques sont donc massivement utilisées afin d'effectuer un premier tri et de passer de plusieurs millions de candidats à seulement quelques milliers. Les candidats ayant passés avec succès les différents tests (toxicité, affinité avec la cible, ...) seront ensuite synthétisés et testés en laboratoire. Une poignée de molécules prometteuses sera enfin testée en essai clinique. C'est seulement à l'issu de ce processus long et coûteux qu'une molécule pourra devenir un médicament. Tous ces phénomènes se déroulent dans le corps humain et donc en solution, le solvant aura donc un rôle clé dans l'ensemble de ces processus.

\section{Solvatation}
La solvatation est le phénomène chimique qui consiste à plonger un composé, le soluté, qu'il soit solide, liquide ou gazeux en solution. Une fois en solution, la stabilité ainsi que le rôle du soluté seront fortement influencés par les molécules de solvant\cite{NickPace_protein_2004, levy_water_2004, Meyer_internal_1992, Ladbury_just_1996, GarcaSosa_hydration_2013, Lemieux_how_1996, Tame_role_1996, Li_effect_2005, Snyder_mechanism_2011, Wang_ligand_2011, Mobley_binding_2009, Barillari2007, Olano_hydration_2004, Bren_individual_2012, Ahmed_bound_2011, VAIANA_molecular_2006, Genheden_accurate_2011, Abel_contribution_2011, Biela_ligand_2012, Stegmann_thermodynamic_2009}. Afin de mieux comprendre et analyser cette influence, deux aspects de la solvatation sont étudiés: (i) l'aspect structural et (ii) l'aspect énergétique. 

\subsection{La structure en solution}
Historiquement, le \textit{drug design} était basé sur la recherche d'une complémentarité de forme entre la cible protéique et le ligand candidat médicament. Cette méthode est appelée \textit{structure-based drug design}\cite{Anderson_process_2003,Zhang_towards_2011,Agrawal_structure_2013}. Dans ce paradigme, la structure de la protéine ainsi que la position des molécules de solvant autour d'elle étaient donc indispensables afin de sélectionner et d'optimiser au mieux les candidats médicaments. Devant la quantité croissante de structures disponibles, la \textit{Protein Data Bank}\cite{Bruno_crystallography_2017, Berman_protein_2000} (PDB) a vu le jour en 1971. La PDB est une base de données collaborative des structures de composés biologiques expérimentalement résolues par RMN\cite{Montelione_recommendations_2013}, rayon X\cite{Read_new_2011} ou encore par microscopie électronique\cite{Henderson_outcome_2012}. À ce jour, plus de 130 000 structures sont disponibles. Cependant, malgré une croissance exponentielle du nombre de structures disponibles, il est récemment apparu que l'étude de la structure ne permettait pas une image complète et précise de ces phénomènes \cite{Henry_structure_2001}.

\subsection{Les énergies liées à la solvatation}
Le développement récent de calorimètres hautes performances permet aujourd'hui d'accéder à une vue énergétique complète de systèmes biologiques\cite{Chaires_Calorimetry_2008, Garbett_thermodynamic_2012, Klebe_applying_2015} et ainsi de venir compléter les données structurales. Il est aujourd'hui par exemple possible, pour une cible donnée, d'acquérir les données énergétiques complètes de plusieurs dizaines de milliers de composés. Parmi les énergies étudiées en drug-design, nous nous intéresserons dans ce rapport aux énergies libres de solvatation ainsi qu'aux énergies libres de liaison.

\subsubsection{L'énergie libre de solvatation}
\begin{figure}[ht]
  \center
  \begin{tikzpicture}
  \tikzstyle{lien}=[->,>=stealth,rounded corners=5pt,thick]
  \node[inner sep=0pt] (prot) at (0,0)
      {\setlength{\fboxrule}{1pt}%
      {\fbox{\includegraphics[width=.25\textwidth]{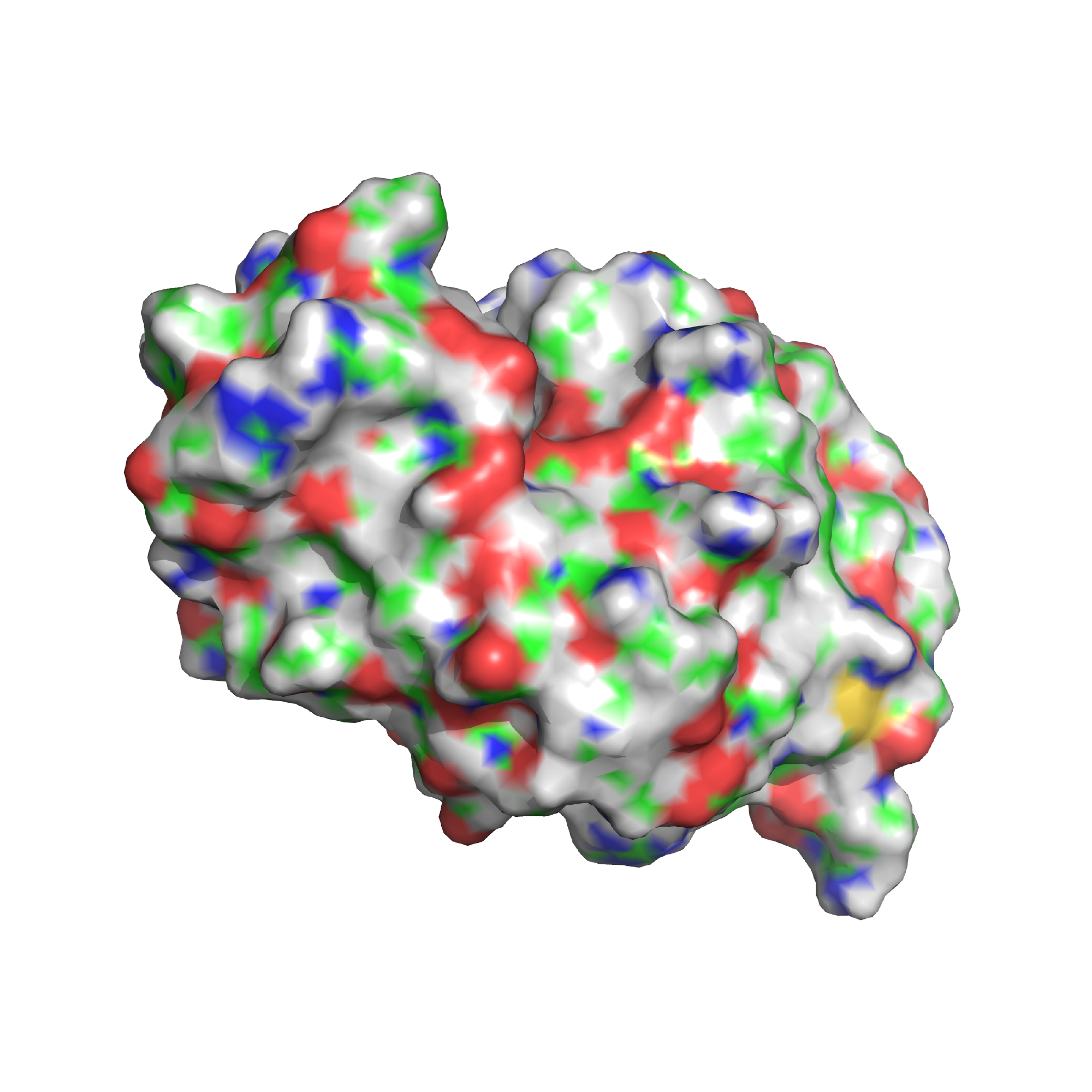}}}%
};
  \node[inner sep=0pt] (water) at (6,0)
      {\setlength{\fboxrule}{1pt}%
      {\fbox{\includegraphics[width=.25\textwidth]{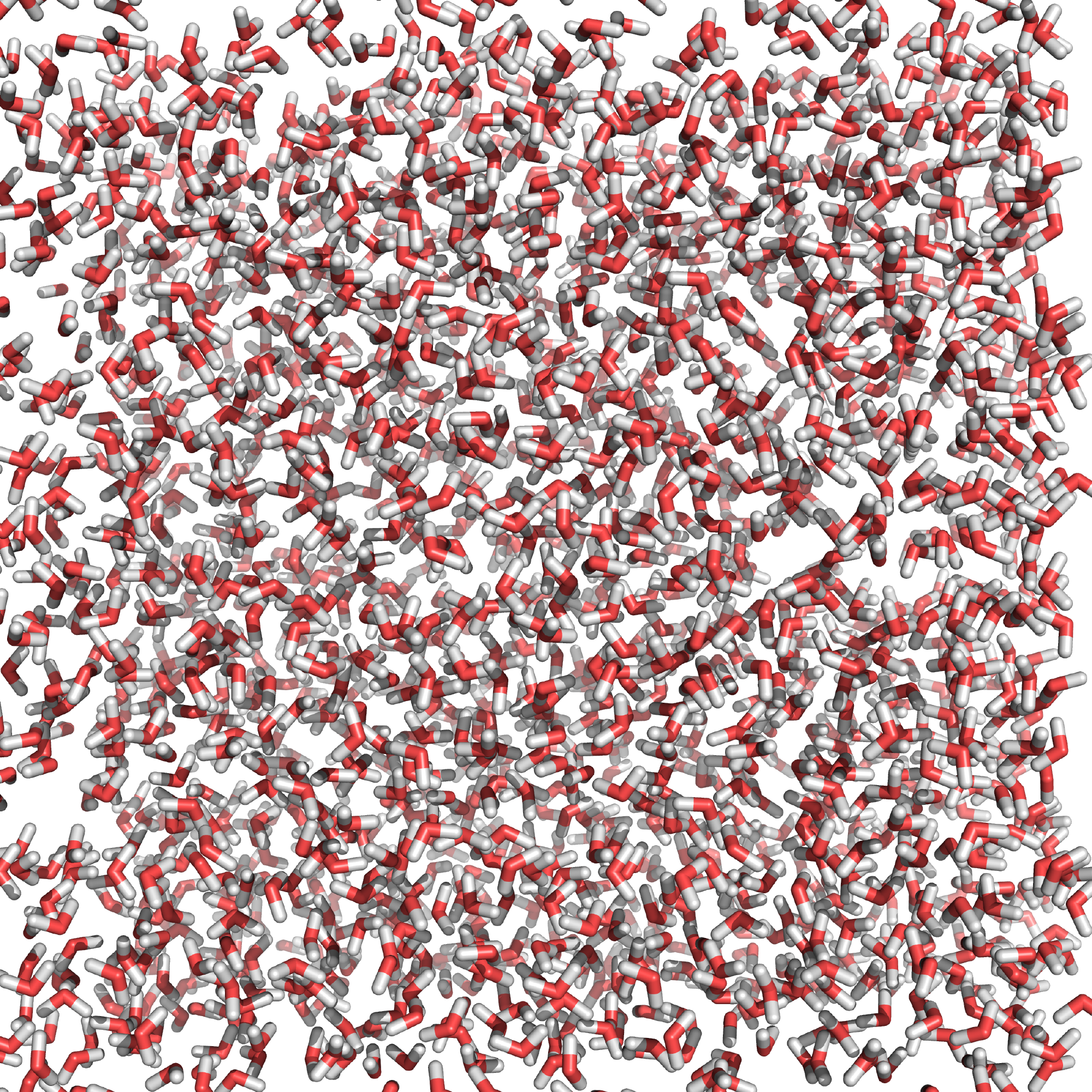}}}%
};
  \node[inner sep=0pt] (prot_water) at (12,0)
      {\setlength{\fboxrule}{1pt}%
      {\fbox{\includegraphics[width=.25\textwidth]{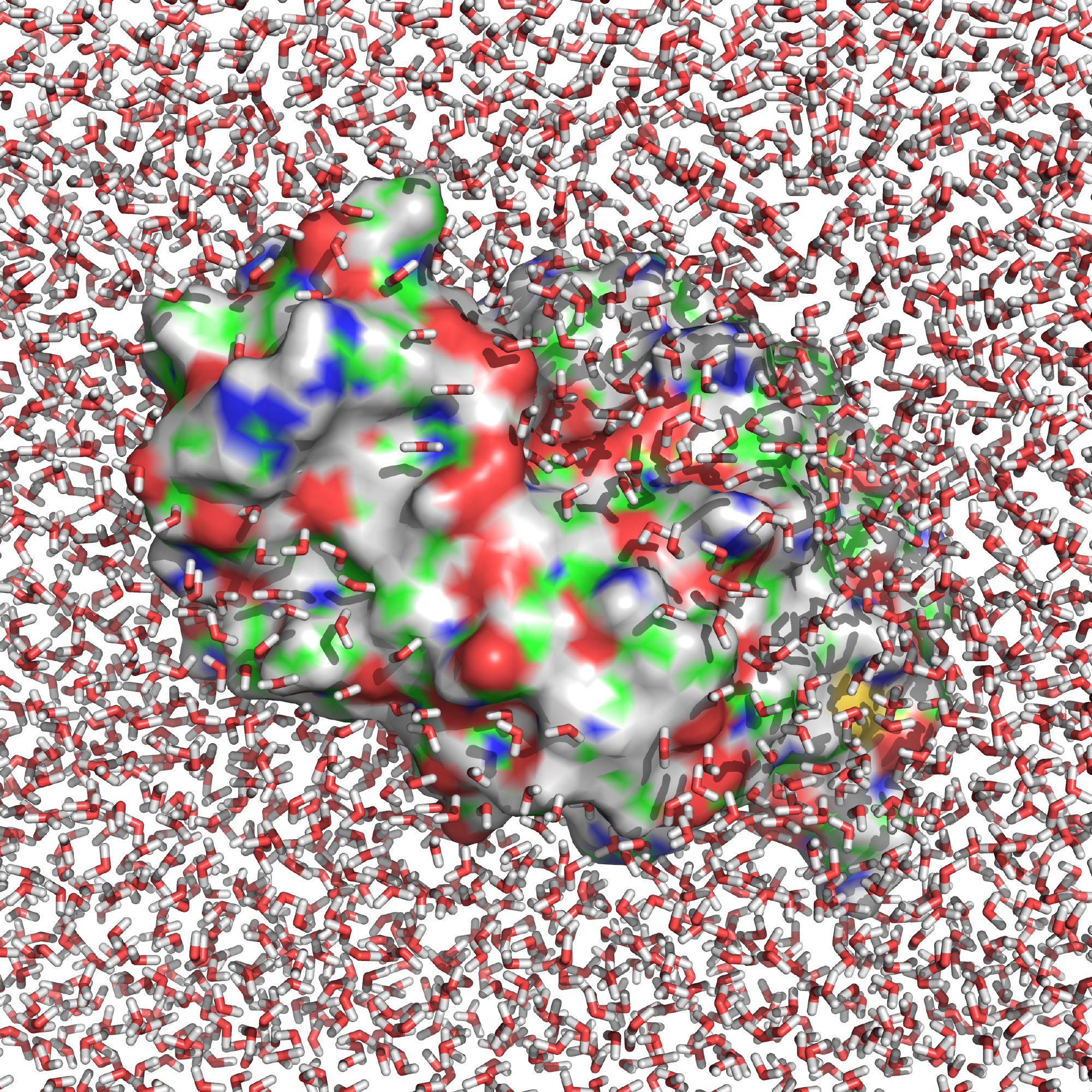}}}%
};

  \draw (2.5,0) node[right] {+};

  \draw[-latex] (water) -- (prot_water);
  
  \end{tikzpicture}
      \caption{Solvatation d'une protéine.}
      \label{fig:solvatation_def}
\end{figure}

La première étape nécessaire à tout phénomène en solution est la solvatation. À ce niveau, on considère deux catégories de composés: les composés hydrophiles, qui aiment l'eau (solvophiles pour un solvant arbitraire) et les composés hydrophobes, qui n'aiment pas l'eau (solvophobe pour un solvant arbitraire). Il est possible de différencier ces composés en mesurant expérimentalement ou en prédisant numériquement leurs énergies libres de solvatation. L'énergie libre de solvatation correspond à l'énergie nécessaire au transfert de notre soluté depuis le vide jusqu'en solution. En d'autres termes, sur la figure \ref{fig:solvatation_def}, elle correspond à la différence d'énergie libre entre le système final (soluté en solution) et le système initial (soluté dans le vide + boîte d'eau). 

La solvation des composés hydrophiles, stabilisés dans l'eau, sera spontanée. Leurs énergies libres de solvatation seront donc négatives. Au contraire, les composés hydrophobes nécessiteront l'apport d'énergie (chauffage, agitation, ...) afin de permettre leur dissolution. Leurs énergies libres de solvatation seront donc positives. 

\subsubsection{L'énergie libre de liaison}
L'énergie libre de liaison correspond à la différence  entre l'énergie libre d'un complexe formé par deux composés (comme une protéine cible et un ligand) et l'énergie libre de ces deux composés séparés. Pour qu'un médicament soit efficace, il doit se fixer à la protéine cible afin d'en altérer la fonction. Lors de l'optimisation d'un candidat médicament, la valeur la plus faible possible d'énergie libre de liaison sera donc recherchée afin de favoriser l'affinité entre ces deux composés.

\section{Les simulations numériques}
Malgré l'évolution des techniques expérimentales, l'acquisition de données structurelles et énergétiques reste longue et coûteuse. Il n'est donc pas possible de traiter systématiquement l'ensemble des millions de composés disponibles au début du processus de sélection. Les simulations numériques sont donc utilisées pour faire un premier tri. Cependant la façon de considérer le solvant au travers de telles simulations reste un challenge actuel: En effet, par définition, les molécules de solvant sont prépondérantes dans la boîte de simulation ce qui fait que la majorité du temps de calcul leur est consacré. Elles représentent donc le facteur limitant de la simulation. Afin de permettre un choix entre précision et rapidité, plusieurs types de représentations ont été proposées pour le solvant\cite{Skyner_review_2015, Reddy_free_2014, brown_free_2010} (voir tableau \ref{tab:temps_calculs}).

\begin{table}[ht]
  \centering
  \begin{tabular}{l | c c | c c}
    \hline & \\[-1em]\hline
    {} & \multicolumn{2}{|c|}{DM/MC} & MDFT & {}\\
    \hline
    Solvant    & explicite & implicite & hybride & {} \\
    \hline
    \multirow{2}{*}{Rapidité}   & \raisebox{-0.3\height}{\includegraphics[width=0.03\textwidth]{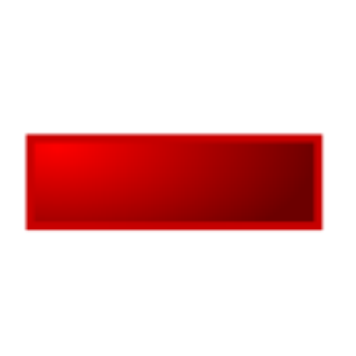}} & \raisebox{-0.3\height}{\includegraphics[width=0.03\textwidth]{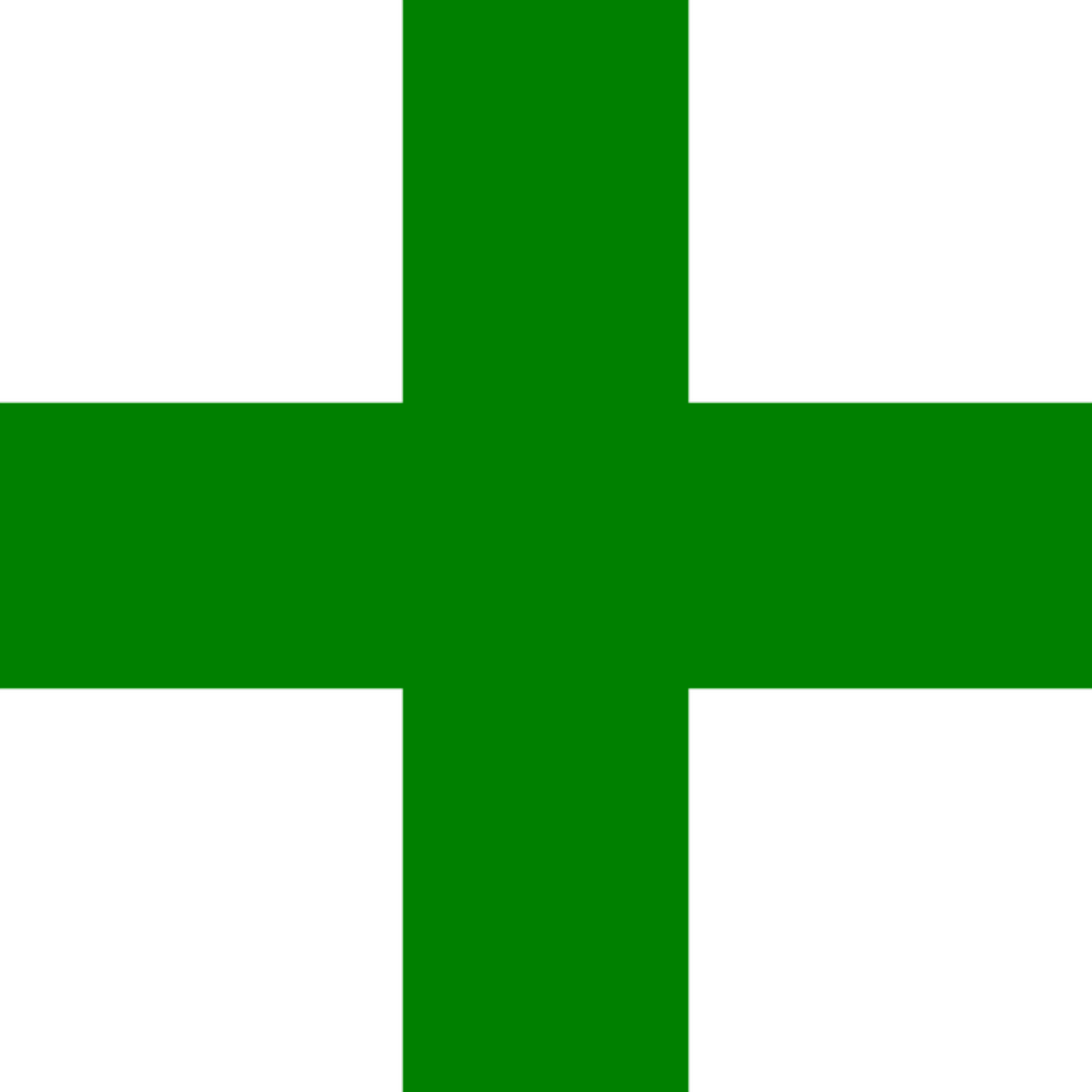}} & \raisebox{-0.3\height}{\includegraphics[width=0.02\textwidth]{chapters/introduction/images/plus.pdf}} & {} \\
    {}   & \raisebox{0.3\height}{({\raise.17ex\hbox{$\scriptstyle\mathtt{\sim}$}} jours)} & \raisebox{0.3\height}{({\raise.17ex\hbox{$\scriptstyle\mathtt{\sim}$}} secondes)} & \raisebox{0.3\height}{({\raise.17ex\hbox{$\scriptstyle\mathtt{\sim}$}} minutes)} & {} \\
    \hline
    Précision  & \raisebox{-0.2\height}{\includegraphics[width=0.03\textwidth]{chapters/introduction/images/plus.pdf}} & \raisebox{-0.2\height}{\includegraphics[width=0.03\textwidth]{chapters/introduction/images/moins.pdf}} & \raisebox{-0.3\height}{\includegraphics[width=0.02\textwidth]{chapters/introduction/images/plus.pdf}} & {} \\
    
  \hline \multicolumn{5}{c}{} \\[-1em]\hline
  \end{tabular}
  \caption{Avantages et inconvénients principaux des différents types de solvant utilisés.}
  \label{tab:temps_calculs}  
\end{table}

\subsection{Les méthodes explicites}
Les méthodes explicites représentent chaque molécule de solvant du système. Au prix de temps de calcul importants, ces méthodes sont actuellement les plus précises. Les logiciels de simulation moléculaire de type dynamique moléculaire (MD) ou Monte Carlo (MC), couplés à une représentation explicite du solvant, permettent d'obtenir avec précision la structure du solvant ainsi que l'énergie libre de solvatation du composé. La structure à l'équilibre du solvant s'obtient à l'issue d'une unique simulation. Elle correspond à la moyenne statistique des positions des atomes de solvant au cours de la simulation. Plus la simulation est longue, plus l’échantillonnage des conformations est important et meilleure sera la statistique. Une bonne précision nécessite donc une simulation longue et donc un temps de calcul important.

Le calcul de l'énergie libre de solvatation nécessite de coupler les simulations explicites à des méthodes comme l'intégration thermodynamique (TI) ou encore de perturbation d'énergie libre (FEP)\cite{Skyner_review_2015, Hansen_Practical_2014, Christ_basic_2009}. Dans le cas de l'intégration thermodynamique, par exemple, afin de simuler une transition lente de l'état initial à l'état final, une vingtaine de simulations de dynamique moléculaire (MD) ou monte carlo (MC) sont lancées. Chacune d'entre elle représente un état intermédiaire de la transformation étudiée. Une fois l'ensemble des simulations terminées, l'énergie libre de la transformation étudiée correspond à la somme des moyennes de la différence d'énergie potentielle entre deux états voisins. Elle s'écrit sous la forme
\begin{eqnarray}
\Delta F(A \rightarrow B) =  \int_0^1 \left\langle U_B(\lambda) - U_A(\lambda) \right\rangle_{\lambda} d\lambda
\end{eqnarray}
\noindent avec $\lambda$ la coordonnée de réaction permettant la transition entre l'état initial ($\lambda$=0) et l'état final ($\lambda$=1).
Ces méthodes nécessitent de nombreuses simulations et multiplient d'autant le temps de calcul.

\subsection{Les méthodes implicites}
Pour dépasser les limites imposées par une représentation explicite du solvant, des méthodes rapides, basées sur une représentation implicite du solvant ont été proposées\cite{Skyner_review_2015}. Elles représentent le solvant sous la forme d'un milieu diélectrique continue polarisable (PCM). Le manque de détails moléculaires comme les liaisons hydrogènes ou la gène stérique ne permet cependant pas un calcul rigoureux des contributions entropiques. Malgré cela, une bonne paramétrisation leur a permis un développement rapide et efficace, avec parfois des prédictions d'énergies libres en bon accord avec les simulations numériques explicites pour des temps de calcul inférieurs de plusieurs ordres de grandeurs. Les deux méthodes implicites majeures sont PBSA et GBSA. Ces méthodes très simplifiées fournissent un résultat quasi instantanément. Pour cela, la partie électrostatique est prise en charge en résolvant soit, l'équation de Poisson (PB) pour le modèle de Poisson-Boltzmann, soit l'équation de Born généralisée (GB) pour le modèle du même nom\cite{Skyner_review_2015}. L'hydrophobicité (la création de la cavité) est quant à elle prise en compte via la surface accessible au solvant (SA). L'énergie libre de solvatation est ensuite déduite de ces deux termes.  Ces deux méthodes, couplées à des énergies de mécanique moléculaire, donnent lieu à des méthodes populaires du calcul de l'énergie libre de liaison en solution. Ces méthodes, MM/PBSA et MM/GBSA\cite{kollman_calculating_2000,srinivasan_continuum_1998,genheden_mm/pbsa_2015} seront développées dans le chapitre \ref{chap:applications}.

Contrairement aux méthodes explicites, ces méthodes ne fournissent aucune information sur l'organisation du solvant.

\subsection{Les méthodes hybrides}
Les méthodes hybrides, en particulier basées sur la théorie des liquides, constituent une 3$^{\text{ème}}$ approche qui allient la vitesse des méthodes implicites à la précision des méthodes explicites. Ces méthodes traitent le solvant sous forme statistique directement à l'équilibre ce qui, contrairement aux méthodes explicites, nous affranchit d'un échantillonnage de l'espace des conformations et permet ainsi un gain de plusieurs ordres de grandeurs en temps de calcul. La première méthode de ce type à avoir été proposée est la théorie des équations intégrales \cite{hansen_theory_2006,gray_theory_1984,hirata_molecular_2003}, avec dans un premier temps une représentation atomistique du soluté et du solvant. Les équations intégrales restent cependant difficiles, sensibles aux instabilités numériques et, d'après nos connaissances, limitées aux systèmes en 1 et 2 dimensions à l'exception des développements de Belloni et al.\cite{Puibasset_bridge_2012, belloni_unpublished} qui permettent l'étude de systèmes en quasi 3 dimensions. Ce secteur reste cependant un champ de recherche ouvert. Une approximation de cette méthode a également été développée, le \textit{Reference interaction-site model} RISM\cite{chandler_optimized_1972}, puis dérivée et adaptée aux solutés complexes en 3 dimensions 3DRISM\cite{Chandler_censity_1986,Kovalenko_self_1999,beglov_integral_1997, du_solvation_2000,  luchko_three-dimensional_2010}. RISM et 3DRISM ont connu un grand succès car elles permettent de prédire les énergies libres de solvatation et les profils du solvant avec une précision acceptable, comme montré récemment sur des petites molécules neutres\cite{roy_predicting_2017, du_solvation_2000, kovalenko_potential_1999, kovalenko_hydration_2000, johnson_small_2016}, des bio-molécules \cite{sindhikara_analysis_2013,imai_hydration_2006,kiyota_new_2011,phongphanphanee_molecular_2010} et même des ions \cite{phongphanphanee_potential_2009, kovalenko_potentials_2000}. Quoi qu'il en soit, ces méthodes considèrent les molécules de solvant comme un ensemble de sites corrélés entre eux, ce qui est schématiquement incorrect.

\subsubsection{La théorie de la fonctionnelle de la densité moléculaire}
La théorie de la fonctionnelle de la densité moléculaire\cite{jeanmairet_molecular_2013-1,jeanmairet_classical_2015,Jeanmairet_introduction_2014,jeanmairet_hydration_2014,levesque_solvation_2012} (MDFT) est une autre approche de la théorie des liquides. Elle a des connexions fortes avec la théorie des équations intégrales, mais est beaucoup moins sensible aux instabilités numériques car elle est basée sur la minimisation d'une fonctionnelle d'un problème variationnel. Le développement d'une fonctionnelle correcte reste cependant difficile et constitue un projet de recherche en cours. La MDFT nous fournit en quelques secondes seulement (quelques minutes pour les plus gros composés), pour des solutés complexes en 3D, deux paramètres essentiels à la compréhension des phénomènes ayant lieu en solution: l'énergie libre de solvatation et le profil de solvatation. Le travail présenté dans ce manuscrit est basé sur cette théorie et son code associé. Le chapitre suivant est dédié à la description de cette théorie.

\subsection{Quelques exemples d'application en \textit{drug design}}
De nombreuses méthodes utilisées en \textit{drug design} comme la modélisation par homologie, le docking, ou encore par exemple la recherche de pharmacophores, nécessitent une structure précise de la protéine cible. Cette information est donc capitale au développement d'un nouveau médicament. 

L'énergie libre de solvatation peut également être dérivée en de nombreuses autres grandeurs utiles en \textit{drug design}. Dans les cas des médicaments administrables par voie orale, Lipinski et al.\cite{Lipinski_lead_2004} ont défini la \textit{régle des 5} qui comporte un ensemble de 4 critères qu'elles doivent respecter. Si une petite molécule ne respecte pas l'ensemble de ces 4 règles, ses chances qu'elles deviennent un jour un médicament oral sont très faibles. Pour respecter ces critères, cette molécule doit posséder au maximum 5 donneurs de liaison hydrogène, au maximum 10 accepteurs de liaison hydrogène, une masse moléculaire inférieure à 550 daltons et un logP inférieur à 5. Si les 3 premiers paramètres peuvent être calculés directement, ce n'est pas le cas du dernier. Le logP correspond au logarithme du rapport entre la solubilité du composé dans l'eau et dans l'octanol. Il peut donc être dérivé des énergies libres de solvatation de ce composé dans ces deux solvants. L'énergie libre de solvatation, couplée à des calculs de mécanique moléculaire permet également de simplifier le calcul de l'énergie libre de liaison. La méthode MM/PBSA\cite{genheden_mm/pbsa_2015} est décrite et dérivée en MM/MDFT dans le chapitre \ref{chap:applications}. Enfin on peut citer également le calcul du logBBB (coefficient de partition entre le cerveau et le sang). Dans le cas des maladies neurologiques, le médicament doit pouvoir atteindre le cerveau et donc traverser la barrière hémato-encéphalique. Pour cela, le logBBB doit être compris entre -1 et 0,3\cite{Vilar_prediction_2010}. Comme l'ont montré Lombardo et al\cite{Lombardo_computation_1996}, ce paramètre peut également être dérivé de la valeur de l'énergie libre de solvatation.

Il est également possible de combiner la structure du solvant et l'énergie libre du système, comme le propose aujourd'hui le logiciel watermap\cite{abel_role_2008,Young_motifs_2007}. En effet, si l'on connaît la valeur d'énergie libre de chaque molécule de solvant proche du site de liaison, il est ensuite possible d'optimiser le candidat médicament afin que l'un de ses groupements se substitue aux molécules les plus énergétiques. Cette optimisation permet de diminuer l'énergie libre du système et ainsi de favoriser la création de la liaison. 

Dans ce paragraphe nous ne présentons qu'une petite partie des possibilités qu'offre une représentation rapide et efficace de la solvatation comme le propose MDFT. Mon projet de thèse consiste à effectuer le premier pas vers toutes ces applications en adaptant la théorie ainsi que son implémentation aux systèmes biologiques.

\clearpage
\strut
\vspace{10\baselineskip}

\boitemagique{A retenir}{
Dans ce chapitre nous introduisons le contexte de cette thèse et présentons l'état de l'art des méthodes de solvatation. Nous montrons également en quoi l'énergie libre de solvatation et la structure du solvant sont omniprésentes tout au long du développement et de l'optimisation d'un médicament. 
}

\clearemptydoublepage
\chapter{MDFT: la théorie de la fonctionnelle de la densité moléculaire}
\label{chap:theorie}

\boitemagique{Objectif}{
Dans ce chapitre nous décrivons la théorie de la fonctionnelle de la densité moléculaire dans l'approximation HNC, son implémentation ainsi que quelques corrections de cette approximation.
}

La théorie de la fonctionnelle de la densité moléculaire (MDFT) permet l'étude de la solvatation de composés de n'importe quelle taille et n'importe quelle forme à l'échelle moléculaire. Cette théorie et son code associé, permettent, en quelques secondes seulement, (i) de calculer l'énergie libre de solvatation et (ii) de générer une carte détaillée en 3 dimensions de la densité ainsi que de l'orientation du solvant autour du soluté.

L'origine de la théorie de la fonctionnelle de la densité (DFT) réside dans le développement d'une fonctionnelle $\mathcal{F}[\rho\left(\boldsymbol{r},\Omega \right)]$. Cette fonctionnelle a pour variable la densité du solvant $\rho\left(\boldsymbol{r},\Omega \right)$, en chaque point de l'espace $\boldsymbol{r}$ et pour chaque orientation $\Omega$ de la molécule de solvant. La fonctionnelle est construite comme la différence entre le grand potentiel du soluté en solution et le grand potentiel du solvant homogène de densité $\rho_{0}$. Par définition, la valeur de la fonctionnelle au minimum correspond donc à l'énergie libre de solvatation du soluté étudié. 

Sans approximation pour le moment, la fonctionnelle est découpée en trois parties: la partie idéale, la partie extérieure et la partie d'excès\cite{evans_density_2009,henderson_fundamentals_1992}. 
\begin{eqnarray}
\mathcal{F} = \mathcal{F}_\mathrm{id} + \mathcal{F}_\mathrm{ext} + \mathcal{F}_\mathrm{exc}
\label{eq:fonctionnelle}
\end{eqnarray}
La partie idéale, représente l’entropie d'information du système, et s'écrit
\begin{eqnarray}
\mathcal{F}_\mathrm{id}&=&\mathrm{k_B}T\int\mathrm{d}\boldsymbol{r}\mathrm{d}\Omega \rho\left(\boldsymbol{r},\Omega \right)\ln\left(\frac{\rho\left(\boldsymbol{r},\Omega \right)}{\rho_0}\right)-\Delta\rho\left(\boldsymbol{r},\Omega \right)
\label{eq:fonctionnelle:id}
\end{eqnarray}
\noindent avec $T$ la température, $\mathrm{k_B}$ la constante de Boltzmann et donc $\mathrm{k_B}T$ l'énergie thermique, $\Delta\rho\left(\boldsymbol{r},\Omega \right)=\rho\left(\boldsymbol{r},\Omega \right)-\rho_0$ la densité d'excès par rapport à la densité bulk de référence $\rho_0$. La seconde partie, la partie extérieure, représente le potentiel d'interaction $\phi\left(\boldsymbol{r},\Omega \right)$ entre le soluté et le solvant. Elle s'écrit:
\begin{eqnarray}
\mathcal{F}_\mathrm{ext}&=&\int\mathrm{d}\boldsymbol{r}\mathrm{d}\Omega\rho\left(\boldsymbol{r},\Omega \right)\phi\left(\boldsymbol{r},\Omega \right)
\label{eq:fonctionnelle:ext}
\end{eqnarray}
\noindent avec
\begin{eqnarray}
\phi\left(\boldsymbol{r},\Omega \right) = \sum\limits_{i=1}^{\mathrm{n}_\mathrm{sv}}\sum\limits_{j=1}^{\mathrm{n}_\mathrm{su}} v_{ij}(|\boldsymbol{r}+\boldsymbol{S}_i(\Omega)+\boldsymbol{r}_j|)
\end{eqnarray}
\noindent avec $\mathrm{n}_\mathrm{sv}$ le nombre de sites du solvent, $\mathrm{n}_\mathrm{su}$ le nombre de sites du solute et $\boldsymbol(S)_i$ le vecteur reliant l'origine du solvant au site i. Le potentiel d'interaction correspond à la somme des interactions électrostatiques et des interactions de Lennard-Jones ou toute autre interaction potentielle. 

\section{L'approximation HNC}

Enfin, la partie d'excès correspond à la corrélation entre les molécules de solvant. La version exacte de cette partie, correspond au développement de Taylor infini autour de la densité bulk liquide de référence, soit pour l'eau $\rho_0$=1kg.L$^{-1}$. Afin d'en permettre son calcul, des approximations doivent être considérées. Nous considérons ici uniquement le premier et le second ordres du développement:
\begin{eqnarray}
\mathcal{F}_\mathrm{exc} = -\frac{\mathrm{k_B}T}{2}\int\mathrm{d}\boldsymbol{r}\mathrm{d}\Omega \Delta\rho\left(\boldsymbol{r}, \Omega   \right) \gamma \left(\boldsymbol{r},\Omega\right)  + \mathcal{O}(\Delta\rho^{3})
\label{eq:fonctionnelle:exc}
\end{eqnarray}
\noindent avec
\begin{eqnarray}
\gamma \left(\boldsymbol{r},\Omega\right) = \int\mathrm{d}\boldsymbol{r}^\prime\mathrm{d}\Omega^\prime\  c\left(\boldsymbol{r}-\boldsymbol{r}^\prime,\Omega,\Omega^\prime \right) \Delta\rho\left(\boldsymbol{r}^\prime, \Omega^\prime \right)
\end{eqnarray}
\noindent avec $c\left(\boldsymbol{r}-\boldsymbol{r}^\prime,\Omega,\Omega^\prime \right) $ la fonction de corrélation directe entre deux molécules de solvant qui dépend de la distance entre ces molécules et de leurs orientations relatives l'une par rapport à l'autre. La fonction de corrélation directe pour un solvant homogène à température et pression données est issue de longues simulations de dynamique moléculaire ou de Monte Carlo corrigées des effets de taille finie\cite{Puibasset_bridge_2012, belloni_unpublished}.

Cette approximation, bien connue, est nommée approximation HNC (hyper-Netted Chain)\cite{hansen_theory_2006}.

\section{L'implémentation}
Une fois la fonctionnelle décrite, il est nécessaire de la minimiser. En effet, par définition, le minimum de la fonctionnelle correspond à l'énergie libre de solvatation. Dans le même temps, le minimum est atteint lorsqu'en tout point de l'espace, la densité du solvant équivaut à sa densité dite à l'équilibre.
\begin{eqnarray}
\min(\mathcal{F}[\rho\left(\boldsymbol{r},\Omega \right)]) = \mathcal{F}[\rho_{eq}\left(\boldsymbol{r},\Omega \right)])= \Delta G_{solv}
\end{eqnarray}
Cette minimisation a été implémentée dans un code en Fortran moderne du même nom: MDFT.

\subsection{La discrétisation}
Comme il n'est numériquement pas possible de travailler avec un système continu infini, le soluté est étudié dans un système fini, discret et périodique. Il existe deux niveaux de discrétisation du système. Le premier, spatial, découpe l'espace sur une grille homogène. Le second niveau, angulaire\cite{ding_cea-01564512}, permet de limiter le nombre d'orientations étudiées. Il est actuellement possible de choisir entre vitesse et précision en faisant varier le nombre d'angles étudiés de 18 à 726 à travers un paramètre nommé $\mathrm{m}_\mathrm{max}$. L'équivalence entre le nombre d'angles et la valeur de ce paramètre traduit des quadratures bien connues de type Gauss-Legendre\cite{abbott_tricks_2005}. Cette équivalence est disponible dans le tableau \ref{tab:mmax};

\begin{table}[ht]
 \centering
  \begin{tabular}{l | c}
    \hline \multicolumn{2}{c}{} \\[-1em]\hline
    $\mathrm{m}_\mathrm{max}$ & nombre d'orientations \\
    \hline
    1  & 18 \\
    2  & 75 \\
    3  & 196 \\
    4  & 405 \\
    5  & 726 \\
    \hline \multicolumn{2}{c}{} \\[-1em]\hline
  \end{tabular}
  \caption[\'Equivalence entre le paramètre $\mathrm{m}_\mathrm{max}$ et le nombre d'angles.]{\'Equivalence entre le paramètre $\mathrm{m}_\mathrm{max}$ et le nombre d'angles considérés lors de la minimisation.}
  \label{tab:mmax}  
\end{table}


\subsection{Les convolutions}
Un des avantages majeurs de MDFT par rapport aux autres méthodes est sa rapidité. La partie idéale et la partie d'excès sont locales ce qui rend leur temps de calcul linéaire, proportionnel à $N_O N_V$. La partie qui nécessite le plus de temps et qui est donc limitante dans ce calcul est la partie d'excès qui est, elle, non locale. Afin de diminuer fortement le temps de calcul de cette partie et par conséquence le temps de calcul global, nous utilisons la propriété suivante des convolutions:
\begin{eqnarray}
f*g = \mathrm{FT}^{-1} [ \mathrm{FT}(f) . \mathrm{FT}(g) ]
\end{eqnarray}
La convolution de deux fonctions peut être calculée comme la transformée de Fourier inverse du produit point à point de la transformée de Fourier de ces deux fonctions. Pour rappel, la fonction $\gamma$ est la convolution entre les fonctions $\Delta\rho$ et $c$. Cette méthode est donc applicable mais ne permet cependant pas à elle seule de diminuer le temps de calcul. Cette propriété a donc été couplée à l'utilisation des FFT (Fast Fourier Transform) et en particulier de la librairie FFTW3 pour la partie spatiale et de FGSHT (fast generalized spherical harmonic transform) récemment proposé par Ding et al\cite{ding_thesis} pour la partie angulaire.
Les auteurs\cite{ding_thesis} ont montré que le couplage de ces deux méthodes permet d'obtenir des temps de calcul du même ordre de grandeurs pour chacune des 3 parties de la fonctionnelle.

\subsection{Le minimiseur}
Le minimiseur utilisé pour minimiser notre fonctionnelle est L-BFGS (Limited-memory Broyden-Fletcher-Goldfarb-Shanno)\cite{Byrd_lbfgs_1995}. L-BFGS correspond à une version de BFGS\cite{bfgs_2006} optimisée pour les problèmes composés de nombreuses variables comme c'est le cas de la MDFT. Contrairement à BFGS qui conserve une approximation de la hessienne sous la forme d'une matrice dense, L-BFGS conserve uniquement quelques vecteurs représentatifs ainsi que l'historique sur quelques pas de la minimisation. L-BFGS nécessite en entrée, l'ensemble des variables à minimiser ainsi que le gradient de la fonctionnelle $\nabla F[\rho\left(\boldsymbol{r}_i,\Omega_j\right)]$ défini comme:
\begin{eqnarray}
\nabla F[\rho(\boldsymbol{r}_i,\Omega_j)] &=& \mathrm{k_B}T \ln(\frac{\rho(\boldsymbol{r}_i,\Omega_j)}{\rho_0}) \\
&+& \phi(\boldsymbol{r}_i,\Omega_j) \nonumber \\
&-& \mathrm{k_B}T \gamma(\boldsymbol{r}_i,\Omega_j) \nonumber
\end{eqnarray}
Le calcul du gradient de chaque partie est détaillé en annexe \ref{chap:annexes:grad}.

\section{Au-delà de HNC}
Pour aller plus loin que la théorie HNC, et ainsi corriger l'approximation faite dans la fonctionnelle d'excès, il est possible (i) d'appliquer des corrections à posteriori ou (ii) d'approximer le terme $\mathcal{O}(\Delta\rho^{3})$ via l'ajout d'un quatrième terme à notre fonctionnelle, une fonctionnelle de bridge.

\subsection{Les corrections à posteriori}
Les corrections à posteriori interviennent après la minimisation. Elles permettent donc uniquement de corriger la valeur de l'énergie libre de solvatation mais n'ont aucun impact sur la carte de densité du solvant. Des corrections de différents types ont été développées et sont actuellement utilisées dans la MDFT.

\subsubsection{Les corrections de pression}
Parmi ces corrections, deux permettent de corriger la pression du système. Pour rappel, l'approximation HNC correspond au premier ordre du développement de Taylor autour de la densité liquide. La phase gazeuse du solvant n'est donc pas représentée, ce qui entraîne une forte surestimation de la pression du système soit 10 000 bar. Nous décrivons de façon détaillé ce problème dans le chapitre \ref{chap:bridge}. Sergiievskyi et al. \cite{sergiievskyi_solvation_2015,sergiievskyi_pressure_2015} ont proposé une correction ad-hoc rigoureuse basée sur la théorie des liquides: la correction \textit{PC}. Au moment de ce développement, la théorie MDFT n'était pas encore au niveau HNC. Elle correspondrait aujourd'hui à une approximation de HNC avec $\mathrm{m}_\mathrm{max}$=1. Les auteurs ont de ce fait également proposé une correction empirique, \textit{PC+}, qui améliorait les résultats\cite{misin_salting-out_2016, misin_hydration_2016, misin_communication:_2015}. Nous montrerons dans le chapitre \ref{chap:BDD} que la correction \textit{PC+} n'est plus adaptée à la théorie dans l'approximation HNC. Nous proposerons également une alternative à ces corrections dans le chapitre \ref{chap:bridge} sous la forme d'un bridge gros gain.


\subsection{Les fonctionnelles de bridge}
Si l'on veut corriger à la fois l'énergie libre de solvatation et la densité du solvant, il est nécessaire de modifier la fonctionnelle. Pour cela, un quatrième terme nommé fonctionnelle de bridge est introduit dans la partie d'excès. Nous obtenons ainsi:
\begin{eqnarray}
\mathcal{F}_\mathrm{exc} = \mathcal{F}_\mathrm{exc}^{HNC} + \mathcal{F}_\mathrm{b}
\end{eqnarray}
Différentes formes pour la fonctionnelle de bridge ont été proposées ces dernières années\cite{levesque_scalar_2012,jeanmairet_molecular_2013,jeanmairet_molecular_2015}. Malheureusement, comme il a été montré dans un papier à venir, aucun de ces bridges ne permet une représentation du système totalement satisfaisante du point de vue thermodynamique. Dans la suite de ce rapport nous proposerons un nouveau bridge qui autorise la création d'une phase gazeuse de l'eau et permet ainsi de représenter de façon correcte la tension de surface de l'eau ainsi que la pression du système.

\clearpage
\strut
\vspace{10\baselineskip}

\boitemagique{A retenir}{
Danc ce chapitre, nous présentons la théorie de la fonctionnelle de la densité moléculaire.
Les récents développements de Ding et al ont permis de porter cette théorie au niveau de l'approximation HNC.
Nous présentons également les différentes corrections associées à cette théorie.
Malheureusement, aucune de ces corrections ne permet de reproduire l'ensemble des propriétés thermodynamiques de nos systèmes.
Dans le chapitre suivant, nous présentons une nouvelle correction adaptée aux systèmes macromoléculaires.
}


\part{Développements théoriques}

\clearemptydoublepage
\chapter{Bridge gros grain}
\label{chap:bridge}

\boitemagique{Objectif}{
L'objectif de ce chapitre est de proposer une fonctionnelle de bridge \textbf{simple} et \textbf{rapide} qui permette de prédire correctement: 
\begin{itemize}
\item Les profils de densité du solvant (g(r))
\item Les énergies libres de solvatation
\end{itemize}
Avec les propriétés thermodynamiques macroscopiques suivantes cohérentes:
\begin{itemize}
\item La bonne tension de surface de l'eau
\item La bonne pression du système
\end{itemize}
}

Comme il a été montré jusqu'ici, l'approximation HNC, y compris corrigée par les bridges décrits dans le chapitre précédent, ne permet pas d'avoir un système thermodynamiquement consistant.

Nous proposons ici un bridge simple et efficace numériquement, basé sur une densité gros-grain, qui prend en compte le démouillage en permettant la quasi-coexistence des phases gazeuse et liquide de l'eau.
Ce bridge permet donc de retrouver la consistance thermodynamique tout en améliorant les rdfs et les énergies libres de solvatation en échange d'un coût de calcul négligeable.

\section{Le démouillage}
\begin{figure}[ht]
    \center    
  \begin{tikzpicture}
    \begin{axis}[
            xlabel= r (\AA),
            ylabel= densité radiale (kg.L$^{-1}$),
            xmin = 10, xmax = 18,
            ymin = 0, ymax = 2,
            no markers,
            legend style = {draw = none, cells={anchor=west}}
      ]

      \addplot[mark=none, black, very thick, domain=10:20, samples=400] {demouillage(-13.5,1)};
    \end{axis}
  \end{tikzpicture}
    \caption[Représentation du démouillage autour d'une sphère hydrophobe.]{Exemple de densité radiale réaliste autour d'une sphère hydrophobe de rayon R=10 \AA\ en fonction de la distance à la sphère. On attend une densité proche de celle du gaz au contact et la densité bulk de référence de l'eau loin de la sphère.}
    \label{fig:demouillage}
\end{figure}
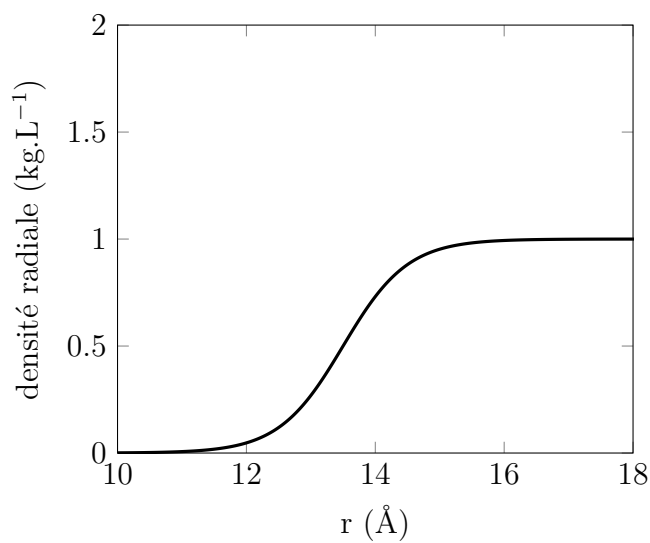

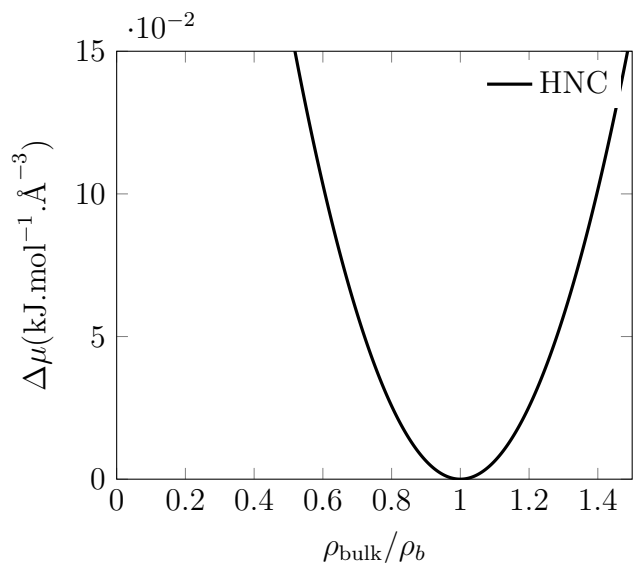
\begin{figure}[ht]
    \center
  \begin{tikzpicture}
    \begin{axis}[
            xlabel= $\rho_{\mathrm{bulk}}/\rho_b$,
            ylabel= $\Delta \mu (\mathrm{kJ.mol}^{-1}.\text{\AA}^{-3}) $,
            xmin = 0, xmax = 1.5,
            ymin = 0, ymax = 0.15,
            scaled y ticks={base 10:2},
            legend style = {draw = none, cells={anchor=west}}
      ]
      \addplot+[mark=none, black, very thick] file {chapters/bridge/datas/fonctionnelles/fonctionnelle.csv};
      \legend{HNC}
    \end{axis}
  \end{tikzpicture}
    \caption[Potentiel chimique d'une molécule de solvant en fonction de sa densité.]{Potentiel chimique d'une molécule de solvant en fonction de la densité $\rho_{\mathrm{bulk}}$ dans l'approximation HNC. $\rho_b$ est la densité bulk de référence de l'eau SPC/E à pression et température standards (1 $\mathrm{kg.L}^{-1}$).}
    \label{fig:fonctionelle_HNC}
\end{figure}

Lorsque de gros composés hydrophobes sont plongés en solution, on observe une transition lente d'une densité quasi-nulle (à la surface du soluté) à la densité bulk (loin du soluté) (voir figure \ref{fig:demouillage}). Ce phénomène est appelé démouillage. Malheureusement, comme on le voit sur la figure \ref{fig:fonctionelle_HNC}, plus on s'éloigne de la densité bulk et plus l'énergie libre d'une unité de volume (soit le potentiel chimique d'une molécule du solvant) augmente. Les phases de faible densité normalement attendues sont donc trop défavorisées pour exister. En réalite, à température ambiante et pression atmosphérique, les phases gazeuses et liquides de l'eau ont la propriété d'être en quasi-coexistence. En effet, comme on le voit sur la figure \ref{fig:diagramme_phase_eau}, à température ambiante et pression atmosphérique, l'eau est proche de la phase gaz. Le potentiel chimique est donc également proche de celui de la phase gaz. Dans l'approximation HNC, les phases de faible densité sont remplacées par des zones de plus forte densité, ce qui entraîne une surestimation forte de la pression du fluide. Comme c'était déjà le cas pour le bridge du 3$^{\text{ème}}$ ordre proposé par Jeanmairet et al\cite{jeanmairet_molecular_2013}, nous allons dans un premier temps fixer l'énergie libre de la phase gaz à la même valeur que celle de la phase liquide, soit proche de zéro. Afin d'augmenter la flexibilité du modèle et ainsi de nous permettre d'obtenir une tension de surface correcte, nous ajoutons un nouveau terme d'ordre 4. Afin de calibrer ces nouveaux termes, nous avons effectué une étude paramétrique (décrite plus bas). Le nombre de calculs étant très important, nous avons développé une version spéciale de MDFT adaptée à cette étude: MDFT à symétrie sphérique.

\begin{center}
    \captionsetup{type=figure}
	\includegraphics[width=0.8\textwidth]{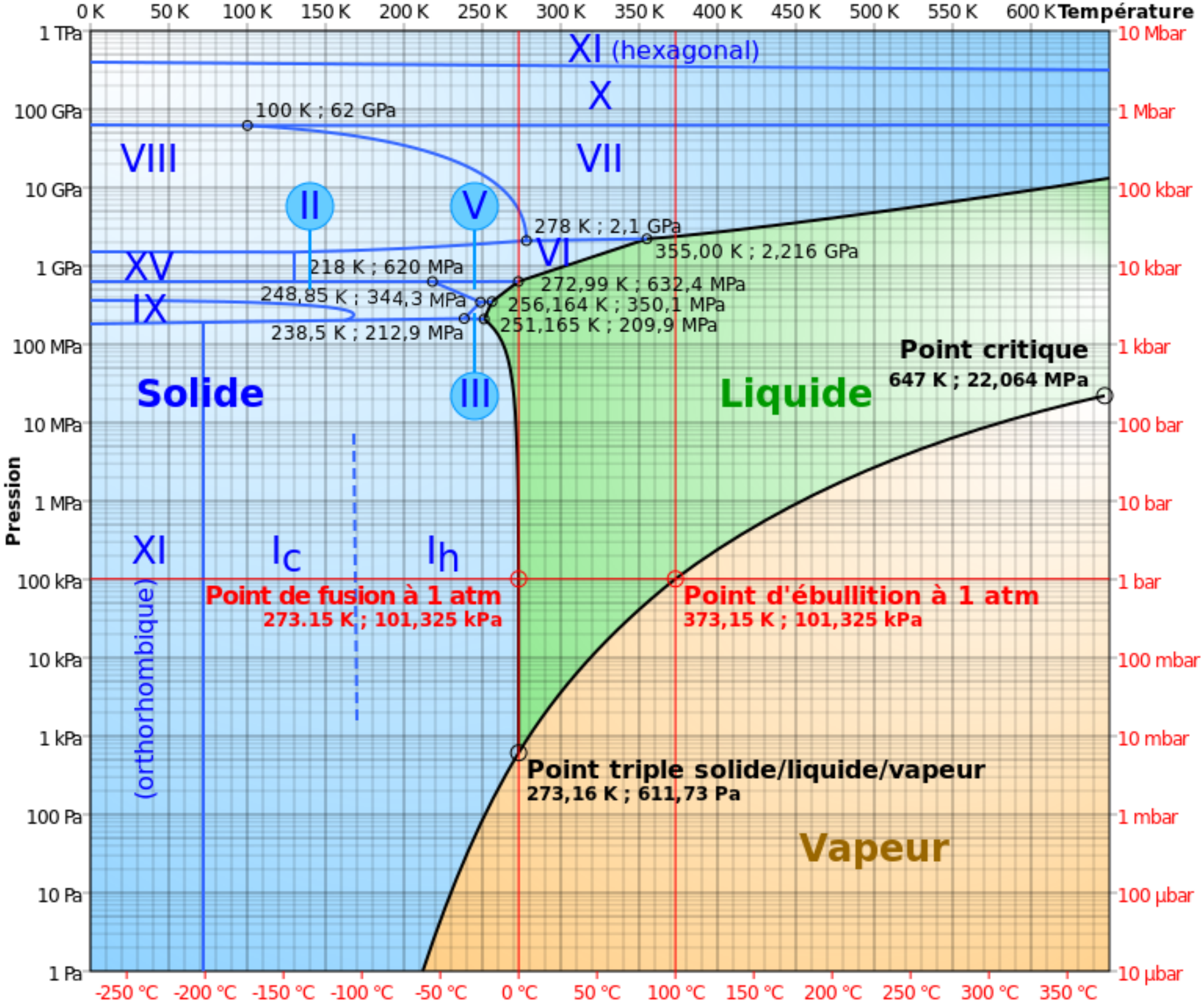}
	\captionof{figure}[Diagramme de phase de l'eau.]{Diagramme de phase de l'eau. Il existe une transition liquide-gaz proche des conditions standards. Cette proximité entraîne la quasi-coexistance des deux phases dans ces conditions. crédits: Olivier Descout}
    \label{fig:diagramme_phase_eau}
\end{center}

\pgfmathdeclarefunction{demouillage}{2}{%
    \pgfmathparse{0.5*(tanh(x+#1)+#2)}%
}

\section{MDFT: version à symétrie sphérique}
Afin de diminuer fortement le temps nécessaire à l'étude paramétrique, nous avons développé une version à symétrie sphérique. Cette version simplifiée repose sur deux approximations:
\begin{itemize}
\item Les solutés étudiés sont neutres
\item Les orientations du solvant sont ignorées
\end{itemize}

La première approximation que nous avons fait est de considérer uniquement des composés neutres.
En effet, les charges permettent la création de liaisons hydrogènes fortes entre l'eau et le soluté ce qui a pour effet de fortement stabiliser ce dernier et ainsi de le rendre hydrophile.
Les composés hydrophobes, entraînant du démouillage sont donc généralement neutres.
Contrairement aux solutés chargés, les solutés neutres forment uniquement des liaisons de Van Der Waals.
Dans le cas du modèle d'eau SPC/E, qui posséde un seul site Lennard-Jones sur l'oxygène, soit en son centre, le soluté n'a aucune inflence directe sur l'orientation des molécules d'eau.
De plus, Jeanmairet et al.\cite{jeanmairet_molecular_2013, jeanmairet_molecular_2016} ont montré que le couplage entre la densité et la polarisation de l'eau est négligeable.
La seconde approximation majeure que nous faisons est donc que chaque orientation du solvant est équiprobable. Cela nous permet ainsi de nous affranchir des angles et de ne considérer qu'une moyenne angulaire de la densité en chaque point de grille.

La symétrie sphérique entraîne une autre différence importante. Au contraire de la version 3D, périodique, les systèmes à symétrie sphérique contiennent par définition un unique soluté dans un solvant infini.

Afin de paramètriser ce nouveau bridge, nous nous concentrons dans un premier temps sur nos molécules modèles: des sphères de Lennard-Jones neutres, avec un solvant sous la forme d'un point et donc sans orientation. L’interaction dépend donc uniquement de la distance entre le soluté et la molécule de solvant. En d'autres termes, tous les points se trouvant sur une sphère centrée sur notre système seront parfaitement identiques. Ces systèmes sont dits à symétrie sphériques. Afin de ne pas minimiser inutilement de nombreuses variables identiques, car équidistantes du centre, nous avons développé une version adaptée de MDFT. Cette version, contrairement à la version 3D, nous autorise à représenter le système sous la forme d'un vecteur 1D de densités partant du centre de notre soluté et allant jusqu'à l'infini et non plus par une grille en 3D. Basé sur ce constat, nous avons adapté la théorie et nous l'avons implémentée dans une version spécifique de MDFT.

\subsection{La théorie}
Comme nous l'avons décrit précedemment, nous nous affranchissons des orientations du solvant. Cela revient, dans la définition de la fonctionnelle (voir équation \ref{eq:fonctionnelle}) à remplacer $\int\mathrm{d}\Omega\rho(\boldsymbol{r}, \Omega)$ par $\rho(r)$. Dans le cas de coordonnées cartésiennes, le découpage de l'espace se fait sous forme de voxels, alors que dans le cas de coordonnées sphériques, le découpage se fait sous forme de coquilles. L'intégration $\int\mathrm{d}\boldsymbol{r}$ est donc remplacée par $\int\mathrm{dV}_{\mathrm{coquille}}(r)$ avec $\mathrm{dV}_{\mathrm{coquille}}(r)=4 \pi r^2 \mathrm{dr}$, $\mathrm{dr}$ étant l'épaisseur de la coquille soit numériquement, l'espacement entre deux points. Les fonctionnelles idéale et externe, centrées sur l'origine, sont les suivantes:
\begin{eqnarray}
\mathcal{F}_\mathrm{id}(r)&=&\mathrm{k_B}T\int\mathrm{dV}_{\mathrm{coquille}}(r) [ \rho\left(r \right)\ln\left(\frac{\rho\left(r \right)}{\rho_0}\right)-\rho\left(r \right)+\rho_0 ],\\
\mathcal{F}_\mathrm{ext}(r)&=&\int\mathrm{dV}_{\mathrm{coquille}}(r)\rho\left(r \right)\phi\left(r \right)\\
\end{eqnarray}
Au contraire des deux premiers termes, la fonctionnelle d'excès n'est pas centrée sur l'origine. Nous ne pouvons donc pas nous placer directement dans des coordonnées sphériques. Nous réécrivons donc dans un premier temps cette partie de la fonctionnelle sous la forme d'une convolution, ce qui nous donne :
\begin{eqnarray}
\mathcal{F}_\mathrm{exc}(r) &=& \mathcal{F}_\mathrm{hnc} + \mathcal{F}_\mathrm{b}
\end{eqnarray}
Avec
\begin{eqnarray}
\mathcal{F}_\mathrm{hnc}&=& -\frac{\mathrm{k_B}T}{2}\int\mathrm{d}\boldsymbol{r} \Delta\rho\left(\boldsymbol{r} \right)  \int\mathrm{d}\boldsymbol{r}^\prime c_s\left(\left|\boldsymbol{r}-\boldsymbol{r}^\prime\right| \right) \Delta\rho\left(\boldsymbol{r}^\prime \right) + \mathcal{F}_\mathrm{b}\\
						 &=& -\frac{\mathrm{k_B}T}{2}\int\mathrm{d}\boldsymbol{r} [ \Delta\rho\left(\boldsymbol{r} \right)  *\gamma(\boldsymbol{r}) ] + \mathcal{F}_\mathrm{b}
\end{eqnarray}
 avec $\gamma(\boldsymbol{r}) = \int\mathrm{d}\boldsymbol{r}\mathrm{d}\boldsymbol{r}^\prime c_s\left(\left|\boldsymbol{r}-\boldsymbol{r}^\prime\right| \right) \Delta\rho\left(\boldsymbol{r}^\prime  \right)$ et $c_s$ la contribution à symétrie sphérique de la fonction de corrélation directe totale. À condition d'utiliser une convolution adaptée, nous sommes ici autorisés à réécrire cette partie de la fonctionnelle dans les coordonnées sphériques. Nous obtenons ainsi:
\begin{eqnarray}
\mathcal{F}_\mathrm{hnc} &=& -\frac{\mathrm{k_B}T}{2} \int \mathrm{d} r (\Delta\rho\left(r \right)  \ast (\gamma(r)) + \mathcal{F}_\mathrm{b}
\end{eqnarray}
Une résolution rapide et simple de ce terme est décrite plus loin.

\subsection{Implémentation}
Une version de cette théorie a été implémentée dans un code de 2800 lignes de C++ objet haute performance. Comme nous l'avons décrit ci-dessus, l'espace est représenté par un ensemble régulier de densités allant du centre du système jusqu'à l'infini. Numériquement, nous avons limité le vecteur à un ensemble de densités réparties entre l'origine et une distance suffisamment loin pour ne plus être influencée par le soluté. Cette limite, ainsi que l'espacement entre deux points, sont fixés par l'utilisateur.
La minimisation est effectuée à l'aide de la librairie L-BFGS\cite{zhu_algorithm_1997}.

\subsubsection{Transformées de Hankel}
Comme nous l'avons décrit dans la section précédente, la fonctionnelle d'excès peut être calculée en utilisant la propriété des convolutions suivante: La convolution de deux fonctions correspond à la transformée de fourier inverse ($\mathrm{FT}^{-1}$) du produit point à point de la transformée de Fourier ($\mathrm{FT}$) des deux fonctions.
\begin{eqnarray}
f \ast g  = \mathrm{FT}^{-1}(\mathrm{FT}(f)\cdot\mathrm{FT}(g))
\end{eqnarray}
Dans notre système à symétrie sphérique, les transformées de Fourier 3D se réécrivent comme des transformées de Hankel. Nous pouvons ainsi directement résoudre cette partie de la fonctionnelle en coordonnées sphériques.

Afin d'adapter au mieux la transformée de Hankel à nos besoins, nous l'avons réimplémentée. Les formules utilisées pour la transformée de Hankel directe (HT) et la transformée de Hankel inverse ($\mathrm{HT}^{-1}$) sont les suivantes:

\begin{eqnarray}
\mathrm{HT}[f](k) &=& 4\pi\int \mathrm{dr}\ f(r)\frac{\sin(kr)}{kr}r^2\\
\mathrm{HT}^{-1}[f](k) &=& \frac{1}{3\pi^2}\int \mathrm{dk}\ f(k)\frac{\sin(kr)}{kr}k^2
\end{eqnarray}
À cette étape, nous minimisons donc la fonctionnelle dans l'approximation HNC.

\subsubsection{Mise en cache partielle de la transformée de Hankel}
Malgré la puissance des machines de calcul actuelles, certaines opérations restent longues à effectuer. C'est le cas par exemple des exponentielles, ou encore des cosinus et sinus que nous utilisons massivement dans le calcul des transformées de Hankel.
Lors de la minimisation, seules les valeurs des densités changent. Leur position, r dans l'espace réel et k dans l'espace réciproque, ne sont pas modifiées. Les valeurs de $4\pi\frac{\sin(kr)}{kr}r^2$ ne sont donc pas non plus modifiées. Afin de diminuer fortement le temps de calcul nécessaire à cette partie, nous avons mis ces valeurs en cache. En d'autres termes, nous les générons une fois au début de la minimisation, nous les stockons en mémoire, puis nous les réutilisons à chaque pas de minimisation. De plus, ces valeurs étant toujours appelées dans le même ordre, nous les stockons en mémoire de manière contiguë, ce qui autorise la vectorisation de cette boucle de calcul et minimise le temps nécessaire au rapatriement des données. Nous bénéficions ainsi du maximum de la puissance de calcul disponible. Nous réécrivons donc la transformée de Hankel sous la forme:
\begin{eqnarray}
\mathrm{HT}[f](k) &=& \sum \mathrm{dr}\ f(r) . \mathrm{cache}(kr)
\end{eqnarray}
Avec cache(kr)=$4\pi\frac{\sin(kr)}{kr}r^2$

\subsubsection{Temps de calcul}
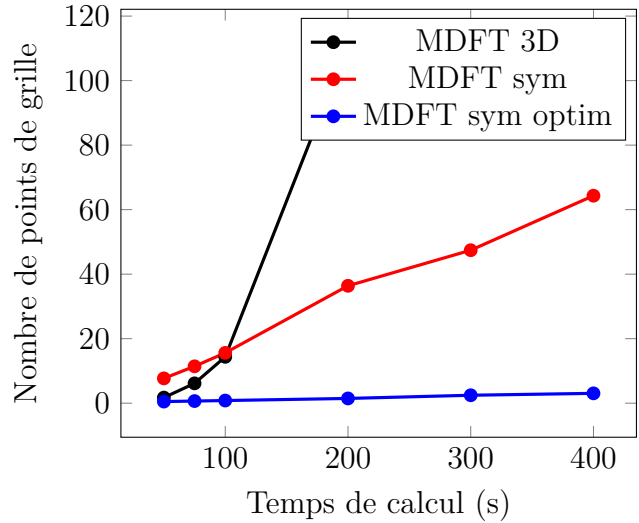
\begin{figure}[ht]
  \centering
  \begin{tikzpicture}
    \begin{axis}[
        xlabel=Temps de calcul (s),
        ylabel=Nombre de points de grille
    ]
    
    \addplot[mark=*, black, very thick] plot coordinates {
        (50,  1.73)
        (75,  6.15)
        (100, 14.37)
        (200, 111.05)
};
    \addplot[mark=*, red, very thick] plot coordinates {
        (50,  7.69)
        (75,  11.43)
        (100, 15.60)
        (200, 36.39)
        (300, 47.42)
        (400, 64.32)
};
    \addplot[mark=*, blue, very thick] plot coordinates {
        (50,  0.50)
        (75,  0.66)
        (100, 0.82)
        (200, 1.47)
        (300, 2.46)
        (400, 3.05)
};

    \legend{MDFT 3D\\MDFT sym\\MDFT sym optim\\}

    \end{axis}
  \end{tikzpicture}
  \caption[Temps de calcul nécessaire à la simulation de la solvatation d'un atome de méthane unifié.]{Temps de calcul nécessaire à la simulation de la solvatation d'un atome de méthane unifié en fonction du nombre de points de grille dans chaque dimension. Le temps nécessaire à la version 3D est représenté en noir, le temps nécessaire à la version à symétrie sphérique sans optimisation est représenté en rouge et avec optimisation en bleu.}
  \label{fig:temps_calcul_methane_versions}
\end{figure}

Afin d'évaluer les performances des différentes implémentations de MDFT, nous avons simulé la solvatation d'un méthane unifié (une boule Lennard-Jones) dans l'eau SPC/E, avec la version 3D puis avec la version à symétrie sphérique, avec et sans mise en cache.
La figure \ref{fig:temps_calcul_methane_versions} représente le temps de calcul nécessaire pour ces 3 versions en fonction du nombre de points de grille dans chaque direction.
Le temps de calcul dépend uniquement du nombre de points de grille et non de la taille du système, nous avons donc fixé, dans tous les cas, la largeur du système à 20 \AA.
Pour 200 points de grille, on voit que la version 3D a besoin de 1 min 51 sec pour compléter la minimisation, alors que la version à symétrie sphérique nécessite uniquement 36 sec.
La mise en cache des transformées de Hankel fait descendre ce temps à seulement 1,47 sec.
Nous divisons, dans ce cas, le temps de calcul par plus de 75.
De plus, nous voyons que les approximations adaptées à la spécificité de notre étude (soluté neutre, solvant sans angle) nous permettent d'atteindre des tailles de boîte inaccessibles avec la version 3D.
Nous minimisons par exemple facilement des systèmes de quelques centaines d'\AA\ avec une précision de 10 points/\AA.

Il existe bien sûr d'autres optimisations possibles comme l'utilisation de transformées de Hankel rapides mais vu le temps de calcul largement convenable, nous avons fait le choix de nous arrêter ici afin de ne pas rendre le code source illisible. Une fois cette version opérationnelle nous avons pu l'utiliser afin de développer notre nouveau bridge.

\section{Le bridge Gros Grain}
A l'aide de la version à symétrie sphérique de la MDFT, nous avons développé un nouveau bridge: le bridge gros grain. Ce bridge doit permettre une amélioration de la prédiction de l'énergie libre de solvatation et des profils de solvant de façon simple et rapide en ajoutant de la consistance thermodynamique au système. Pour cela nous devons, d'une part, reproduire une tension de surface correcte et, d'autre part, corriger la pression du système, qui est largement surestimée dans l'approximation HNC, en rendant possible la coexistence liquide-vapeur. 

\subsection{La tension de surface}
La tension de surface, notée $\gamma$, correspond à l'énergie nécessaire pour créer une unité de surface d'une interface liquide-gaz.

L'énergie libre de solvatation d'une bulle peut être exprimée en fonction de sa surface et de son volume sous la forme:

\begin{equation} \label{eq:energie_libre_terme_volume_surface}
\Delta G_{solv}= \mathrm{A} V_{\mathrm{sphere}} + \gamma S_{\mathrm{sphere}} 
\end{equation}

\noindent Or on sait que physiquement, pour des sphères de petits rayons, le terme proportionnel au volume est prépondérant.
Au contraire, pour les sphères de rayons importants, lorsque leurs profils peuvent être assimilés à des murs plats, c'est le terme en surface qui devient prépondérant soit:
\begin{equation}
\lim\limits_{r_{sphere} \to \infty} \Delta G_{solv} = \gamma S_{\mathrm{sphere}} 
\end{equation}
Que l'on réécrit:
\begin{equation}
\lim\limits_{r_{sphere} \to \infty} \frac{\Delta G_{solv}}{S_{\mathrm{sphere}}} = \gamma 
\end{equation}
Le rapport entre l'énergie libre de solvatation et la surface d'une sphère dure de grand diamètre correspond à la tension de surface du solvant, ici de l'eau.
Par la suite nous tracerons donc le rapport entre l'énergie libre de solvatation et la surface des sphères, soit $\frac{\Delta G_{solv}}{S_{\mathrm{sphere}}}$, afin de nous assurer que la tension de surface tend bien vers celle de l'eau.
Nous avons choisi comme référence la tension de surface de l'eau SPC/E et non celle de l'eau réelle afin de rester cohérent avec le modèle utilisé par MDFT.
Comme on l'a montré précédemment, la densité de la phase gazeuse tend vers 0. Pour faciliter le calcul numérique, la bulle de gaz est remplacée par une bulle de vide. Cela revient à faire l'approximation suivante: $\gamma_{liq-gaz} = \gamma_{liq-vide}$

\subsection{Définition du bridge gros grain}
Pour construire notre bridge, nous partons de l'approximation HNC. Cette approximation peut être interprétée comme un développement de Taylor à l'ordre deux de la fonctionnelle d'excès autour de la densité de référence $\rho_{0}$, la densité de l'eau liquide.
Dans cette approximation, comme on le voit sur la figure \ref{fig:fonctionelle_HNC}, plus on s'éloigne de la densité liquide et plus le potentiel chimique du solvant augmente. La phase gaz n'existe donc pas, ce qui à pour conséquence de surestimer fortement la pression du fluide.
Afin de corriger ce phénomène, nous ajoutons un terme d'ordre 3 qui permet de rendre cohérentes les énergies libres de solvatation des deux phases.

D'après la théorie de Landau-Ginzburg\cite{Ginzburg2009}, la hauteur de la courbe entre les deux phases est directement liée à la tension de surface. Nous ajoutons donc un terme d'ordre 4 qui s'annule en $\rho=0$ et $\rho=\rho_0$ afin d'autoriser la modification de cette hauteur sans modifier les deux phases précédemment ajustées.

L'avantage majeur de la MDFT est sa vitesse. Des termes à 3 et 4 corps demanderaient un temps de calcul trop important pour rester concurrentiel par rapport aux méthodes explicites. Afin de rendre le temps de calcul négligeable, nous reprenons une idée de Tarazona et al.\cite{tarazona_free-energy_1985}, qui consiste à remplacer des termes à 3 ou 4 corps par de simples puissances d'ordre 3 et 4 d'une densité gros grain notée $\bar{\rho}(\boldsymbol{r})$ définie comme

\begin{equation} \label{eq:convolution_gros_grain}
\bar{\rho}(\boldsymbol{r}) = \int \mathrm{d}\boldsymbol{r}^\prime \rho(\boldsymbol{r}^\prime) K\left(\left|\boldsymbol{r}-\boldsymbol{r}^\prime\right|\right) = \rho\ast K \left(\boldsymbol{r} \right)
\end{equation}

Le choix du noyau de convolution K sera décrit plus bas. Nous obtenons donc un bridge de la forme suivante:

\begin{equation} \label{eq:fbridge_2}
F_{\mathrm{b}}[\bar{\rho}(\boldsymbol{r})]=A\int\Delta\bar{\rho}(\boldsymbol{r})^3d\boldsymbol{r}+B\int\bar{\rho}(\boldsymbol{r})^2\Delta\bar{\rho}(\boldsymbol{r})^4d\boldsymbol{r}
\end{equation}

\subsection{\'Etude paramétrique}
À ce niveau, nous disposons d'un bridge que nous pouvons ajuster au travers de 3 paramètres: $A$, $B$ et le choix du noyau de convolution. Le premier paramètre $A$ est résolu analytiquement de façon à annuler le potentiel chimique de la phase gaz, soit $F[\rho_{liq}]=F[\rho_{gas}]=0$ . Nous obtenons ainsi $A=\mathrm{k_B}T(\frac{1}{\rho_0^2} - \frac{\int c(\boldsymbol{r}) d\boldsymbol{r}}{2\rho_0})$ (en $kJ.mol^{-1}.\text{\AA}^{6}$). Les deux autres paramètres, ont été déterminés à l'aide d'une étude paramétrique.

\subsubsection{Choix du noyau de convolution gros grain}
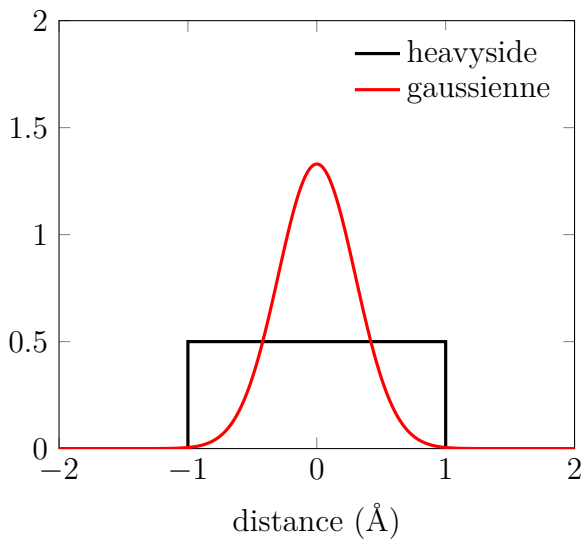
\begin{figure}[ht]
    \center    
  \begin{tikzpicture}
    \begin{axis}[
            xlabel= distance (\AA),
            xmin = -2, xmax = 2,
            ymin = 0, ymax = 2,
            no markers,
            legend style = {draw = none, cells={anchor=west}}
      ]
      \addplot[mark=none, black, very thick] plot coordinates {
        (-2,  0)
        (-1,  0)
        (-1,  0.5)
        (1,  0.5)
        (1, 0)
        (2, 0)
};
      
      \addplot[mark=none, red, very thick, domain=-2:2, samples=400] {gauss(0,0.3)};
      \legend{heavyside, gaussienne}
    \end{axis}
  \end{tikzpicture}
    \caption[Noyaux de convolution utilisés dans l'étude paramétrique permettant la définition du bridge gros grain.]{Noyaux de convolution utilisés dans l'étude paramétrique permettant la définition du bridge gros grain. En noir le heavyside et en rouge la gaussienne.}
    \label{fig:noyaux}
\end{figure}

Il existe différents noyaux de convolution permettant l'obtention d'une densité gros grain. Notre choix s'est dans un premier temps naturellement porté vers le plus simple, un heavyside (en noir sur la figure \ref{fig:noyaux}). Notre étude a permis d'éliminer rapidement ce noyau de convolution. Il n'existait aucune combinaison de paramètres permettant de reproduire correctement la tension de surface (résultats non présentés ici).

\pgfmathdeclarefunction{gauss}{2}{%
    \pgfmathparse{1/(#2*sqrt(2*pi))*exp(-((x-#1)^2)/(2*#2^2))}%
}

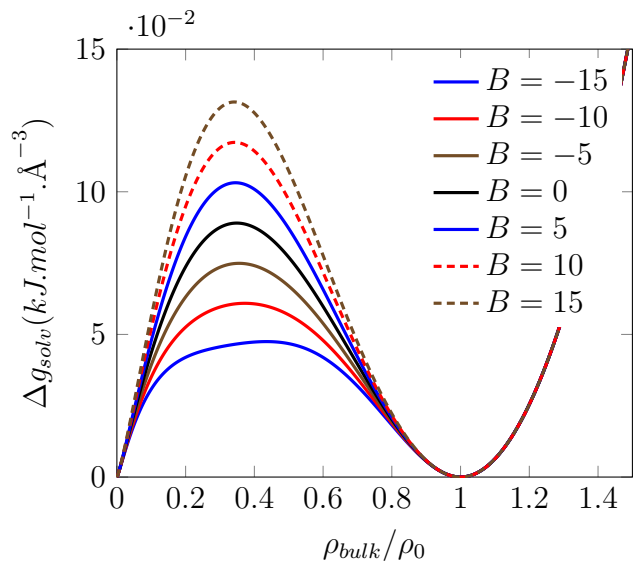
\begin{figure}[ht]
    \center
  \begin{tikzpicture}
    \begin{axis}[
           xlabel= $\rho_{bulk}/\rho_0$,
            ylabel= $\Delta g_{solv} (kJ.mol^{-1}.\text{\AA}^{-3}) $,
            xmin = 0, xmax = 1.5,
            ymin = 0, ymax = 0.15,
            scaled y ticks={base 10:2},
            legend style = {draw = none, cells={anchor=west}}
      ]
      \addplot+[mark=none, very thick] file {chapters/bridge/datas/fonctionnelles/fonctionnelle_-15.0.csv};
      \addplot+[mark=none, very thick] file {chapters/bridge/datas/fonctionnelles/fonctionnelle_-10.0.csv};
      \addplot+[mark=none, very thick] file {chapters/bridge/datas/fonctionnelles/fonctionnelle_-5.0.csv};
      \addplot+[mark=none, very thick] file {chapters/bridge/datas/fonctionnelles/fonctionnelle_0.0.csv};
      \addplot+[mark=none, very thick] file {chapters/bridge/datas/fonctionnelles/fonctionnelle_5.0.csv};
      \addplot+[mark=none, very thick] file {chapters/bridge/datas/fonctionnelles/fonctionnelle_10.0.csv};
      \addplot+[mark=none, very thick] file {chapters/bridge/datas/fonctionnelles/fonctionnelle_15.0.csv};
      \legend{$B=-15$, $B=-10$, $B=-5$, $B=0$, $B=5$, $B=10$, $B=15$}
    \end{axis}
  \end{tikzpicture}
    \caption[\'Energie libre d'une unité de volume du solvant homogène en fonction du paramètre B.]{\'Energie libre d'une unité de volume du solvant homogène de densité $\rho_\mathrm{bulk}$ en fonction du paramètre $B$ ($10^{-8} \mathrm{kJ.mol}^{-1}.\text{\AA}^{15}$). $\rho_0$ est la densité bulk de l'eau SPC/E à pression et température standards (1$kg.L^{-1}$).}
    \label{fig:etude_B}
\end{figure}

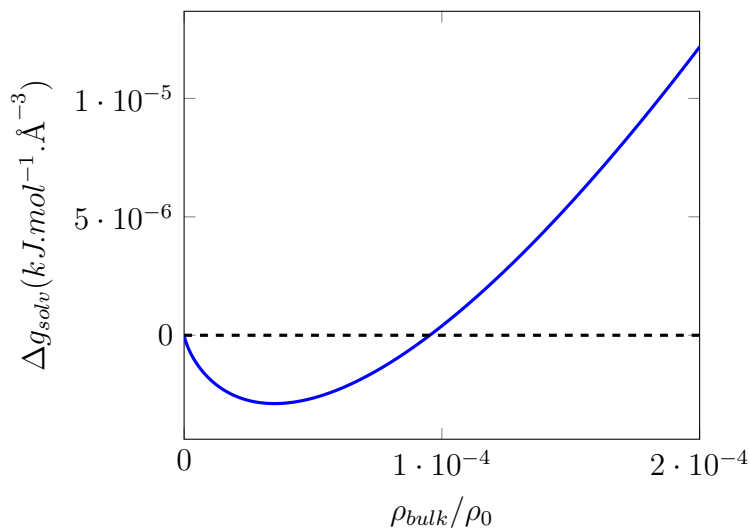
\begin{figure}[ht]
    \center
  \begin{tikzpicture}
    \begin{axis}[
           xlabel= $\rho_{bulk}/\rho_0$,
            ylabel= $\Delta g_{solv} (kJ.mol^{-1}.\text{\AA}^{-3}) $,
            scaled x ticks={base 10:0},
            scaled y ticks={base 10:0},
            xtick = {0, 0.0001, 0.0002},
			xmin = 0, xmax = 0.0002,
            legend style = {draw = none, cells={anchor=west}}
      ]
      \addplot+[mark=none, very thick] file {chapters/bridge/datas/fonctionnelles/fonctionnelle_0.0_zoom.csv};
      \addplot[mark=none, dashed, black, very thick] coordinates {(0, 0) (0.0002, 0)};
    \end{axis}
  \end{tikzpicture}
    \caption[Zoom de l'\'energie libre d'une unité de volume du solvant homogène autour de l'origine.]{Zoom de l'\'energie libre d'une unité de volume du solvant homogène autour de l'origine. On observe un nouveau minimum correspondant à la phase gazeuse du solvant.}
    \label{fig:etude_B_zoom}
\end{figure}

Nous nous sommes ensuite intéressés à la gaussienne (en rouge sur la figure \ref{fig:noyaux}).
Ce noyau est défini par deux paramètres, sa largeur à mi hauteur $\sigma_{gauss}$ (\AA) et un pré-facteur définissant sa hauteur.
Afin de conserver une cohérence entre la densité et la densité gros grain, le pré-facteur est choisi de façon à obtenir une aire sous la courbe de la gaussienne toujours égale à 1.
\`A ce niveau, nous disposons donc de deux paramètres, $B$ et $\sigma_{gauss}$.
Dans un premier temps, nous avons cherché les limites de $B$ qui ont un impact direct sur la forme de la courbe de potentiel chimique du fluide homogène.
Comme on le voit sur la figure \ref{fig:etude_B}, une valeur de $B$ inférieure à $-15.10^{-8}\  \mathrm{kJ.mol}^{-1}.\text{\AA}^{15}$ ou supérieure à $15.10^{-8} \mathrm{kJ.mol}^{-1}.\text{\AA}^{15}$ déforme la courbe.
Nous avons donc choisi de concentrer notre étude sur des valeurs de $B$ allant de $-15.10^{-8}$ à $15.10^{-8} \mathrm{kJ.mol}^{-1}.\text{\AA}^{15}$.
La figure \ref{fig:etude_B_zoom} correspond à un zoom de la fonctionnelle autour de l'origine. On voit la création d'un second minimum proche de zéro qui confirme la présence d'une phase gazeuse. On voit également que le minimum est très légèrement négatif. Ce résultat est attendu car on impose $\rho[0]=0$ et on sait que $\frac{\delta F}{\delta \rho}<0$ (voir annexe \ref{chap:annexes:grad}).

\subsubsection{\'Exploration des paramètres $B$ et $\sigma_{gauss}$}
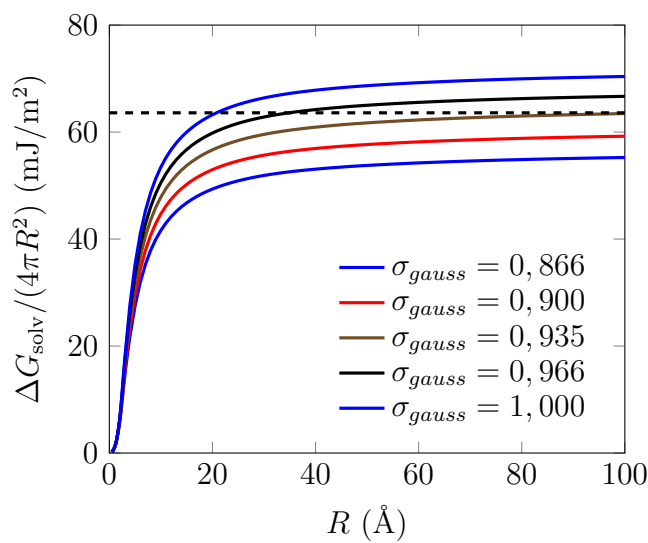
\begin{figure}[ht]
\center
    \begin{tikzpicture}
      \begin{axis}[
          xlabel= $R$ (\AA),
          ylabel= $\Delta G_{\textrm{solv}}/(4\pi R^2)$ (mJ/m$^2$), 
          xmin = 0, xmax = 100,
          ymin = 0, ymax = 80,
          legend style = {draw = none, at={(0.95,0.05)},anchor=south east}
          ]
        \addplot+[mark=none, very thick] table[x index=0, y index=4] {chapters/bridge/datas/free_energy_by_volume/results_HS_0.866_15.csv};
        \addplot+[mark=none, very thick] table[x index=0, y index=4] {chapters/bridge/datas/free_energy_by_volume/results_HS_0.900_15.csv};
        \addplot+[mark=none, very thick] table[x index=0, y index=4] {chapters/bridge/datas/free_energy_by_volume/results_HS_0.935_15.csv};
        \addplot+[mark=none, very thick] table[x index=0, y index=4] {chapters/bridge/datas/free_energy_by_volume/results_HS_0.966_15.csv};
        \addplot+[mark=none, very thick] table[x index=0, y index=4] {chapters/bridge/datas/free_energy_by_volume/results_HS_1.000_15.csv};
        \addplot[mark=none, dashed, black, very thick] coordinates {(0, 63.6) (100, 63.6)};
        \legend{{$\sigma_{gauss}=0,866$}, {$\sigma_{gauss}=0,900$}, {$\sigma_{gauss}=0,935$}, {$\sigma_{gauss}=0,966$}, {$\sigma_{gauss}=1,000$}}
      \end{axis}
    \end{tikzpicture}
	\caption[\'Energie libre de solvatation d'une sphère dure divisée par sa surface en fonction de son rayon  pour différentes valeurs de $\sigma_{gauss}$.]{\'Energie libre de solvatation d'une sphère dure divisée par sa surface en fonction de son rayon  pour différentes valeurs de $\sigma_{gauss}$ (en \AA), largeur de la gaussienne servant à la convolution. $B$ est fixé ici à $15.10^{-8} kJ.mol^{-1}.\text{\AA}^{15}$. La valeur de référence en pointillé est la valeur de la tension de surface de l'eau SPC/E soit 63,3 mJ.m$^{-2}$\cite{vega_surface_2007}.}
    \label{fig:etude_sigma}
\end{figure}

Pour chaque valeur de ce paramètre B, nous avons tracé l'énergie libre de solvatation d'une sphère dure divisée par son volume en fonction de son rayon.
Nous avons ensuite ajusté la valeur de $\sigma_{gauss}$ pour obtenir une tension de surface correcte (voir figure \ref{fig:etude_sigma}).

Nous avons ainsi obtenu, pour chaque valeur de $B$ sélectionnée, la valeur de $\sigma_{gauss}$ reproduisant correctement la tension de surface $\gamma$ de l'eau. Ces valeurs sont disponibles dans le tableau \ref{tab:parametres_bridge}.

\begin{table}[ht]
 \centering
  \begin{tabular}{l || c c c c c c c}
    \hline
    $B$ ($10^{-8} \mathrm{kJ.mol}^{-1}.\text{\AA}^{15}$)  & -15 & -10 & -5 & 0 & 5 & 10 & 15  \\
    \hline
     $\sigma_{gauss}$ (\AA)  & 1,177 & 1,110 & 1,061 & 1,021 & 0,989 & 0,960 & 0,935  \\
    \hline
  \end{tabular}
  \caption{Couples de paramètres permettant d'obtenir la bonne tension de surface de l'eau.}
  \label{tab:parametres_bridge}  
\end{table}

Afin de sélectionner le meilleur couple de paramètres, nous disposons de différentes références, que ce soit de structure de solvant ou d'énergie libre de solvatation.

\subsubsection{\'Energie libre de solvatation de sphères dures}
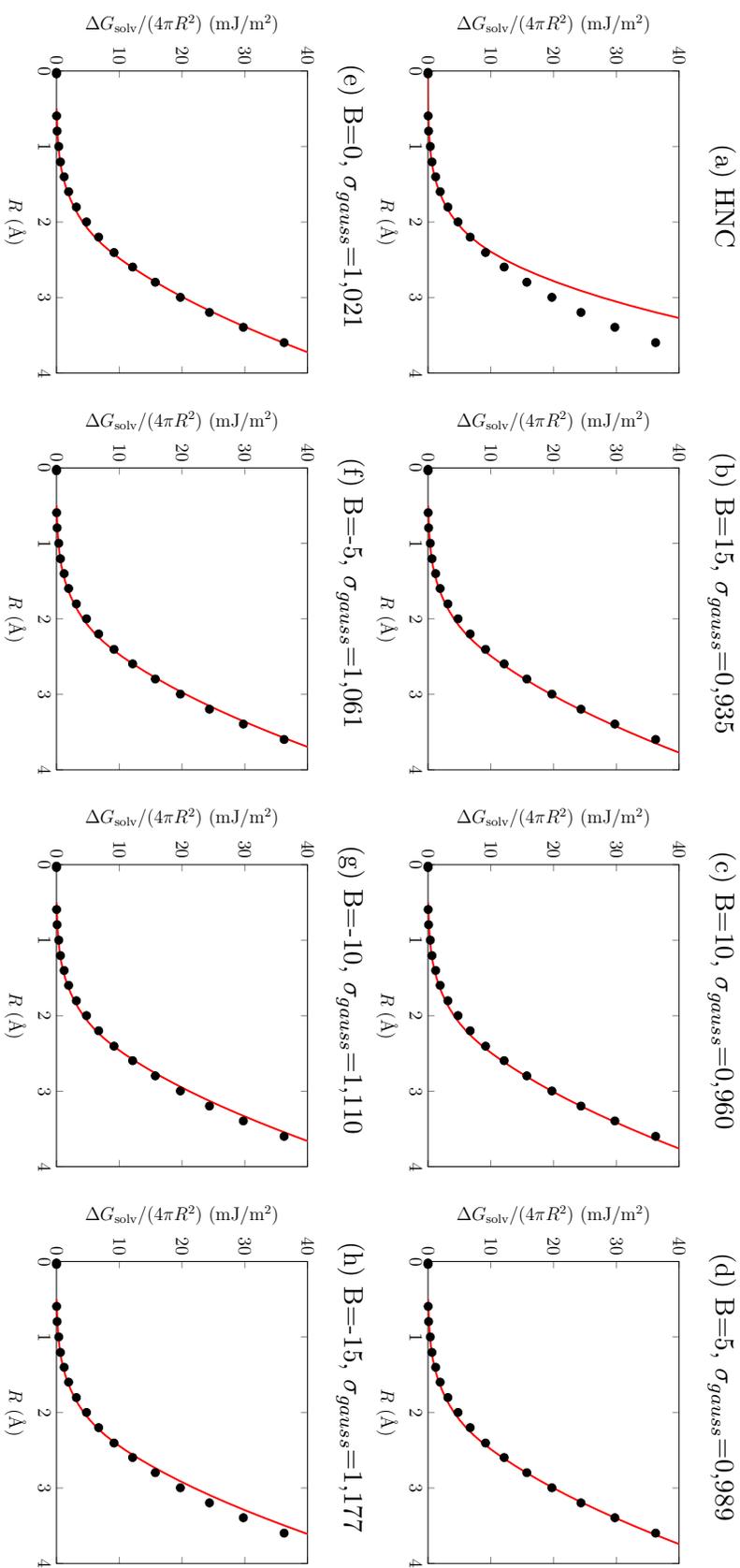
\begin{sidewaysfigure}
\begin{figure}[H]
\centering
\begin{subfigure}{.24\textwidth}
  \centering
    \caption{HNC}
    \resizebox{\linewidth}{!}{
    \begin{tikzpicture}
        	\begin{axis}[
                xlabel= $R$ (\AA),
          		ylabel= $\Delta G_{\textrm{solv}}/(4\pi R^2)$ (mJ/m$^2$), 
    	        restrict x to domain=0:5,
    	        restrict y to domain=0:50,
            	xmin = 0,
        	    xmax = 4,
                ymin = 0,
	            ymax = 40,
        	    legend style = {draw = none, cells={anchor=west}}
    	        ]
            	\addplot+[mark=none, red, very thick] table[x index=0, y index=2]{chapters/bridge/datas/free_energy_by_volume/results_HS_HNC.csv};
                \addplot+[only marks,mark=*,mark options={scale=1.3, black, fill=black},text mark as node=true] table[x index=0, y index=1] {chapters/bridge/datas/free_energy_by_volume/hummer_ref.csv};
        	\end{axis}
    \end{tikzpicture}
}
  \end{subfigure}
\begin{subfigure}{.24\textwidth}
  \centering
    \caption{B=15, $\sigma_{gauss}$=0,935}
    \resizebox{\linewidth}{!}{
    \begin{tikzpicture}
        	\begin{axis}[
                xlabel= $R$ (\AA),
          		ylabel= $\Delta G_{\textrm{solv}}/(4\pi R^2)$ (mJ/m$^2$), 
    	        restrict x to domain=0:5,
    	        restrict y to domain=0:50,
            	xmin = 0,
        	    xmax = 4,
                ymin = 0,
	            ymax = 40,
        	    legend style = {draw = none, cells={anchor=west}}
    	        ]
            	\addplot+[mark=none, red, very thick] table[x index=0, y index=2]{chapters/bridge/datas/free_energy_by_volume/results_HS_0.935_15.csv};
                \addplot+[only marks,mark=*,mark options={scale=1.3, black, fill=black},text mark as node=true] table[x index=0, y index=1] {chapters/bridge/datas/free_energy_by_volume/hummer_ref.csv};
        	\end{axis}
    \end{tikzpicture}
}
  \end{subfigure}
\begin{subfigure}{.24\textwidth}
  \centering
    \caption{B=10, $\sigma_{gauss}$=0,960}
    \resizebox{\linewidth}{!}{
    \begin{tikzpicture}
        	\begin{axis}[
                xlabel= $R$ (\AA),
          		ylabel= $\Delta G_{\textrm{solv}}/(4\pi R^2)$ (mJ/m$^2$), 
    	        restrict x to domain=0:5,
    	        restrict y to domain=0:50,
            	xmin = 0,
        	    xmax = 4,
                ymin = 0,
	            ymax = 40,
        	    legend style = {draw = none, cells={anchor=west}}
    	        ]
            	\addplot+[mark=none, red, very thick] table[x index=0, y index=2]{chapters/bridge/datas/free_energy_by_volume/results_HS_0.960_10.csv};
                \addplot+[only marks,mark=*,mark options={scale=1.3, black, fill=black},text mark as node=true] table[x index=0, y index=1] {chapters/bridge/datas/free_energy_by_volume/hummer_ref.csv};
        	\end{axis}
    \end{tikzpicture}
}
  \end{subfigure}
\begin{subfigure}{.24\textwidth}
  \centering
    \caption{B=5, $\sigma_{gauss}$=0,989}
    \resizebox{\linewidth}{!}{
    \begin{tikzpicture}
        	\begin{axis}[
                xlabel= $R$ (\AA),
          		ylabel= $\Delta G_{\textrm{solv}}/(4\pi R^2)$ (mJ/m$^2$), 
    	        restrict x to domain=0:5,
    	        restrict y to domain=0:50,
            	xmin = 0,
        	    xmax = 4,
                ymin = 0,
	            ymax = 40,
        	    legend style = {draw = none, cells={anchor=west}}
    	        ]
            	\addplot+[mark=none, red, very thick] table[x index=0, y index=2]{chapters/bridge/datas/free_energy_by_volume/results_HS_0.989_5.csv};
                \addplot+[only marks,mark=*,mark options={scale=1.3, black, fill=black},text mark as node=true] table[x index=0, y index=1] {chapters/bridge/datas/free_energy_by_volume/hummer_ref.csv};
        	\end{axis}
    \end{tikzpicture}
}
  \end{subfigure}
\begin{subfigure}{.24\textwidth}
  \centering
    \caption{B=0, $\sigma_{gauss}$=1,021}
    \resizebox{\linewidth}{!}{
    \begin{tikzpicture}
        	\begin{axis}[
                xlabel= $R$ (\AA),
          		ylabel= $\Delta G_{\textrm{solv}}/(4\pi R^2)$ (mJ/m$^2$), 
    	        restrict x to domain=0:5,
    	        restrict y to domain=0:50,
            	xmin = 0,
        	    xmax = 4,
                ymin = 0,
	            ymax = 40,
        	    legend style = {draw = none, cells={anchor=west}}
    	        ]
            	\addplot+[mark=none, red, very thick] table[x index=0, y index=2]{chapters/bridge/datas/free_energy_by_volume/results_HS_1.021_0.csv};
                \addplot+[only marks,mark=*,mark options={scale=1.3, black, fill=black},text mark as node=true] table[x index=0, y index=1] {chapters/bridge/datas/free_energy_by_volume/hummer_ref.csv};
        	\end{axis}
    \end{tikzpicture}
}
  \end{subfigure}
\begin{subfigure}{.24\textwidth}
  \centering
    \caption{B=-5, $\sigma_{gauss}$=1,061}
    \resizebox{\linewidth}{!}{
    \begin{tikzpicture}
        	\begin{axis}[
                xlabel= $R$ (\AA),
          		ylabel= $\Delta G_{\textrm{solv}}/(4\pi R^2)$ (mJ/m$^2$), 
    	        restrict x to domain=0:5,
    	        restrict y to domain=0:50,
            	xmin = 0,
        	    xmax = 4,
                ymin = 0,
	            ymax = 40,
        	    legend style = {draw = none, cells={anchor=west}}
    	        ]
            	\addplot+[mark=none, red, very thick] table[x index=0, y index=2]{chapters/bridge/datas/free_energy_by_volume/results_HS_1.061_-5.csv};
                \addplot+[only marks,mark=*,mark options={scale=1.3, black, fill=black},text mark as node=true] table[x index=0, y index=1] {chapters/bridge/datas/free_energy_by_volume/hummer_ref.csv};
        	\end{axis}
    \end{tikzpicture}
}
  \end{subfigure}
\begin{subfigure}{.24\textwidth}
  \centering
    \caption{B=-10, $\sigma_{gauss}$=1,110}
    \resizebox{\linewidth}{!}{
    \begin{tikzpicture}
        	\begin{axis}[
                xlabel= $R$ (\AA),
          		ylabel= $\Delta G_{\textrm{solv}}/(4\pi R^2)$ (mJ/m$^2$), 
    	        restrict x to domain=0:5,
    	        restrict y to domain=0:50,
            	xmin = 0,
        	    xmax = 4,
                ymin = 0,
	            ymax = 40,
        	    legend style = {draw = none, cells={anchor=west}}
    	        ]
            	\addplot+[mark=none, red, very thick] table[x index=0, y index=2]{chapters/bridge/datas/free_energy_by_volume/results_HS_1.110_-10.csv};
                \addplot+[only marks,mark=*,mark options={scale=1.3, black, fill=black},text mark as node=true] table[x index=0, y index=1] {chapters/bridge/datas/free_energy_by_volume/hummer_ref.csv};
        	\end{axis}
    \end{tikzpicture}
}
  \end{subfigure}
\begin{subfigure}{.24\textwidth}
  \centering
    \caption{B=-15, $\sigma_{gauss}$=1,177}
    \resizebox{\linewidth}{!}{
    \begin{tikzpicture}
        	\begin{axis}[
                xlabel= $R$ (\AA),
          		ylabel= $\Delta G_{\textrm{solv}}/(4\pi R^2)$ (mJ/m$^2$), 
    	        restrict x to domain=0:5,
    	        restrict y to domain=0:50,
            	xmin = 0,
        	    xmax = 4,
                ymin = 0,
	            ymax = 40,
        	    legend style = {draw = none, cells={anchor=west}}
    	        ]
            	\addplot+[mark=none, red, very thick] table[x index=0, y index=2]{chapters/bridge/datas/free_energy_by_volume/results_HS_1.177_-15.csv};
                \addplot+[only marks,mark=*,mark options={scale=1.3, black, fill=black},text mark as node=true] table[x index=0, y index=1] {chapters/bridge/datas/free_energy_by_volume/hummer_ref.csv};
        	\end{axis}
    \end{tikzpicture}
}
  \end{subfigure}
    \caption[\'Energie libre de solvatation d'une sphère dure divisée par sa surface en fonction de son rayon pour différentes valeurs de B.]{\'Energie libre de solvatation d'une sphère dure divisée par sa surface en fonction de son rayon calculée dans l'approximation HNC puis avec les différents paramètres du bridge. Les calculs ont été effectués pour chaque couple de paramètres $B$ ($10^{-8} \mathrm{kJ.mol}^{-1}.\text{\AA}^{15}$) et $\sigma_{gauss}$ (\AA). Ces courbes en rouge, sont comparées à des valeurs de référence (sphères noires) calculées par Monte-Carlo\cite{hummer_information_1996}}
    \label{fig:comparaison_hummer}
\end{figure}
\end{sidewaysfigure}

Dans un premier temps, nous comparons, pour chaque jeu de paramètres, l'\'energie libre de solvatation d'une sphère dure divisée par sa surface aux valeurs de références calculées par Monte Carlo\cite{hummer_information_1996} (voir figure \ref{fig:comparaison_hummer}). À cause du temps de calcul nécessaire, Hummer et al\cite{hummer_information_1996}, se sont limités à une dizaine de points pour des sphères de rayon inférieur à 4 \AA.

On voit que l'approximation HNC, surestime les énergies libres de solvatation. Notre bridge, corrige fortement cet écart, quelque soit le couple de paramètres choisi. Les meilleurs paramètres ici semblent être pour un $B$ autour de $5.10^{-8}\  \mathrm{kJ.mol}^{-1}.\text{\AA}^{15}$. 

\boitesimple{Les \'energies libres de solvatation de sphères dures les plus précises sont obtenues pour un $B$ autour de $5.10^{-8}\  \mathrm{kJ.mol}^{-1}.\text{\AA}^{15}$. Chacun des paramètres proposés sont acceptables à ce niveau.}

\subsubsection{\'Energie libre de solvatation de molécules modèles}
Nous avons ensuite calculé l'énergie libre de solvatation de molécules modèles avec MDFT, pour les différents paramètres du bridge, et par dynamique moléculaire. Nos molécules modèles sont le méthane unifié et les gaz rares: Néon, Argon, Krypton, Xénon. L'ensemble de ces résultats est disponible dans le tableau \ref{tab:energie_libre_molecules_models}. Les paramètres Lennard-Jones de ces composés sont disponibles dans le tableau \ref{tab:param_lj}.

\begin{table}[ht]
  \centering
  \begin{tabular}{l c c c c c c c c c c c}
   \hline & \\[-1em]\hline
    Méthode    & Exp   & DM    & HNC    &      &       &       & MDFT  &       &       & \\
    \hline
    $B$          &       &       &        & -15   & -10   & -5    & 0     & 5     & 10    & 15     \\
    $\sigma_{gauss}$   &       &       &        & 1,177 & 1,110 & 1,061 & 1,021 & 0,989 & 0,960 & 0,935  \\
    \hline
    Méthane    &       &  9,23 & 28,14  & 15,97 & 14,55 & 13,70 & 13,05 & 12,62 & 12,25 & \textbf{11,99}  \\
    Néon       & 10,36 & 11,73 & 19,04  & 14,89 & 14,25 & 13,83 & 13,53 & 13,34 & 13,19 & \textbf{13,10}  \\
    Argon      &  8,40 &  8,61 & 22,89  & 14,16 & 13,12 & 12,42 & 11,89 & 11,55 & 11,25 & \textbf{11,01}  \\
    Krypton    &  6,96 &  8,03 & 26,41  & 14,54 & 13,28 & 12,44 & 11,80 & 11,38 & 11,00 & \textbf{10,74}  \\
    Xénon       &  6,06 &  6,47 & 31,23 & 15,02 & 13,50 & 12,48 & 11,71 & 11,20 & 10,74 & \textbf{10,33}  \\
    \hline & \\[-1em]\hline
  \end{tabular}
  \caption[\'Energie libre de solvatation du méthane unifié et des gaz rares.]{Valeurs de l'énergie libre de solvatation (en kJ.mol$^{-1}$) du méthane unifié et des gaz rares: Argon, Xénon, Krypton, Néon pour les différents paramètres possibles de la fonctionnelle de bridge gros grain. Les valeurs les plus proches de notre référence, la dynamique moléculaire, sont en gras.}
  \label{tab:energie_libre_molecules_models}  
\end{table}

\begin{table}[ht]
  \centering
  \begin{tabular}{l c c}
   \hline & \\[-1em]\hline
    Soluté    & $\sigma_{LJ}$ (\AA) & $\epsilon_{LJ}$ (kJ.mol$^{-1}$) \\
    \hline
      méthane & 3,73000 & 1,2300 \\
	  néon & 3,03500 & 0,15432 \\
	  argon & 3,41500 & 1,03931 \\
	  krypton & 3,67500 & 1,4051 \\
	  xenon & 3,97500 & 1,7851 \\
    \hline & \\[-1em]\hline
  \end{tabular}
  \caption[Paramètres Lennard-Jones du méthane unifié et des gaz rares utilisés dans nos simulations.]{Paramètres Lennard-Jones du méthane unifié et des gaz rares utilisés dans nos simulations.}
  \label{tab:param_lj}  
\end{table}

On remarque que l'approximation HNC surestime fortement le calcul de l'énergie libre de solvatation de plusieurs dizaines de $\mathrm{kJ.mol}^{-1}$. Le bridge gros grain permet de considérablement réduire cet écart, à quelques $\mathrm{kJ.mol}^{-1}$ seulement.
On remarque également que plus $B$ est grand et donc plus la barrière de transition entre la phase liquide et gaz est importante, et meilleures sont les prédictions d'énergies libres de solvatation. 

\boitesimple{Les meilleures énergies libres de solvatation de nos molécules modèles sont obtenues avec B=$15.10^{-8}\  \mathrm{kJ.mol}^{-1}.\text{\AA}^{15}$  et $\sigma_{gauss}$ = 0,935 \AA}

\subsubsection{Structure du solvant autour de sphères dures}
\begin{sidewaysfigure}
\begin{figure}[H]
\centering
\begin{subfigure}{.24\textwidth}
  \centering
    \caption{HNC}
    \resizebox{\linewidth}{!}{
      \begin{tikzpicture}
          \begin{axis}[
              xlabel= r ($\text{\AA}$),
              ylabel= g(r),
              xmin = 0,
              xmax = 16,
              ymin = 0,
              ymax = 8,
              legend style = {draw = none, cells={anchor=west}}
              ]
              \addplot+[mark=none, black, very thick, solid] table[x index=0,y index=1] {chapters/bridge/datas/g_of_r/data_HS_HNC/g_HS_HNC_2_4_6_8_10.csv};
              \addplot+[mark=none, blue, very thick, dashed] file {chapters/bridge/datas/g_of_r/chandler_ref/chandler_2.csv};
              \addplot+[mark=none, black, very thick, solid] table[x index=0,y index=2] {chapters/bridge/datas/g_of_r/data_HS_HNC/g_HS_HNC_2_4_6_8_10.csv};
              \addplot+[mark=none, blue, very thick, dashed] file {chapters/bridge/datas/g_of_r/chandler_ref/chandler_4.csv};
              \addplot+[mark=none, black, very thick, solid] table[x index=0,y index=3] {chapters/bridge/datas/g_of_r/data_HS_HNC/g_HS_HNC_2_4_6_8_10.csv};
              \addplot+[mark=none, blue, very thick, dashed] file {chapters/bridge/datas/g_of_r/chandler_ref/chandler_6.csv};
              \addplot+[mark=none, black, very thick, solid] table[x index=0,y index=4] {chapters/bridge/datas/g_of_r/data_HS_HNC/g_HS_HNC_2_4_6_8_10.csv};
              \addplot+[mark=none, blue, very thick, dashed] file {chapters/bridge/datas/g_of_r/chandler_ref/chandler_8.csv};
              \addplot+[mark=none, black, very thick, solid] table[x index=0,y index=5] {chapters/bridge/datas/g_of_r/data_HS_HNC/g_HS_HNC_2_4_6_8_10.csv};
              \addplot+[mark=none, blue, very thick, dashed] file {chapters/bridge/datas/g_of_r/chandler_ref/chandler_10.csv};
              \legend{MDFT(HNC), MC}
          \end{axis}
      \end{tikzpicture}
}
  \end{subfigure}
\begin{subfigure}{.24\textwidth}
  \centering
    \caption{B=15, $\sigma_{gauss}$=0,935}
    \resizebox{\linewidth}{!}{
      \begin{tikzpicture}
          \begin{axis}[
              xlabel= r ($\text{\AA}$),
              ylabel= g(r),
              xmin = 0,
              xmax = 16,
              ymin = 0,
              ymax = 8,
              legend style = {draw = none, cells={anchor=west}}
              ]
              \addplot+[mark=none, red, very thick, solid] table[x index=0,y index=1] {chapters/bridge/datas/g_of_r/data_HS_HNC_bridge_0.935_15/g_HS_HNC_bridge_2_4_6_8_10.csv};
              \addplot+[mark=none, blue, very thick, dashed] file {chapters/bridge/datas/g_of_r/chandler_ref/chandler_2.csv};
              \addplot+[mark=none, red, very thick, solid] table[x index=0,y index=2] {chapters/bridge/datas/g_of_r/data_HS_HNC_bridge_0.935_15/g_HS_HNC_bridge_2_4_6_8_10.csv};
              \addplot+[mark=none, blue, very thick, dashed] file {chapters/bridge/datas/g_of_r/chandler_ref/chandler_4.csv};
              \addplot+[mark=none, red, very thick, solid] table[x index=0,y index=3] {chapters/bridge/datas/g_of_r/data_HS_HNC_bridge_0.935_15/g_HS_HNC_bridge_2_4_6_8_10.csv};
              \addplot+[mark=none, blue, very thick, dashed] file {chapters/bridge/datas/g_of_r/chandler_ref/chandler_6.csv};
              \addplot+[mark=none, red, very thick, solid] table[x index=0,y index=4] {chapters/bridge/datas/g_of_r/data_HS_HNC_bridge_0.935_15/g_HS_HNC_bridge_2_4_6_8_10.csv};
              \addplot+[mark=none, blue, very thick, dashed] file {chapters/bridge/datas/g_of_r/chandler_ref/chandler_8.csv};
              \addplot+[mark=none, red, very thick, solid] table[x index=0,y index=5] {chapters/bridge/datas/g_of_r/data_HS_HNC_bridge_0.935_15/g_HS_HNC_bridge_2_4_6_8_10.csv};
              \addplot+[mark=none, blue, very thick, dashed] file {chapters/bridge/datas/g_of_r/chandler_ref/chandler_10.csv};
              \legend{MDFT(HNC+B), MC}
          \end{axis}
      \end{tikzpicture}
}
  \end{subfigure}
\begin{subfigure}{.24\textwidth}
  \centering
    \caption{B=10, $\sigma_{gauss}$=0,960}
    \resizebox{\linewidth}{!}{
      \begin{tikzpicture}
          \begin{axis}[
              xlabel= r ($\text{\AA}$),
              ylabel= g(r),
              xmin = 0,
              xmax = 16,
              ymin = 0,
              ymax = 8,
              legend style = {draw = none, cells={anchor=west}}
              ]
              \addplot+[mark=none, red, very thick, solid] table[x index=0,y index=1] {chapters/bridge/datas/g_of_r/data_HS_HNC_bridge_0.960_10/g_HS_HNC_bridge_2_4_6_8_10.csv};
              \addplot+[mark=none, blue, very thick, dashed] file {chapters/bridge/datas/g_of_r/chandler_ref/chandler_2.csv};
              \addplot+[mark=none, red, very thick, solid] table[x index=0,y index=2] {chapters/bridge/datas/g_of_r/data_HS_HNC_bridge_0.960_10/g_HS_HNC_bridge_2_4_6_8_10.csv};
              \addplot+[mark=none, blue, very thick, dashed] file {chapters/bridge/datas/g_of_r/chandler_ref/chandler_4.csv};
              \addplot+[mark=none, red, very thick, solid] table[x index=0,y index=3] {chapters/bridge/datas/g_of_r/data_HS_HNC_bridge_0.960_10/g_HS_HNC_bridge_2_4_6_8_10.csv};
              \addplot+[mark=none, blue, very thick, dashed] file {chapters/bridge/datas/g_of_r/chandler_ref/chandler_6.csv};
              \addplot+[mark=none, red, very thick, solid] table[x index=0,y index=4] {chapters/bridge/datas/g_of_r/data_HS_HNC_bridge_0.960_10/g_HS_HNC_bridge_2_4_6_8_10.csv};
              \addplot+[mark=none, blue, very thick, dashed] file {chapters/bridge/datas/g_of_r/chandler_ref/chandler_8.csv};
              \addplot+[mark=none, red, very thick, solid] table[x index=0,y index=5] {chapters/bridge/datas/g_of_r/data_HS_HNC_bridge_0.960_10/g_HS_HNC_bridge_2_4_6_8_10.csv};
              \addplot+[mark=none, blue, very thick, dashed] file {chapters/bridge/datas/g_of_r/chandler_ref/chandler_10.csv};
              \legend{MDFT(HNC+B), MC}
          \end{axis}
      \end{tikzpicture}
}
  \end{subfigure}
\begin{subfigure}{.24\textwidth}
  \centering
    \caption{B=5, $\sigma_{gauss}$=0,989}
    \resizebox{\linewidth}{!}{
      \begin{tikzpicture}
          \begin{axis}[
              xlabel= r ($\text{\AA}$),
              ylabel= g(r),
              xmin = 0,
              xmax = 16,
              ymin = 0,
              ymax = 8,
              legend style = {draw = none, cells={anchor=west}}
              ]
              \addplot+[mark=none, red, very thick, solid] table[x index=0,y index=1] {chapters/bridge/datas/g_of_r/data_HS_HNC_bridge_0.989_5/g_HS_HNC_bridge_2_4_6_8_10.csv};
              \addplot+[mark=none, blue, very thick, dashed] file {chapters/bridge/datas/g_of_r/chandler_ref/chandler_2.csv};
              \addplot+[mark=none, red, very thick, solid] table[x index=0,y index=2] {chapters/bridge/datas/g_of_r/data_HS_HNC_bridge_0.989_5/g_HS_HNC_bridge_2_4_6_8_10.csv};
              \addplot+[mark=none, blue, very thick, dashed] file {chapters/bridge/datas/g_of_r/chandler_ref/chandler_4.csv};
              \addplot+[mark=none, red, very thick, solid] table[x index=0,y index=3] {chapters/bridge/datas/g_of_r/data_HS_HNC_bridge_0.989_5/g_HS_HNC_bridge_2_4_6_8_10.csv};
              \addplot+[mark=none, blue, very thick, dashed] file {chapters/bridge/datas/g_of_r/chandler_ref/chandler_6.csv};
              \addplot+[mark=none, red, very thick, solid] table[x index=0,y index=4] {chapters/bridge/datas/g_of_r/data_HS_HNC_bridge_0.989_5/g_HS_HNC_bridge_2_4_6_8_10.csv};
              \addplot+[mark=none, blue, very thick, dashed] file {chapters/bridge/datas/g_of_r/chandler_ref/chandler_8.csv};
              \addplot+[mark=none, red, very thick, solid] table[x index=0,y index=5] {chapters/bridge/datas/g_of_r/data_HS_HNC_bridge_0.989_5/g_HS_HNC_bridge_2_4_6_8_10.csv};
              \addplot+[mark=none, blue, very thick, dashed] file {chapters/bridge/datas/g_of_r/chandler_ref/chandler_10.csv};
              \legend{MDFT(HNC+B), MC}
          \end{axis}
      \end{tikzpicture}
}
  \end{subfigure}
\begin{subfigure}{.24\textwidth}
  \centering
    \caption{B=0, $\sigma_{gauss}$=1,021}
    \resizebox{\linewidth}{!}{
      \begin{tikzpicture}
          \begin{axis}[
              xlabel= r ($\text{\AA}$),
              ylabel= g(r),
              xmin = 0,
              xmax = 16,
              ymin = 0,
              ymax = 8,
              legend style = {draw = none, cells={anchor=west}}
              ]
              \addplot+[mark=none, red, very thick, solid] table[x index=0,y index=1] {chapters/bridge/datas/g_of_r/data_HS_HNC_bridge_1.021_0/g_HS_HNC_bridge_2_4_6_8_10.csv};
              \addplot+[mark=none, blue, very thick, dashed] file {chapters/bridge/datas/g_of_r/chandler_ref/chandler_2.csv};
              \addplot+[mark=none, red, very thick, solid] table[x index=0,y index=2] {chapters/bridge/datas/g_of_r/data_HS_HNC_bridge_1.021_0/g_HS_HNC_bridge_2_4_6_8_10.csv};
              \addplot+[mark=none, blue, very thick, dashed] file {chapters/bridge/datas/g_of_r/chandler_ref/chandler_4.csv};
              \addplot+[mark=none, red, very thick, solid] table[x index=0,y index=3] {chapters/bridge/datas/g_of_r/data_HS_HNC_bridge_1.021_0/g_HS_HNC_bridge_2_4_6_8_10.csv};
              \addplot+[mark=none, blue, very thick, dashed] file {chapters/bridge/datas/g_of_r/chandler_ref/chandler_6.csv};
              \addplot+[mark=none, red, very thick, solid] table[x index=0,y index=4] {chapters/bridge/datas/g_of_r/data_HS_HNC_bridge_1.021_0/g_HS_HNC_bridge_2_4_6_8_10.csv};
              \addplot+[mark=none, blue, very thick, dashed] file {chapters/bridge/datas/g_of_r/chandler_ref/chandler_8.csv};
              \addplot+[mark=none, red, very thick, solid] table[x index=0,y index=5] {chapters/bridge/datas/g_of_r/data_HS_HNC_bridge_1.021_0/g_HS_HNC_bridge_2_4_6_8_10.csv};
              \addplot+[mark=none, blue, very thick, dashed] file {chapters/bridge/datas/g_of_r/chandler_ref/chandler_10.csv};
              \legend{MDFT(HNC+B), MC}
          \end{axis}
      \end{tikzpicture}
}
  \end{subfigure}
\begin{subfigure}{.24\textwidth}
  \centering
    \caption{B=-5, $\sigma_{gauss}$=1,061}
    \resizebox{\linewidth}{!}{
      \begin{tikzpicture}
          \begin{axis}[
              xlabel= r ($\text{\AA}$),
              ylabel= g(r),
              xmin = 0,
              xmax = 16,
              ymin = 0,
              ymax = 8,
              legend style = {draw = none, cells={anchor=west}}
              ]
              \addplot+[mark=none, red, very thick, solid] table[x index=0,y index=1] {chapters/bridge/datas/g_of_r/data_HS_HNC_bridge_1.061_-5/g_HS_HNC_bridge_2_4_6_8_10.csv};
              \addplot+[mark=none, blue, very thick, dashed] file {chapters/bridge/datas/g_of_r/chandler_ref/chandler_2.csv};
              \addplot+[mark=none, red, very thick, solid] table[x index=0,y index=2] {chapters/bridge/datas/g_of_r/data_HS_HNC_bridge_1.061_-5/g_HS_HNC_bridge_2_4_6_8_10.csv};
              \addplot+[mark=none, blue, very thick, dashed] file {chapters/bridge/datas/g_of_r/chandler_ref/chandler_4.csv};
              \addplot+[mark=none, red, very thick, solid] table[x index=0,y index=3] {chapters/bridge/datas/g_of_r/data_HS_HNC_bridge_1.061_-5/g_HS_HNC_bridge_2_4_6_8_10.csv};
              \addplot+[mark=none, blue, very thick, dashed] file {chapters/bridge/datas/g_of_r/chandler_ref/chandler_6.csv};
              \addplot+[mark=none, red, very thick, solid] table[x index=0,y index=4] {chapters/bridge/datas/g_of_r/data_HS_HNC_bridge_1.061_-5/g_HS_HNC_bridge_2_4_6_8_10.csv};
              \addplot+[mark=none, blue, very thick, dashed] file {chapters/bridge/datas/g_of_r/chandler_ref/chandler_8.csv};
              \addplot+[mark=none, red, very thick, solid] table[x index=0,y index=5] {chapters/bridge/datas/g_of_r/data_HS_HNC_bridge_1.061_-5/g_HS_HNC_bridge_2_4_6_8_10.csv};
              \addplot+[mark=none, blue, very thick, dashed] file {chapters/bridge/datas/g_of_r/chandler_ref/chandler_10.csv};
              \legend{MDFT(HNC+B), MC}
          \end{axis}
      \end{tikzpicture}
}
  \end{subfigure}
\begin{subfigure}{.24\textwidth}
  \centering
    \caption{B=-10, $\sigma_{gauss}$=1,110}
    \resizebox{\linewidth}{!}{
      \begin{tikzpicture}
          \begin{axis}[
              xlabel= r ($\text{\AA}$),
              ylabel= g(r),
              xmin = 0,
              xmax = 16,
              ymin = 0,
              ymax = 8,
              legend style = {draw = none, cells={anchor=west}}
              ]
              \addplot+[mark=none, red, very thick, solid] table[x index=0,y index=1] {chapters/bridge/datas/g_of_r/data_HS_HNC_bridge_1.110_-10/g_HS_HNC_bridge_2_4_6_8_10.csv};
              \addplot+[mark=none, blue, very thick, dashed] file {chapters/bridge/datas/g_of_r/chandler_ref/chandler_2.csv};
              \addplot+[mark=none, red, very thick, solid] table[x index=0,y index=2] {chapters/bridge/datas/g_of_r/data_HS_HNC_bridge_1.110_-10/g_HS_HNC_bridge_2_4_6_8_10.csv};
              \addplot+[mark=none, blue, very thick, dashed] file {chapters/bridge/datas/g_of_r/chandler_ref/chandler_4.csv};
              \addplot+[mark=none, red, very thick, solid] table[x index=0,y index=3] {chapters/bridge/datas/g_of_r/data_HS_HNC_bridge_1.110_-10/g_HS_HNC_bridge_2_4_6_8_10.csv};
              \addplot+[mark=none, blue, very thick, dashed] file {chapters/bridge/datas/g_of_r/chandler_ref/chandler_6.csv};
              \addplot+[mark=none, red, very thick, solid] table[x index=0,y index=4] {chapters/bridge/datas/g_of_r/data_HS_HNC_bridge_1.110_-10/g_HS_HNC_bridge_2_4_6_8_10.csv};
              \addplot+[mark=none, blue, very thick, dashed] file {chapters/bridge/datas/g_of_r/chandler_ref/chandler_8.csv};
              \addplot+[mark=none, red, very thick, solid] table[x index=0,y index=5] {chapters/bridge/datas/g_of_r/data_HS_HNC_bridge_1.110_-10/g_HS_HNC_bridge_2_4_6_8_10.csv};
              \addplot+[mark=none, blue, very thick, dashed] file {chapters/bridge/datas/g_of_r/chandler_ref/chandler_10.csv};
              \legend{MDFT(HNC+B), MC}
          \end{axis}
      \end{tikzpicture}
}
  \end{subfigure}
\begin{subfigure}{.24\textwidth}
  \centering
    \caption{B=-15, $\sigma_{gauss}$=1,177}
    \resizebox{\linewidth}{!}{
      \begin{tikzpicture}
          \begin{axis}[
              xlabel= r ($\text{\AA}$),
              ylabel= g(r),
              xmin = 0,
              xmax = 16,
              ymin = 0,
              ymax = 8,
              legend style = {draw = none, cells={anchor=west}}
              ]
              \addplot+[mark=none, red, very thick, solid] table[x index=0,y index=1] {chapters/bridge/datas/g_of_r/data_HS_HNC_bridge_1.177_-15/g_HS_HNC_bridge_2_4_6_8_10.csv};
              \addplot+[mark=none, blue, very thick, dashed] file {chapters/bridge/datas/g_of_r/chandler_ref/chandler_2.csv};
              \addplot+[mark=none, red, very thick, solid] table[x index=0,y index=2] {chapters/bridge/datas/g_of_r/data_HS_HNC_bridge_1.177_-15/g_HS_HNC_bridge_2_4_6_8_10.csv};
              \addplot+[mark=none, blue, very thick, dashed] file {chapters/bridge/datas/g_of_r/chandler_ref/chandler_4.csv};
              \addplot+[mark=none, red, very thick, solid] table[x index=0,y index=3] {chapters/bridge/datas/g_of_r/data_HS_HNC_bridge_1.177_-15/g_HS_HNC_bridge_2_4_6_8_10.csv};
              \addplot+[mark=none, blue, very thick, dashed] file {chapters/bridge/datas/g_of_r/chandler_ref/chandler_6.csv};
              \addplot+[mark=none, red, very thick, solid] table[x index=0,y index=4] {chapters/bridge/datas/g_of_r/data_HS_HNC_bridge_1.177_-15/g_HS_HNC_bridge_2_4_6_8_10.csv};
              \addplot+[mark=none, blue, very thick, dashed] file {chapters/bridge/datas/g_of_r/chandler_ref/chandler_8.csv};
              \addplot+[mark=none, red, very thick, solid] table[x index=0,y index=5] {chapters/bridge/datas/g_of_r/data_HS_HNC_bridge_1.177_-15/g_HS_HNC_bridge_2_4_6_8_10.csv};
              \addplot+[mark=none, blue, very thick, dashed] file {chapters/bridge/datas/g_of_r/chandler_ref/chandler_10.csv};
              \legend{MDFT(HNC+B), MC}
          \end{axis}
      \end{tikzpicture}
}
  \end{subfigure}
    \caption[Fonctions de distribution radiale autour de petites sphères dures.]{Fonctions de distribution radiale autour de sphères dures de rayons 2,0, 4,0, 6,0, 8,0 et 10,0 $\text{\AA}$. Les rdfs ont été calculés avec MDFT dans l'approximation HNC puis pour chaque couple de paramètres $B$ ($.10^{-8} \mathrm{kJ.mol}^{-1}.\text{\AA}^{15}$) et $\sigma_{gauss}$ (\AA) de notre bridge. Les références ont été calculées par Monte-Carlo\cite{huang_hydrophobic_2002} (pointillés bleu).}
    \label{fig:g_r_HS}
\end{figure}
\end{sidewaysfigure}
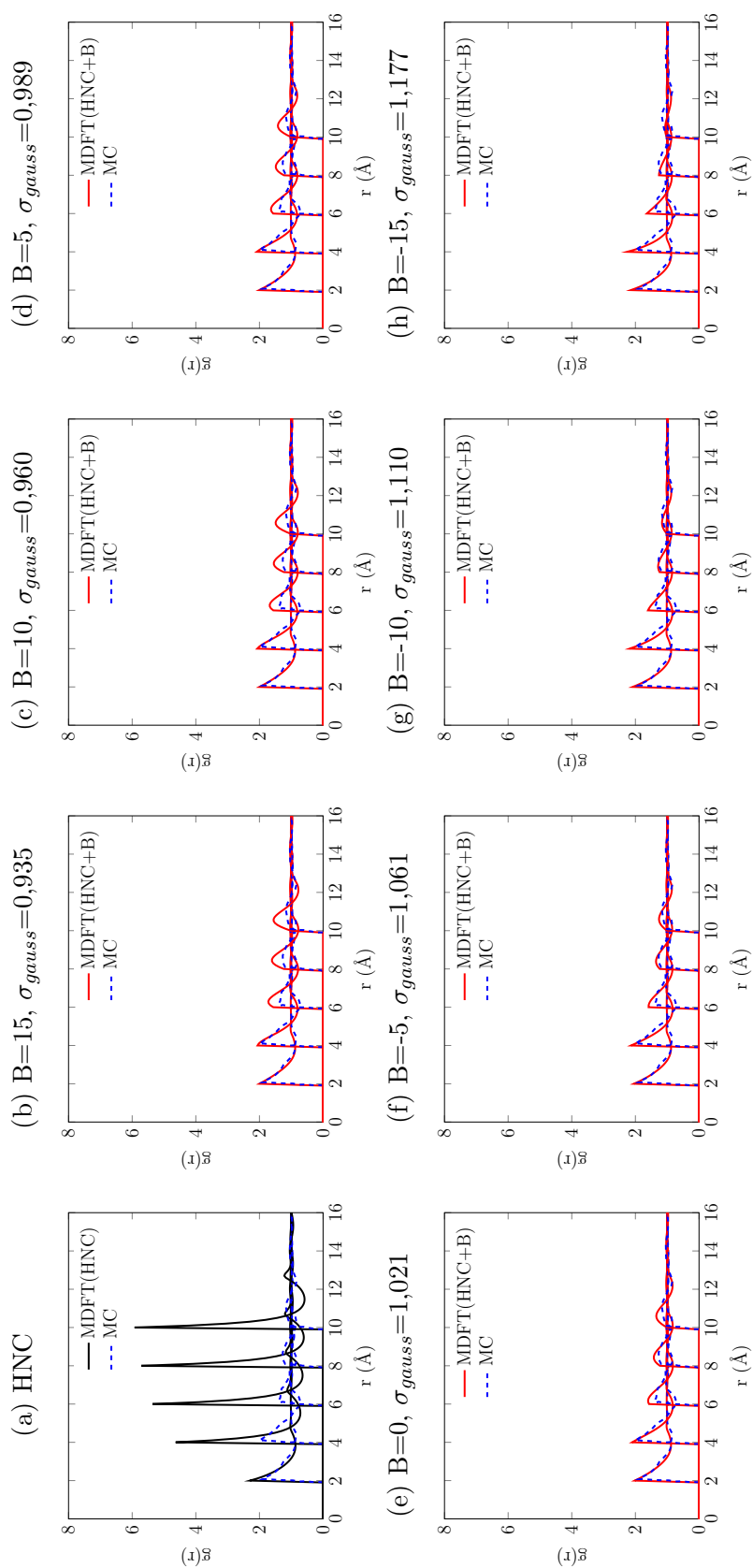

Nous nous sommes également intéressés aux structures de solvant, dans un premier temps, autour de sphères dures (voir figure \ref{fig:g_r_HS}). Les profils calculés pour chacun des paramètres ont été comparés à des profils de référence calculés par Monte Carlo et publiés par Chandler et al.\cite{huang_hydrophobic_2002}. \'Etant donné les temps de calcul nécessaires à la production de ces références, ils se sont limités à des sphères dures de 2, 4, 6, 8 et 10 \AA.

Les références, calculées par Monte Carlo, pour des sphères de diamètre croissant, montrent un premier pic qui augmente avant de diminuer. Cette diminution traduit la présence de démouillage. Dans l'approximation HNC, les premiers pics sont tous fortement surestimés. Ils augmentent jusqu'à tendre vers une valeur seuil. Le bridge corrige ce défaut en autorisant le démouillage dans le système et ce, quel que soit le couple de paramètres choisi. 

\boitesimple{Les meilleurs profils sont obtenus pour les valeurs $B$ = $-5.10^{-8}\ \mathrm{kJ.mol}^{-1}.\text{\AA}^{15}$  et $\sigma_{gauss}$ = 1,061 \AA. Les autres paramètres fournissent des résultats légèrement plus éloignés de la référence mais restent cependant acceptables.}

\subsubsection{Structure du solvant autour de molécules modèles}
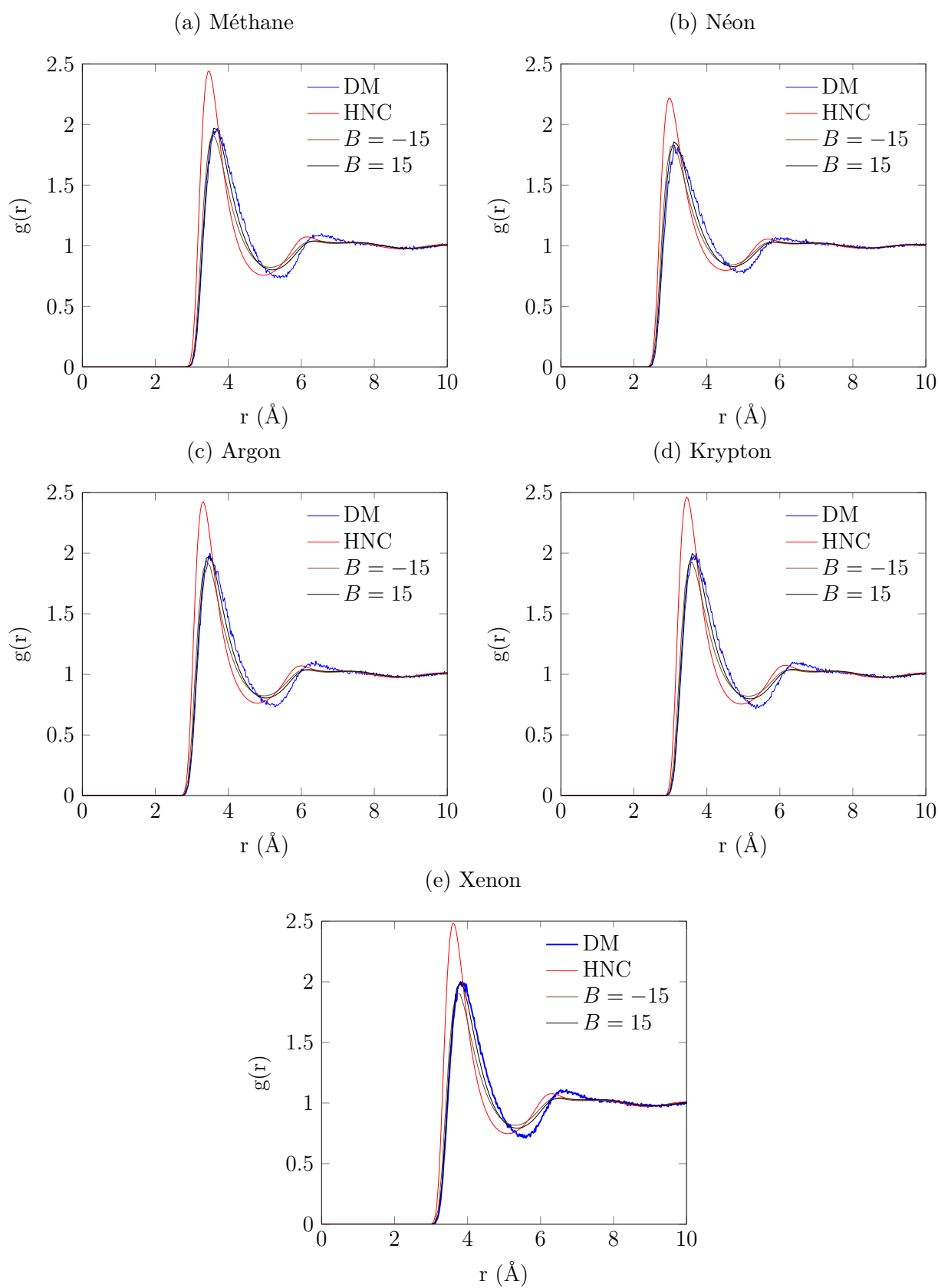
\begin{figure}[ht]
  \centering
  \begin{subfigure}{.5\textwidth}
    \caption{Méthane}
    \resizebox{\linewidth}{!}{
     \begin{tikzpicture}
        \begin{axis}[
            xlabel= r ($\text{\AA}$), ylabel= g(r),
            xmin = 0, xmax = 10,
            ymin = 0, ymax = 2.5,
            legend style = {draw = none, cells={anchor=west}}
            ]
            \addplot+[mark=none, smooth, tension=1] table [x expr={10*\thisrowno{0}}, y index=1] {chapters/bridge/datas/g_of_r/data_MD/g_methane.csv};
            \addplot+[mark=none] file {chapters/bridge/datas/g_of_r/data_HNC/g_methane.csv};
            \addplot+[mark=none] file {chapters/bridge/datas/g_of_r/data_bridge_1.177_-15/g_methane.csv};
            \addplot+[mark=none] file {chapters/bridge/datas/g_of_r/data_bridge_0.935_15/g_methane.csv};
            \legend{DM, HNC, $B=-15$, $B=15$}
        \end{axis}
    \end{tikzpicture}
}
  \end{subfigure}%
  \begin{subfigure}{.5\textwidth}
    \caption{Néon}
    \resizebox{\linewidth}{!}{
     \begin{tikzpicture}
        \begin{axis}[
            xlabel= r ($\text{\AA}$), ylabel= g(r),
            xmin = 0, xmax = 10,
            ymin = 0, ymax = 2.5,
            legend style = {draw = none, cells={anchor=west}}
            ]
            \addplot+[mark=none, smooth, tension=1] table [x expr={10*\thisrowno{0}}, y index=1] {chapters/bridge/datas/g_of_r/data_MD/g_neon.csv};
            \addplot+[mark=none] file {chapters/bridge/datas/g_of_r/data_HNC/g_neon.csv};
            \addplot+[mark=none] file {chapters/bridge/datas/g_of_r/data_bridge_1.177_-15/g_neon.csv};
            \addplot+[mark=none] file {chapters/bridge/datas/g_of_r/data_bridge_0.935_15/g_neon.csv};
            \legend{DM, HNC, $B=-15$, $B=15$}
        \end{axis}
    \end{tikzpicture}
}
  \end{subfigure}
  \begin{subfigure}{.5\textwidth}
    \caption{Argon}
    \resizebox{\linewidth}{!}{
     \begin{tikzpicture}
        \begin{axis}[
            xlabel= r ($\text{\AA}$), ylabel= g(r),
            xmin = 0, xmax = 10,
            ymin = 0, ymax = 2.5,
            legend style = {draw = none, cells={anchor=west}}
            ]
            \addplot+[mark=none, smooth, tension=1] table [x expr={10*\thisrowno{0}}, y index=1] {chapters/bridge/datas/g_of_r/data_MD/g_argon.csv};
            \addplot+[mark=none] file {chapters/bridge/datas/g_of_r/data_HNC/g_argon.csv};
            \addplot+[mark=none] file {chapters/bridge/datas/g_of_r/data_bridge_1.177_-15/g_argon.csv};
            \addplot+[mark=none] file {chapters/bridge/datas/g_of_r/data_bridge_0.935_15/g_argon.csv};
            \legend{DM, HNC, $B=-15$, $B=15$}
        \end{axis}
    \end{tikzpicture}
}
  \end{subfigure}%
  \begin{subfigure}{.5\textwidth}
    \caption{Krypton}
    \resizebox{\linewidth}{!}{
     \begin{tikzpicture}
        \begin{axis}[
            xlabel= r ($\text{\AA}$), ylabel= g(r),
            xmin = 0, xmax = 10,
            ymin = 0, ymax = 2.5,
            legend style = {draw = none, cells={anchor=west}}
            ]
            \addplot+[mark=none, smooth, tension=1] table [x expr={10*\thisrowno{0}}, y index=1] {chapters/bridge/datas/g_of_r/data_MD/g_krypton.csv};
            \addplot+[mark=none] file {chapters/bridge/datas/g_of_r/data_HNC/g_krypton.csv};
            \addplot+[mark=none] file {chapters/bridge/datas/g_of_r/data_bridge_1.177_-15/g_krypton.csv};
            \addplot+[mark=none] file {chapters/bridge/datas/g_of_r/data_bridge_0.935_15/g_krypton.csv};
            \legend{DM, HNC, $B=-15$, $B=15$}
        \end{axis}
    \end{tikzpicture}
}
  \end{subfigure}
  \begin{subfigure}{.5\textwidth}
    \caption{Xenon}
    \resizebox{\linewidth}{!}{
     \begin{tikzpicture}
        \begin{axis}[
            xlabel= r ($\text{\AA}$), ylabel= g(r),
            xmin = 0, xmax = 10,
            ymin = 0, ymax = 2.5,
            legend style = {draw = none, cells={anchor=west}}
            ]
            \addplot+[mark=none,, smooth, tension=1, thick] table [x expr={10*\thisrowno{0}}, y index=1] {chapters/bridge/datas/g_of_r/data_MD/g_xenon.csv};
            \addplot+[mark=none] file {chapters/bridge/datas/g_of_r/data_HNC/g_xenon.csv};
            \addplot+[mark=none] file {chapters/bridge/datas/g_of_r/data_bridge_1.177_-15/g_xenon.csv};
            \addplot+[mark=none] file {chapters/bridge/datas/g_of_r/data_bridge_0.935_15/g_xenon.csv};
            \legend{DM, HNC, $B=-15$, $B=15$}
        \end{axis}
    \end{tikzpicture}
}
  \end{subfigure}
  \caption[Fonctions de distribution radiale autour du méthane et des gaz rares.]{Fonctions de distribution radiale autour du méthane et des gaz rares. Les rdfs ont été calculées avec MDFT dans l'approximation HNC (en noir) et les deux paramètres extrêmes disponibles pour notre bridge (en marron et noir). Les références (en pointillées bleu) ont été calculées par dynamique moléculaire.}
  \label{fig:g_of_r_molecules_modeles}
\end{figure}

La dernier résultat observé est le profil du solvant autour de nos molécules modèles (voir figure \ref{fig:g_of_r_molecules_modeles}): le méthane unifié et les gaz rares: Néon, Argon, Krypton, Xénon. Les références ont été produites par Dynamique moléculaire. Afin de ne pas rendre ces images illisibles, nous avons choisi de ne mettre que les paramètres extrêmes, soit B=$-15.10^{-8} \mathrm{kJ.mol}^{-1}.\text{\AA}^{15}$ et B=$15.10^{-8} \mathrm{kJ.mol}^{-1}.\text{\AA}^{15}$ et leurs $\sigma_{gauss}$ correspondantes.

Dans l'approximation HNC, la MDFT surestime fortement le premier pic de solvatation et décale le premier creux. Pour chacune de nos molécules modèles, notre bridge, quelque soit le couple de paramètres choisi, corrige fortement les profils de solvatation de nos molécules modèles. Les paramètres $B$ = $15.10^{-8} \mathrm{kJ.mol}^{-1}.\text{\AA}^{15}$ et $\sigma_{gauss}$ = 0,935 \AA\ sont les plus efficaces, quelque soit la molécule étudiée.

\boitesimple{Les meilleurs profils de solvatation de nos molécules modèles sont obtenus pour les valeurs $B$ = $15.10^{-8} \mathrm{kJ.mol}^{-1}.\text{\AA}^{15}$ et $\sigma_{gauss}$ = 0,935 \AA.}

\subsection{Conclusion}
Afin de choisir les meilleurs paramètres de ce modèle, nous avons comparé les résultats pour chaque couple B, $\sigma_{gauss}$, à des références d'énergie libre de solvatation et de profils de solvant d'une part sur des sphères dures, indispensables au développement de ce bridge, et d'autre part sur des molécules modèles plus réalistes: le méthane, et les gaz rares.
On voit que tous ces paramètres améliorent considérablement l'ensemble des résultats par rapport à l'approximation HNC. Sur les sphères dures, les paramètres n'influencent que peu les résultats. Des écarts plus importants apparaissent lors de l'étude de nos molécules modèles. Sur ces molécules, les paramètres $B$ = $15.10^{-8} \mathrm{kJ.mol}^{-1}.\text{\AA}^{15}$ et $\sigma_{gauss}$ = 0,935 \AA\ apparaissent comme le meilleur compromis. Le bridge retenu est donc:

\boitesimplelarge{
\begin{equation} \label{eq:fbridge_complete}
F_{\mathrm{b}}[\bar{\rho}(\boldsymbol{r})]=\mathrm{k_B}T(\frac{1}{\rho_0^2} - \frac{\int c(\boldsymbol{r}) d\boldsymbol{r}}{2\rho_0})\int\Delta\bar{\rho}(\boldsymbol{r})^3d\boldsymbol{r}+15.10^{-8}\int\bar{\rho}(\boldsymbol{r})^2\Delta\bar{\rho}(\boldsymbol{r})^4d\boldsymbol{r}
\end{equation} 
}

Le gradient de cette fonctionnelle de bridge est disponible en annexe \ref{sec:annexes:grad:bridge}.

\section{Implémentation 3D}
Une fois le bridge développé, nous l'avons implémenté dans la version 3D de MDFT. Nous avons, dans un premier temps, comparé les résultats obtenus sur nos molécules modèles avant de nous intéresser à des molécules d’intérêt biologique: des protéines.

\begin{table}[ht]
  \centering
  \begin{tabular}{l c c c c c c c}
    \hline & \\[-1em]\hline
    compound   & exp  & DM & MDFT & MDFT & MDFT  & MDFT \\
               &      &    & (HNC)  & (HNC+PC)  & (HNC+B) & (HNC+B3D) \\
    \hline
    Méthane    &       &  9,23 & 28,14 & 7,73 & 11,99 & 12,25 \\
    Néon       & 10,36 & 11,73 & 19,04 & 7,59 & 13,10 & 11,26 \\
    Argon      &  8,40 &  8,61 & 22,89 & 7,26 & 11,01 & 10,92 \\
    Krypton    &  6,96 &  8,03 & 26,41 & 7,51 & 10,74 & 10,99 \\
    Xenon      &  6,06 &  6,47 & 31,23 & 9,81 & 10,33 & 11,26 \\
    \hline & \\[-1em]\hline
  \end{tabular}
  \caption[\'Energie libre de solvatation du méthane et des gaz rares.]{\'Energie libre de solvatation du méthane et des gaz rares en $\mathrm{kJ}.\mathrm{mol}^{-1}$ avec MDFT dans l'approximation HNC et avec le bridge. Les valeurs sont comparées à nos références calculées par Dynamique moléculaire et aux valeurs expérimentales\cite{straatsma_free_1986}.}
  \label{tab:deltag_1D_3D}  
\end{table}

Comme nous le voyons dans le tableau \ref{tab:deltag_1D_3D}, les énergies libres de solvatation produites par la version à symétrie sphérique et la version 3D de MDFT sont proches. Il n'est cependant pas étonnant d'observer de légères différences entre les résultats fournis par la version 2D et la version à symétrie sphérique car ces versions diffèrent sur plusieurs points. 

\begin{itemize}
\item La version en 3D est périodique, contrairement à la version à symétrie sphérique
\item Contrairement à la verion 3D, la version à symétrie sphérique ignore l'orientation du solvant
\end{itemize}

De plus, par définition, la version à symétrie sphérique permet une finesse de grille bien plus importante que la version 3D. Les calculs ont été effectués avec un écart de 0,5 \AA\ entre deux points de grille dans la version 3D contre un écart de seulement 0,1 \AA\ dans la version à symétrie sphérique. Ces différences peuvent entraîner des différences dans le calcul de la densité gros grain car la largeur de la gaussienne à mi hauteur utilisée comme noyau de convolution est proche de l'écart entre deux points en 3D.

\begin{figure}[ht]
\centering
 \begin{subfigure}{.5\textwidth}
    \caption{Méthane}
    \resizebox{\linewidth}{!}{
     \begin{tikzpicture}
        \begin{axis}[
            xlabel= r ($\text{\AA}$), ylabel= g(r),
            xmin = 0, xmax = 10,
            ymin = 0, ymax = 2.5,
            legend style = {draw = none, cells={anchor=west}}
            ]
            \addplot+[mark=none, smooth, tension=1, thick] table [x expr={10*\thisrowno{0}}, y index=1] {chapters/bridge/datas/g_of_r/data_MD/g_methane.csv};
            \addplot+[mark=none, thick] file {chapters/bridge/datas/g_of_r/data_HNC/g_methane.csv};
            \addplot+[mark=none, thick] file {chapters/bridge/datas/g_of_r/data_bridge_0.935_15/g_methane.csv};
            \addplot+[mark=none, thick] file {chapters/bridge/datas/g_of_r/data_HNC_bridge_3D/g_methane.csv};
            \legend{DM, HNC, HNC+B, HNC+B (3D)}
        \end{axis}
    \end{tikzpicture}
}
  \end{subfigure}%
 \begin{subfigure}{.5\textwidth}
    \caption{Néon}
    \resizebox{\linewidth}{!}{
     \begin{tikzpicture}
        \begin{axis}[
            xlabel= r ($\text{\AA}$), ylabel= g(r),
            xmin = 0, xmax = 10,
            ymin = 0, ymax = 2.5,
            legend style = {draw = none, cells={anchor=west}}
            ]
            \addplot+[mark=none, smooth, tension=1, thick] table [x expr={10*\thisrowno{0}}, y index=1] {chapters/bridge/datas/g_of_r/data_MD/g_neon.csv};
            \addplot+[mark=none, thick] file {chapters/bridge/datas/g_of_r/data_HNC/g_neon.csv};
            \addplot+[mark=none, thick] file {chapters/bridge/datas/g_of_r/data_bridge_0.935_15/g_neon.csv};
            \addplot+[mark=none, thick] file {chapters/bridge/datas/g_of_r/data_HNC_bridge_3D/g_neon.csv};
            \legend{DM, HNC, HNC+B, HNC+B (3D)}
        \end{axis}
    \end{tikzpicture}
}
  \end{subfigure}
 \begin{subfigure}{.5\textwidth}
    \caption{Argon}
    \resizebox{\linewidth}{!}{
     \begin{tikzpicture}
        \begin{axis}[
            xlabel= r ($\text{\AA}$), ylabel= g(r),
            xmin = 0, xmax = 10,
            ymin = 0, ymax = 2.5,
            legend style = {draw = none, cells={anchor=west}}
            ]
            \addplot+[mark=none, smooth, tension=1, thick] table [x expr={10*\thisrowno{0}}, y index=1] {chapters/bridge/datas/g_of_r/data_MD/g_argon.csv};
            \addplot+[mark=none, thick] file {chapters/bridge/datas/g_of_r/data_HNC/g_argon.csv};
            \addplot+[mark=none, thick] file {chapters/bridge/datas/g_of_r/data_bridge_0.935_15/g_argon.csv};
            \addplot+[mark=none, thick] file {chapters/bridge/datas/g_of_r/data_HNC_bridge_3D/g_argon.csv};
            \legend{DM, HNC, HNC+B, HNC+B (3D)}
        \end{axis}
    \end{tikzpicture}
}
  \end{subfigure}%
 \begin{subfigure}{.5\textwidth}
    \caption{Krypton}
    \resizebox{\linewidth}{!}{
     \begin{tikzpicture}
        \begin{axis}[
            xlabel= r ($\text{\AA}$), ylabel= g(r),
            xmin = 0, xmax = 10,
            ymin = 0, ymax = 2.5,
            legend style = {draw = none, cells={anchor=west}}
            ]
            \addplot+[mark=none, smooth, tension=1, thick] table [x expr={10*\thisrowno{0}}, y index=1] {chapters/bridge/datas/g_of_r/data_MD/g_krypton.csv};
            \addplot+[mark=none, thick] file {chapters/bridge/datas/g_of_r/data_HNC/g_krypton.csv};
            \addplot+[mark=none, thick] file {chapters/bridge/datas/g_of_r/data_bridge_0.935_15/g_krypton.csv};
            \addplot+[mark=none, thick] file {chapters/bridge/datas/g_of_r/data_HNC_bridge_3D/g_krypton.csv};
            \legend{DM, HNC, HNC+B, HNC+B (3D)}
        \end{axis}
    \end{tikzpicture}
}
  \end{subfigure}
 \begin{subfigure}{.5\textwidth}
    \caption{Xenon}
    \resizebox{\linewidth}{!}{
     \begin{tikzpicture}
        \begin{axis}[
            xlabel= r ($\text{\AA}$), ylabel= g(r),
            xmin = 0, xmax = 10,
            ymin = 0, ymax = 2.5,
            legend style = {draw = none, cells={anchor=west}}
            ]
            \addplot+[mark=none, smooth, tension=1, thick] table [x expr={10*\thisrowno{0}}, y index=1] {chapters/bridge/datas/g_of_r/data_MD/g_xenon.csv};
            \addplot+[mark=none, thick] file {chapters/bridge/datas/g_of_r/data_HNC/g_xenon.csv};
            \addplot+[mark=none, thick] file {chapters/bridge/datas/g_of_r/data_bridge_0.935_15/g_xenon.csv};
            \addplot+[mark=none, thick] file {chapters/bridge/datas/g_of_r/data_HNC_bridge_3D/g_xenon.csv};
            \legend{DM, HNC, HNC+B, HNC+B (3D)}
        \end{axis}
    \end{tikzpicture}
}
  \end{subfigure}
  \caption[Fonctions de distribution radiale autour du méthane et des gaz rares en 3D.]{Fonctions de distribution radiale autour du méthane et des gaz rares. Les rdfs ont été calculés avec les versions de MDFT à symétrie sphérique et 3D, dans l'approximation HNC et avec le bridge gros grain. Les références ont été calculées par Dynamique moléculaire.}
    \label{fig:g_of_r_methane_3D}
\end{figure}
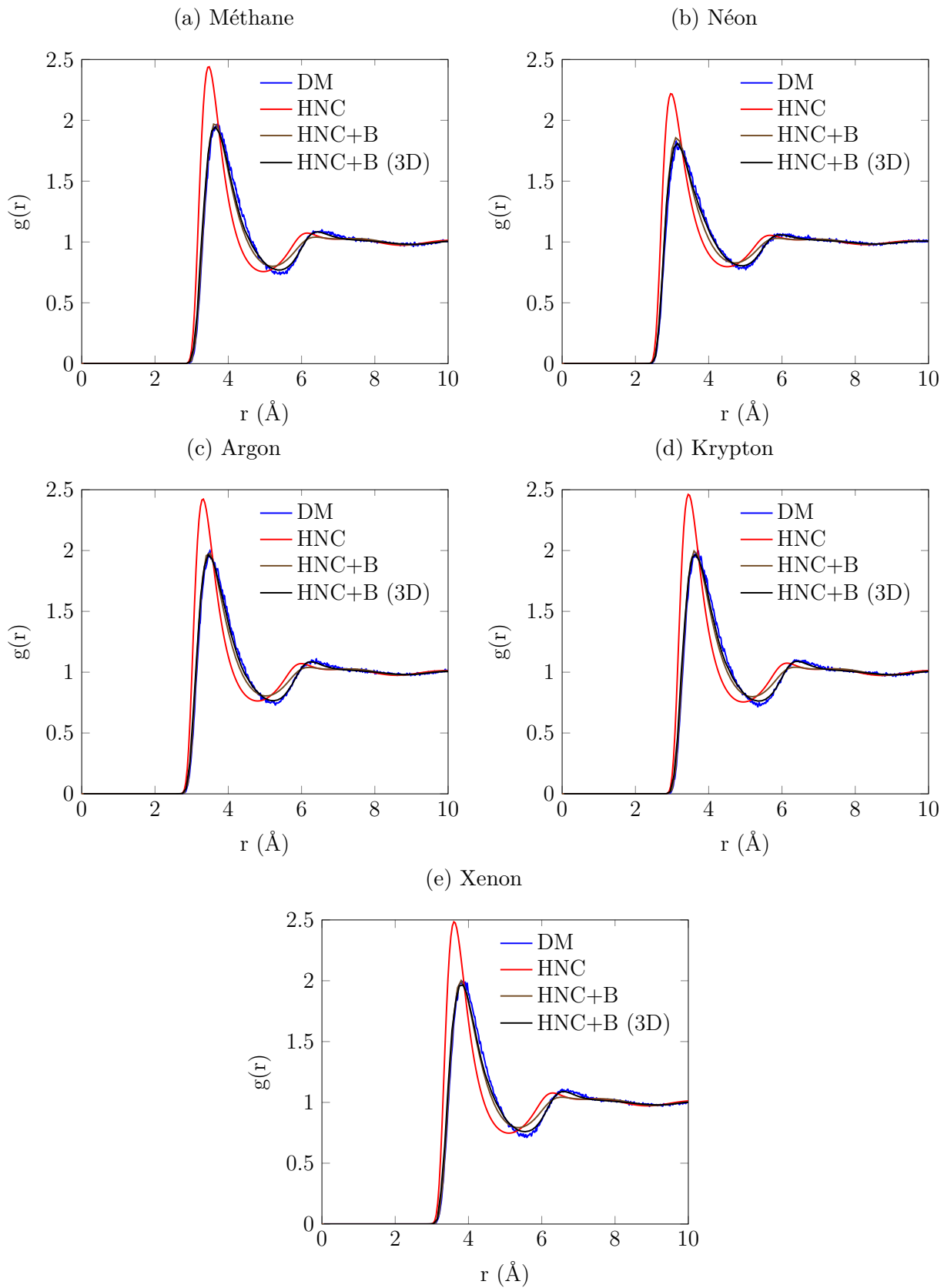

En ce qui concerne les profils de solvant, on voit sur la figure \ref{fig:g_of_r_methane_3D} que la version 3D est bien plus précise que la version à symétrie sphérique. Les profils fournis sont quasi-identiques à ceux de référence issus de dynamique moléculaire.


\subsubsection{Système biologique}

Enfin, nous avons étudié la prédiction de profils de solvatation autour de plus grosses molécules et en particulier autour d'une protéine. Cet exemple sera traité et analysé plus en détails dans le chapitre \ref{chap:applications}.

\clearpage
\strut
\vspace{10\baselineskip}

\boitemagique{A retenir}{
Dans ce chapitre nous décrivons le développement d'un nouveau bridge \textbf{simple} et \textbf{rapide}. Ce bridge, basé sur une densité gros grain ajoute de la consistance thermodynamique à nos modèles en reproduisant une pression et une tension de surface correcte. Il améliore également fortement le calcul de l'énergie libre de solvatation et la prédiction de structures de solvant, sur les petites molécules mais également sur des molécules plus grosses comme des systèmes biologiques. 
}

\part{Développements numériques}

\clearemptydoublepage
\chapter{Développements numériques}
\label{chap:numerique}

\boitemagique{Objectif}{Dans ce chapitre, nous décrivons les développements numériques ainsi que les optimisations nécessaires à l'application de MDFT sur des systèmes biologiques.}

Afin de porter MDFT vers des applications biologiques, certains développements numériques ont été nécessaires. En effet ces systèmes imposent, de par leur taille, de nouvelles contraintes (taille de la boîte de simulation, mémoire nécessaire, ...) et rendent certaines parties du calcul bloquantes alors qu'elles étaient jusque là négligeables. C'est par exemple le cas avec le calcul du potentiel extérieur. Dans ce chapitre nous décrivons dans un premier temps le processus JUBE que nous avons mis en place afin de suivre et évaluer les différentes évolutions. Nous faisons ensuite une revue non exhaustive des améliorations les plus importantes. Le développement haute performance (HPC) est un aspect important de cette thèse. Le choix a cependant été fait de ne pas expliciter chaque terme de ce chapitre. Ces aspects sont décrits plus en détails dans les rapports EoCoE de MDFT.

\section{Reproductibilité}
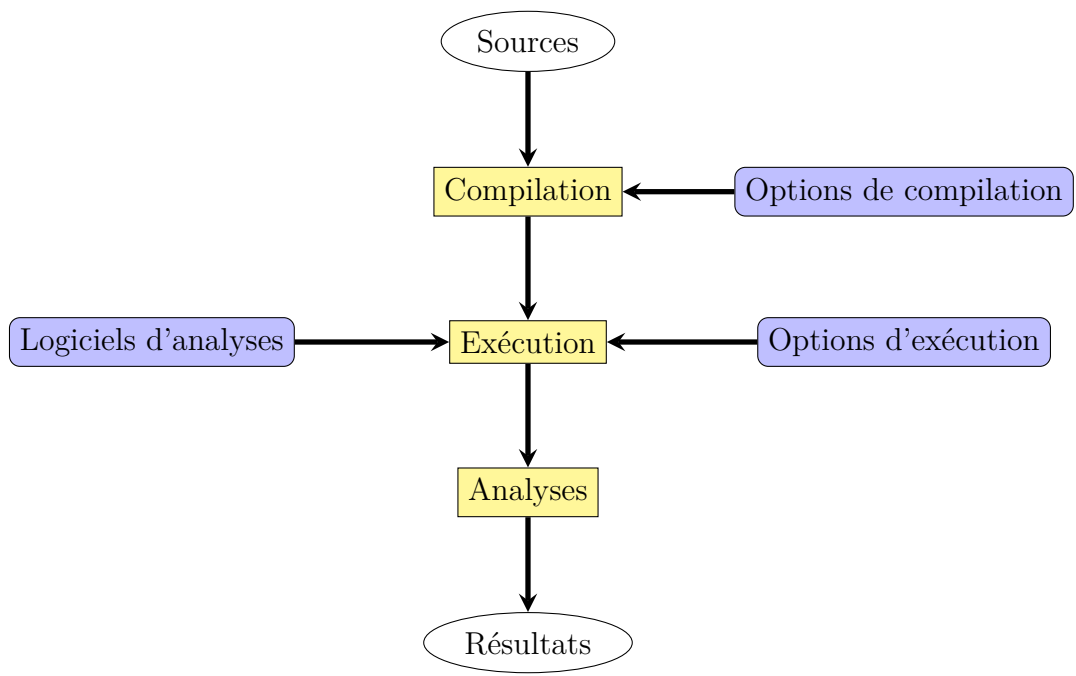
\begin{figure}[ht]
  \center
  \begin{tikzpicture}
  
    \tikzstyle{files}=[ellipse,draw,text=black]
    \tikzstyle{instruct}=[rectangle,draw,fill=yellow!50]
    \tikzstyle{test}=[diamond, aspect=2.5,thick,draw=blue,fill=yellow!50,text=blue]
    \tikzstyle{es}=[rectangle,draw,rounded corners=4pt,fill=blue!25]
    \tikzstyle{suite}=[->,>=stealth,line width=2pt,rounded corners=4pt]

    \node[files] (debut) at (0,0) {Sources};
    \node[es] (compilationOptions) at (5,-2) {Options de compilation};
    \node[instruct] (compilation) at (0,-2) {Compilation};
    \node[es] (executionOptions) at (5,-4) {Options d'exécution};
    \node[es] (softLink) at (-5,-4) {Logiciels d'analyses};
    \node[instruct] (execution) at (0,-4) {Exécution};
    \node[instruct] (analyse) at (0,-6) {Analyses};
    \node[files] (resultats) at (0,-8) {Résultats};
 
    \draw[suite] (debut) -- (compilation);
    \draw[suite] (compilationOptions) -- (compilation);
    \draw[suite] (compilation) -- (execution);
    \draw[suite] (executionOptions) -- (execution);
    \draw[suite] (softLink) -- (execution);
    \draw[suite] (execution) -- (analyse);
    \draw[suite] (analyse) -- (resultats);

  \end{tikzpicture}
  \caption{Processus d'exécution d'un logiciel de la récupération des sources à l'analyse des résultats.}
  \label{fig:JUBE_process}
\end{figure}

La reproductibilité est une problématique récurrente lors de développement, de la modification ou de l'optimisation de logiciels de calcul. En effet, la comparaison de deux mesures liées à l’exécution d'un logiciel n'est pas pertinente si nous n'avons pas la certitude que les deux exécutions ont eu lieu strictement dans les mêmes conditions: options de compilation, environnement, options d’exécution, etc (voir image \ref{fig:JUBE_process}).

Afin de contrôler cette chaîne d’exécution, et de pouvoir relancer le même calcul des semaines, des mois plus tard, nous avons mis en place l’outil de gestion de flux JUBE\cite{Luhrs_parallel_2016,Galonska_advances_2012}.

\subsection{JUBE}
JUBE est un logiciel écrit en python et développé au \textit{Jülich Supercomputing Centre} qui permet d'une part, d'automatiser toutes les étapes nécessaires au lancement de cas tests et à leurs analyses et, d'autre part, de conserver un historique des exécutions précédentes et des résultats obtenus. Il allie la robustesse nécessaire à la reproductibilité et la flexibilité permettant d'adapter les cas tests et les mesures aux évolutions de MDFT. Un ensemble d'options (entièrement automatisables) nous permet d'étudier différentes métriques de MDFT. En effet, après une modification, qu'il s'agisse d'un changement global d'algorithme ou de la plus minime des optimisations, nous souhaitons dans un premier temps nous assurer que les résultats scientifiques (\'energie libre de solvatation, ...) sont inchangés, puis nous souhaitons étudier l'évolution des comportements informatiques (temps d'exécution, quantité de mémoire utilisée, ...).

Ce script a plusieurs objectifs:

\subsubsection{Les cas tests}
Il doit permettre de lancer simplement un ou plusieurs des 3 cas tests suivants:

\begin{table}[ht]
  \begin{center}
    \begin{tabular}{c c c c c}
      \hline & \\[-1em]\hline
      Nom & solute & Nombre de points de grille & taille de la boîte (\AA) & $\mathrm{m}_\mathrm{max}$  \\
      \hline
      petit & Pyridine & 64 & 20 & 3  \\
      moyen & lysosyme & 128 & 25 & 3  \\
      grand & lysosyme & 256 & 20 & 5  \\
      \hline & \\[-1em]\hline%
    \end{tabular}
  \end{center}
  \caption{Récapitulatif des 3 cas tests accessibles via JUBE.}
  \label{tab:JUBE_bench_cases}  
\end{table}

\subsubsection{L'environnement}
JUBE permet également l’exécution de MDFT sur différentes machines en s'adaptant à leurs environnements spécifiques. En effet, contrairement à des ordinateurs personnels, les super-calculateurs disposent généralement d'un système de gestion de tâches comme SLURM ou encore d'un système de gestion de modules qui permettent de charger des logiciels ou librairies indispensables à la bonne compilation/exécution de MDFT (GFORTRAN, FFTW3, JUBE, ...).

\subsubsection{Les options de compilation}
Afin de ne pas être impacté par les calculs précédents, les sources sont récupérées et recopiées depuis le dépôt distant (github\footnote{\url{https://github.com/}}) et recompilées pour chaque nouveau calcul. Afin d'étudier l'évolution au cours des modifications, il est possible de spécifier la version via le numéro de commit et/ou la branche à utiliser.

Les options de compilation s'adaptent également aux mesures effectuées. Si l'on prend l'exemple du calcul du taux de vectorisation, deux calculs sont lancés. Le premier, avec l'option de compilation empêchant la vectorisation nous sert de référence, le second, sans cette option, permet l'évaluation du taux de vectorisation. Il en est de même pour les options d'exécution.

\subsubsection{Les options d'exécution}
Afin de ne pas avoir un nombre infini de cas, la majorité des options est gérée via les 3 cas tests décrits précédemment. Il existe cependant des options indépendantes du cas étudié comme le nombre de processeurs utilisés ou encore le nombre d’itérations du calcul de la fonctionnelle effectuées.

Il est également possible à cette étape de coupler MDFT aux logiciels d'analyses suivants:
\begin{itemize}
\item darshan
\item scorep
\item scalasca
\item papi
\item VTune
\item valgrind
\end{itemize}
Cette thèse n'étant pas orientée vers le HPC, nous n'entrerons pas dans le détail du rôle et de l’exécution de ces différents logiciels.

\subsubsection{Les grandeurs mesurées}
L'ensemble des métriques extraites grâce aux logiciels listés ci-dessus sont décrits ci-dessous:

\begin{itemize}
\item[$\bullet$] Les métriques globaux:
  \begin{itemize}
    \item \textbf{Temps d’exécution}: Le temps réel d'exécution est fourni par MDFT et exprimé en secondes. Il correspond au temps nécessaire à l'exécution de MDFT.
    \item $\mathbf{\Delta G_{\mathrm{solv}}}$ : L’énergie libre de solvatation prédite par MDFT et exprimée en $\mathrm{kJ.mol}^{-1}$.
    \item \textbf{Nombre d'itérations} : Correspond au nombre d'itérations nécessaires à MDFT pour minimiser le système étudié.
  \end{itemize}
  \vspace*{1.5ex}%

\item[$\bullet$] Les métriques OpenMP:
  \begin{itemize}
  \item \textbf{Répartition de charge} : La répartition de la charge, exprimée en \%. Elle permet d'évaluer le déséquilibre entre les différents threads OpenMP.
  \item \textbf{Temps OpenMP} : Exprimé en seconde, il correspond à la durée passée dans les parties du code parallélisées en OpenMP.
  \item \textbf{Ratio OpenMP} : Exprimé en pourcentage, le ratio OpenMP correspond au rapport entre le temps OpenMP et le temps total d'exécution. Un code séquentiel, a un ratio de 0, alors qu'un code entièrement parallélisé en OpenMP aura le ratio maximum 1. 
  \end{itemize}
  \vspace*{1.5ex}%

\item[$\bullet$] Les métriques liées à la mémoire:
  \begin{itemize}
  \item \textbf{Empreinte mémoire}: L'empreinte mémoire correspond à la quantité maximum de mémoire utilisée lors du calcul. Elle est exprimée en Go.
  \item \textbf{Intensité d'utilisation du cache}: L'intensité d'utilisation du cache est exprimée en \% et correspond à la fraction des données directement disponibles en cache. Lors de la création ou de l'utilisation d'une variable, celle ci est copiée de la RAM vers le cache. Le cache est une mémoire restreinte, proche des processeurs, ce qui en rend l'accès très rapide. Lorsque le cache arrive à saturation, les variables les plus anciennes en sont supprimées et seront accessibles uniquement dans la RAM. Lors d'un accès à une variable, le processus vérifie dans un premier temps si elle est toujours dans le cache. C'est à ce taux de succès que correspond l'intensité d'utilisation du cache. Un ratio important correspond à un accès mémoire plus rapide et donc à un temps d'exécution plus faible.
  \end{itemize}
  \vspace*{1.5ex}%

\item[$\bullet$] Les métriques liées à l'utilisation des processeurs:
  \begin{itemize}
  \item \textbf{IPC}: L'IPC correspond au nombre d'instruction par cycle. Plus ce nombre est important et plus MDFT exploite la puissance fournie par le processeur.
  \item \textbf{Temps d'exécution sans vectorisation}: Temps d’exécution d'un calcul sans vectorisation exprimé en sec. La vectorisation est désactivée à l'aide de l'option -fno-tree-vectorize pour Gfortran ou des options -no-simd et -no-vec pour ICC. 
  \item \textbf{Efficacité de la vectorisation}: L'efficacité de la vectorisation est mesurée en calculant le ratio entre le temps d'exécution avec et sans vectorisation. Plus ce nombre est important et plus MDFT exploite la puissance fournie par le processeur.
  \end{itemize}
  \vspace*{1.5ex}%

\end{itemize}

Le script JUBE décrit ci-dessus, nous permet ainsi de suivre et de quantifier les évolutions de MDFT tout en nous assurant une constance dans les résultats fournis.

\section{Optimisations}
Lors de l'exécution de MDFT sur des systèmes biologiques, certaines parties dépendants du nombre d'atomes du soluté sont apparues limitantes alors que leurs temps d'exécution étaient négligeables jusque là. Le script JUBE précédemment décrit nous a permis d'identifier facilement ces parties comme par exemple le module qui calcule les forces Lennard-Jones ou encore le minimiseur. En collaboration avec la Maison de La Simulation (CEA), MDFT a également été parallélisé en OpenMP.

\subsection{Le module Lennard Jones}
Lors de l’initialisation du système, et en particulier du calcul du $\mathrm{V}_\mathrm{ext}$, un module calcul l’interaction Lennard-Jones entre chaque atome du soluté et chaque atome d'eau pour chaque point de grille et chaque orientation. Dans sa version naïve, le nombre de calculs effectués par ce module est de $N_{\mathrm{voxels}}\mathrm{x}N_{\mathrm{orientations}}\mathrm{x}N_{\mathrm{atomes\ du\ solute}}\mathrm{x}N_{\mathrm{atomes\ du\ solvant}}$. Les systèmes biologiques, composés de plusieurs milliers d'atomes, nécessitent une grande boîte de simulation. La quantité de calcul de ce module croît dont très rapidement. L'optimisation de ce calcul a eu lieu en deux étapes décrites ci-dessous.

\subsubsection{Ajout d'une distance limite}
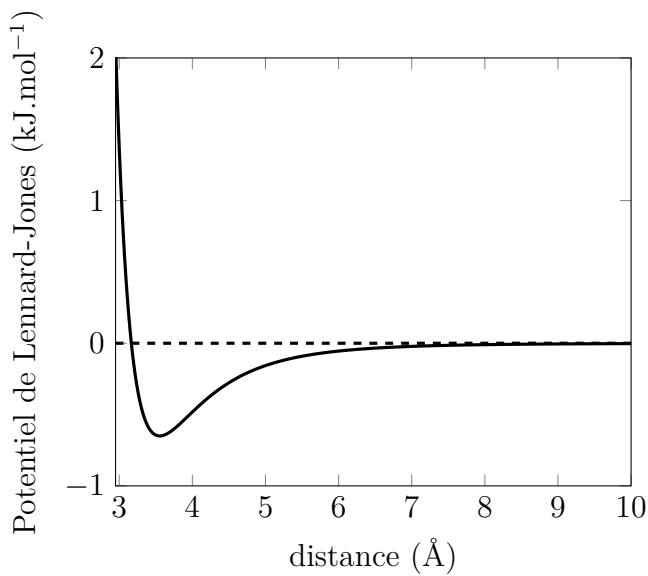
\begin{figure}[ht]
    \center    
  \begin{tikzpicture}
    \begin{axis}[
            xlabel= distance (\AA),
            ylabel= Potentiel de Lennard-Jones (kJ.mol$^{-1}$),
            xmin = 2.95, xmax = 10,
            ymin = -1, ymax = 2,
            no markers,
            legend style = {draw = none, cells={anchor=west}}
      ]
      \addplot[mark=none, black, very thick, domain=2.95:10, samples=400] {lj(0.65,3.166)};
      \addplot[mark=none, black, very thick, dashed] plot coordinates {
        (2.95,  0)
        (10,  0)
};
    \end{axis}
  \end{tikzpicture}
    \caption{Exemple du potentiel de Lennard-Jones entre deux atomes d'oxygènes de l'eau SPC/E.}
    \label{fig:lj}
\end{figure}

Comme on le voit sur la figure \ref{fig:lj}, le potentiel de Lennard-Jones tend rapidement vers 0. Il n'est donc pas nécessaire de calculer ce potentiel pour des molécules trop distantes. Nous avons donc ajouté une distance limite, configurable par l'utilisateur, au delà de laquelle le potentiel n'est plus calculé. Malheureusement, une distance limite simple, sphérique, ne permet qu'un gain limité car nous sommes toujours obligés de calculer la distance entre chaque atome du soluté et chaque atome d'eau pour chaque point de grille et chaque orientation pour ensuite la comparer à notre distance limite. Nous avons donc fait le choix de sous-espaces cubiques de tailles égales à la valeur de la limite. En échange de quelques calculs supplémentaires, dans les angles, il n'est plus nécessaire de tester chaque distance. En effet, une représentation cubique permet de limiter les boucles de calcul à cette zone.

\pgfmathdeclarefunction{lj}{2}{%
    \pgfmathparse{4*#1*((#2/x)^12-(#2/x)^6)}%
}

\subsubsection{Mise en cache}
Malgré la puissance actuelle des ordinateurs, le calcul de fonctions trigonométriques reste coûteux en temps. Dans une implémentation naïve, il est nécessaire de reconstruire la position de chaque atome du solvant, pour chaque atome du soluté, pour chaque point de grille et pour chaque orientation. Si l'on considère l'étude de l'une des plus petites protéines existantes, le lysosyme (1960 atomes), dans une boîte divisée en 64 points dans chaque direction et pour une valeur de $\mathrm{m}_\mathrm{max}$=1, il est nécessaire d'effectuer $64^3\mathrm{x}3\mathrm{x}18\mathrm{x}1960$ soit environ 27,7 milliards reconstructions de position. Nous avons donc stocké en mémoire l'ensemble des positions relatives de chaque atome de solvant par rapport au point de grille étudié pour chaque orientation. Pour le même système, le nombre de reconstructions est donc aujourd'hui de 3x18 soit seulement 54. Le nombre de calculs est ainsi divisé par plus de 500 millions.

\subsubsection{Performances}
Si l'on reprend l'exemple du lysosyme (1960 atomes), dans une boîte de 32 \AA\ de coté divisée en 64 points dans chaque direction et pour une valeur de $\mathrm{m}_\mathrm{max}$=1, avant optimisation le module Lennard-Jones était complété en 1 h 47 min. Grâce à l'ensemble de ces optimisations, le même calcul est aujourd'hui complété en moins de 6 sec. Ces optimisations ont permi de diviser le temps dédié à ce module par plus de 1000.

\subsection{Le minimiseur: steepest descent}
Comme nous l'avons décrit dans le chapitre \ref{chap:theorie}, le minimiseur nativement implémenté dans MDFT est L-BFGS. Ce minimiseur est le meilleur compromis pour des systèmes comportants de nombreuses variables comme c'est le cas avec MDFT. Cependant, le nombre de variables reste limité. En effet, il stocke en mémoire un tableau de taille $2mn + 5n + 11mm + 8m$ avec n le nombre de variables à minimiser et m le nombre de pas d'historique que l'on conserve. En fortran, les entiers sont codés sur 32 bits. Ils sont donc limités à un maximum de $2^{32}$ soit 2,29.10$^9$. Si l'on réduit l'historique au minimum, soit 1, la taille du tableau est de $7n+16$. Notre minimiseur est donc limité aux systèmes de moins de $\frac{2,29e9-16}{7}$ soit 6,14$^{8}$ variables. 

Dans notre cas, le nombre de variables est égal au nombre de points de grille multiplié par le nombre d'orientations dans chaque direction, soit $N_g^3 \mathrm{x} N_o$ avec $N_g$ le nombre de points de grille dans chaque dimension et $N_o$ le nombre d'orientations. La taille des boîtes de simulation est donc limitée à $\sqrt[3]{\frac{2^{32}-7}{7N_o}}$. La taille de boîte maximale autorisée en fonction de la valeur de $\mathrm{m}_\mathrm{max}$ est décrite dans le tableau \ref{tab:taille_boîte_max}.

\begin{table}[ht]
 \centering
  \begin{tabular}{l | c | c}
    \hline \multicolumn{3}{c}{} \\[-1em]\hline
    $\mathrm{m}_\mathrm{max}$ & Nombre d'orientations & Nombre de points de grilles autorisé \\
    \hline
    1  & 18 & 324 \\
    2  & 75 & 201 \\
    3  & 196 & 146 \\
    4  & 405 & 114 \\
    5  & 726 & 94 \\
    \hline \multicolumn{3}{c}{} \\[-1em]\hline
  \end{tabular}
  \caption{Taille de boîte maximum autorisée par L-BFGS en fonction du paramètre $\mathrm{m}_\mathrm{max}$.}
  \label{tab:taille_boîte_max}  
\end{table}

Afin de dépasser ces limites, nous avons donc implémenté un nouveau minimiseur dans MDFT: le steepest descent. Si l'on reprend l'exemple du lysosyme (1960 atomes), dans une boîte de 64 \AA\ de coté divisée en 128 points dans chaque direction, les performances obtenues en fonction de la valeur de $\mathrm{m}_\mathrm{max}$ avec les deux minimiseurs sont regroupées dans le tableau \ref{tab:perf_minimiseurs}.

\begin{table}[ht]
 \centering
  \begin{tabular}{c || c | c | c || c | c | c}
    \hline \multicolumn{3}{c}{} \\[-1em]\hline
         & \multicolumn{3}{c ||}{Mémoire utilisée (Go)} & \multicolumn{3}{c}{Temps de calcul} \\
    \hline
      $\mathrm{m}_\mathrm{max}$ & L-BFGS & SD & gain(\%) & L-BFGS & SD & gain(\%) \\
    \hline
    1  &  1,36 &  0,77 & 43,4 &  2 min 42 &  2 min 24 & 11,1 \\
    2  &  8,71 &  4,29 & 50,7 &  8 min 47 &  8 min 18 &  5,5 \\
    3  & 16,20 &  7,94 & 51,0 & 19 min 50 & 14 min 56 & 24,7 \\
    4  &   /   & 20,38 &  /   &    /      & 18 min 37 &   /  \\
    5  &   /   & 30,07 &  /   &    /      & 22 min 04 &   /  \\
    \hline \multicolumn{3}{c}{} \\[-1em]\hline
  \end{tabular}
  \caption[Comparaison des performances des minimiseurs L-BFGS et \textit{steepest descent}.]{Comparaison des performances des minimiseurs L-BFGS et \textit{steepest descent} dans le cas de la solvatation du lysosyme.}
  \label{tab:perf_minimiseurs}  
\end{table}

Comme on le voit dans ce tableau, en plus de rendre possibles les calculs les plus importants, le temps de calcul nécessaire au \textit{steepest descent} est légèrement inférieur à celui nécessaire à L-BFGS. Cela illuste l'importance de l'optimisation des accès mémoire. De plus, la mémoire utilisée par cette version de \textit{steepest descent} est moitié moins importante que celle utilisée par L-BFGS.

\subsection{La parallélisation OpenMP}
Afin d'améliorer les performances de MDFT et de bénéficier au maximum des architectures actuelles, les boucles les plus coûteuses en temps de calcul ont été parallélisées en OpenMP. Ce travail a été effectué en collaboration avec Yacine Ould-Rouis à la Maison de La Simulation. 

Les deux cadres d'utilisation de MDFT les plus coûteux sont:
\begin{itemize}
\item l'étude de bases de données complètes
\item l'étude d'une macro-molécule
\end{itemize}

Dans le premier cas, le nombre de calcul à lancer peut être largement supérieur au nombre de cœurs disponibles. La parallélisation n'a donc aucun avantage. En effet, à cause de la communication entre les différents processus, le temps de calcul nécessaire pour l’exécution d'un code parrallèle sur n cœurs est toujours supérieur au temps sur un cœur divisé par n. Dans ce cas de figure, il est donc plus intéressant de lancer chaque calcul en série soit sur un seul cœur.

Dans le second cas, c'est le temps de restitution qui est important. La parallélisation permet donc ici un gain de temps considérable. Les temps de calcul en fonction du nombre de cœurs OpenMP sont disponibles dans le tableau \ref{tab:perf_openmp}. Le système étudié est le lysosyme dans une boîte de simulation de 64 \AA\ de coté divisée en 128 points de grille dans chaque direction pour une valeur de $\mathrm{m}_\mathrm{max}$=3.

\begin{table}[ht]
 \centering
  \begin{tabular}{c | c | c}
      Nombre de cœurs & Temps de calcul & gain(\%)\\
    \hline
    sans OpenMP (ref) & 25 min & 0\\
     1 & 31 min 44 & -27,0 \\
     2 & 16 min 34 &  33,4 \\
     4 &  9 min 32 &  61,8 \\
     8 &  5 min 30 &  78,0 \\
    12 &  4 min 03 &  83,8 \\
    24 &  2 min 34 &  89,8 \\
    \hline \multicolumn{3}{c}{} \\[-1em]\hline
  \end{tabular}
  \caption[Temps de calcul en fonction du nombre de cœurs OpenMP.]{Temps de calcul en fonction du nombre de cœurs OpenMP. Le système étudié est le lysosyme dans une boîte de simulation de 64 \AA\ de coté divisée en 128 points de grille dans chaque direction pour une valeur de $\mathrm{m}_\mathrm{max}$=3.}
  \label{tab:perf_openmp}  
\end{table}

On voit que sur un coeur, l'éxecution de MDFT est plus lente avec OpenMP (31min) que sans (25min). Ces résultats montrent que l'utilisation de OpenMP est co\^uteuse en temps de calcul. C'est pourquoi, dans le cas de benchmark, il est plus efficace de lancer les calculs en série. On voit également que l'utilisation de 24 cœurs permet de faire passer le temps de calcul de 25 min à seulement 2 min 34 et ainsi de le diviser par 10. Il aurait bien sûr été possible d'encore plus optimiser la parallélisation de MDFT, cependant le choix a été fait de s’arrêter ici afin que le code ne perde pas en lisibilité.



\clearpage
\strut
\vspace{10\baselineskip}

\boitemagique{A retenir}{Dans ce chapitre, nous avons dans un premier décrit la mise en place d'un outil permettant un suivi simple et efficace de l'évolution de MDFT. Nous avons ensuite présenté les développements numériques et optimisations qui permettent aujourd'hui d'atteindre des tailles de systèmes intéressantes en biologie dans des temps de calcul raisonnables.}

\clearemptydoublepage
\chapter{Base de données}
\label{chap:BDD}

\boitemagique{Objectif}{Dans ce chapitre, nous voulons benchmarker MDFT. Pour cela, nous appliquons MDFT sur une base de données de plus de 600 petites molécules neutres de type médicament. Une étude détaillée de ces résultats ainsi obtenus nous permettra d'orienter les futurs développements de MDFT.}

Dans les chapitres précédents, nous avons testé et validé la théorie et son implémentation sur seulement quelques molécules modèles comme des sphères de Lennard-Jones. Dans ce chapitre, nous allons étudier un espace chimique plus large afin de mettre en évidence les points forts et faibles de MDFT et d'ainsi pouvoir proposer des correctifs adaptés.

\section{La base de données FreeSolv}
Dans ce chapitre nous nous concentrerons sur la base de données FreeSolv\cite{Mobley_small_2009}. Cette base de données est largement utilisée dans l'évaluation de méthodes de calcul d'énergie libres de solvatation. Elle est composée de 643 petites molécules neutres, accompagnées de nombreuses méta-données telles que leurs énergies libres de solvatation expérimentales issues de la littérature, leurs énergies libres de solvatation calculées par dynamique moléculaire et intégration thermodynamique ou encore les groupes chimiques qu'elles possèdent. Les paramètres quantitatifs sont parfois accompagnés de leur incertitude.

\subsection{Les molécules}
Lors de sa première publication, FreeSolv, composée de 504 molécules, était le regroupement de la base de données de Rizzo et al\cite{Rizzo_estimation_2006} et de calculs précédemment effectués par David Mobley et ses collaborateurs. Depuis, David Mobley et al, ont concentré leurs efforts à la nettoyer\cite{Mobley_small_2009, Mobley_treating_2008, Mobley_comparison_2007, Mobley_predictions_2009, Beckstein_prediction_2011, Mobley_alchemical_2011, Mobley_blind_2014, Mobley_freesolv_2014, DuarteRamosMatos_approaches_2017} en supprimant les doublons, en vérifiant dans la littérature les valeurs des énergies libres de solvatation ou encore en essayant de peupler au mieux les zones sous-représentées de l'espace chimique. Le nombre de composés est aujourd'hui de 643. Ces molécules contiennent entre 3 et 44 atomes, pour des énergies libres de solvatation expérimentales entre -25,47 et 3,43 $\mathrm{kcal}.\mathrm{mol}^{-1}$. 

\subsection{Les groupes chimiques}
Comme on le voit dans le tableau \ref{tab:freeSolvGroupNumbers}, un ensemble de 73 groupes chimiques sont représentés. Par exemple, nous comptons 267 composés aromatiques, 88 hétérocycles contre seulement 1 sulfoxyde ou 1 acide sulfurique diester. Parmi les 643 molécules qui composent FreeSolv, seules 38 ne sont classées dans aucun groupe. Cette catégorisation nous permet une analyse en profondeur des points forts et faibles de MDFT en fonction de chaque groupe chimique.

\begin{filecontents*}{chapters/BDD/datas/freesolv_groups.csv}
groupe chimique (en anglais),nombre de molécules
aromatic,267
heterocyclic,88
nombre de groupes,73
aryl chloride,61
carboxylic acid ester,53
alkene,50
phenol or hydroxyhetarene,49
alkyl chloride,37
ketone,36
halogen derivative,33
primary amine,31
primary alcohol,30
dialkyl ether,29
nitro,26
aldehyde,24
primary aromatic amine,21
alkyl aryl ether,20
secondary alcohol,19
alkyl bromide,17
secondary amine,17
tertiary amine,16
carboxylic acid,14
diaryl ether,13
carbonitrile,12
oxo(het)arene,12
secondary aliphatic amine (dialkylamine),11
primary aliphatic amine (alkylamine),10
thiophosphoric acid ester,10
orthocarboxylic acid derivative,10
nitrate,9
alkyl iodide,9
thioether,9
tertiary aliphatic/aromatic amine (alkylarylamine),8
tertiary aliphatic amine (trialkylamine),8
orthoester,8
tertiary carboxylic acid amide,7
aryl fluoride,6
alkyl fluoride,6
aryl bromide,6
alkyne,6
1,2-diol,5
secondary aliphatic/aromatic amine (alkylarylamine),5
thiol (sulfanyl),5
primary carboxylic acid amide,4
alkylthiol,4
aryl iodide,3
carbamic acid ester (urethane),3
phosphoric acid ester,3
tertiary alcohol,3
carboxylic acid imide, N-substituted,2
secondary carboxylic acid amide,2
sulfone,2
acetal,2
thiocarbamic acid ester,2
disulfide,2
hemiacetal,2
sulfonyl halide,1
hemiaminal,1
cation,1
secondary aromatic amine (diarylamine),1
anion,1
sulfonic acid ester,1
carboxylic acid imide, N-unsubstituted,1
sulfoxide,1
ketene acetal or derivative,1
carbamic acid,1
sulfuric acid diester,1
arylthiol (thiophenol),1
phosphonic acid ester,1
urea,1
hydrazine derivative,1
phosphonic acid derivative,1
1,2-aminoalcohol,1
\end{filecontents*}
\csvautolongtable[separator=comma,
                    table head=\caption{Répartition des groupes chimiques présents dans FreeSolv.}\label{tab:freeSolvGroupNumbers}\\\hline
                    \csvlinetotablerow\\\hline
                    \endfirsthead\hline
                    \csvlinetotablerow\\\hline
                    \endhead\hline
                    \endfoot,
                    respect all
                  ]{chapters/BDD/datas/freesolv_groups.csv}


\section{MDFT Database Tool}
Afin d'étudier en profondeur les limites de MDFT, nous avions besoin de lancer de nombreux calculs sur plusieurs centaines de molécules et pour différents paramètres d'entrée. Devant ce cas de figure, deux stratégies ont été envisagées. La première consiste à écrire un script spécifique qui pourra pas être utilisé uniquement par son développeur et uniquement pour la version actuelle de MDFT et du jeu de données. C'est le choix qui avait été fait lors de l'étude d'une version précédente de la théorie MDFT\cite{sergiievskyi_fast_2014, sergiievskyi_solvation_2015}. Nous sommes arrivés aujourd'hui à un niveau de théorie et de performance qui permet et nécessite la répétition d'études plus poussées sur des espaces chimiques variés. Rien que pour ce chapitre, nous avons effectué 3874 calculs (voir tableau \ref{tab:calculs_lances}). De plus, en parallèle de ce projet, un bridge utilisant des outils de \textit{machine learning} a été développé par Sohvi Luukkonen lors de son stage de M2. Cet outil, durant sa phase d'apprentissage nécessite également un nombre très important de calculs MDFT. Nous avons donc suivi une autre stratégie. Nous avons consacré un temps plus important au développement d'un outil d'analyse semi automatique qui soit efficace, générique, qui s'adapte facilement aux modifications de MDFT et du jeu de données et qui puisse s'interfacer facilement à un maximum d'outils. Cet outil nommé \textit{MDFT Database Tool} a été développé en collaboration avec José François durant son stage de seconde année de master sous ma supervision. Il a été écrit en python orienté objet et gère toute la chaîne d'analyse, de la préparation des fichiers à l'analyse des résultats, en un minimum de commandes. Un effort particulier a été consacré à développer un outil adaptable et qui soit facile à prendre en main, à maintenir et à étendre. Pour cela, nous avons externalisé un maximum de paramètres dans des fichiers de configuration. Un format générique, le JSON, est utilisé pour chacun de ces fichiers.

\begin{table}[ht]
  \begin{tabular}{l l l l}
    \hline & \\[-1em]\hline
    Base de Données & $\mathrm{m}_\mathrm{max}$   & correction & Nombre de molécules \\
    \hline
    FreeSolv  & 1 & correction de pression \textit{PC} & 643 \\
    FreeSolv  & 1 & correction de pression \textit{PC+} & 643 \\
    FreeSolv  & 1 & bridge gros grain & 643 \\
    FreeSolv  & 3 & correction de pression \textit{PC} & 643 \\
    FreeSolv  & 3 & correction de pression \textit{PC+} & 643 \\
    FreeSolv  & 3 & bridge gros grain & 643 \\
    \hline
    Ions  & 3 & correction de pression \textit{PC} & 8 \\
    Ions  & 3 & bridge gros grain & 8 \\
    \hline & \\[-1em]\hline
  \end{tabular}
  \caption[Liste des benchmark lancés.]{Liste des calculs nécessaires pour ce chapitre. Au total 3874 calculs ont été lancés et analysés.}
  \label{tab:calculs_lances}  
\end{table}

Le format JSON (voir code \ref{code:json}) est un format générique de stockage et de transfert de l'information. Ce format, très utilisé par les technologies web et mobiles, permet de stocker et transférer facilement et lisiblement des tableaux et dictionnaires de tous types (nombre, texte). Un des avantages majeurs de ce format de données est qu'il est implémenté dans de nombreux langages \footnote{\url{http://www.json.org/}}, il est ainsi facile d'interfacer \textit{MDFT database tool} à d'autres codes, en particulier en fortran et python comme les outils \textit{machine learning} développés au laboratoire.

\begin{lstlisting}[caption={Exemple de fichier json. Ici on décrit un atome, son nom, son symbole, sa position ainsi que ses paramètres de champ de force.}, label={code:json},captionpos=b]
  "atom": {
      "name": "Hydrogen", 
      "epsilon": 0.12552, 
      "charge": 0.06, 
      "coordinates: [1.211, -0.854, 0.045],  
      "sigma": 2.5,
      "symbol" : "H"
}
\end{lstlisting}

\subsection{Description du code}
Comme nous l'avons indiqué ci-dessus, un des objectifs principaux est de faire un outil facile à utiliser et recouvrant à un minimum de commandes. Afin de bénéficier de la puissance de calcul des serveurs actuels, nous ne pouvions descendre en dessous de 4 étapes. Ces étapes sont décrites ci-dessous.

\subsubsection{Préparation des fichiers}
La première commande permet de transformer la base de données initiale en des fichiers interprétables par MDFT. De nombreuses options sont disponibles à cette étape. Il est possible de choisir les paramètres utilisés pour le calcul MDFT comme la taille du buffer d'eau entre les solutés et le bord de la boîte, l'espacement entre deux points de grilles, le nombre d'orientations ($\mathrm{m}_\mathrm{max}$), ou encore le serveur sur lequel seront lancés les calculs. Un descriptif complet des options et des valeurs possibles est disponible dans le tableau \ref{tab:processParameters}.

\begin{table}
  \begin{tabular}{l p{10cm} p{4cm}}
    \hline & \\[-1em]\hline
    Option   & Description  & Valeurs possibles \\
    \hline
    -h  & Affiche la liste des options disponibles & \\
    -db & Distance en \AA\ entre le composé étudié et les bords de la boîte dans chaque direction & $\mathbf{R}^{+*}$ \\
    -dx & Distance en \AA\ entre deux points de grilles & $\mathbf{R}^{+*}$ \\
    -solvent & Solvent & spce/tip3p/acetonitrile\\
    -$\mathrm{m}_\mathrm{max}$ & Paramètres $\mathrm{m}_\mathrm{max}$ permettant de fixer le nombre d'orientations du solvant & $\mathbf{E}[1-5]$ \\
    -T & Température en K & $\mathbf{R}^{+}$ \\
    -sv & Serveur sur lequel les calculs vont être effectués & localhost/abalone \\
    -\,-mdftcommit & Commit à utiliser pour effectuer les calculs MDFT & Commit valide \\
    -\,-mdftpath & Chemin de la version locale de MDFT à utiliser & \\
    -bg & Bridge à utiliser & cgb/wca/3b \\
    -\,-scsf & \textit{Solute charge scale factor}: Préfacteur permettant d'atténuer les charges partielles & $\mathbf{R}[0-1]$ \\
    \hline & \\[-1em]\hline
  \end{tabular}
  \caption[Description des paramètres disponibles lors de la préparation des fichiers par \textit{MDFT Database Tool}.]{Description des paramètres disponibles lors de la préparation des fichiers par \textit{MDFT Database Tool}. Chaque option est accompagnée de sa déscription et des valeurs qu'elle peut prendre.}
  \label{tab:processParameters}  
\end{table}

Si le serveur choisi est autre que localhost, l'ensemble des fichiers créé est automatiquement compressé afin d'en faciliter le transfert.

\subsubsection{\'Exécution de MDFT}
Une fois l'archive transférée et décompressée sur le serveur distant, il suffit d’exécuter le script d'orchestration runAll.sh qui permet de cloner, compiler, et exécuter MDFT sur l'ensemble des molécules de la base de données choisie à l'étape précédente. Pour cette étape, le choix du langage s'est porté sur le bash afin d'être compatible avec la majorité des serveurs de calcul scientifique. Aujourd'hui, il existe de nombreux gestionnaires de queue ou encore de modules, ce qui rend chaque supercalculateur unique. Afin de permettre l'utilisation de n'importe lequel d'entre eux, le fichier d'orchestration runAll.sh est créé à la volée durant l'étape de préparation. Ce fichier est ainsi adapté en fonction du serveur choisi.

\subsubsection{Récupération des résultats}
Après l’exécution de MDFT sur l'ensemble des composés de la chimiothèque choisie, la $3^\text{ème}$ commande permet d'analyser les fichiers de sortie de MDFT et regrouper l'ensemble des résultats dans un fichier JSON unique. Ce fichier compact et unique est ainsi simple à rapatrier sur son ordinateur pour la dernière étape.

\subsubsection{Analyse}
\begin{figure}[ht]
   \begin{subfigure}[b]{0.40\textwidth}
         \includegraphics[width=\textwidth]{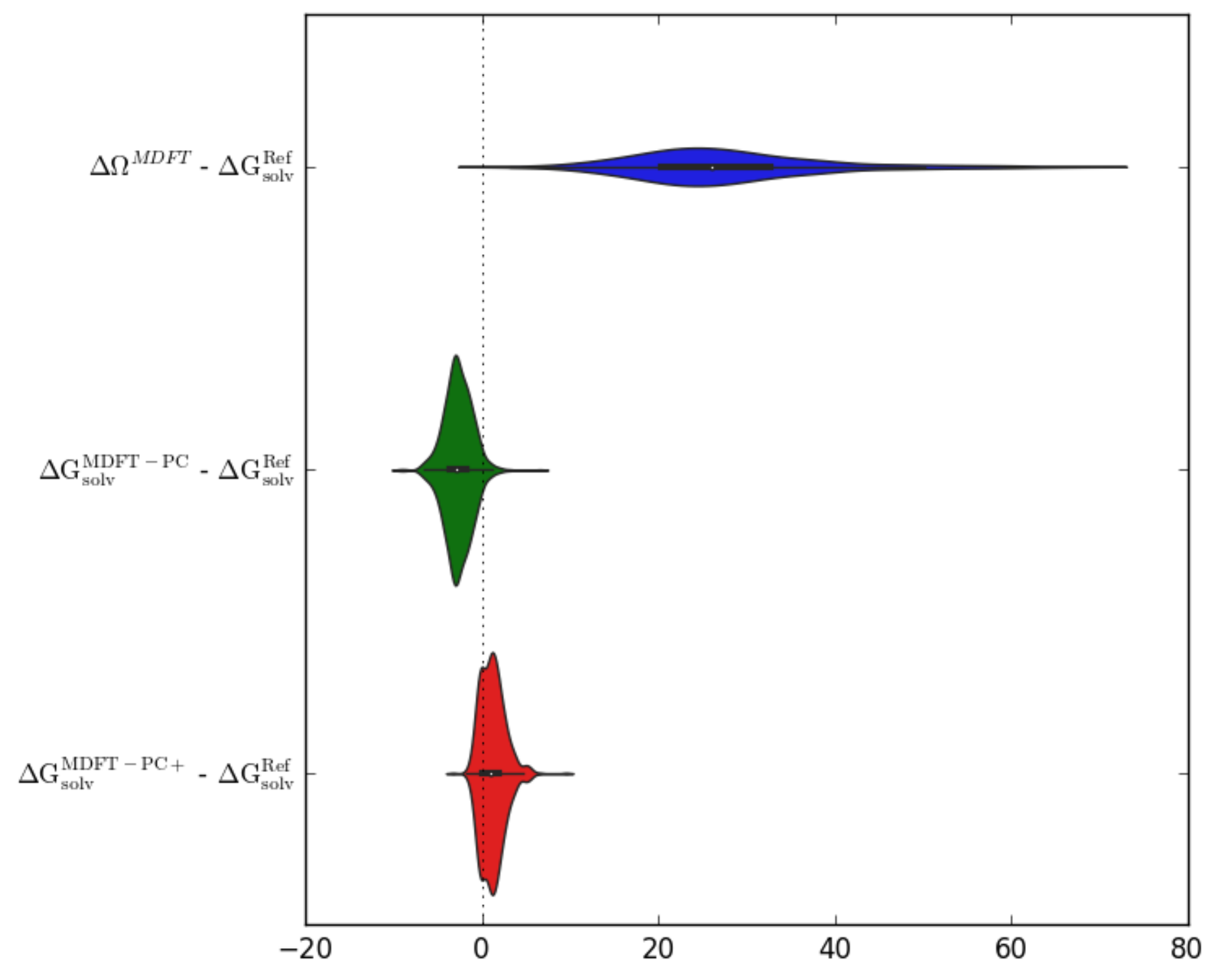}
         \par\bigskip
         \includegraphics[width=\textwidth]{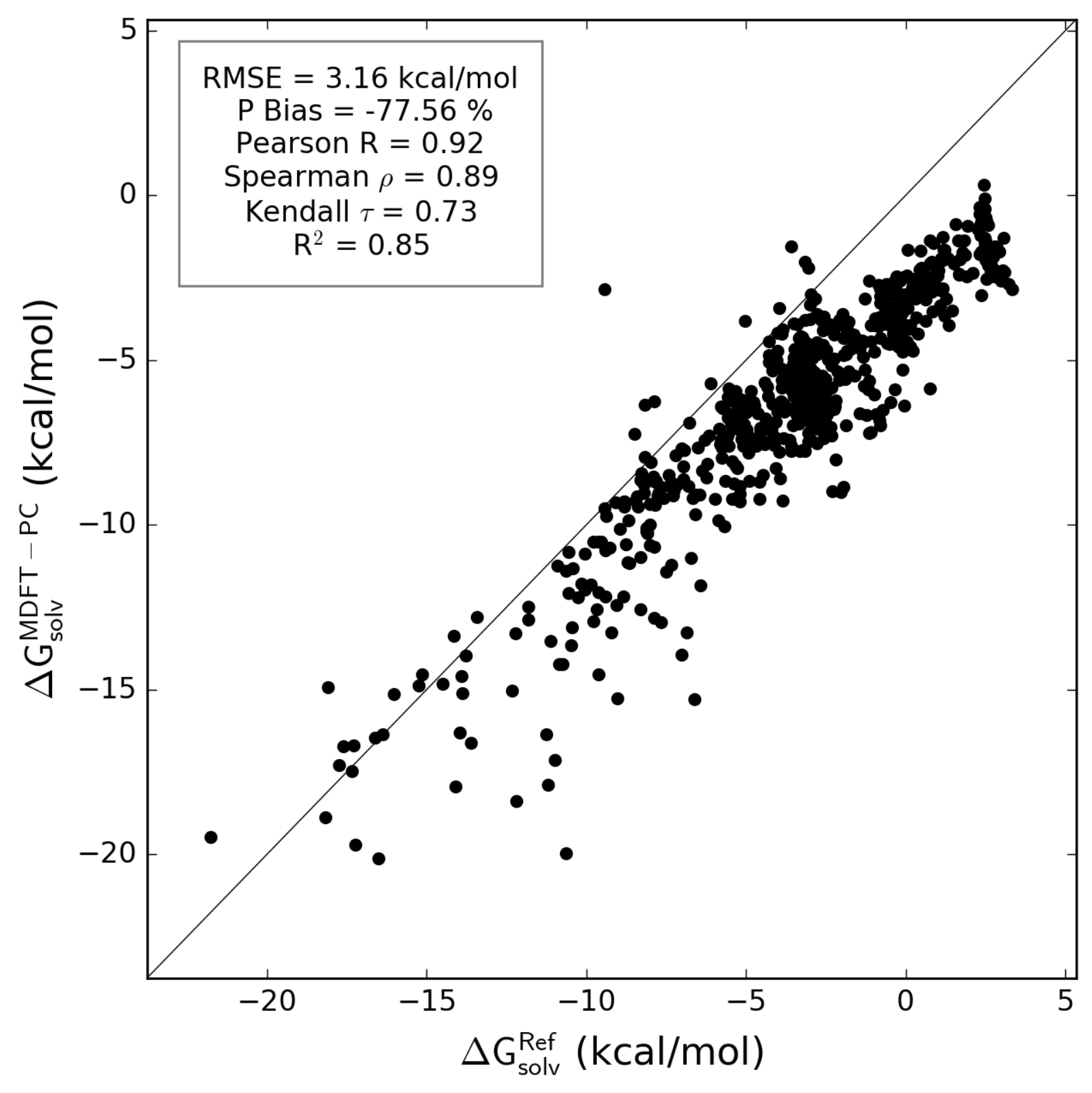}
   \end{subfigure}
   \hspace{5mm}
   \begin{subfigure}[b]{0.59\textwidth}
         \includegraphics[width=\textwidth]{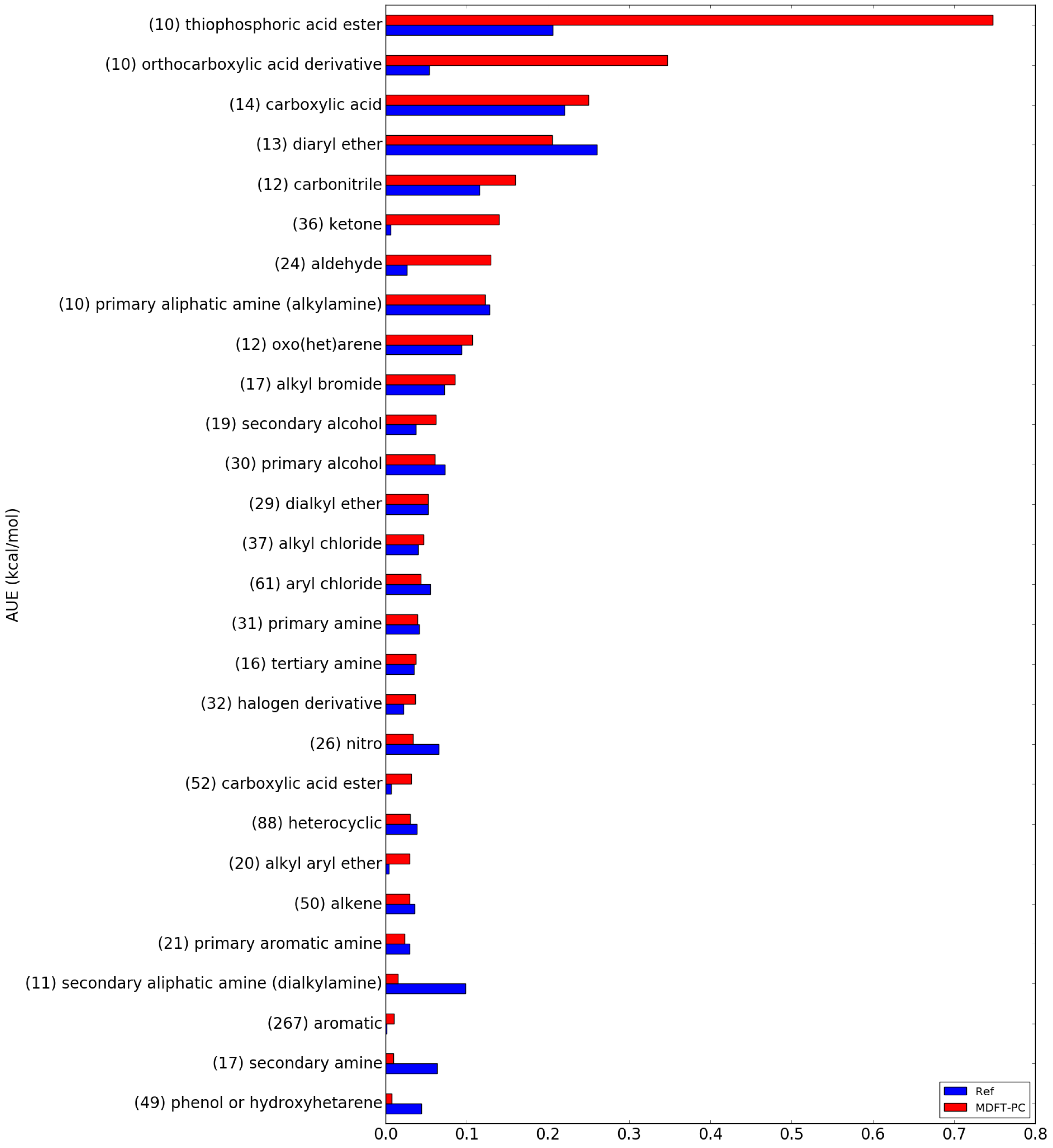}
    \end{subfigure}
  \caption[Exemple de figures d'analyse fournies par \textit{MDFT Database Tool}.]{Exemple de figures d'analyse fournies par \textit{MDFT Database Tool}. En haut à gauche un exemple de violon. En bas à gauche, un exemple de corrélation et à droite un exemple d'analyse par groupe chimique.}
  \label{fig:examples}
\end{figure}

Cette étape permet de transformer les données brutes en figures facilement analysables. Aujourd'hui, il existe 3 types de figures créées (voir figure \ref{fig:examples}):

(i) Des analyses de corrélation qui permettent une comparaison directe entre une méthode et une référence. Afin d'évaluer la corrélation entre les deux méthodes comparées, différentes grandeurs sont calculées et affichées sur l'image: le RMSE, le P Bias, le coefficient R de Pearson, le coefficient $\rho$ de Spearmann, le coefficient $\tau$ de Kendall et enfin le coefficient de corrélation R$^2$. L'ensemble de ces paramètres est détaillé en annexe \ref{chap:annexes:stats}. Dans ce chapitre nous utiliserons le RMSE, qui correspond à l'erreur quadratique moyenne et qui doit donc être le plus bas possible et le coefficient de corrélation R$^2$ qui doit être le plus proche possible de 1.

(ii) des \textit{violons}. Ils permettent de comparer visuellement l'erreur relative de plusieurs méthodes par rapport à une méthode commune de référence. Ils représentent la distribution de l'erreur soit la différence entre la valeur de l'énergie libre analysée et la valeur de référence. La largeur du violon correspond à la fréquence relative de la valeur correspondante. La partie de l'axe plus épaisse au milieu correspond aux limites du premier et troisième quartile.

(iii) des histogrammes qui permettent de visualiser rapidement l'erreur relative de différentes méthodes par rapport à une méthode de référence commune en fonction d'une donnée qualitative comme des groupes chimiques.

\subsection{La modularité du code}
Un effort tout particulier a été porté afin que le code soit le plus modulable et maintenable possible. 

\subsubsection{Les bases de données}
Aujourd'hui, deux formats de base de données sont implémentées dans ce code. Le format gromacs (fichiers gro et top) qui permet l'étude de la base de données FreeSolv et le format JSON qui permet de créer facilement des chimiothèques. Ces chimiothèques peuvent être soit créées par d'autres codes de référence, soit des chimiothèques de référence dans le groupe. Le code a été pensé pour qu'il soit facile d'ajouter de nouveaux formats de base de données. Il suffit pour cela uniquement d'implémenter ou de lier le \textit{parser} adapté.

Chaque base de données est ensuite décrite dans un fichier de configuration: lien git ou local, format, valeurs présentes etc..

\subsubsection{Les supercalculateurs}
Actuellement, il est possible de lancer MDFT sur sa propre machine, ou sur le cluster du pôle théorie, nommé "abalone". Nous avons développé ce logiciel de façon à ce qu'il soit très simple d'ajouter un nouvel ordinateur cible. Pour cela, deux fichiers suffisent: un fichier de paramètres en format json, ainsi qu'un fichier d'exemple d’exécution.

\subsubsection{Les images d'analyse}
De même, la gestion des images est entièrement externalisée. Un fichier de paramètres regroupe l'ensemble des descripteurs de chaque graphique, comme le type, les valeurs, les labels ou encore l'unité dans lequel il doit apparaître. Les valeurs extraites sont automatiquement converties dans l'unité choisie et affichées conformément à tous ces paramètres. Pour ajouter des nouvelles images, il suffit d'ajouter une nouvelle entrée dans ce fichier de paramètres.

\subsubsection{Les logiciels à étudier}
Il est pour l'instant possible de lancer uniquement MDFT. Il est cependant relativement simple d'adapter ce code à d'autres logiciels. Un utilisateur peut vouloir se comparer à d'autres méthodes, comme des simulations de dynamique moléculaire ou de Monte Carlo. Il suffit pour cela, d'implémenter la classe nécessaire à l'écriture des fichier d'entrée de cette méthode et d'adapter légèrement le script d’exécution.

\section{Résultats}
\subsection{Les corrections de pressions}\label{sec:corrections_pression}
Pour rappel, l'approximation HNC engendre une forte surestimation de la pression du système dans les conditions standard de pression et température. Sergiievskyi et al. \cite{sergiievskyi_solvation_2015,sergiievskyi_pressure_2015} ont proposés une correction ad-hoc rigoureuse basée sur la théorie des liquides: la correction \textit{PC} (voir chapitre \ref{chap:theorie}). Au moment de ce développement, la théorie MDFT n'était pas encore au niveau HNC. Elle correspondrait aujourd'hui à une approximation de HNC avec $\mathrm{m}_\mathrm{max}$=1. Les auteurs ont également proposé une correction empirique, \textit{PC+}, qui améliore considérablement les résultats sans interprétation physique claire. L'étude de base de données nous permet d'étudier l'efficacité de ces deux corrections. Pour cela, nous avons tracé la distribution de l'écart entre les valeurs calculées par MDFT et les valeurs de référence calculées par dynamique moléculaire. Ces calculs ont été lancés sans bridge.

\begin{center}
\begin{figure}
   \begin{subfigure}[b]{0.8\textwidth}
       \centering
       \resizebox{\linewidth}{!}{
         \includegraphics[width=\textwidth]{chapters/BDD/images/freesolv_1/error_distribution_calc_all.pdf}
}
       \caption{$\mathrm{m}_\mathrm{max}$=1}
       \label{fig:distrib_error_PC_PCPlus:mmax1}
    \end{subfigure}
   \begin{subfigure}[b]{0.8\textwidth}
       \centering
       \resizebox{\linewidth}{!}{
         \includegraphics[width=\textwidth]{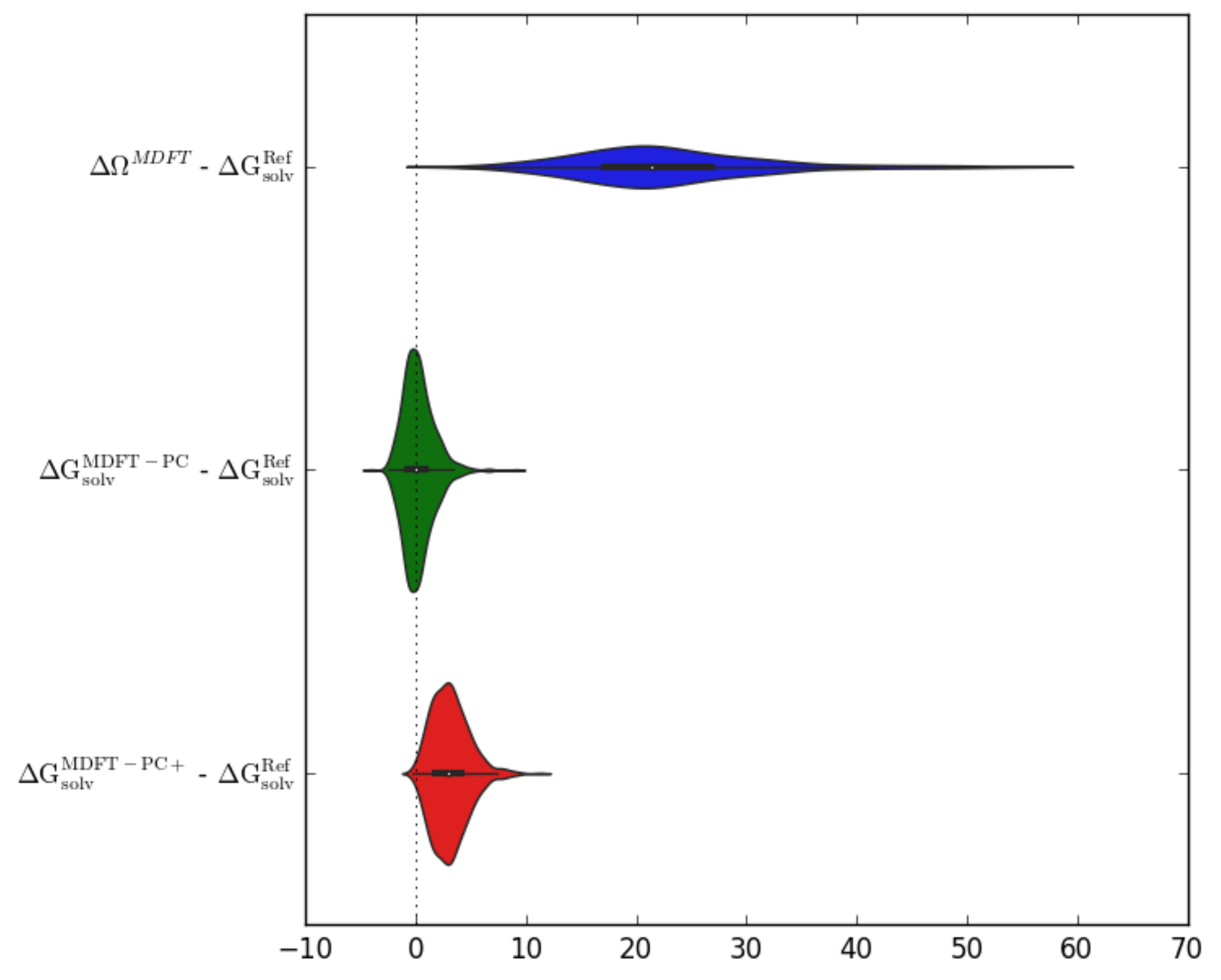}
}
       \caption{$\mathrm{m}_\mathrm{max}$=3}
       \label{fig:distrib_error_PC_PCPlus:mmax3}
    \end{subfigure}
  \caption[Distribution de l'écart entre l'énergie libre de solvatation calculée par MDFT et par dynamique moléculaire sur la base de données FreeSolv.]{Distribution de l'écart entre l'énergie libre de solvatation calculée par MDFT et par dynamique moléculaire sur la base de données FreeSolv. En bleu, MDFT sans corrélation de pression, en vert MDFT avec la correction de pression \textit{PC} et en rouge, MDFT avec la correction de pression \textit{PC+}, pour $\mathrm{m}_\mathrm{max}$=1 et 3.}
  \label{fig:distrib_error_PC_PCPlus}
\end{figure}
\end{center}

On voit sur la figure \ref{fig:distrib_error_PC_PCPlus} que la MDFT, sans correction de pression, surestime fortement les énergies libres de solvatation, et ce quelque soit la valeur de $\mathrm{m}_\mathrm{max}$ choisie. Dans tous les cas, les deux corrections de pression \textit{PC} et \textit{PC+} améliorent les valeurs et diminuent cet écart à seulement quelques $\mathrm{kJ.mol}^{-1}$. On voit également que pour $\mathrm{m}_\mathrm{max}$=1, la correction \textit{PC} a tendance à sous-estimer les valeurs d'énergies libres de solvatation. Dans ces conditions, mimant au mieux le niveau de théorie disponible au moment du développement de ces corrections, \textit{PC+} améliore bien les résultats. Pour un niveau de théorie supérieur, $\mathrm{m}_\mathrm{max}$=3, la correction \textit{PC} propose des résultats plus précis que la correction \textit{PC+}.

\boitesimple{
Au travers de cette étude, nous avons montré l'efficacité de la correction de pression \textit{PC} quelque soit le niveau de la théorie utilisée. Nous avons également montré que la correction de pression empirique \textit{PC+} permet de corriger simplement les approximations engendrées par l'utilisation d'un $\mathrm{m}_\mathrm{max}$=1.
}

\subsection{Le bridge gros grain}
Dans le chapitre précédent, nous avons proposé un nouveau bridge gros grain. Nous avons montré que ce bridge améliore considérablement les structures de solvatation. Afin d'étudier la précision de la prédiction des énergies libres de solvatation, nous avons calculé ces valeurs pour $\mathrm{m}_\mathrm{max}$=1 et $\mathrm{m}_\mathrm{max}$=3 avec et sans le bridge gros grain. Pour chaque jeu de paramètres, nous avons ensuite étudié la corrélation entre les valeurs calculées par MDFT et par dynamique moléculaire.

\begin{sidewaysfigure}
\begin{figure}[H]
\centering
   \begin{subfigure}[b]{0.30\textwidth}
       \centering
       \caption{$\mathrm{m}_\mathrm{max}$=1}
       \resizebox{\linewidth}{!}{
         \includegraphics[width=\textwidth]{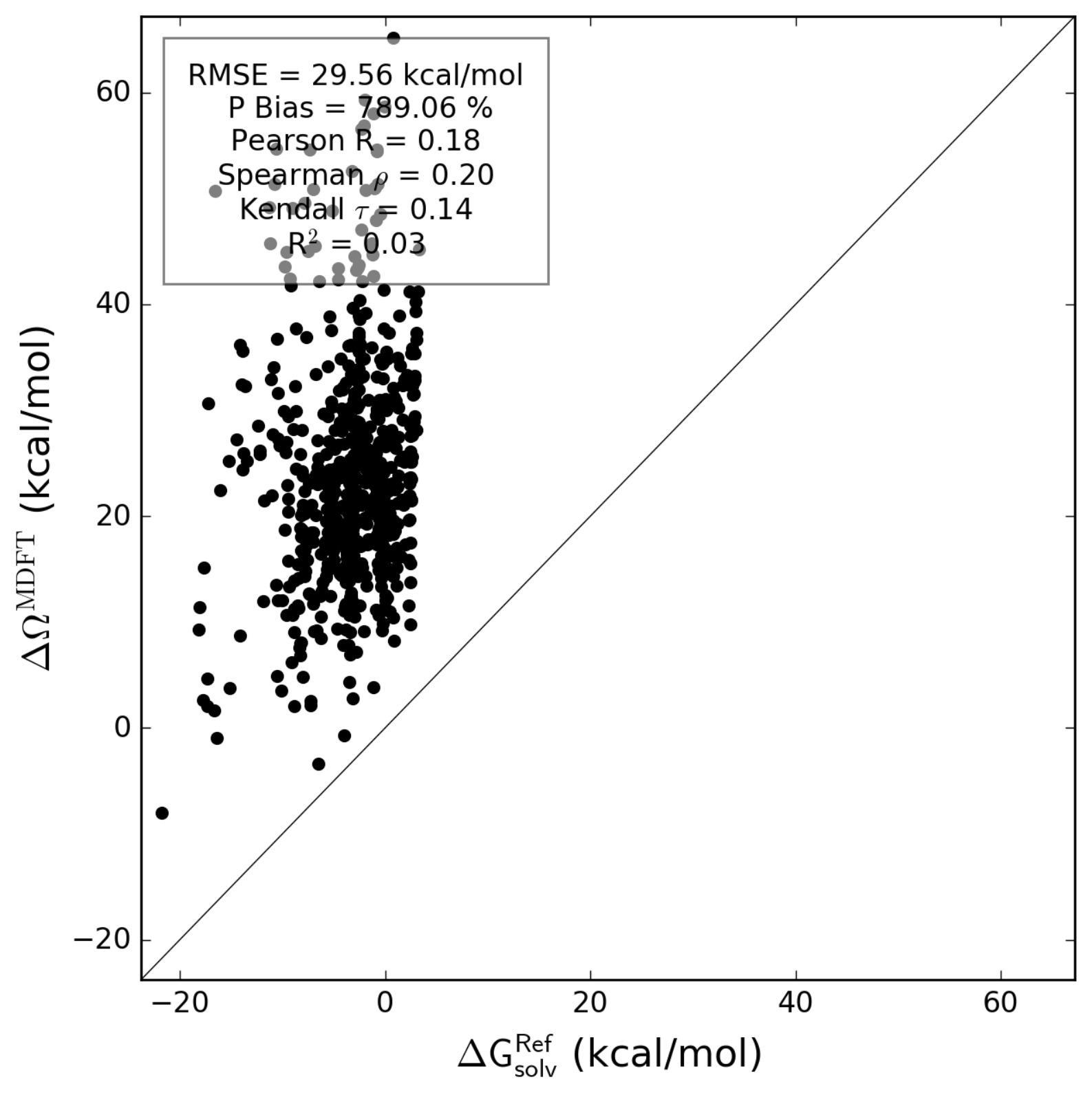}
}
       \label{fig:correlation_avec_sans_cgb:mmax1}
    \end{subfigure}    
   \begin{subfigure}[b]{0.30\textwidth}
       \centering
       \caption{$\mathrm{m}_\mathrm{max}$=1,PC}
       \resizebox{\linewidth}{!}{
         \includegraphics[width=\textwidth]{chapters/BDD/images/freesolv_1/correlation__mdft_energy_pc__vs__calc.pdf}
}
       \label{fig:correlation_avec_sans_cgb:mmax1_pc}
    \end{subfigure}
   \begin{subfigure}[b]{0.30\textwidth}
       \centering
       \caption{$\mathrm{m}_\mathrm{max}$=1,cgb}
       \resizebox{\linewidth}{!}{
         \includegraphics[width=\textwidth]{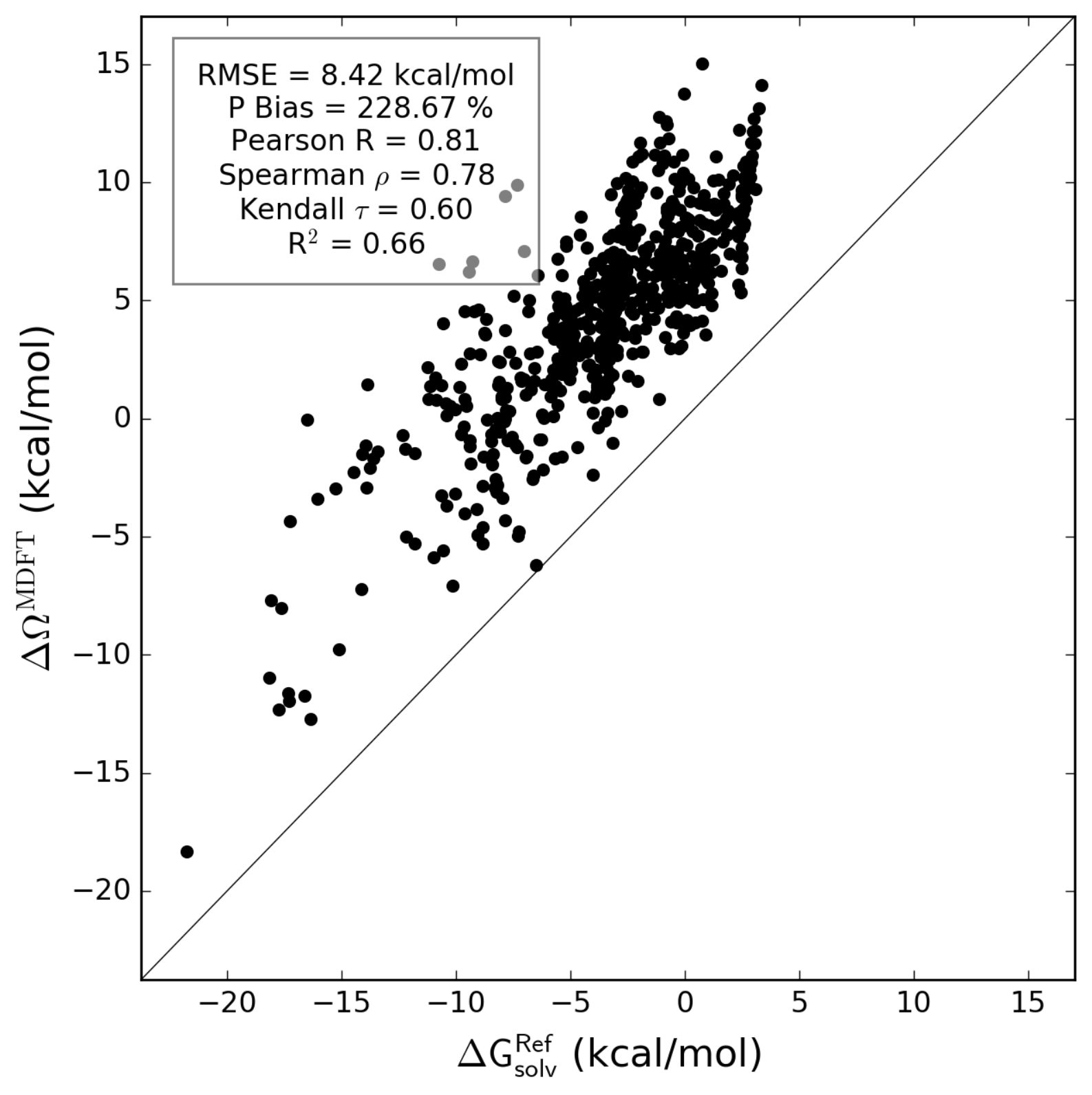}
}
       \label{fig:correlation_avec_sans_cgb:mmax1_cgb}
    \end{subfigure}

   \begin{subfigure}[b]{0.30\textwidth}
       \centering
       \caption{$\mathrm{m}_\mathrm{max}$=3}
       \resizebox{\linewidth}{!}{
         \includegraphics[width=\textwidth]{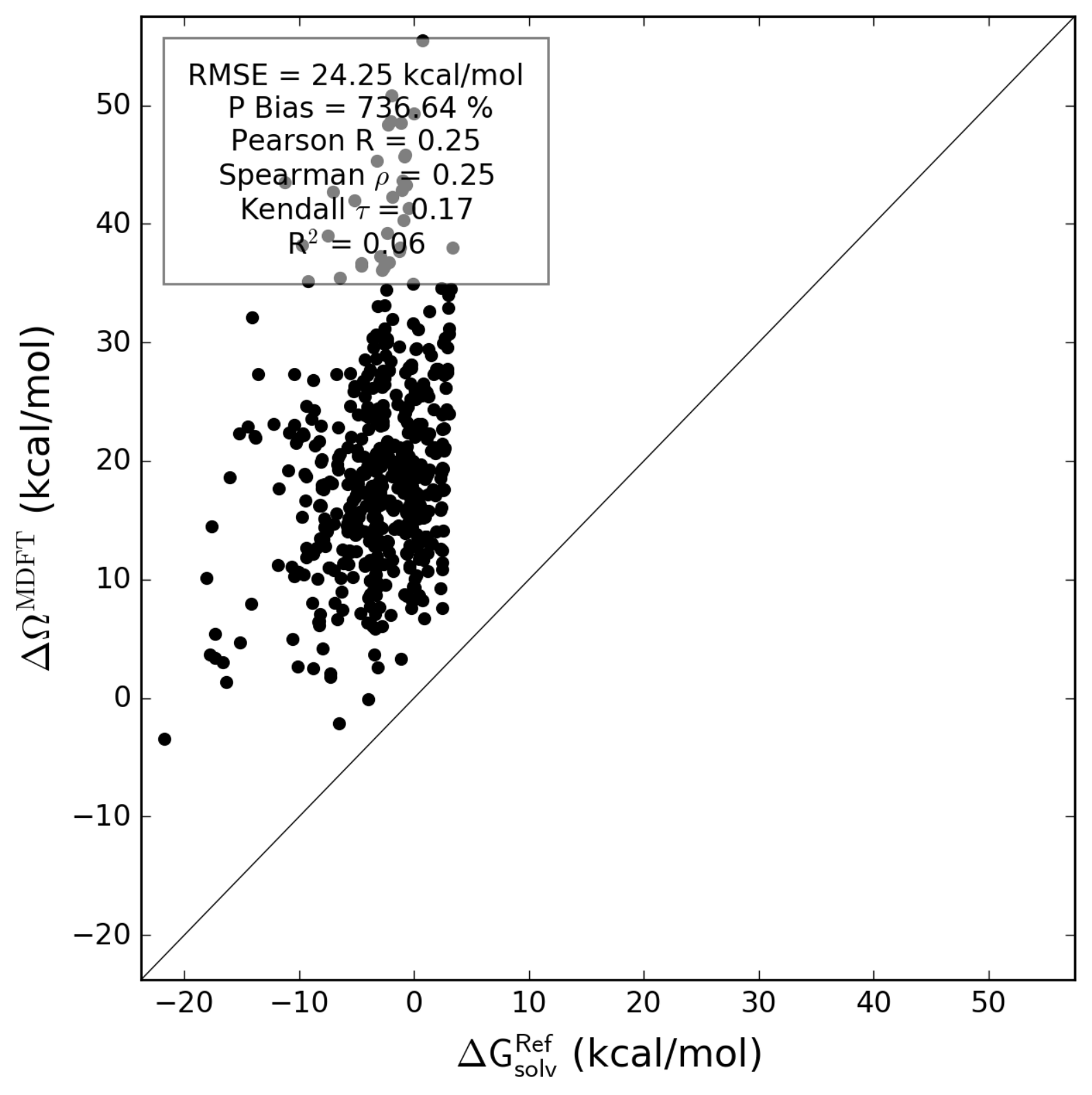}
}
       \label{fig:correlation_avec_sans_cgb:mmax3}
    \end{subfigure}    
   \begin{subfigure}[b]{0.30\textwidth}
       \centering
       \caption{$\mathrm{m}_\mathrm{max}$=3,PC}
       \resizebox{\linewidth}{!}{
         \includegraphics[width=\textwidth]{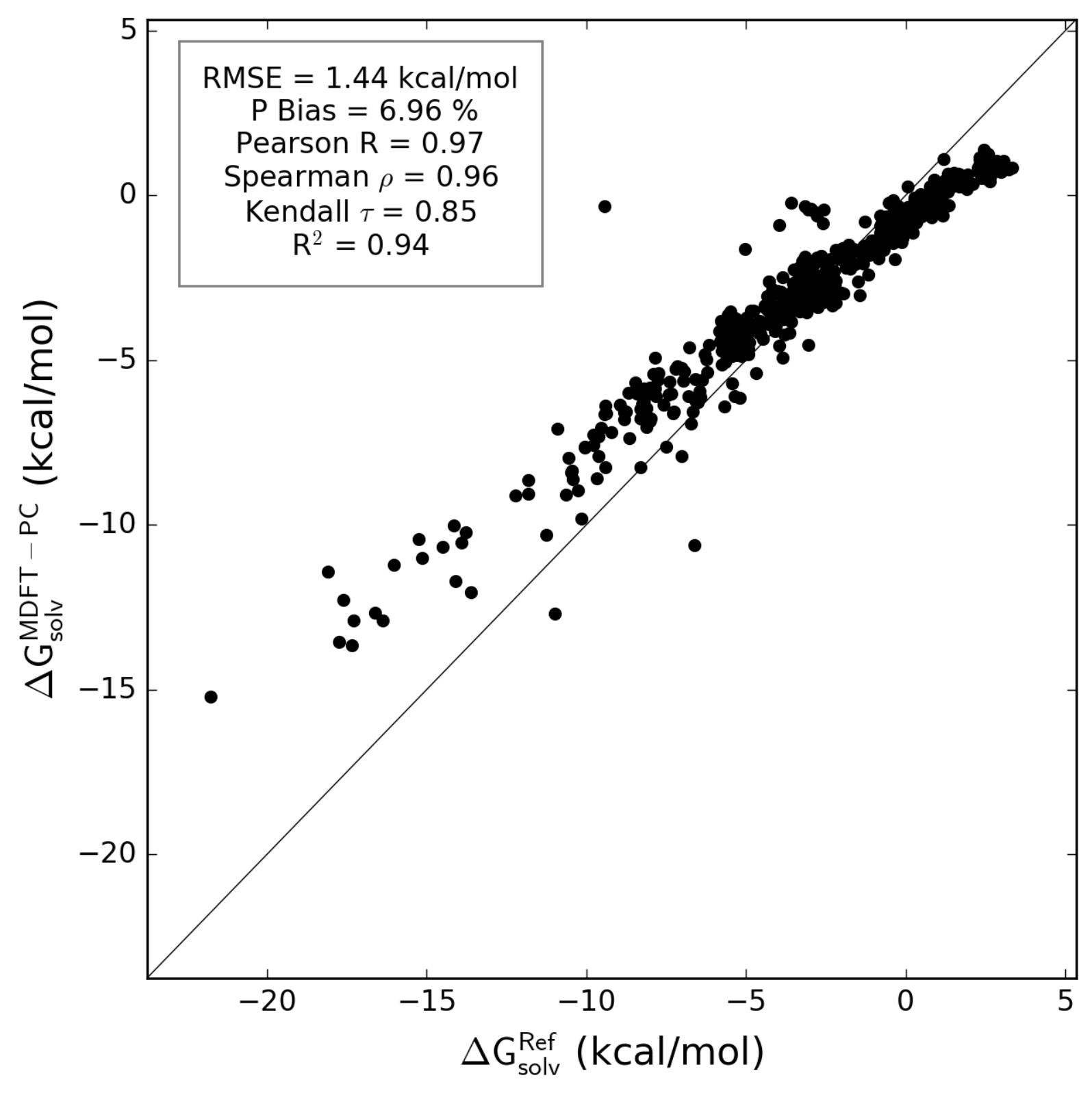}
}
       \label{fig:correlation_avec_sans_cgb:mmax3_pc}
    \end{subfigure}
   \begin{subfigure}[b]{0.30\textwidth}
       \centering
       \caption{$\mathrm{m}_\mathrm{max}$=3,cgb}
       \resizebox{\linewidth}{!}{
         \includegraphics[width=\textwidth]{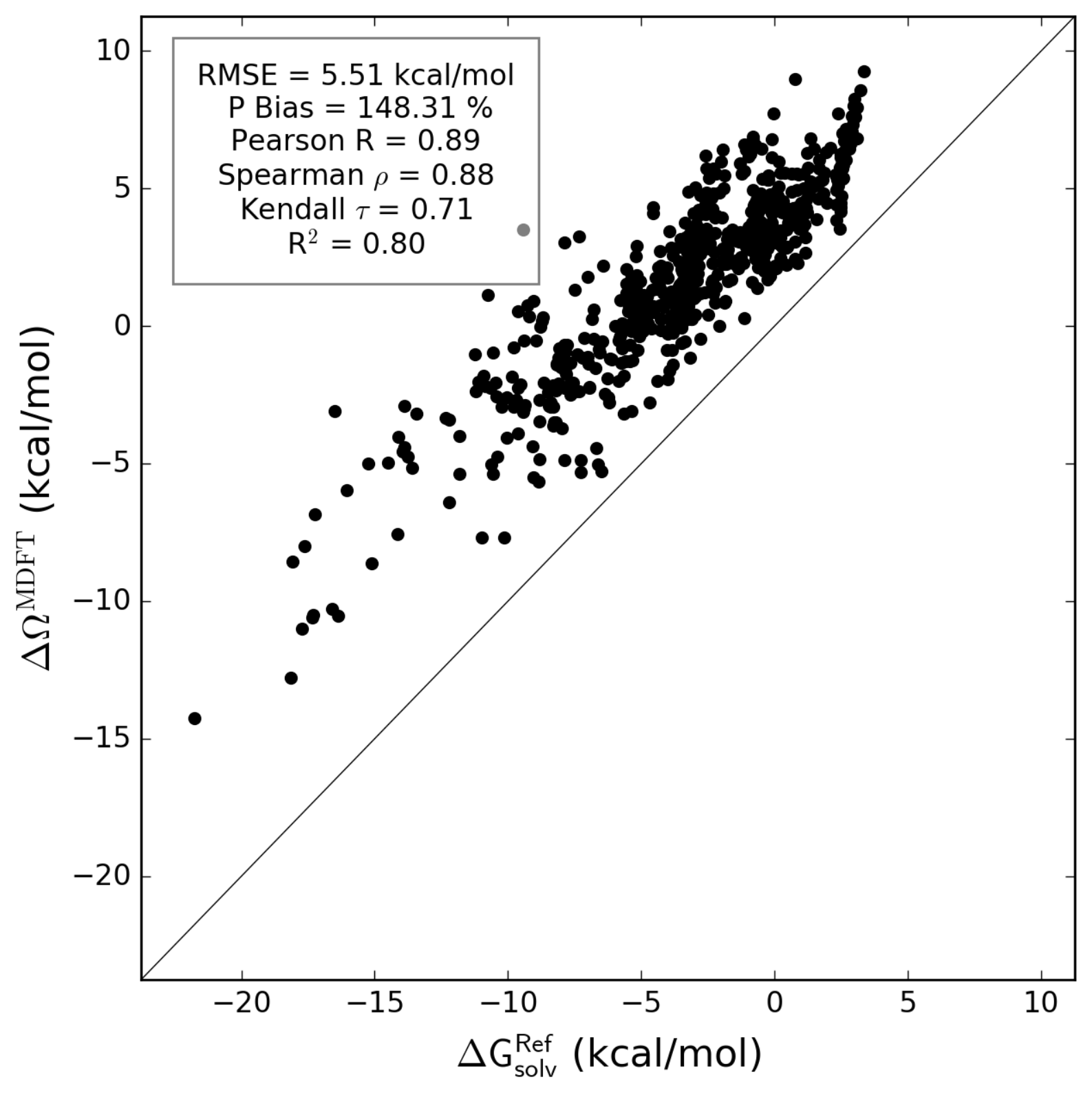}
}
       \label{fig:correlation_avec_sans_cgb:mmax3_cgb}
    \end{subfigure}

  \caption[Corrélation entre les valeurs d'énergies libres de solvatation calculées par MDFT et par dynamique moléculaire pour les composés de la base de données FreeSolv.]{Corrélation entre les valeurs d'énergies libres de solvatation calculées par MDFT et par dynamique moléculaire pour les composés de la base de données FreeSolv. Sur la première ligne $\mathrm{m}_\mathrm{max}$=1, alors que sur la seconde $\mathrm{m}_\mathrm{max}$=3. La première colonne correspond à un calcul MDFT sans correction, la seconde à MDFT avec la correction de pression \textit{PC} et la dernière MDFT avec le bridge gros grain.}
  \label{fig:correlation_avec_sans_cgb}
\end{figure}
\end{sidewaysfigure}

A ce niveau nous disposons de plusieurs possibilités d'amélioration de l'approximation HNC. Soit la pression ad-hoc de pression \textit{PC}, soit le bridge gros grain. Ces deux corrections améliorent fortement les résultats. Comme on peut le voir sur les images \ref{fig:correlation_avec_sans_cgb:mmax1} et \ref{fig:correlation_avec_sans_cgb:mmax3}, la corrélation est faible entre les valeurs calculées dans l'approximation HNC et les valeurs de référence obtenues par dynamique moléculaire. 
Le bridge gros grain ainsi que la correction \textit{PC} améliorent fortement la corrélation de ces données. Avec le bridge, nous obtenons ainsi, pour $\mathrm{m}_\mathrm{max}$=1 $\mathrm{R}^2$=0,66 et pour $\mathrm{m}_\mathrm{max}$=3 $\mathrm{R}^2$=0,80. Avec la correction \textit{PC}, nous obtenons, pour $\mathrm{m}_\mathrm{max}$=1 $\mathrm{R}^2$=0,85 et pour $\mathrm{m}_\mathrm{max}$=3 $\mathrm{R}^2$=0,94. L'ensemble des coefficients de corrélation $\mathrm{R}^2$ est disponible dans le tableau \ref{tab:correlation}.
\begin{table}[ht]
  \begin{center}
    \begin{tabular}{c c}
      \hline & \\[-1em]\hline
       fonctionnelle  & R²  \\
      \hline
       $\mathrm{m}_\mathrm{max}$ 1      & 0,03  \\
       $\mathrm{m}_\mathrm{max}$ 1 PC   & 0,85  \\
       $\mathrm{m}_\mathrm{max}$ 1 cgb  & 0,66  \\
       $\mathrm{m}_\mathrm{max}$ 3      & 0,06  \\
       $\mathrm{m}_\mathrm{max}$ 3 PC   & 0,94  \\
       $\mathrm{m}_\mathrm{max}$ 3 cgb  & 0,80  \\
      \hline & \\[-1em]\hline%
    \end{tabular}
  \end{center}
  \caption[Coefficient de corrélation des énergies libres de solvatation calculées par MDFT par rapport aux valeurs calculées par DM.]{Coefficient de corrélation des énergies libres de solvatation calculées par différentes approximations de MDFT par rapport aux valeurs de référence calculées par dynamique moléculaire. On rappelle que plus la valeur est proche et meilleure sera la corrélation.}
  \label{tab:correlation}  
\end{table}

Quelque-soit la valeur du paramètre $\mathrm{m}_\mathrm{max}$, notre bridge gros grain améliore les résultats par rapport à l'approximation HNC, tout en fournissant des structures de solvatation proches des références (voir chapitre \ref{chap:bridge}). Ces valeurs d'énergies restent cependant moins précises que celles obtenues avec la correction de pression \textit{PC}.

\boitesimple{
Cette étude nous a permis d'évaluer les différentes corrections dont nous disposons: la correction ad-hoc \textit{PC} et le bridge gros grain. Nous avons ainsi montré que ces deux corrections améliorent fortement la prédiction d'énergies libres de solvatation par rapport à l'approximation HNC seule avec un léger avantage pour la correction de pression \textit{PC}.\\
En fonction des études, nous avons donc le choix entre la correction de pression  \textit{PC}, plus précise au niveau des énergies libres de solvatation, et le bridge gros grain, plus précis en ce qui concerne les profils de solvatation.
}

\subsection{Analyse par groupe chimique}
\begin{figure}[ht]
  \centering
  \resizebox{\linewidth}{!}{
  	\fbox{\includegraphics[width=\textwidth]{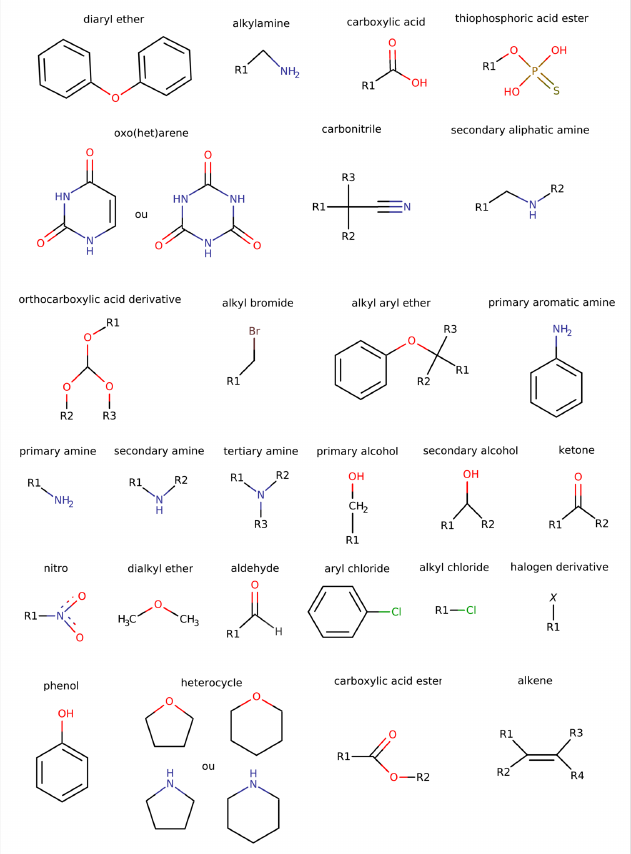}}
}
  \caption[Représentation en 2 dimensions des groupes chimiques de la base de données FreeSolv étudiés.]{Représentation en 2 dimensions des groupes chimiques étudiés dans ce chapitre. \protect\footnotemark}
  \label{fig:groupes_chimiques}
\end{figure}

Pour nous permettre de mieux comprendre les points forts et points faibles de MDFT, nous avons étudié des énergies libres de solvatation en fonction des différents groupes chimiques présents dans la base de données FreeSolv. Afin de ne pas être influencés par la sous représentation de certains groupes chimiques (voir tableau \ref{tab:freeSolvGroupNumbers}), nous ignorons dans ce chapitre les résultats obtenus pour des groupes étant composés de moins de 10 molécules. Une représentation 2D de ces groupes est disponible en figure \ref{fig:groupes_chimiques}. Dans un premier temps nous avons calculé et affiché l'erreur relative moyenne pour chaque groupe avec MDFT et la correction de pression \textit{PC} et avec MDFT et le bridge gros grain.

Nous rappelons au lecteur que les calculs MDFT et dynamique moléculaire utilisent le même champ de force soit GAFF couplé à AM1-BCC pour les charges. Il existe cependant quelques différences notables. La première est le modèle d'eau. En effet MDFT utilise le modèle SPC/E alors que les calculs en dynamique moléculaire utilisent le modèle TIP3P. De plus, la dynamique moléculaire est effectuée sur des molécules flexibles ce qui n'est pour l'instant pas le cas de MDFT. 

Si le champ de force et la théorie étaient idéaux, les erreurs moyennes pour chaque groupe et chaque méthode devraient être nulles. La dynamique moléculaire est une méthode considérée comme exacte. Pour un groupe donné, si l'erreur moyenne calculée par dynamique moléculaire s'écarte de zéro, cela signifie donc que le champ de force n'est pas optimal pour le calcul d'énergies libres de solvatation de molécules dans cette zone de l'espace chimique. Au contraire, un écart entre la dynamique moléculaire et la MDFT indique que la théorie de la MDFT n'est pas optimisée pour le groupe chimique en question. Ainsi, si pour un groupe donné, la MDFT est plus précise que la dynamique moléculaire, ces résultats seraient obtenus par chance mais traduiraient en réalité un défaut de notre théorie. 

C'est dans le but de reproduire les énergies libres de solvatation expérimentales, que Sohvi Luukkonen, développe un bridge \textit{machine learning}. Il permet de corriger à la fois les approximations de la MDFT et du champ de force (résultats non présentés dans ce rapport).

\footnotetext{Image réalisée avec le logiciel MarvinSketch 17.17.0 , 2017, ChemAxon (\url{http://www.chemaxon.com}).}

\begin{figure}[ht]
  \centering
  \resizebox{\linewidth}{!}{
  	\fbox{\includegraphics[width=\textwidth]{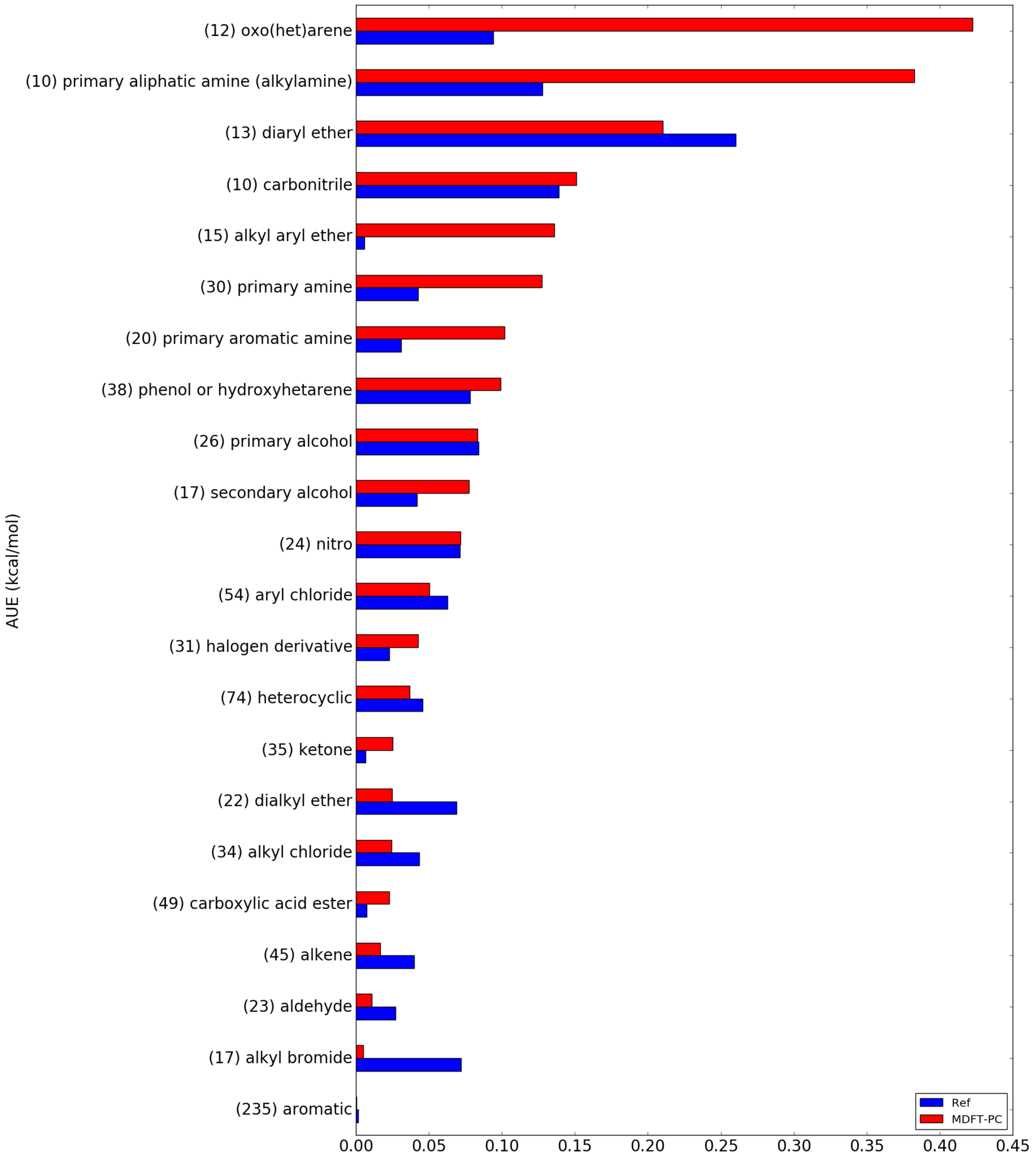}}
}
  \caption[Erreur absolue moyenne pour chaque groupe chimique de la base données FreeSolv calculée par MDFT avec la correction \textit{PC}.]{Erreur absolue moyenne calculée pour chaque groupe chimique de la base de données FreeSolv, par rapport aux valeurs expérimentales. En bleu, les résultats de dynamique moléculaire et en rouge ceux de MDFT pour $\mathrm{m}_\mathrm{max}$=3 avec la correction de pression \textit{PC}. Nous n'affichons ici que les groupes comportant 10 molécules ou plus.}
  \label{fig:AUE:mmax3}
\end{figure}

\begin{figure}[ht]
  \centering
  \resizebox{\linewidth}{!}{
  	\fbox{\includegraphics[width=\textwidth]{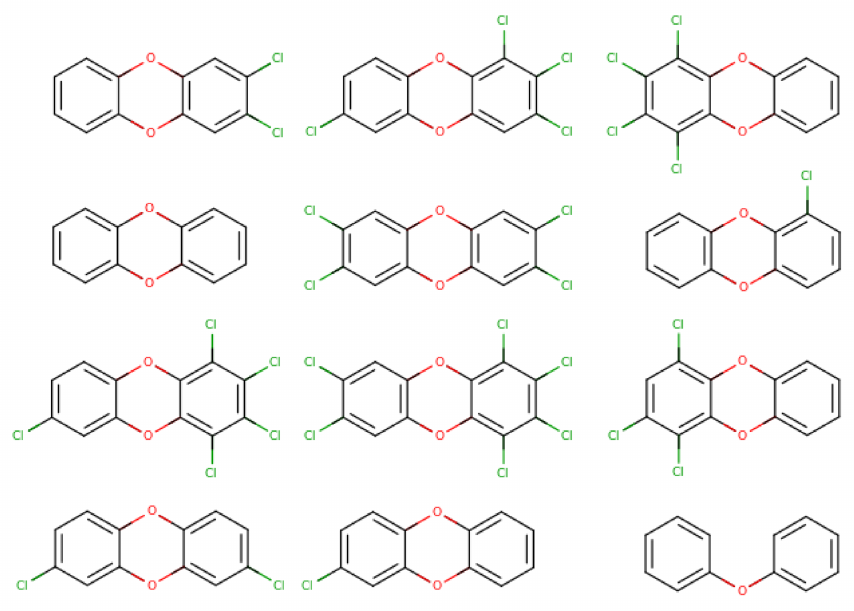}}
}
  \caption{Représentation en 2 dimensions des molécules composant le groupe des diaryl ethers.}
  \label{fig:diaryl_ether}
\end{figure}

Dans un premier temps, on voit que la dynamique moléculaire et donc le champ de force est moins précis pour les diaryl ether, les alkylamines ou encore les carbonitriles. Ces résultats sont cependant à prendre avec précaution car ces groupes peuvent être peuplés par des molécules similaires comme c'est le cas pour les éthers de diaryle (voir figure \ref{fig:diaryl_ether}). En effet, la base de données FreeSolv est issue de la littérature et il est fréquent que des molécules soient étudiées par séries. Dans le cas des diaryle ether, cette série de molécules, correspond à la fraction d'une série plus importante qui a été utilisée lors du challenge SAMPL3\cite{Geballe_sampl_2012}. Ces déséquilibres n'ont cependant aucun impact sur la comparaison de la dynamique moléculaire et de la MDFT car le biais est identique dans les deux cas.

Cette étude nous permet également de confirmer deux points faibles de MDFT: l'hydrophobicité (voir chapitre \ref{chap:bridge}) et les charges partielles\cite{ding_thesis}. En effet, les différences les plus importantes entre les deux méthodes sont obtenues pour les oxo(het)arenes (charges partielles), les amines primaires aliphatiques(charges partielles), les alkyl aryl éthers(charges partielles, hydrophobe), les amines primaires(charges partielles) et les amines primaires aromatiques (charges partielles, hydrophobe). 

Enfin, on voit que MDFT avec le bridge gros grain, nous donne globalement des erreurs plus importantes avec une erreur maximale autour de 0,85 kcal.mol$^{-1}$ contre 0,43 kcal.mol$^{-1}$ pour MDFT avec la correction de pression \textit{PC}. Ces résultats confirment ceux précédemment obtenus dans le paragraphe \ref{sec:corrections_pression}. 

\boitesimple{
Dans cette partie, nous avons mis en évidence des groupes chimiques qui exacerbent les faiblesses de MDFT dans l'approxmation HNC. Ces groupes sont principalement hydrophobes ou contiennent des charges partielles importantes. L'hydrophobocité ayant été traitée dans le chapitre précédent, nous étudions dans la suite de ce chapitre une base de données d'ions afin de mieux comprendre l'impact des charges sur MDFT au niveau HNC.
}

\begin{figure}[ht]
  \centering
  \resizebox{\linewidth}{!}{
         \fbox{\includegraphics[width=\textwidth]{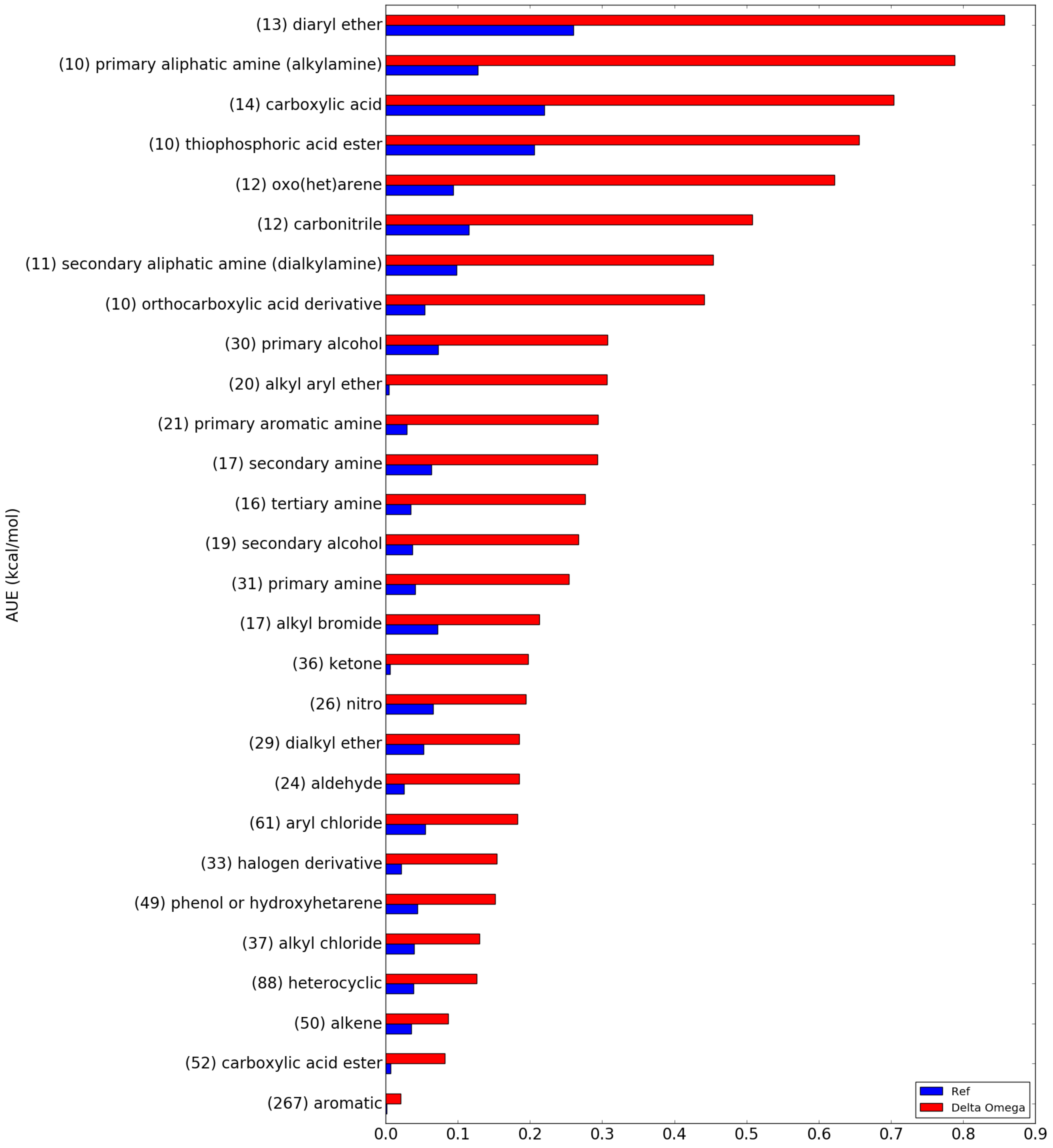}}
}
  \caption[Erreur absolue moyenne pour chaque groupe chimique de la base données FreeSolv calculée par MDFT avec le bridge gros grain.]{Erreur absolue moyenne calculée pour chaque groupe chimique de la base données FreeSolv, par rapport aux valeurs expérimentales. En bleu, les résultats de dynamique moléculaire et en rouge ceux de MDFT pour $\mathrm{m}_\mathrm{max}$=3 avec le bridge gros grain. Nous n'affichons ici que les groupes comportant 10 molécules ou plus.}
  \label{fig:AUE:mmax3_cgb}
\end{figure}

\section{Les ions}
En plus de l'analyse de bases de données officielles comme FreeSolv, \textit{MDFT Database Tool} nous permet d'étudier simplement et efficacement un ensemble de molécules d’intérêt pharmaceutique. Nous nous en sommes donc servi afin d'étudier l'impact de la charge sur les résultats de MDFT au niveau HNC. Notre jeu de données est composé de 4 cations: Li$^+$, Na$^+$, K$^+$, Cs$^+$ et de 4 anions F$^-$, Cl$^-$, Br$^-$, I$^-$.

\subsection{Les énergies libres de solvatation}
Dans un premier temps nous avons étudié l'énergie libre de solvatation de ces ions. Pour cela, nous avons utilisé les paramètres (voir tableau \ref{tab:param_lj_horinek}) proposés par Horinek et al.\cite{Horinek_rational_2009}.

\begin{table}[ht]
  \centering
  \begin{tabular}{l c c c c}
   \hline & \\[-1em]\hline
    Ion    & $\sigma_{LJ\ ion}$ (\AA) & $\epsilon_{LJ\ ion}$ (kJ.mol$^{-1}$) & $\sigma_{LJ\ ion-eau}$ (\AA) & $\epsilon_{LJ\ ion-eau}$ (kJ.mol$^{-1}$) \\
    \hline
      F$^-$	 & 3,434 & 4,654.10$^{-1}$ & 3,30 & 0,55 \\
      Cl$^-$ & 4,394 & 4,160.10$^{-1}$ & 3,78 & 0,52 \\
      Br$^-$ & 4,834 & 2,106.10$^{-1}$ & 4,00 & 0,37 \\
      I$^-$  & 5,334 & 1,575.10$^{-1}$ & 4,25 & 0,32 \\
      Li$^+$ & 2,874 & 6,154.10$^{-4}$ & 3,02 & 0,02 \\
      Na$^+$ & 3,814 & 6,154.10$^{-4}$ & 3,49 & 0,02 \\
      K$^+$  & 4,534 & 6,154.10$^{-4}$ & 3,85 & 0,02 \\
      Cs$^+$ & 5,174 & 6,154.10$^{-4}$ & 4,17 & 0,02 \\
    \hline & \\[-1em]\hline
  \end{tabular}
  \caption[Paramètres Lennard-Jones des ions utilisés dans nos calculs d'énergies libres de solvatation.]{Paramètres Lennard-Jones des ions utilisés dans nos calculs d'énergies libres de solvatation. $\sigma_{LJ\ ion}$ et $\epsilon_{LJ\ ion}$ correspondent aux paramètres Lennard-Jones des ions. $\sigma_{LJ\ ion-eau}$ et $\epsilon_{LJ\ ion-eau}$  correspondent aux paramètres d'intéraction Lennard-Jones entre l'ion étudié et l'oxygène de l'eau SPC/E. La règle de mélange de Lorentz-Berthelot\cite{Lorentz_Ueber_1881} a été utilisée.}
  \label{tab:param_lj_horinek}  
\end{table}

\begin{figure}[ht]
   \centering
   \begin{subfigure}[b]{0.5\textwidth}
       \centering
       \resizebox{\linewidth}{!}{
         \includegraphics[width=\textwidth]{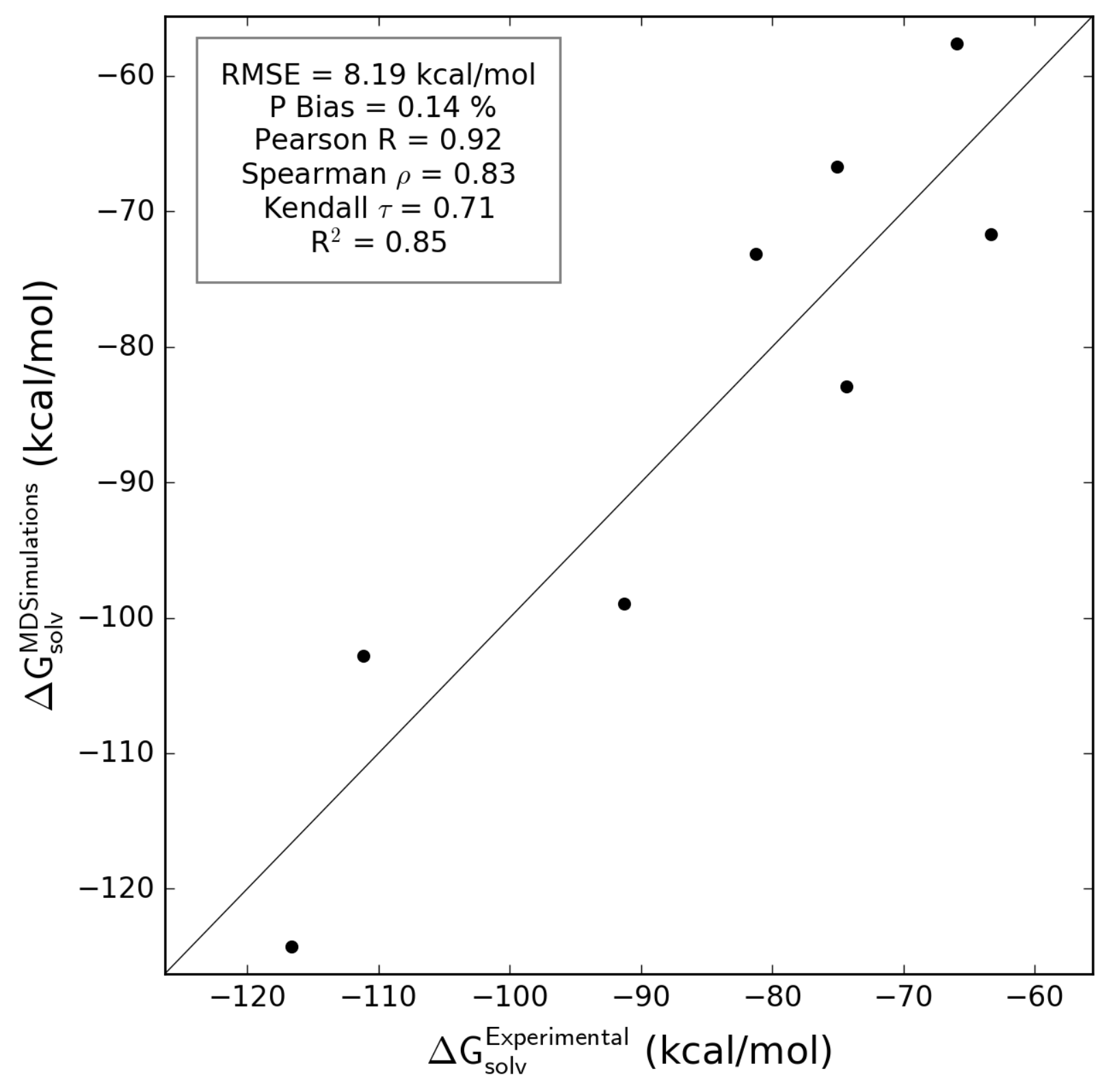}
}
    \end{subfigure}
   \begin{subfigure}[b]{0.49\textwidth}
       \centering
       \resizebox{\linewidth}{!}{
         \includegraphics[width=\textwidth]{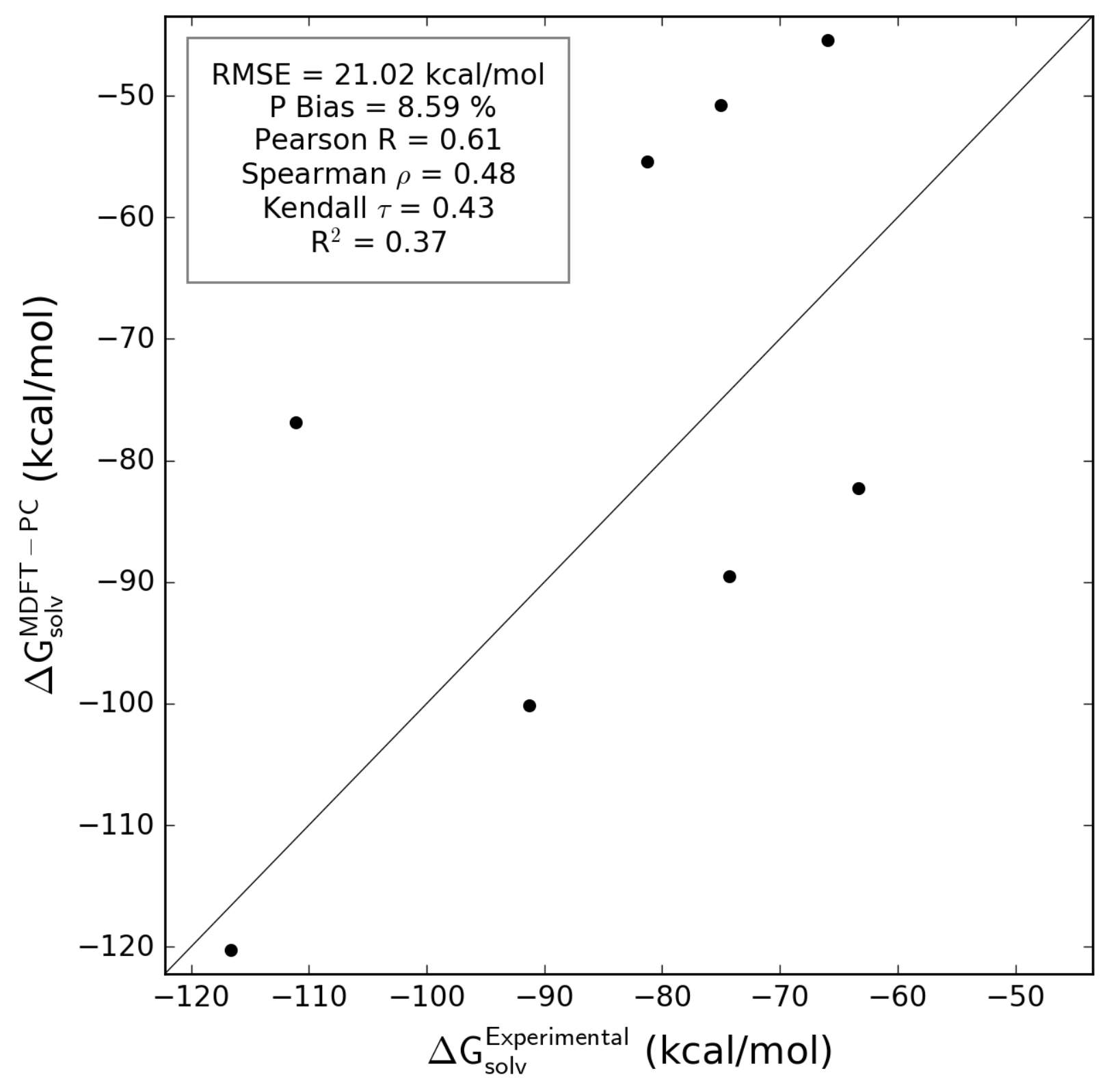}
}
    \end{subfigure}
  \caption[Corrélation des énergies libres de solvatation calculées par rapport aux valeurs expérimentales pour les ions.]{Corrélation de l'énergie libre de solvatation calculée par MDFT, par dynamique moléculaire et expérimentale pour un ensemble de 4 anions et 4 cations.}
  \label{fig:correlation_ions}
\end{figure}

Ces paramètres ont été optimisés de façon à supprimer l'erreur systématique des méthodes utilisées.
Pour cela, les auteurs ont considéré la différence entre l'énergie libre de solvatation du composé étudié et celle d'un composé de référence (ici l'ion Cl$^-$). Cette différence, notée $\Delta\Delta G_{\text{solv}}$ s'exprime:
\begin{eqnarray}
\Delta\Delta G_{\text{solv}} = \Delta G_{\text{solv}} + \frac{\text{q}}{\text{e}} \Delta G_{\text{solv Cl}^{-}}
\end{eqnarray}
\noindent avec q la charge du soluté et e la charge élémentaire. En effet, comme on le voit sur la figure \ref{fig:correlation_ions}, le biais entre les simulations et l'expérience dépend de la charge des composés étudiés. Le terme $\frac{q}{e}$ permet donc de moduler la valeur de référence en fonction du signe de la charge de l'ion étudié. Dans le cas des anions, cela revient à calculer l'énergie libre relative par rapport à celle de l'ion Chlorure. Dans le cas des cations, cela revient par contre à calculer l'énergie libre du sel qu'il formerait avec le Chlorure dans l'hypothèse d'un sel infiniment dilué. Dans ce dernier cas, on voit que les erreurs expérimentales et théoriques du cations compensent celles de l'anions.

Dans un premier temps nous avons comparé les énergies libres de solvatation calculées par MDFT et par dynamique moléculaire\cite{Horinek_rational_2009} aux valeurs expérimentales\cite{Marcus_simple_1994, Noyes_thermodynamics_1962}.

Comme attendu, une erreur systématique, différente pour les deux méthodes, appara\^it. L'énergie libre de solvatation des cations est systématiquement sous-estimée alors que celle des anions est systématiquement sur-estimée.

\begin{figure}[ht]
   \centering
   \begin{subfigure}[b]{0.5\textwidth}
       \centering
       \resizebox{\linewidth}{!}{
         \includegraphics[width=\textwidth]{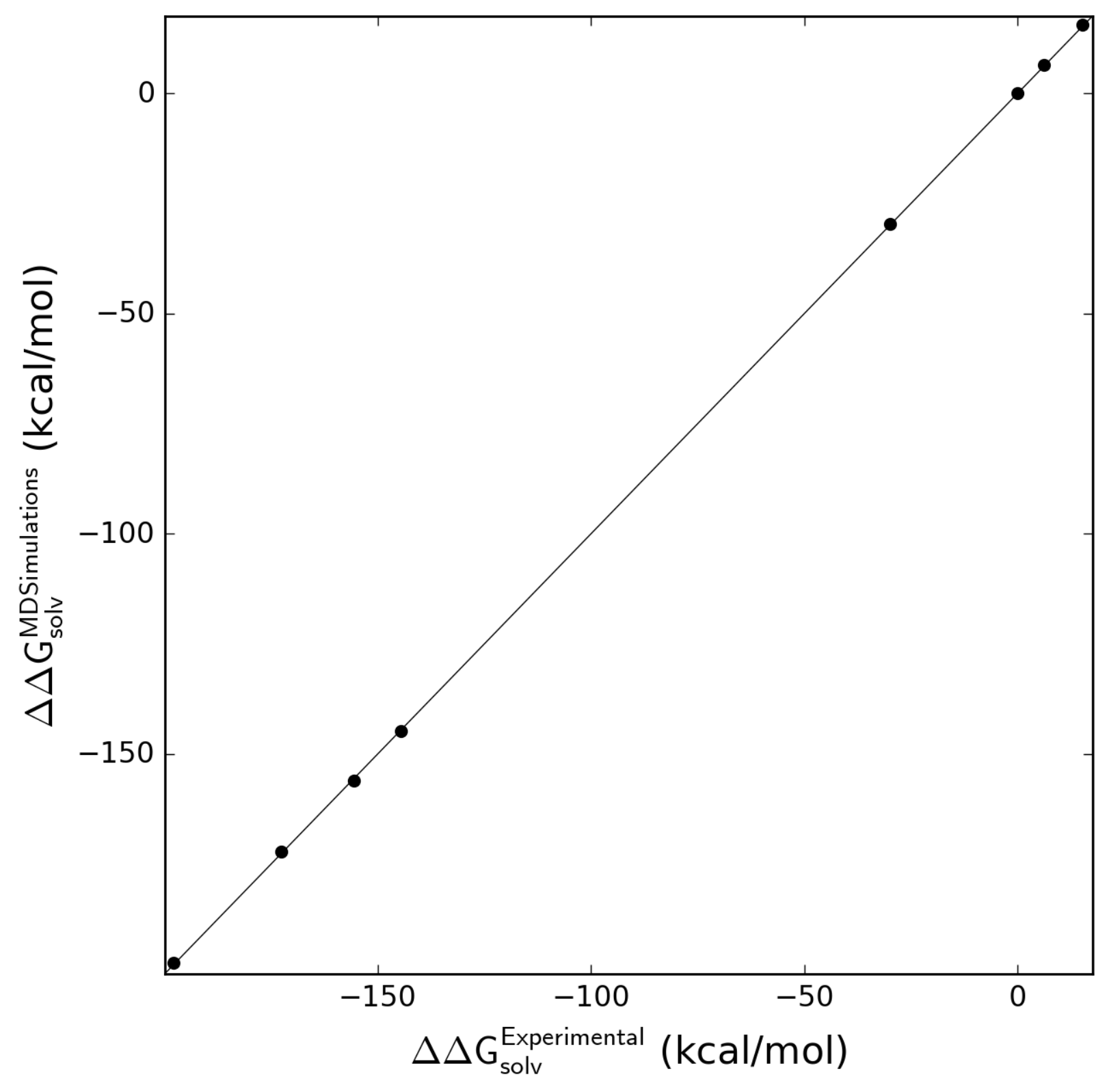}
}
    \end{subfigure}
   \begin{subfigure}[b]{0.49\textwidth}
       \centering
       \resizebox{\linewidth}{!}{
         \includegraphics[width=\textwidth]{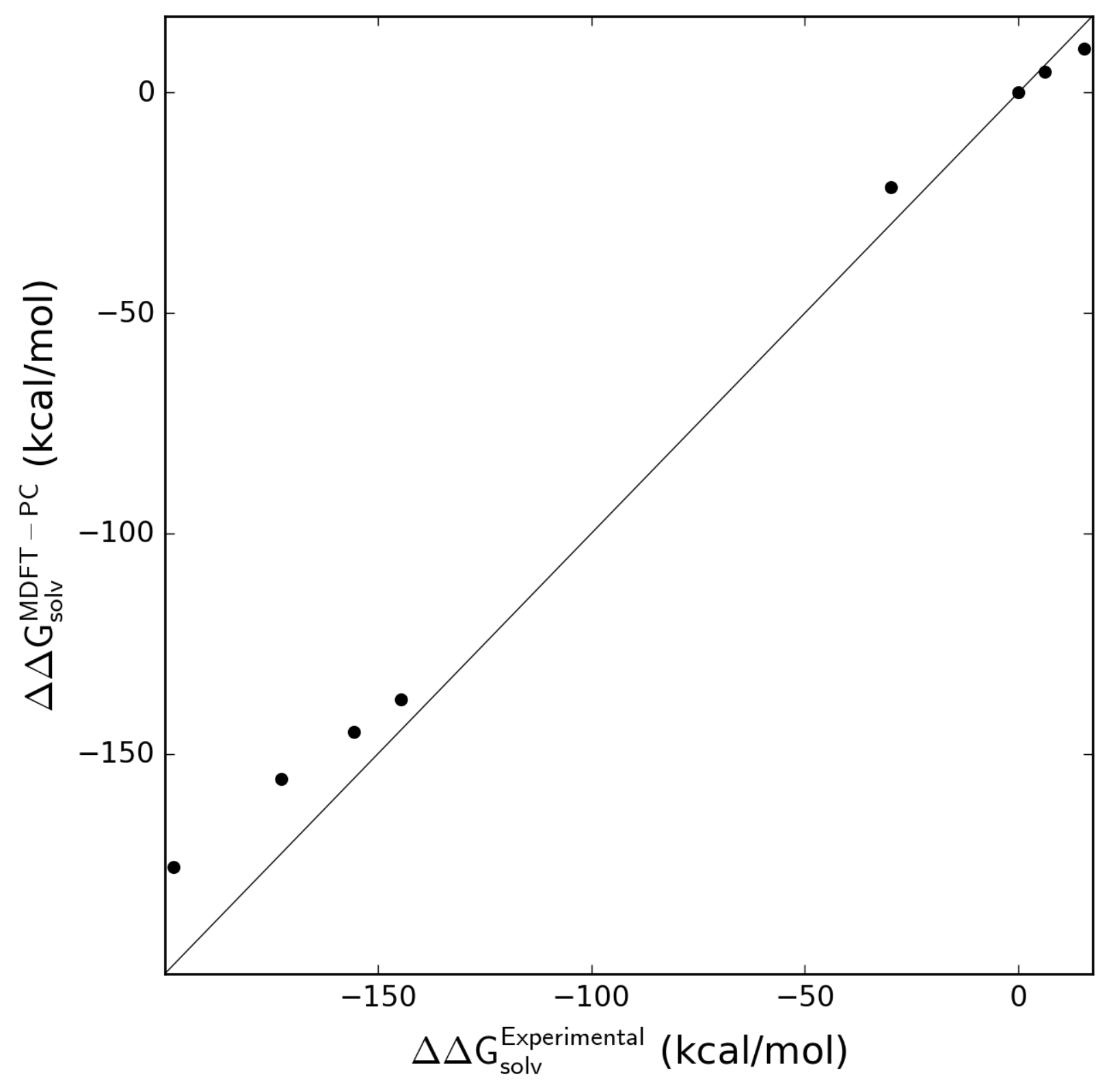}
}
    \end{subfigure}
  \caption[Corrélation des énergies libres de solvatation relatives calculées par rapport aux valeurs expérimentales pour les ions.]{Corrélation des énergies libres de solvatation relatives calculées par rapport aux valeurs expérimentales pour un ensemble de 4 anions et 4 cations. L'énergie libre de solvation du Chlorure est soustraite à celle des anions et ajoutée à celles des cations.}
  \label{fig:correlation_ions_delta}
\end{figure}
Pour nous placer dans les mêmes conditions que celles utilisées par Horinek et al. lors de l'optimisation des paramètres, nous avons tracé la différence d'énergie libre de solvatation par rapport au composé de référence pour la MDFT et la dynamique moléculaire en fonction des valeurs expérimentales.
Comme on le voit sur la figure \ref{fig:correlation_ions_delta} la corrélation est parfaite pour les calculs de dynamique moléculaire. Nous rappelons au lecteur que les paramètres de simulation ont été optimisés par Horinek et al afin d'obtenir ce résultat.
On voit également que les résultats MDFT sont proches des résultats expérimentaux. 

À ce niveau, ces résultats confirment qu'il existe un biais dépendant de la charge du composé étudiant. Ils ne permettent cependant pas de déterminer si l'erreur systématique dépend de la valeur de la charge ou uniquement de son signe. Afin d'aller plus loin, l'étape suivante sera d'étudier une base de données d'ions polyvalents et ainsi le lien entre la valence et l'erreur.

\subsection{La structures du solvant autour des ions}

\subsubsection{La structure du solvant}
Dans un second temps, nous avons étudié la structure du solvant autour de ces ions. Pour cela, nous avons comparé la fonction de distribution radiale ainsi que la polarisation du solvant autour de ces 8 ions à des structures de référence.
Les calculs MDFT ont été effectués avec des paramètres OPLS\cite{jorgensen_development_1996} pour lesquels nous disposions de calcul MD de référence\cite{zhao_molecular_2011} (voir tableau \ref{tab:param_lj_ions_daniel}).

\begin{table}[ht]
  \centering
  \begin{tabular}{l c c}
   \hline & \\[-1em]\hline
    Ion    & $\sigma_{LJ}$ (\AA) & $\epsilon_{LJ}$ (kJ.mol$^{-1}$) \\
    \hline
      F$^-$	 & 4,03 & 0,042 \\
      Cl$^-$ & 4,034 & 0,418 \\
      Br$^-$ & 4,58 & 0,45 \\
      I$^-$  & 4,92 & 0,67 \\
      Li$^+$ & 2,44 & 0,013 \\
      Na$^+$ & 2,584 & 0,4185 \\
      K$^+$  & 2,93 & 0,76 \\
      Cs$^+$ & 3,53 & 1,50 \\
    \hline & \\[-1em]\hline
  \end{tabular}
  \caption{Paramètres Lennard-Jones des ions utilisés dans nos calculs de structure du solvant.}
  \label{tab:param_lj_ions_daniel}  
\end{table}

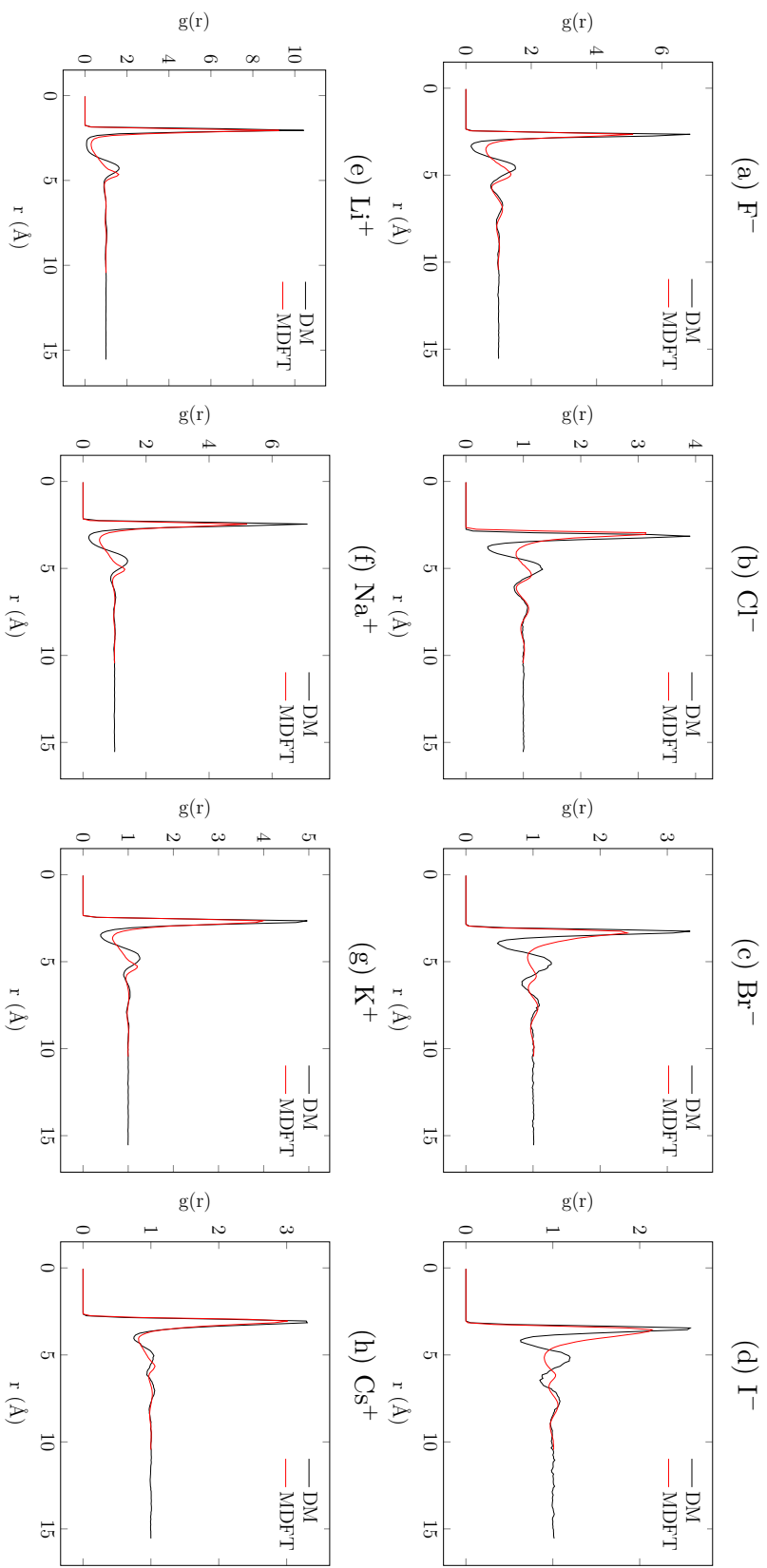
\begin{sidewaysfigure}
\begin{figure}[H]
  \centering
  \begin{subfigure}{.24\textwidth}
    \caption{F$^-$}
    \resizebox{\linewidth}{!}{
     \begin{tikzpicture}
        \begin{axis}[
            xlabel= r ($\text{\AA}$), ylabel= g(r),
            legend style = {draw = none, cells={anchor=west}}
            ]
            \addplot+[mark=none, black] file {chapters/BDD/datas/ions/MD/rdf/F.csv};
            \addplot+[mark=none, red] file {chapters/BDD/datas/ions/MDFT/rdf/F.csv};
            \legend{DM, MDFT}
        \end{axis}
    \end{tikzpicture}
	}
  \end{subfigure}%
  \begin{subfigure}{.24\textwidth}
    \caption{Cl$^-$}
    \resizebox{\linewidth}{!}{
     \begin{tikzpicture}
        \begin{axis}[
            xlabel= r ($\text{\AA}$), ylabel= g(r),
            legend style = {draw = none, cells={anchor=west}}
            ]
            \addplot+[mark=none, black] file {chapters/BDD/datas/ions/MD/rdf/Cl.csv};
            \addplot+[mark=none, red] file {chapters/BDD/datas/ions/MDFT/rdf/Cl.csv};
            \legend{DM, MDFT}
        \end{axis}
    \end{tikzpicture}
	}
  \end{subfigure}%
  \begin{subfigure}{.24\textwidth}
    \caption{Br$^-$}
    \resizebox{\linewidth}{!}{
     \begin{tikzpicture}
        \begin{axis}[
            xlabel= r ($\text{\AA}$), ylabel= g(r),
            legend style = {draw = none, cells={anchor=west}}
            ]
            \addplot+[mark=none, black] file {chapters/BDD/datas/ions/MD/rdf/Br.csv};
            \addplot+[mark=none, red] file {chapters/BDD/datas/ions/MDFT/rdf/Br.csv};
            \legend{DM, MDFT}
        \end{axis}
    \end{tikzpicture}
	}
  \end{subfigure}%
  \begin{subfigure}{.24\textwidth}
    \caption{I$^-$}
    \resizebox{\linewidth}{!}{
     \begin{tikzpicture}
        \begin{axis}[
            xlabel= r ($\text{\AA}$), ylabel= g(r),
            legend style = {draw = none, cells={anchor=west}}
            ]
            \addplot+[mark=none, black] file {chapters/BDD/datas/ions/MD/rdf/I.csv};
            \addplot+[mark=none, red] file {chapters/BDD/datas/ions/MDFT/rdf/I.csv};
            \legend{DM, MDFT}
        \end{axis}
    \end{tikzpicture}
	}
  \end{subfigure}
  \begin{subfigure}{.24\textwidth}
    \caption{Li$^+$}
    \resizebox{\linewidth}{!}{
     \begin{tikzpicture}
        \begin{axis}[
            xlabel= r ($\text{\AA}$), ylabel= g(r),
            legend style = {draw = none, cells={anchor=west}}
            ]
            \addplot+[mark=none, black] file {chapters/BDD/datas/ions/MD/rdf/Li.csv};
            \addplot+[mark=none, red] file {chapters/BDD/datas/ions/MDFT/rdf/Li.csv};
            \legend{DM, MDFT}
        \end{axis}
    \end{tikzpicture}
	}
  \end{subfigure}%
  \begin{subfigure}{.24\textwidth}
    \caption{Na$^+$}
    \resizebox{\linewidth}{!}{
     \begin{tikzpicture}
        \begin{axis}[
            xlabel= r ($\text{\AA}$), ylabel= g(r),
            legend style = {draw = none, cells={anchor=west}}
            ]
            \addplot+[mark=none, black] file {chapters/BDD/datas/ions/MD/rdf/Na.csv};
            \addplot+[mark=none, red] file {chapters/BDD/datas/ions/MDFT/rdf/Na.csv};
            \legend{DM, MDFT}
        \end{axis}
    \end{tikzpicture}
	}
  \end{subfigure}%
  \begin{subfigure}{.24\textwidth}
    \caption{K$^+$}
    \resizebox{\linewidth}{!}{
     \begin{tikzpicture}
        \begin{axis}[
            xlabel= r ($\text{\AA}$), ylabel= g(r),
            legend style = {draw = none, cells={anchor=west}}
            ]
            \addplot+[mark=none, black] file {chapters/BDD/datas/ions/MD/rdf/K.csv};
            \addplot+[mark=none, red] file {chapters/BDD/datas/ions/MDFT/rdf/K.csv};
            \legend{DM, MDFT}
        \end{axis}
    \end{tikzpicture}
	}
  \end{subfigure}%
  \begin{subfigure}{.24\textwidth}
    \caption{Cs$^+$}
    \resizebox{\linewidth}{!}{
     \begin{tikzpicture}
        \begin{axis}[
            xlabel= r ($\text{\AA}$), ylabel= g(r),
            legend style = {draw = none, cells={anchor=west}}
            ]
            \addplot+[mark=none, black] file {chapters/BDD/datas/ions/MD/rdf/Cs.csv};
            \addplot+[mark=none, red] file {chapters/BDD/datas/ions/MDFT/rdf/Cs.csv};
            \legend{DM, MDFT}
        \end{axis}
    \end{tikzpicture}
	}
  \end{subfigure}
  \caption[Fonctions de distribution radiale autour d'ions.]{Fonctions de distribution radiale autour des ions Li$^+$, Na$^+$, K$^+$, Cs$^+$, F$^-$, Cl$^-$, Br$^-$ et I$^-$. Les calculs MDFT (en rouge) sont comparés aux calculs de référence (en noir).}
  \label{fig:g_of_r_ions}
\end{figure}
\end{sidewaysfigure}

Comme on le voit sur la figure \ref{fig:g_of_r_ions}, la MDFT, pour l'ensemble des 8 ions étudiés, prédit correctement la position du premier pic en sous-estimant cependant sa hauteur. Les pics et creux suivants sont, en plus d'être sous-estimés, légèrement décalés.

\subsubsection{La polarisation}
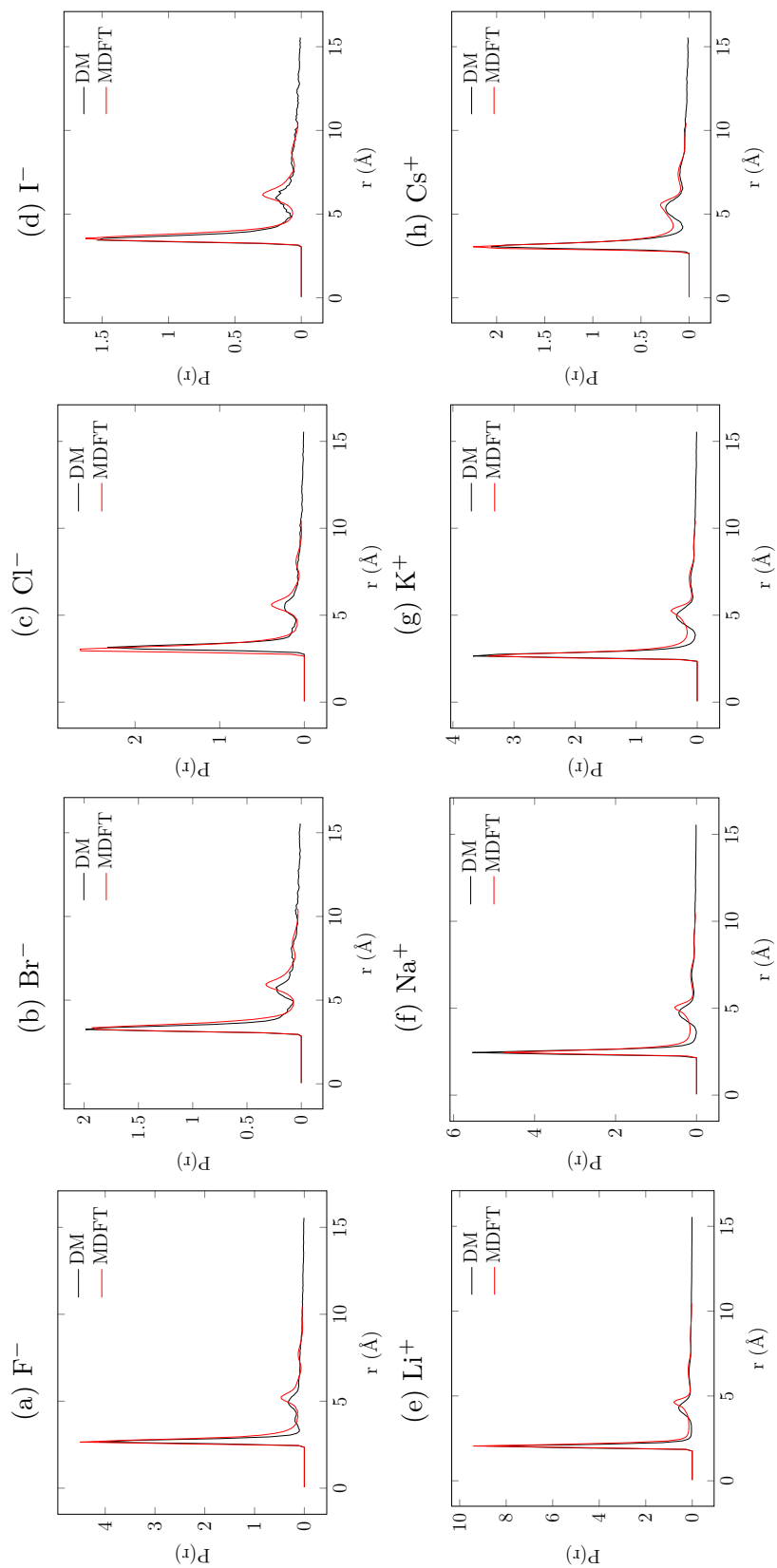
\begin{sidewaysfigure}
\begin{figure}[H]
  \centering
  \begin{subfigure}{.24\textwidth}
    \caption{F$^-$}
    \resizebox{\linewidth}{!}{
     \begin{tikzpicture}
        \begin{axis}[
            xlabel= r ($\text{\AA}$), ylabel= P(r),
            legend style = {draw = none, cells={anchor=west}}
            ]
            \addplot+[mark=none, black] table [x index=0, y expr={-1*\thisrowno{1}}] {chapters/BDD/datas/ions/MD/polarisation/F.csv};
            \addplot+[mark=none, red] table [x index=0, y expr={8*0.5*3.14*3.14*\thisrowno{1}}] {chapters/BDD/datas/ions/MDFT/polarisation/F.csv};
            \legend{DM, MDFT}
        \end{axis}
    \end{tikzpicture}
	}
  \end{subfigure}%
  \begin{subfigure}{.24\textwidth}
    \caption{Br$^-$}
    \resizebox{\linewidth}{!}{
     \begin{tikzpicture}
        \begin{axis}[
            xlabel= r ($\text{\AA}$), ylabel= P(r),
            legend style = {draw = none, cells={anchor=west}}
            ]
            \addplot+[mark=none, black] table [x index=0, y expr={-1*\thisrowno{1}}] {chapters/BDD/datas/ions/MD/polarisation/Br.csv};
            \addplot+[mark=none, red] table [x index=0, y expr={8*0.5*3.14*3.14*\thisrowno{1}}] {chapters/BDD/datas/ions/MDFT/polarisation/Br.csv};
            \legend{DM, MDFT}
        \end{axis}
    \end{tikzpicture}
	}
  \end{subfigure}%
  \begin{subfigure}{.24\textwidth}
    \caption{Cl$^-$}
    \resizebox{\linewidth}{!}{
     \begin{tikzpicture}
        \begin{axis}[
            xlabel= r ($\text{\AA}$), ylabel= P(r),
            legend style = {draw = none, cells={anchor=west}}
            ]
            \addplot+[mark=none, black] table [x index=0, y expr={-1*\thisrowno{1}}] {chapters/BDD/datas/ions/MD/polarisation/Cl.csv};
            \addplot+[mark=none, red] table [x index=0, y expr={8*0.5*3.14*3.14*\thisrowno{1}}] {chapters/BDD/datas/ions/MDFT/polarisation/Cl.csv};
            \legend{DM, MDFT}
        \end{axis}
    \end{tikzpicture}
	}
  \end{subfigure}%
  \begin{subfigure}{.24\textwidth}
    \caption{I$^-$}
    \resizebox{\linewidth}{!}{
     \begin{tikzpicture}
        \begin{axis}[
            xlabel= r ($\text{\AA}$), ylabel= P(r),
            legend style = {draw = none, cells={anchor=west}}
            ]
            \addplot+[mark=none, black] table [x index=0, y expr={-1*\thisrowno{1}}] {chapters/BDD/datas/ions/MD/polarisation/I.csv};
            \addplot+[mark=none, red] table [x index=0, y expr={8*0.5*3.14*3.14*\thisrowno{1}}] {chapters/BDD/datas/ions/MDFT/polarisation/I.csv};
            \legend{DM, MDFT}
        \end{axis}
    \end{tikzpicture}
	}
  \end{subfigure}
  \begin{subfigure}{.24\textwidth}
    \caption{Li$^+$}
    \resizebox{\linewidth}{!}{
     \begin{tikzpicture}
        \begin{axis}[
            xlabel= r ($\text{\AA}$), ylabel= P(r),
            legend style = {draw = none, cells={anchor=west}}
            ]
            \addplot+[mark=none, black] file {chapters/BDD/datas/ions/MD/polarisation/Li.csv};
            \addplot+[mark=none, red] table [x index=0, y expr={8*0.5*3.14*3.14*\thisrowno{1}}] {chapters/BDD/datas/ions/MDFT/polarisation/Li.csv};
            \legend{DM, MDFT}
        \end{axis}
    \end{tikzpicture}
	}
  \end{subfigure}%
  \begin{subfigure}{.24\textwidth}
    \caption{Na$^+$}
    \resizebox{\linewidth}{!}{
     \begin{tikzpicture}
        \begin{axis}[
            xlabel= r ($\text{\AA}$), ylabel= P(r),
            legend style = {draw = none, cells={anchor=west}}
            ]
            \addplot+[mark=none, black] file {chapters/BDD/datas/ions/MD/polarisation/Na.csv};
            \addplot+[mark=none, red] table [x index=0, y expr={8*0.5*3.14*3.14*\thisrowno{1}}] {chapters/BDD/datas/ions/MDFT/polarisation/Na.csv};
            \legend{DM, MDFT}
        \end{axis}
    \end{tikzpicture}
	}
  \end{subfigure}%
  \begin{subfigure}{.24\textwidth}
    \caption{K$^+$}
    \resizebox{\linewidth}{!}{
     \begin{tikzpicture}
        \begin{axis}[
            xlabel= r ($\text{\AA}$), ylabel= P(r),
            legend style = {draw = none, cells={anchor=west}}
            ]
            \addplot+[mark=none, black] file {chapters/BDD/datas/ions/MD/polarisation/K.csv};
            \addplot+[mark=none, red] table [x index=0, y expr={8*0.5*3.14*3.14*\thisrowno{1}}] {chapters/BDD/datas/ions/MDFT/polarisation/K.csv};
            \legend{DM, MDFT}
        \end{axis}
    \end{tikzpicture}
	}
  \end{subfigure}%
  \begin{subfigure}{.24\textwidth}
    \caption{Cs$^+$}
    \resizebox{\linewidth}{!}{
     \begin{tikzpicture}
        \begin{axis}[
            xlabel= r ($\text{\AA}$), ylabel= P(r),
            legend style = {draw = none, cells={anchor=west}}
            ]
            \addplot+[mark=none, black] file {chapters/BDD/datas/ions/MD/polarisation/Cs.csv};
            \addplot+[mark=none, red] table [x index=0, y expr={8*0.5*3.14*3.14*\thisrowno{1}}] {chapters/BDD/datas/ions/MDFT/polarisation/Cs.csv};
            \legend{DM, MDFT}
        \end{axis}
    \end{tikzpicture}
	}
  \end{subfigure}
  \caption[Polarisation radiale autour d'ions.]{Polarisation radiale autour des ions Li$^+$, Na$^+$, K$^+$, Cs$^+$, F$^-$, Cl$^-$, Br$^-$ et I$^-$. Les calculs MDFT (en rouge) sont comparés aux calculs de référence (en noir).}
  \label{fig:polarisation_ions}
\end{figure}
\end{sidewaysfigure}

La polarisation P($\boldsymbol{r}$) permet de déterminer s'il existe une orientation préférentielle du solvant au point $\boldsymbol{r}$ de l'espace. Elle se calcule:
\begin{eqnarray}
\mathrm{P}(\boldsymbol{r})=\int\mathrm{d}\boldsymbol{\Omega}\rho\left(\boldsymbol{r},\boldsymbol{\Omega} \right)\boldsymbol{\Omega}
\end{eqnarray}
Nous en traçons ensuite la norme en fonction de la distance. Dans le cas d'une orientation prépondérante, la norme est importante, contrairement au cas d'une distribution angulaire homogène pour laquelle la norme serait nulle.
Comme on le voit sur la figure \ref{fig:polarisation_ions}, la polarisation prédite par MDFT est en accord avec celle de référence calculée par dynamique moléculaire.

\clearpage
\boitesimple{Dans le paragraphe précédent, nous avons montré que MDFT dans l'approximation HNC est moins précis pour les groupes chimiques contenant des charges partielles importantes. Cette étude, sur une base de données de 8 ions, nous permet de montrer que la MDFT prédit des énergies libres de solvatation ainsi que des structures de solvant (rdfs et polarisation) autour d'ions monovalents satisfaisantes. Afin d'aller plus loin, il sera nécessaire de transposer cette étude à un ensemble d'ions polyvalents.}

\clearpage
\strut
\vspace{10\baselineskip}

\boitemagique{A retenir}{
Dans ce chapitre nous proposons un outil simple et efficace qui permet une analyse complète et reproductible de MDFT sur de larges chimiothèques. Nous avons dans un premier temps montré que la correction \textit{PC+}, plus adaptée au niveau actuel de la théorie, devait être abandonnée au profit de la correction \textit{PC}. Nous avons ensuite mis en avant certaines zones de l'espace chimique (charge partielles fortes, groupements hydrophobes) pour lequel MDFT dans l'approximation HNC manque encore de précision. Cette étude nous permet ainsi d'orienter les futurs développements théoriques de MDFT.
}

\part{Applications}

\clearemptydoublepage
\chapter{Applications}
\label{chap:applications}

\boitemagique{Objectif}{
Dans ce chapitre nous présentons deux exemples d'applications de MDFT sur des systèmes biologiques. Le premier exemple permet d'évaluer la qualité de la prédiction de la structure de solvatation et le second permet dévaluer l'autre pan de MDFT, soit la prédiction des énergies libres de solvatation.
}

Jusqu'ici, nous avons montré la façon dont nous avons étendu la théorie derrière MDFT et adapté le code afin de permettre l'étude de systèmes biologiques. Pour rappel, MDFT permet la prédiction (i) des structures du solvant et (ii) des énergies libres de solvatation de systèmes complexes. Dans ce chapitre, nous présentons deux exemples d'applications sur des systèmes biologiques, le premier exemple permet d'évaluer la qualité de la prédiction de la structure de solvatation autour de macromolécules et le second permet d'évaluer la prédiction des énergies libres de liaison et donc de solvatation de complexes protéines-ligands.

\clearpage
\section{Application 1: Peut on retrouver les molécules d'eau cristallographiques?}
Certaines molécules d'eau jouent un rôle important dans la stabilité et le rôle des systèmes biologiques. Ces molécules interagissent entre elles ainsi qu'avec les systèmes biologiques via des liaisons hydrogènes. Comme il a été montré par Papoian et al. \cite{papoian_water_2004}, l'eau joue un rôle dans le repliement et la liaison des protéines via des interactions à courte portée mais également via des liaisons longues portées. Ces molécules sont donc indispensables à la bonne compréhension des différents processus biologiques ayant lieu dans le corps humain. Il existe différentes approches, expérimentales ou théoriques, permettant de détecter ces molécules.

Expérimentalement, ces molécules d'eau sont liées aux systèmes biologiques via un réseau de liaisons hydrogènes. Lors de la cristallisation d'une protéine par exemple, ces liaisons figent une partie des molécules de solvant, ce qui les rend alors détectables lors de la résolution expérimentale de la structure en 3 dimensions de tels systèmes. À cause des conditions nécessaires à la cristallisation \cite{wlodawer_advanced_2017} (agent précipitant, pH, température, ...), certaines liaisons vont être favorisées et d'autres vont être affaiblies. Les molécules d'eau expérimentalement détectées ne correspondent qu'en partie à celles que l'on retrouverait dans les conditions du laboratoire. Une fois publiées, ces structures sont, pour une majorité d'entre elles, ajoutées à la \textit{Protein Data Bank}\cite{pdb_2011} (PDB). La PDB est la base de données collaborative de référence pour les structures expérimentales de composés biologiques.

Il existe également différentes méthodes théoriques permettant la prédiction des molécules d'eau. Azuara et al. proposent par exemple leur logiciel Aquasol \cite{azuara_pdb_hydro_2006} disponible en ligne et gratuit \footnote{\url{http://lorentz.dynstr.pasteur.fr/suny/index.php?id0=aquasol}}. Cette méthode hybride, basée sur la résolution de l'équation de Poisson Boltzmann permet de prédire une densité en eau autour d'une macromolécule. Il existe également des méthodes explicites comme celle proposée par Schrödinger à travers son logiciel WaterMap\cite{abel_role_2008, Young_motifs_2007}. Afin de prédire la position des molécules d'eau, une dynamique moléculaire de 10 ns est effectuée puis une carte de densité en eau est générée. Les molécules d'eau sont ensuite reconstruites en partant des maximums locaux de la densité.

Cette étude à pour but d'évaluer la capacité de MDFT à retrouver les molécules d'eau cristallographiques. En effet, comme nous l'avons décrit ci-dessus ces molécules sont intégrées à des réseaux de liaisons hydrogènes ce qui limite fortement leur mouvement. En d'autres termes, ces molécules peuvent être considérées comme fixes et par conséquent la probabilité de trouver une molécule d'eau à cet endroit est élevée. Ces molécules doivent donc se trouver dans des zones de forte densité ou de forte probabilité de présence prédites par MDFT. Dans cette partie, nous comparons dans un premier temps, les résultats obtenus par MDFT et ceux obtenus par dynamique moléculaire, notre référence, sur des systèmes complexes. Dans un second temps nous vérifions l'adéquation entre les zones de forte probabilité de présence fournies par ces deux méthodes et la position des molécules d'eau expérimentales. Nous allions ainsi la rapidité des méthodes implicites comme Aquasol à la précision des méthodes explicites comme la dynamique moléculaire.

\subsection{Les systèmes étudiés}
Afin de mener cette étude, la protéine \textit{Streptomyces Erythraeus Trypsin}, composée de 227 acides aminés, et issue de la PDB sous le code 4M7G, a été sélectionnée pour la qualité de sa résolution expérimentale (0,81 \AA).

\begin{figure}[ht]
   \centering
   \fbox{\includegraphics[width=\textwidth]{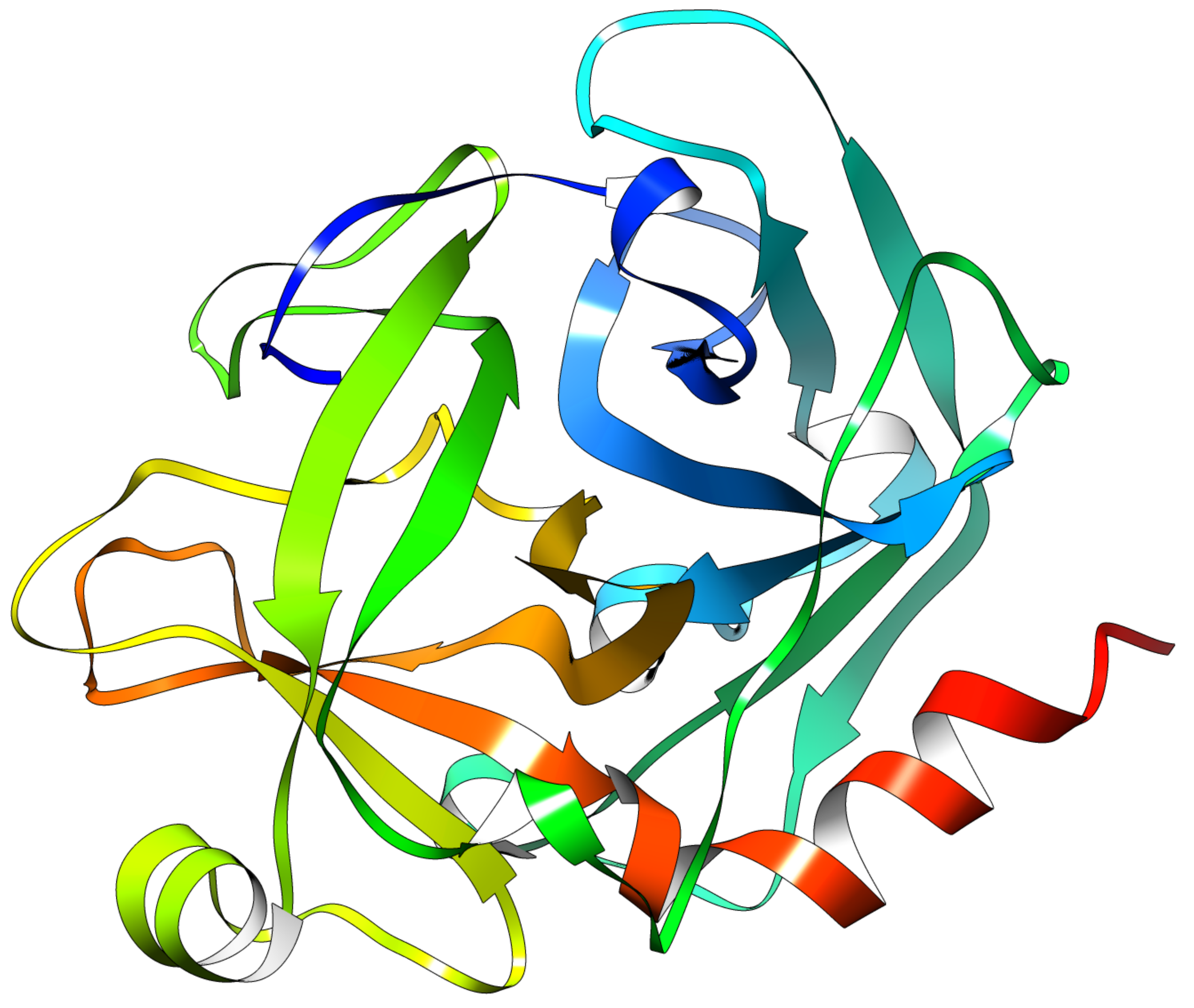}}
   \caption{Représentation en structure secondaire de la protéine 4M7G.}
   \label{fig:struct_secondaires}
\end{figure}

\subsection{Protocole}

\subsubsection{Récupération et nettoyage de fichiers PDB}
\begin{figure}[ht]
  \center
  \begin{tikzpicture}
  \tikzstyle{lien}=[->,>=stealth,rounded corners=5pt,thick]
  \node[inner sep=0pt] (protein_water) at (0,0)
      {\includegraphics[width=.25\textwidth]{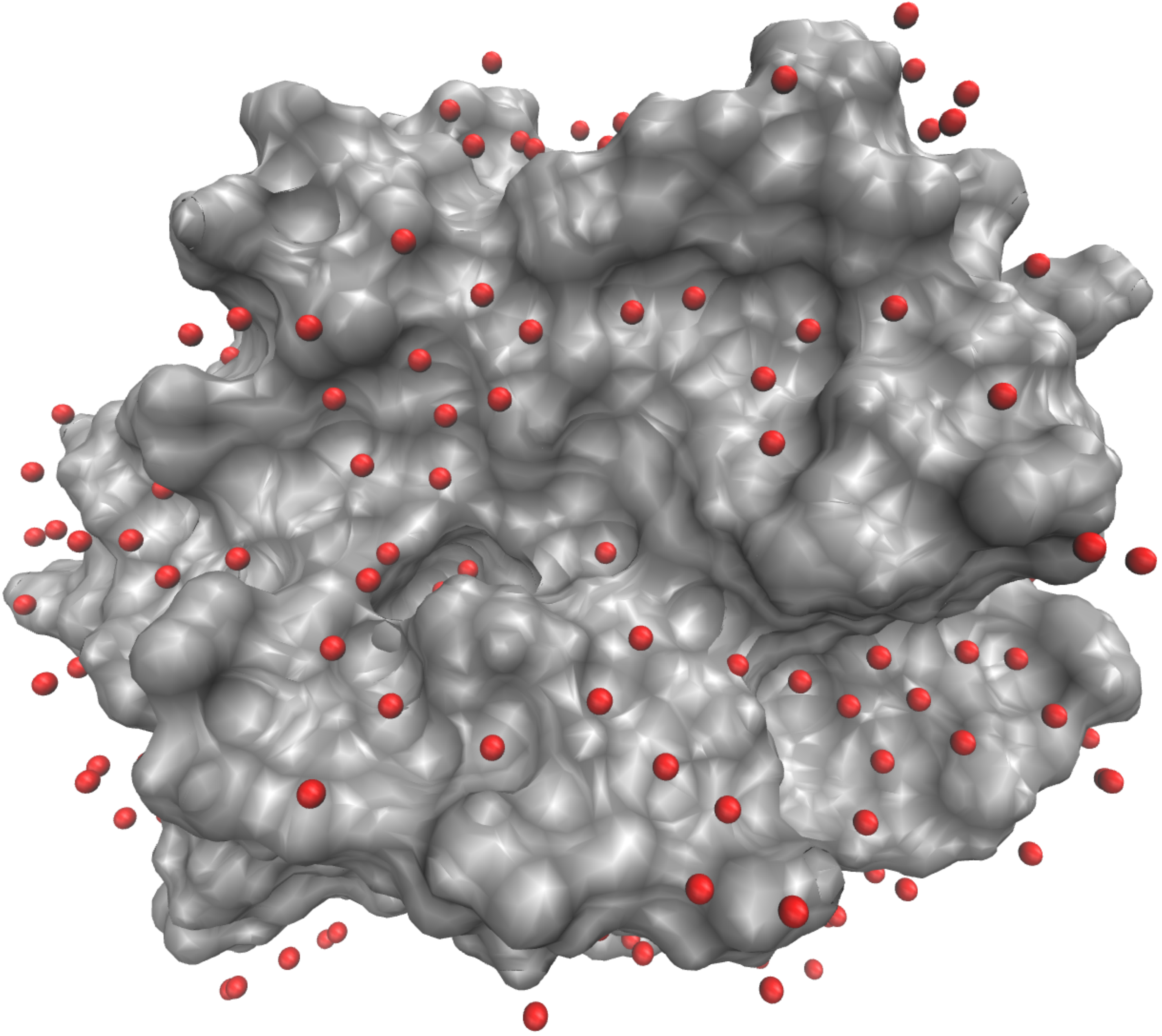}};
  \node[inner sep=0pt] (water) at (6,0)
      {\includegraphics[width=.30\textwidth]{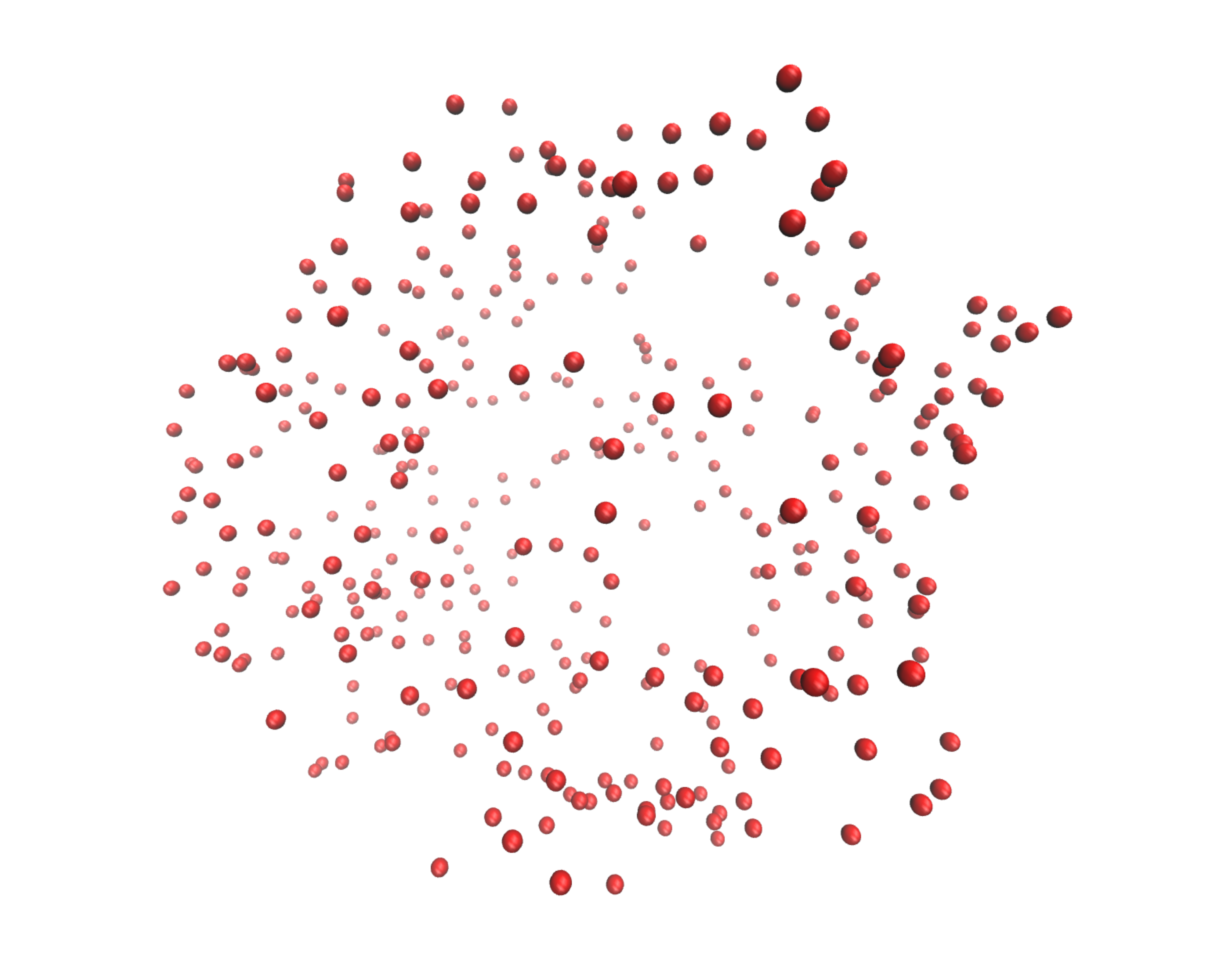}};
  \node[inner sep=0pt] (protein) at (12,0)
      {\includegraphics[width=.25\textwidth]{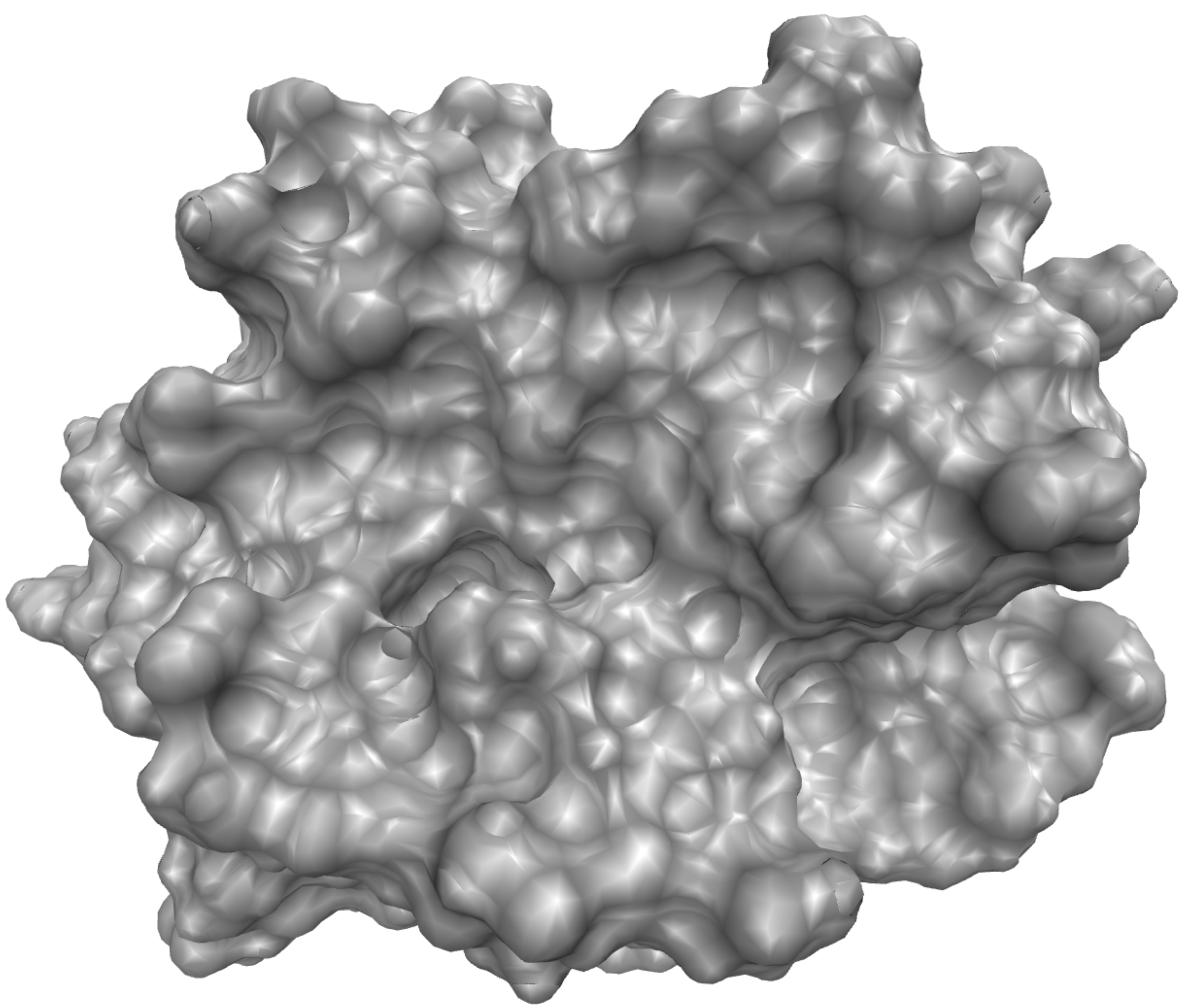}};
  \node[inner sep=0pt] (protein_water_2) at (0,-6)
      {\includegraphics[width=.25\textwidth]{chapters/Applications/images/water_molecules_grey/prot_eau.pdf}};
  \node[inner sep=0pt] (protein_gromacs_water) at (6,-6)
      {\includegraphics[width=.25\textwidth]{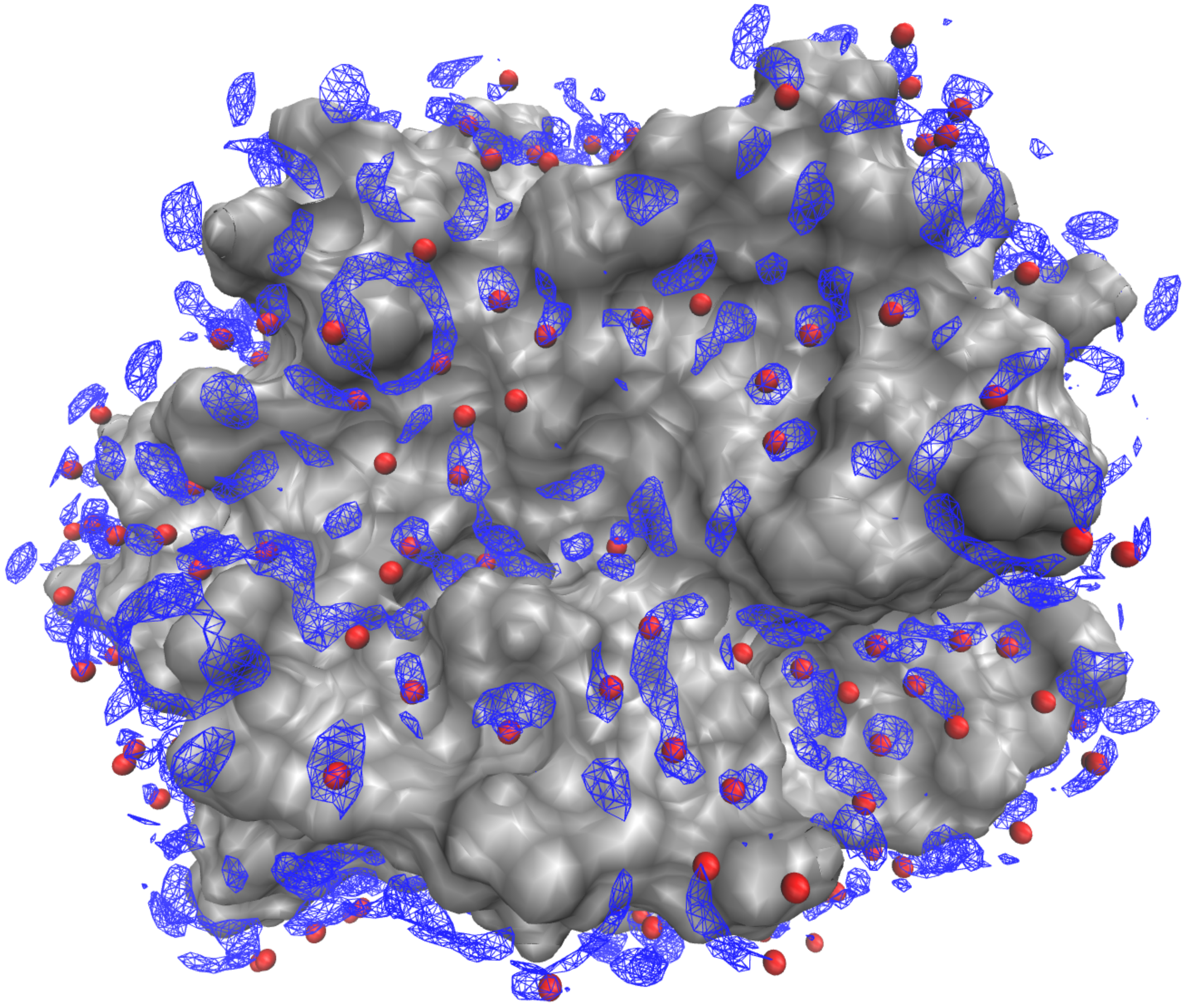}};
  \node[inner sep=0pt] (protein_mdft_water) at (12,-6)
      {\includegraphics[width=.25\textwidth]{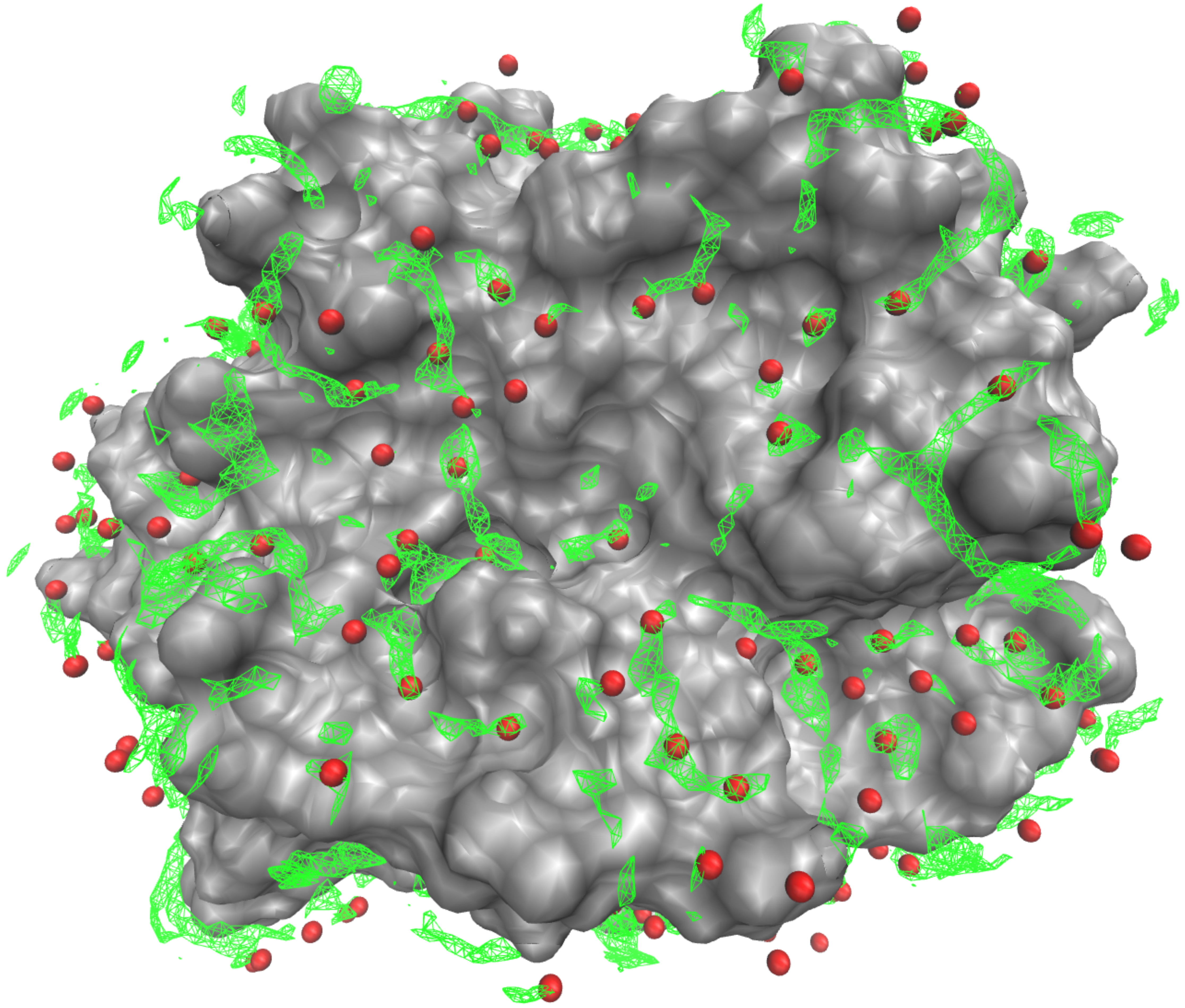}};

  \draw[line width=5pt,red] (7.5, 1.5)--(4.5,-1.5);
  \draw[line width=5pt,red] (7.5,-1.5)--(4.5, 1.5);

  \draw (8.5,0) node[right] {+};

  \node[draw] (MDFT) at (12,-3) {MDFT};
  \node[draw] (MD) at (6,-3) {MD};

  \draw[-latex] (protein_water) -- (protein_water_2);
  \draw[-latex] (protein_water) -- (water);
  
  \draw[-latex] (protein.south) -- (MDFT.north);
  \draw[-latex] (protein.south) |- (6,-2) -- (MD.north);
  
  \draw[-latex] (MDFT.south) -- (protein_mdft_water.north);
  \draw[-latex] (MD.south) -- (protein_gromacs_water.north);
  
  \end{tikzpicture}
      \caption[Protocole de détection des molécules d'eau autour de la protéine 4M7G.]{Protocole de détection des molécules d'eau autour de la protéine 4M7G. La protéine est représentée en surface. Les sphères rouges correspondent aux molécules d'eau cristallographiques. Les zones de forte probabilité de présence proposées par dynamique moléculaire sont en bleu et celles proposées par MDFT sont en vert.}
      \label{fig:water_molecule_protocol}
\end{figure}

Afin de ne pas être influencé par les résultats expérimentaux, nous avons suivi le protocole (voir image \ref{fig:water_molecule_protocol}) suivant: La structure 3D de notre molécule a été téléchargée depuis le site de la PDB\footnote{\url{http://www.rcsb.org/pdb/explore.do?structureId=4m7g}} sous le code 4M7G. Ce fichier comporte la structure 3D de la protéine ainsi que d'éventuels ions ou molécules d'eau détectés expérimentalement. Les molécules d'eau ont été supprimées du fichier pdb afin de lancer les calculs MDFT et DM. Il n'y a donc à ce stade plus aucune trace de molécules d'eau expérimentales. Notre système est ensuite préparé en utilisant le champ de force OPLS/AA\cite{jorgensen_opls_1988} et le modèle d'eau SPC/E\cite{berendsen_missing_1987}. Pour chacune des méthodes (MDFT et DM), nous avons laissé un espace de 10 \AA\ entre les bords de notre boîte de simulation et le bord de notre protéine. Chaque simulation a été effectuée à 298,15 K.

\subsubsection{Dynamique moléculaire}
Un calcul de référence a été lancé en dynamique moléculaire en utilisant le logiciel Gromacs\cite{berendsen_gromacs:_1995}. Après avoir solvaté nos protéines dans de l'eau SPC/E, l'énergie interne du système est minimisée, d'abord par \textit{steepest descent} puis par gradient conjugué. Nous supprimons ainsi tous les \textit{clash} stériques créés lors des étapes de préparation du système et de solvatation. Afin de permettre la comparaison des résultats avec ceux proposés par MDFT, la protéine est rendue rigide en utilisant l'option \textit{freezegrps} proposée par gromacs. Lorsque cette option est activée, les forces appliquées aux atomes de la protéines sont ignorées. Le système est ensuite équilibré à 298,15K et 1 bar en utilisant le barostat Berendsen et le thermostat \textit{V-rescale}. Une fois le système proche de l'équilibre, nous changeons le barostat pour Parinello-Rahman qui est plus précis mais diverge si le système est trop éloigné de l'équilibre. Une fois le système minimisé, nous lançons une simulation NPT de 100 ns. La configuration du système est sauvegardée toutes les 10 ps, nous obtenons ainsi un ensemble 10 000 conformations. Sur 32 coeurs OpenMP/MPI, couplés à deux GPU, une simulation nécessite deux jours de calcul.

\subsubsection{MDFT}
Comme nous l'avons montré dans les chapitres précédents, avec notre bridge gros grain, $\mathrm{m}_\mathrm{max}$=3 est suffisant pour améliorer fortement la prédiction de structures de solvatation autour de petites molécules. Nous avons cependant choisi de lancer la simulation avec $\mathrm{m}_\mathrm{max}$=5 afin que cet exemple serve en même temps de test de robustesse de la nouvelle implémentation de MDFT. En effet, avec un espacement entre chaque point de grille de 0,5 \AA, MDFT minimise plus de $2\mathrm{e}^9$ de variables. Sur 16 cœurs OpenMP, les résultats sont obtenus en seulement 17 min.

\subsubsection{Conversion xtc en cube}
Pour comparer ces deux méthodes, nous avons développé un logiciel permettant de convertir une trajectoire XTC, fournie par gromacs, en une représentation statistique 3D au format CUBE. Ce logiciel, librement accessible sur github\footnote{\url{https://github.com/cgageat/xtc2Cube}}, découpe l'espace sous forme d'une grille ayant les même paramètres que celle utilisée par MDFT (ici 0,5 \AA\ de maille) puis, compte, pour chaque étape de la simulation, le nombre de molécules présentes dans chaque voxel. Ce total est ensuite divisé par le nombre d'étapes de simulation puis par la taille d'un voxel (voir équation \ref{eq:equationXtcToCube}). Nous obtenons ainsi une probabilité de présence en eau à l'équilibre pour chaque voxel, $\mathrm{n}\left(\boldsymbol{r}\right)$, qui est la même que celle obtenue à l'issue d'un calcul MDFT, c'est à dire le g($\boldsymbol{r}$). Cela nous permet ainsi une comparaison directe de ces deux méthodes.

\begin{equation}\label{eq:equationXtcToCube}
\mathrm{n}\left(\boldsymbol{r}\right) = \frac{1}{NV_{\boldsymbol{r}}}\sum\limits_{i=1}^N n(\boldsymbol{r}, i)
\end{equation}

\subsection{Résultats}
\subsubsection{En surface}
Dans un premier temps, nous évaluons l'efficacité de notre bridge en comparant les zones de forte probabilité de présence prédites en utilisant MDFT dans l'approximation HNC (en jaune) puis avec notre bridge (en vert) à notre référence calculée par dynamique moléculaire (en bleu). Comme nous le voyons sur la figure \ref{fig:surface_protein_hnc}, l'approximation HNC surestime fortement la densité ou la probabilité de présence d'une molécule d'eau à de nombreux endroits. Notre bridge corrige cet effet et produit une carte de densité ayant une bien meilleure adéquation avec celle fournie par notre référence. Nous rappelons au lecteur que MDFT est 1 000 fois plus rapide que la dynamique moléculaire pour des calculs de structure de solvatation.

\begin{center}
    \captionsetup{type=figure}
	\fbox{\includegraphics[width=\textwidth]{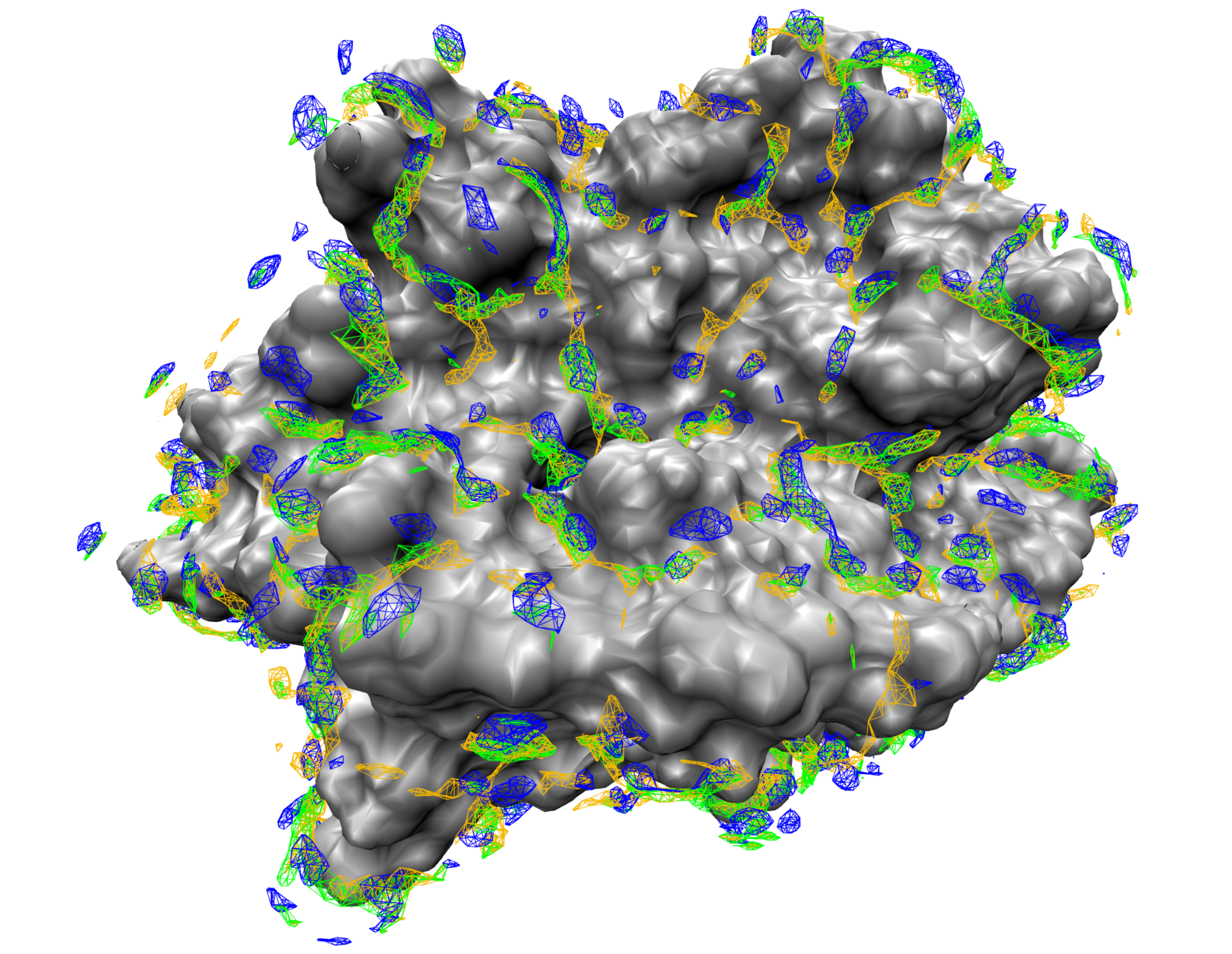}}
	\captionof{figure}[Zones de forte probabilité de présence de l'eau autour de la surface de la protéine 4M7G.]{Zones de forte probabilité de présence de l'eau autour de la surface de la protéine 4M7G. En jaune, les zones prédites par MDFT dans l'approximation HNC et en vert MDFT avec notre nouveau bridge. La référence, en bleu, est calculée par dynamique moléculaire.}
      \label{fig:surface_protein_hnc}
\end{center}

Dans un second temps, nous comparons la position de ces zones de forte probabilité de présence à la position des molécules d'eau expérimentales. On voit sur l'image \ref{fig:surface_water_molecule} que la majorité des molécules d'eau se situent dans les zones de forte probabilité de présence proposées à la fois en dynamique moléculaire et par MDFT avec notre nouveau bridge. Cependant, certaines molécules ne sont retrouvées ni en dynamique moléculaire ni par MDFT. Cette différence peut venir des conditions expérimentales indispensables à la cristallisation et donc à la résolution de la structure 3D. En effet, les résolutions cristallographiques ne sont possibles qu'après une forte modification du système (agent de précipitation, pH, température) qui a pour effet de figer de nouvelles molécules d'eau mais également d'en libérer d'autres. Il est donc attendu qu'il y ait une différence entre les résultats expérimentaux et théoriques.

\begin{center}
    \captionsetup{type=figure}
	\fbox{\includegraphics[width=\textwidth]{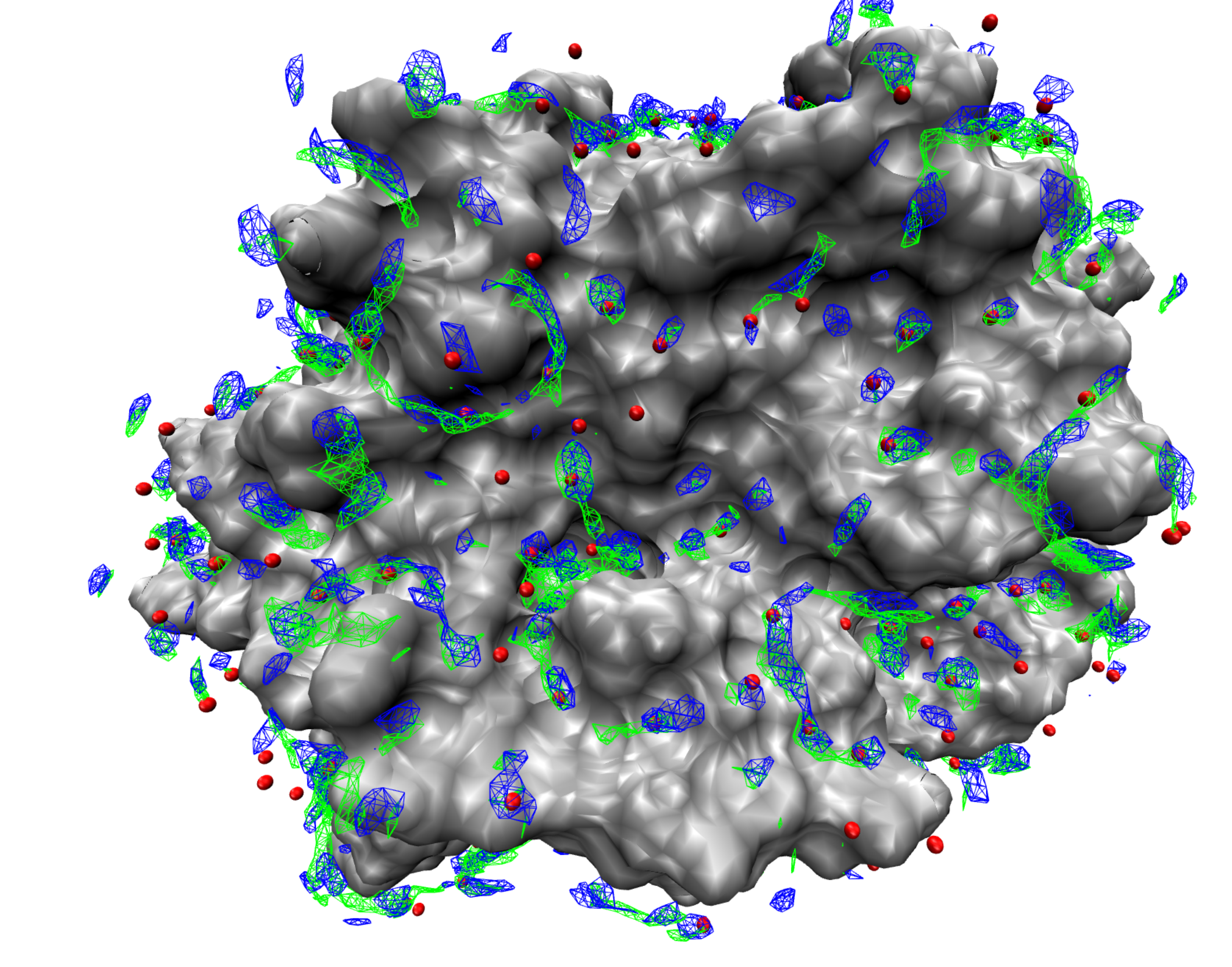}}
	\captionof{figure}[Comparaison des molécules d'eau cristallographiques et des résultats produits en dynamique moléculaire et par MDFT à la surface de la protéine 4M7G.]{Comparaison des molécules d'eau cristallographiques et des résultats produits en dynamique moléculaire et par MDFT avec notre nouveau bridge à la surface de la protéine 4M7G. Les sphères rouges correspondent aux oxygènes des molécules d'eau cristallographiques. Les zones de forte probabilité de présence proposées par dynamique moléculaire sont représentées en bleu et celles proposées par MDFT sont en vert.}
      \label{fig:surface_water_molecule}
\end{center}

\subsubsection{A l'intérieur de la protéine}
La dernière partie de cette étude consiste à étudier notre solvant à l'intérieur des poches de la protéine. Comme on le voit sur la figure \ref{fig:interieur_water_molecule}, MDFT avec notre bridge (en vert) est en accord parfait avec les résultats expérimentaux. MDFT dans l'approximation HNC (en jaune) prédit de nombreuses zones de forte densité qui ne correspondent à aucune molécule d'eau expérimentale comme c'est le cas en surface. De même, la dynamique moléculaire (en bleu) ne trouve pas deux molécules d'eau et en prédit plusieurs inexistantes. Ces résultats, pour la dynamique moléculaire, à l'intérieur de la protéine, s'expliquent par la dépendance des résultats de dynamique moléculaire aux conditions initiales. Durant les premières étapes d'une dynamique moléculaire le système est solvaté. À l'intérieur de la protéine, si une cavité est assez grande, des molécules d'eau y sont placées. En d'autres termes, si une cavité est anormalement grande dans cette configuration, une molécule y sera placée et restera bloquée tout au long de la simulation. A contrario, si une cavité est anormalement petite, aucune molécule d'eau n'y sera placée. Pour corriger cela, la protéine doit s'ouvrir durant la simulation, libérer ou absorber une molécule d'eau, puis se refermer. Ces phénomènes, de l'ordre de la seconde, ne sont pas accessibles en dynamique moléculaire. MDFT propose des résultats directement à l'équilibre et permet donc une meilleure prédiction à l'intérieur des systèmes biologiques.

\clearpage
\begin{center}
    \captionsetup{type=figure}
	\fbox{\includegraphics[width=\textwidth]{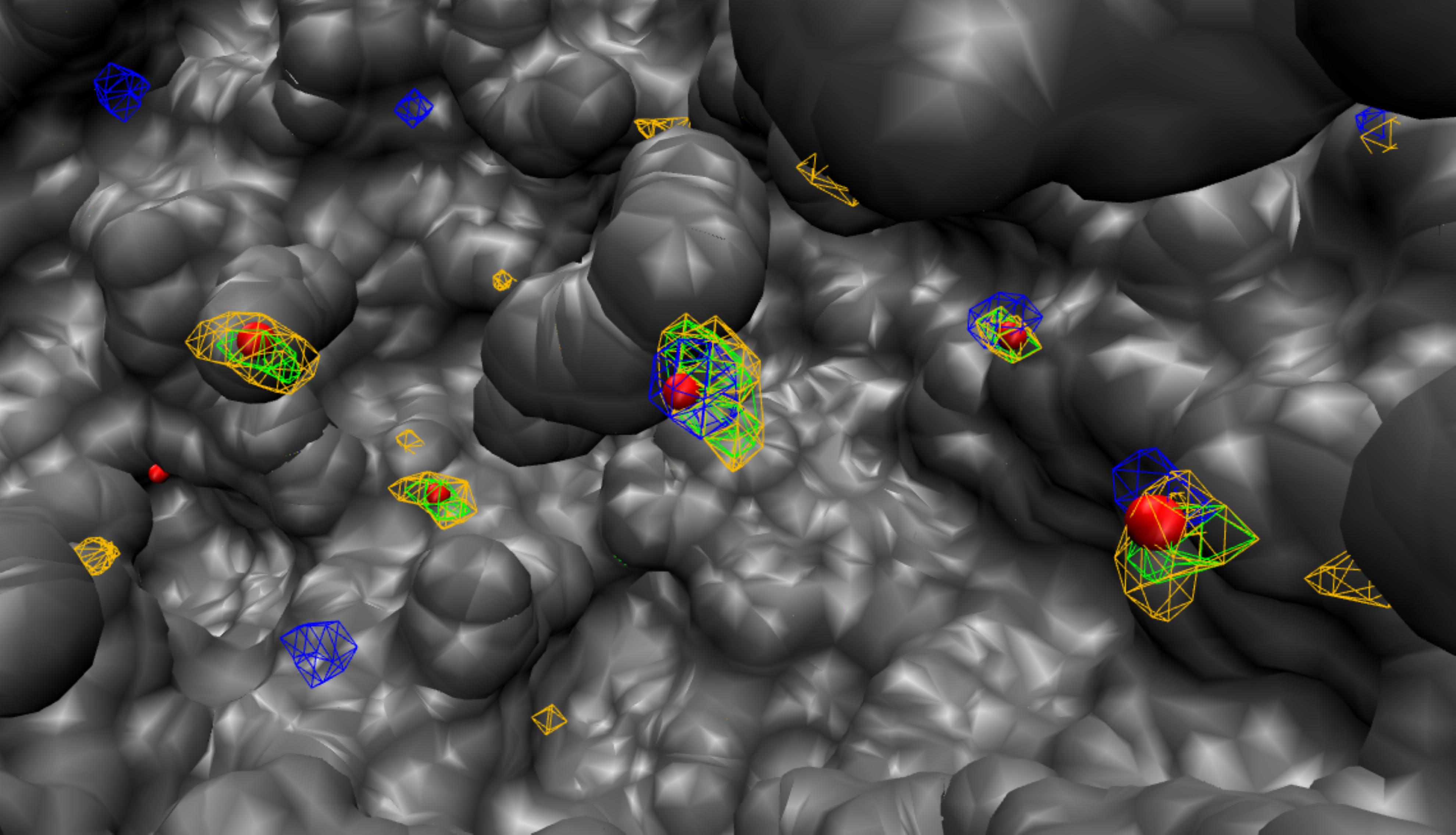}}
	\captionof{figure}[Comparaison des molécules d'eau cristallographiques et des résultats produits en dynamique moléculaire et par MDFT à l'intérieur de 4M7G.]{Comparaison des molécules d'eau cristallographiques et des résultats produits en dynamique moléculaire et par MDFT à l'intérieur de 4M7G. La protéine est représentée en surface. Les sphères rouges correspondent aux molécules d'eau cristallographiques. Les zones de forte probabilité de présence proposées par dynamique moléculaire sont en bleu et celles proposées par MDFT sont en vert.}
      \label{fig:interieur_water_molecule}
\end{center}

\clearpage
\section{application 2: MM-MDFT pour remplacer MM-PBSA}
Lors du développement d'un nouveau médicament, les calculs d'énergie libre de liaison sont indispensables et interviennent à différentes étapes du procédé. Ils permettent par exemple d'évaluer l'affinité d'un petit composé, candidat médicament, avec (i) sa cible thérapeutique et ainsi d'évaluer son efficacité et (ii) avec des cibles secondaires et ainsi d'évaluer le risque d'effets secondaires. Durant les premières étapes, un premier tri large est effectué et la rapidité est favorisée à la précision. La méthode de choix est donc MM-PBSA\cite{genheden_mm/pbsa_2015}. Cette méthode implicite permet une évaluation quasi-instantanée de l'énergie libre de liaison en échange de fortes approximations. Un solvant implicite ne permet par exemple pas de prendre en compte les effets stériques des molécules d'eau ou encore les liaisons hydrogènes qu'elles pourraient former avec le complexe solvaté. Dans cette partie, nous remplaçons MM-PBSA par méthode MM-MDFT. Les résultats MM-PBSA de 46 complexes protéine-ligand, récemment publiés par Chéron et al.\cite{cheron_effect_2017}, sont comparés à ceux obtenus par MM-MDFT.

\subsection{Théorie}
\begin{figure}[ht]
  \center
  \begin{tikzpicture}
  
   \draw (0,0,0)--(3,0,0)--(3,3,0)--(0,3,0)--cycle;
   \draw (4,0,0)--(7,0,0)--(7,3,0)--(4,3,0)--cycle;
   \draw (11,0,0)--(14,0,0)--(14,3,0)--(11,3,0)--cycle;
   \draw[fill=white!70!cyan] (0,6,0)--(3,6,0)--(3,9,0)--(0,9,0)--cycle;
   \draw[fill=white!70!cyan] (4,6,0)--(7,6,0)--(7,9,0)--(4,9,0)--cycle;
   \draw[fill=white!70!cyan] (11,6,0)--(14,6,0)--(14,9,0)--(11,9,0)--cycle;

   \draw[fill=black] (5.4, 1.5) circle (1);
   \draw[fill=black] (12.4,1.5) circle (1);
   \draw[fill=black] (5.4, 7.5) circle (1);
   \draw[fill=black] (12.4,7.5) circle (1);
   \draw[gray,fill=gray] (2.1, 1.5) circle (0.5);
   \draw[white,fill=white] (6.1, 1.5) circle (0.5);
   \draw[gray,fill=gray] (13.1,1.5) circle (0.5);
   \draw[gray,fill=gray] (2.1, 7.5) circle (0.5);
   \draw[white!70!cyan,fill=white!70!cyan] (6.1, 7.5) circle (0.5);
   \draw[gray,fill=gray] (13.1,7.5) circle (0.5);

   \draw[line width=5pt, -latex] (7.5, 1.5) -- (10.5,1.5);
   \draw (8.8,1.4) node[below]{$\Delta G_{\mathrm{liaison}}\ \mathrm{vide}$} ;
   \draw[line width=5pt, -latex] (7.5, 7.5) -- (10.5,7.5);
   \draw (8.8,7.4) node[below]{$\Delta G_{\mathrm{liaison}}\ \mathrm{eau}$} ;

   \draw[line width=5pt, -latex] (1.5, 3.5) -- (1.5,5.5);
   \draw (1.5,4.3) node[right]{$\Delta G_{\mathrm{solv}}\ \mathrm{ligand}$} ;
   \draw[line width=5pt, -latex] (5.5, 3.5) -- (5.5,5.5);
   \draw (5.5,4.3) node[right]{$\Delta G_{\mathrm{solv}}\ $protéine} ;
   \draw[line width=5pt, -latex] (12.5, 3.5) -- (12.5,5.5);
   \draw (12.5,4.3) node[right]{$\Delta G_{\mathrm{solv}}\ \mathrm{complexe}$} ;

  \end{tikzpicture}
    \caption{Cycle thermodynamique utilisé dans le calcul de l'énergie libre de liaison entre une protéine et un ligand par MM-PBSA et MM-MDFT.}
    \label{fig:Energie_libre_de_binding}
\end{figure}
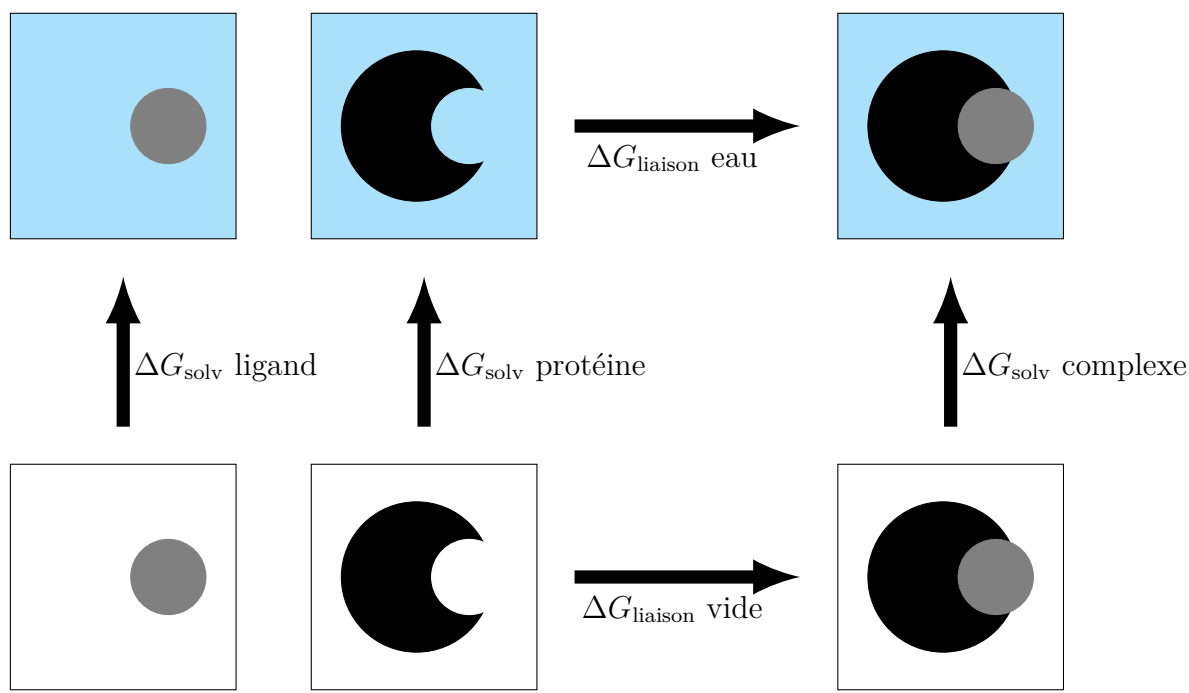

Par définition, le calcul exact d'une énergie libre de liaison, implique un échantillonnage complet du système étudié. Dans le cas d'un complexe rigide protéine-ligand, ce calcul implique l'échantillonnage de toutes les conformations du solvant. Afin de simplifier et d’accélérer les calculs, il est d'usage d'appliquer le cycle thermodynamique (voir figure \ref{fig:Energie_libre_de_binding}) suivant: dans un premier temps, la protéine et le ligand sont désolvatés. Une fois dans le vide, le calcul de l'énergie libre de liaison correspond à la somme des termes inter et intra-moléculaires moyennés sur l'ensemble des conformations possibles. Dans le cas de molécules rigides, les énergies internes des deux composés ne sont pas modifiées lors de la liaison et peuvent donc être ignorées. L'énergie libre de liaison correspond ainsi à la somme des termes intra-moléculaires. Enfin, le complexe nouvellement formé est de nouveau solvaté. Comme nous l'avons détaillé dans le chapitre \ref{chap:introduction}, il existe différentes méthodes permettant le calcul de l'énergie libre de solvatation. Les systèmes biologiques ne permettent pas, de par leur taille, d'utiliser des méthodes explicites telles que l'intégration thermodynamique associées à des simulations de dynamique moléculaire ou de Monte Carlo. Les méthodes implicites telles que \textit{Poisson-Boltzman and Surface-Area} (PBSA) sont donc favorisées.

\subsubsection{MM-PBSA}
MM-PBSA est une méthode basée sur le cycle thermodynamique décrit ci-dessus, qui permet d'estimer, en quelques secondes seulement, l'énergie libre de liaison entre deux molécules dans un solvant implicite. Les énergies libres de solvatation sont calculées par PBSA et l'énergie libre de liaison est calculée par \textit{Molecular Mechanics}. PBSA est basée sur la résolution de l'équation de Poisson-Boltzmann (PB) pour les charges et sur le calcul de la surface accessible au solvant (SA). L’énergie libre de solvatation correspond à la somme de ces deux termes.

\subsubsection{MM-MDFT}
MDFT calcule les énergies libres de solvatation, ce qui nous permet de remplacer les calculs PBSA par des calculs MDFT et ainsi de décliner MM-PBSA en MM-MDFT. Afin de rester comparable, nous avons fait le choix d'une approximation de la théorie MDFT, qui nous propose des résultats dans des temps équivalents à ceux fournis par PBSA, soit de l'ordre de la seconde. Nous avons donc fixé $\mathrm{m}_\mathrm{max}=1$, et testé les corrections de pression PC et PC+ ainsi que le bridge gros grain. Notre objectif ici n'est pas d'être plus rapide que MM-PBSA mais d'ajouter de la précision aux résultats tout en apportant des informations supplémentaires au travers de la structure de solvatation.

\subsection{Les systèmes étudiés}
Dans une étude récente, Chéron et al.\cite{cheron_effect_2017} ont optimisé le calcul MM-PBSA sur un ensemble de 46 complexes protéine-ligand. Afin de créer cette chimiothèque, l'ensemble des 264 complexes impliquant la protéine BACE1 (voir figure \ref{fig:structure_proteine_MMPBSA}) ont été extrait de la base de données PDBBind-CN. Parmi ces 264, seuls 46 étaient de qualité suffisante (résolution < 2,5 \AA) et accompagnés de leur valeur d'énergie libre de solvatation. Dans leur article, Chéron et al. montrent que les calculs MM-PBSA sur ces complexes sont optimums en utilisant le champ de force Amber03 pour la protéine, le modèle d'eau TIP3P et en laissant un espace supérieur à 10 \AA\ entre les molécules le composant et le bord des boîtes de simulation. Les valeurs d'énergies libres de solvatation expérimentales et calculées par MM-PBSA présentées dans la suite de ce chapitre ont toutes été fournies par Nicolas Chéron.

Le code PDB ainsi que les charges des différents complexes sont listés dans le tableau \ref{tab:MMPBSA_systemes}. Une représentation 2D des ligands de chaque complexe est également disponible en figure \ref{fig:structures_MMPBSA}.

\begin{center}
\begin{longtable}{c c c c c}

\hline
Code &  \multicolumn{3}{c}{Charge} & $\mathrm{N}_\mathrm{atomes}$ \\
PDB  & complex & protein & ligand  & Ligand \\ \hline 
\endfirsthead

\multicolumn{5}{c}%
{{\tablename\ \thetable{} -- suite}} \\
\hline Code PDB &  \multicolumn{3}{c}{Charge} & $\mathrm{N}_\mathrm{atomes} \mathrm{Ligand}$ \\
               & complex & protein & ligand  & \\ \hline 
\endhead

\hline \multicolumn{5}{r}{{Suite page suivante}} \\ \hline
\endfoot

\hline \hline
\caption[Description des systèmes utilisés dans l'étude MM-MDFT.]{Description des systèmes utilisés dans l'étude MM-MDFT. Pour chaque système la charge du complexe, de la protéine et du ligand seuls sont indiqués ainsi que le nombre d'atomes composant le ligand.} \label{tab:MMPBSA_systemes} \\
\endlastfoot

      \hline & \\[-1em]\hline
      
      \hline
      1FKN & -10,0 &  -8,0 & -2,0 & 125 \\
      1M4H & -11,0 &  -8,0 & -3,0 & 132 \\
      2FDP &  -9,0 & -10,0 &  1,0 &  83 \\
      2G94 & -10,0 & -10,0 &  0,0 &  99 \\
      2P4J & -10,0 & -10,0 &  0,0 & 104 \\
      2Q11 & -10,0 & -10,0 &  0,0 &  61 \\
      2Q15 &  -9,0 & -10,0 &  1,0 &  78 \\
      2QMG &  -9,0 & -10,0 &  1,0 &  84 \\
      3BRA &  -9,0 & -10,0 &  1,0 &  22 \\
      3BUF &  -9,0 & -10,0 &  1,0 &  25 \\
      3BUG & -10,0 & -11,0 &  1,0 &  28 \\
      3BUH &  -9,0 & -10,0 &  1,0 &  38 \\
      3CKP &  -9,0 & -10,0 &  1,0 &  80 \\
      3I25 & -10,0 & -10,0 &  0,0 & 118 \\
      3KMX &  -9,0 &  -9,0 &  0,0 &  34 \\
      3KMY &  -9,0 &  -9,0 &  0,0 &  29 \\
      3L59 & -14,0 & -14,0 &  0,0 &  31 \\
      3L5B & -10,0 & -10,0 &  0,0 &  40 \\
      3LPI &  -9,0 & -10,0 &  1,0 &  89 \\
      3LPK &  -9,0 & -10,0 &  1,0 &  91 \\
      3RSX & -10,0 & -10,0 &  0,0 &  26 \\
      3RU1 & -10,0 & -10,0 &  0,0 &  48 \\
      3UDH & -10,0 & -11,0 &  1,0 &  27 \\
      3WB4 & -10,0 & -10,0 &  0,0 &  37 \\
      3WB5 & -10,0 & -10,0 &  0,0 &  38 \\
      4B05 &  -9,0 &  -9,0 &  0,0 &  48 \\
      4DJU & -10,0 & -10,0 &  0,0 &  35 \\
      4DJV & -10,0 & -10,0 &  0,0 &  49 \\
      4DJW & -10,0 & -10,0 &  0,0 &  44 \\
      4DJX & -10,0 & -10,0 &  0,0 &  41 \\
      4DJY & -10,0 & -10,0 &  0,0 &  46 \\
      4FRS &  -9,0 &  -9,0 &  0,0 &  42 \\
      4FS4 & -10,0 & -10,0 &  0,0 &  45 \\
      4FSL & -10,0 & -10,0 &  0,0 &  57 \\
      4GID &  -9,0 & -10,0 &  1,0 &  95 \\
      4H1E &  -9,0 &  -9,0 &  0,0 &  54 \\
      4H3F & -10,0 & -10,0 &  0,0 &  55 \\
      4H3G & -10,0 & -10,0 &  0,0 &  52 \\
      4H3I & -10,0 & -10,0 &  0,0 &  55 \\
      4H3J & -10,0 & -10,0 &  0,0 &  52 \\
      4HA5 & -10,0 & -10,0 &  0,0 &  39 \\
      4R8Y &  -9,0 &  -9,0 &  0,0 &  70 \\
      4R91 &  -9,0 & -10,0 &  1,0 &  72 \\
      4R92 &  -9,0 &  -9,0 &  0,0 &  69 \\
      4R93 & -10,0 & -10,0 &  0,0 &  72 \\
      4R95 & -10,0 & -10,0 &  0,0 &  73 \\
\end{longtable}
\end{center}

\begin{center}
    \captionsetup{type=figure}
	\fbox{\includegraphics[width=0.8\textwidth]{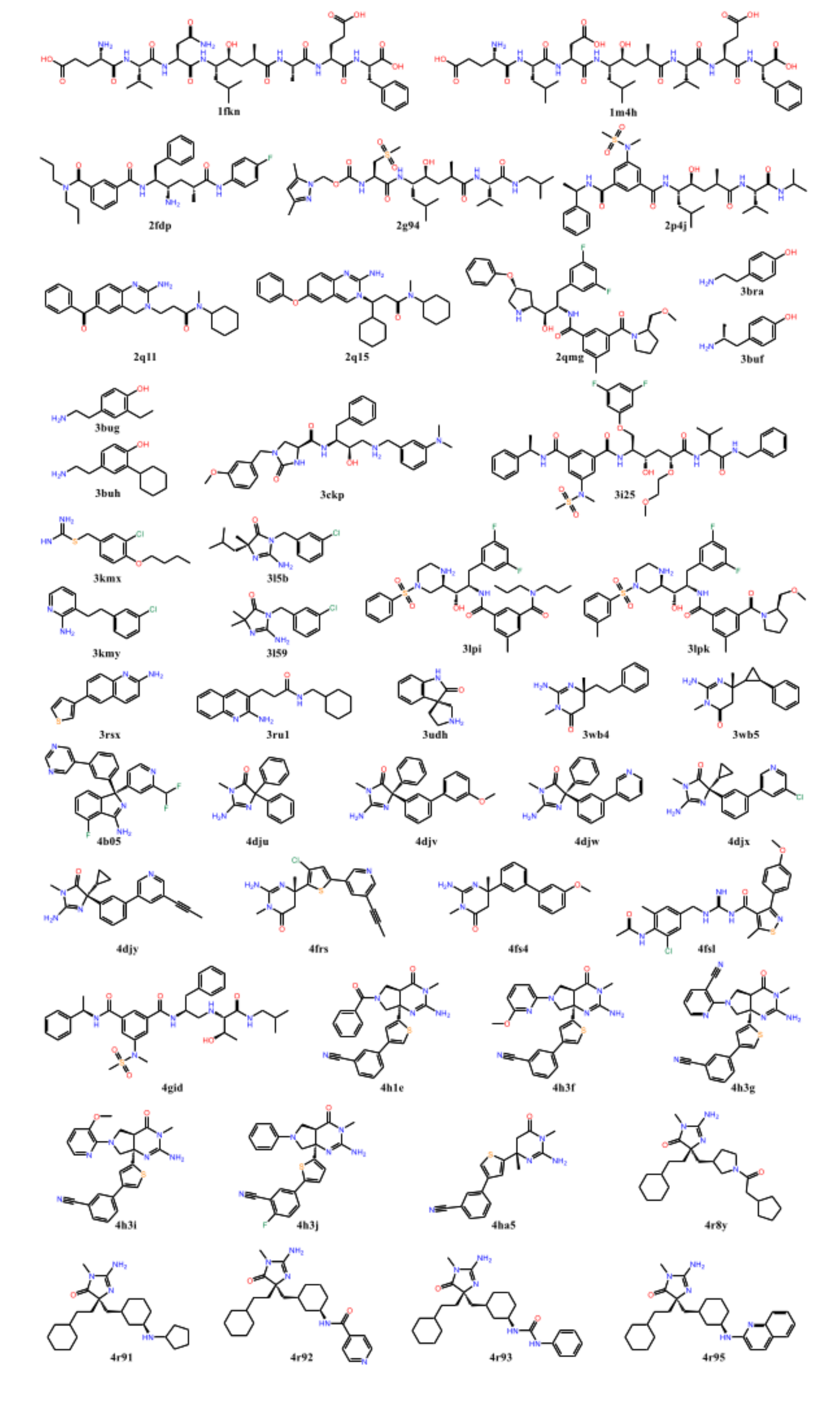}}
	\captionof{figure}[Structure 2D des ligands de chaque complexe utilisés dans l'étude MM-MDFT.]{Structure 2D des ligands de chaque complexe étudié.}
    \label{fig:structures_MMPBSA}
\end{center}

\begin{center}
    \captionsetup{type=figure}
	\fbox{\includegraphics[width=0.75\textwidth]{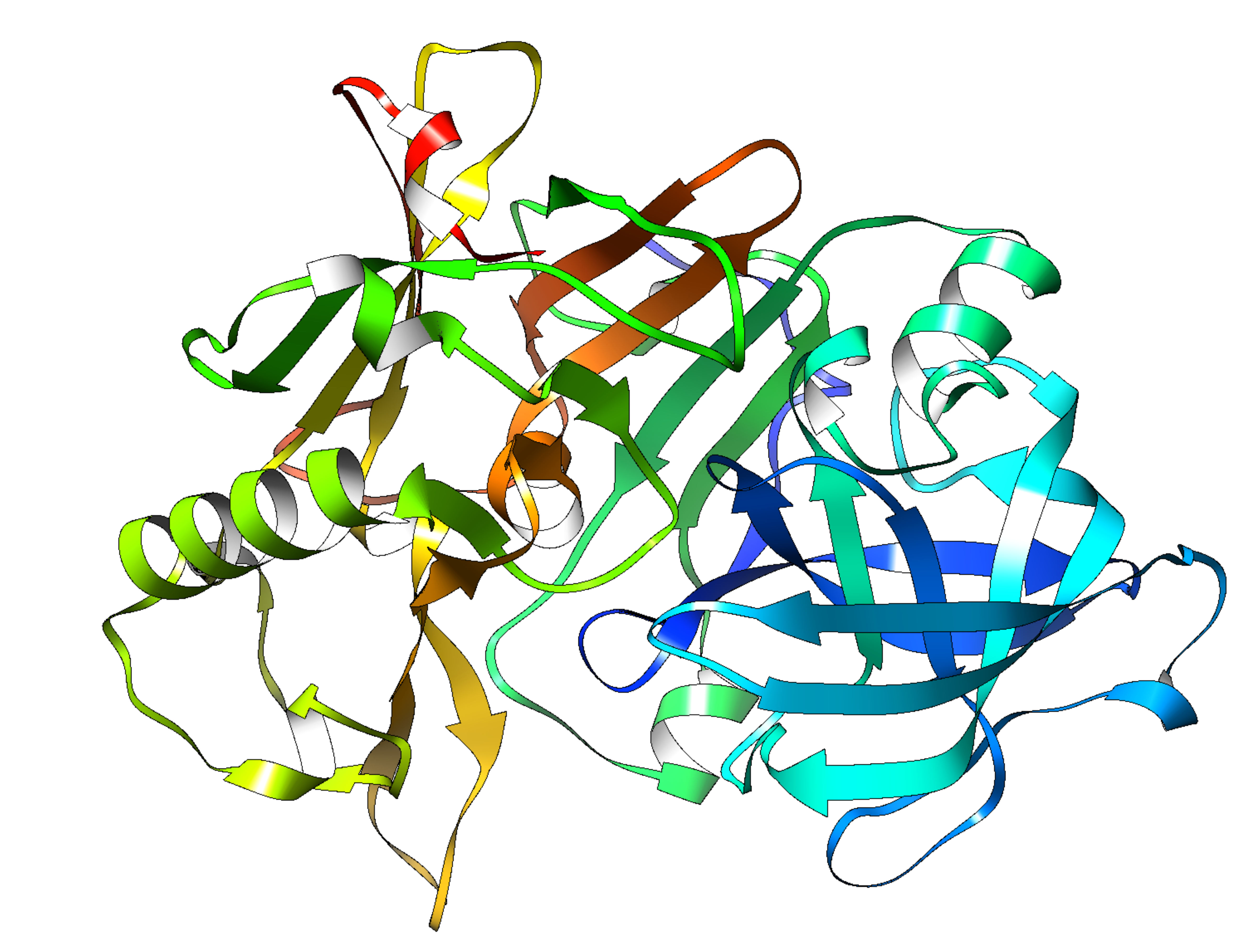}}
	\captionof{figure}[Structure 3D de la protéine BACE1.]{Structure 3D de la protéine BACE1 commune à tous les complexes étudiés.}
    \label{fig:structure_proteine_MMPBSA}
\end{center}

\subsection{Résultats}

\begin{figure}[ht]
    \begin{subfigure}[b]{0.50\textwidth}
        \centering
        \caption{MM-PBSA}
        \resizebox{\linewidth}{!}{
          \begin{tikzpicture}
            \begin{axis}[
                    xlabel= $\Delta \mathrm{G_{liaison}\ exp}\ (\mathrm{kCal.mol}^{-1}$),
                    ylabel= $\Delta \mathrm{G_{liaison}\ calc}\ (\mathrm{kCal.mol}^{-1}$),
                    legend style = {draw = none},
                    xmin = -16, xmax = -2, ymin = -200, ymax = 20,
                    legend entries={$\mathrm{R}^2$=0,66},
                 legend pos=south east]
                 \addlegendimage{only marks,white};
              
              \addplot[only marks,mark=*, black, very thick] table[x index=0, y index=1]{chapters/Applications/datas/MMPBSA/results.csv};
            \end{axis}
          \end{tikzpicture}
}
     \end{subfigure}%
     \begin{subfigure}[b]{0.50\textwidth}
        \centering
        \caption{$\mathrm{m}_\mathrm{max}$ 1 PC}
        \resizebox{\linewidth}{!}{
          \begin{tikzpicture}
            \begin{axis}[
                    xlabel= $\Delta \mathrm{G_{liaison}\ exp}\ (\mathrm{kCal.mol}^{-1}$),
                    ylabel= $\Delta \mathrm{G_{liaison}\ calc}\ (\mathrm{kCal.mol}^{-1}$),
                    legend style = {draw = none},
                    xmin = -16, xmax = -2, ymin = -200, ymax = 20,
              legend entries={$\mathrm{R}^2$=0,61},
                 legend pos=south east]
                 \addlegendimage{only marks,white};
              \addplot[only marks,mark=*, black, very thick] table[x index=0, y index=2]{chapters/Applications/datas/MMPBSA/results.csv};
            \end{axis}
          \end{tikzpicture}
}
     \end{subfigure}
     \begin{subfigure}[b]{0.50\textwidth}
        \centering
        \caption{$\mathrm{m}_\mathrm{max}$ 1 PC+}
        \resizebox{\linewidth}{!}{
          \begin{tikzpicture}
            \begin{axis}[
                    xlabel= $\Delta \mathrm{G_{liaison}\ exp}\ (\mathrm{kCal.mol}^{-1}$),
                    ylabel= $\Delta \mathrm{G_{liaison}\ calc}\ (\mathrm{kCal.mol}^{-1}$),
                    legend style = {draw = none},
                    xmin = -16, xmax = -2, ymin = -200, ymax = 20,
              legend entries={$\mathrm{R}^2$=0,62},
                 legend pos=south east]
                 \addlegendimage{only marks,white};
              \addplot[only marks,mark=*, black, very thick] table[x index=0, y index=3]{chapters/Applications/datas/MMPBSA/results.csv};
            \end{axis}
          \end{tikzpicture}
}
     \end{subfigure}%
     \begin{subfigure}[b]{0.50\textwidth}
        \centering
        \caption{$\mathrm{m}_\mathrm{max}$ 1 + bridge}
        \resizebox{\linewidth}{!}{
          \begin{tikzpicture}
            \begin{axis}[
                    xlabel= $\Delta \mathrm{G_{liaison}\ exp}\ (\mathrm{kCal.mol}^{-1}$),
                    ylabel= $\Delta \mathrm{G_{liaison}\ calc}\ (\mathrm{kCal.mol}^{-1}$),
                    legend style = {draw = none},
                    xmin = -16, xmax = -2, ymin = -200, ymax = 20,
              legend entries={$\mathrm{R}^2$=0,69},
                 legend pos=south east]
                 \addlegendimage{only marks,white};
              \addplot[only marks,mark=*, black, very thick] table[x index=0, y index=4]{chapters/Applications/datas/MMPBSA/results.csv};
            \end{axis}
          \end{tikzpicture}
}
     \end{subfigure}
    \caption[Énergie libre de liaison calculée par MM-PBSA et par MM-MDFT.]{Énergie libre de liaison calculée par MM-PBSA et par MM-MDFT. Les valeurs calculées sont comparées à l'énergie libre de liaison expérimentale pour chacun des complexes étudiés pour différents paramètres de la fonctionnelle.}
    \label{fig:MMPBSA_MMMDFT_complete}
\end{figure}
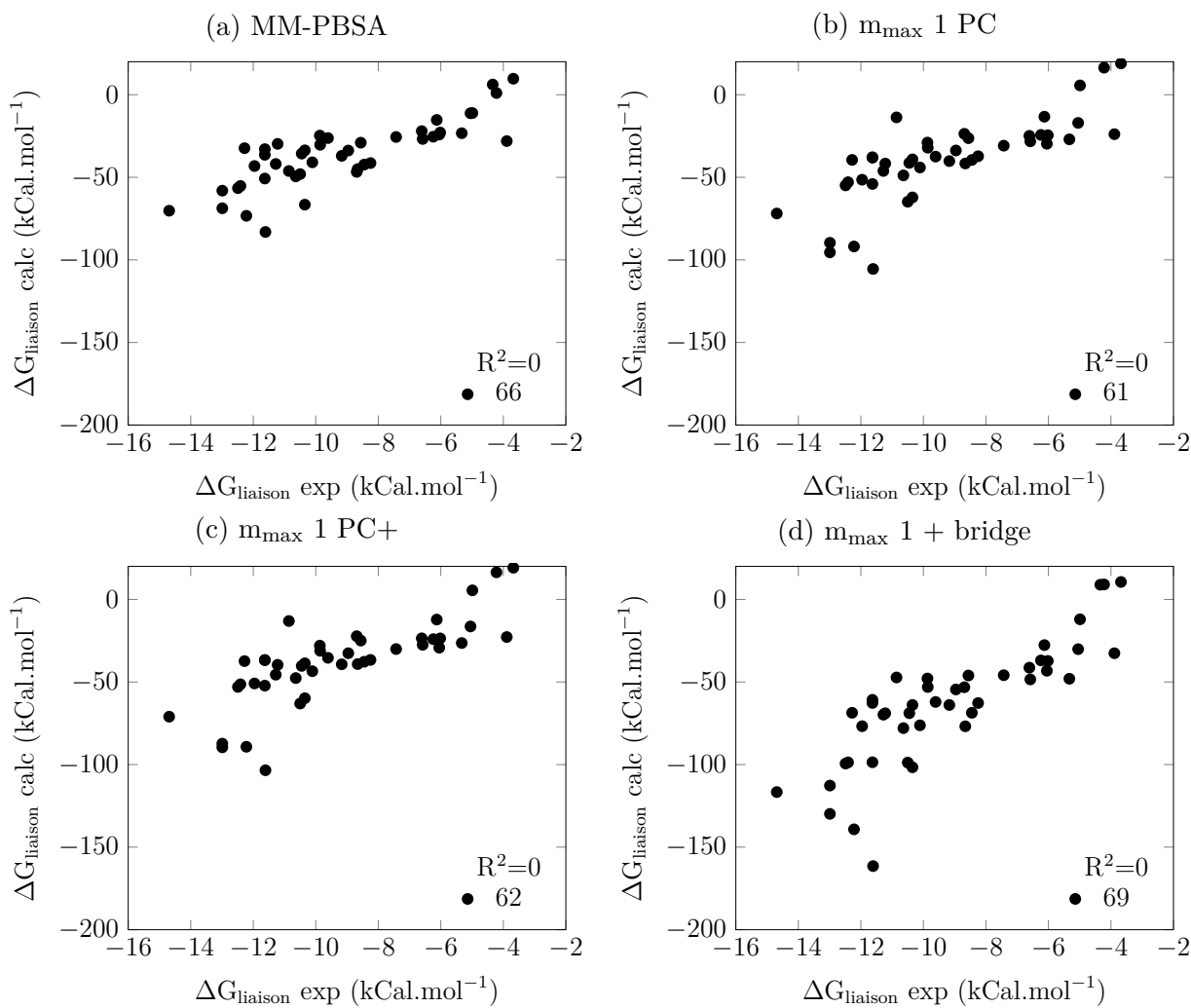

La figure \ref{fig:MMPBSA_MMMDFT_complete} compare les énergies libres de solvatation calculées (a) par MM-PBSA, (b) par MM-MDFT avec la correction de pression PC, (c) par MM-MDFT avec la correction de pression PC et (d) par MM-MDFT avec le bridge gros grain, aux valeurs expérimentales pour chacun des 46 complexes protéine-ligand étudiés. Pour chaque jeu de données, nous avons ensuite calculé le coefficient de corrélation $\mathrm{R}^2$ (voir tableau \ref{tab:MMPBSA_correlation}).

Les résultats obtenus avec MM-MDFT sans bridge, avec la correction de pression PC ou PC+ sont légèrement moins bons ($\mathrm{R}^2$=0,61 et $\mathrm{R}^2$=0,62) que ceux obtenus avec MM-PBSA ($\mathrm{R}^2$=0,66). Au contraire, les résultats obtenus avec $\mathrm{m}_\mathrm{max}=1$ et le bridge gros grain, sont quant à eux légèrement plus corrélés que notre référence MM-PBSA aux valeurs expérimentales. En plus de cette amélioration de la prédiction des valeurs d'énergie libre de liaison, dans le même temps, MM-MDFT nous propose les structures de solvatation de la protéine, du ligand ainsi que du complexe.

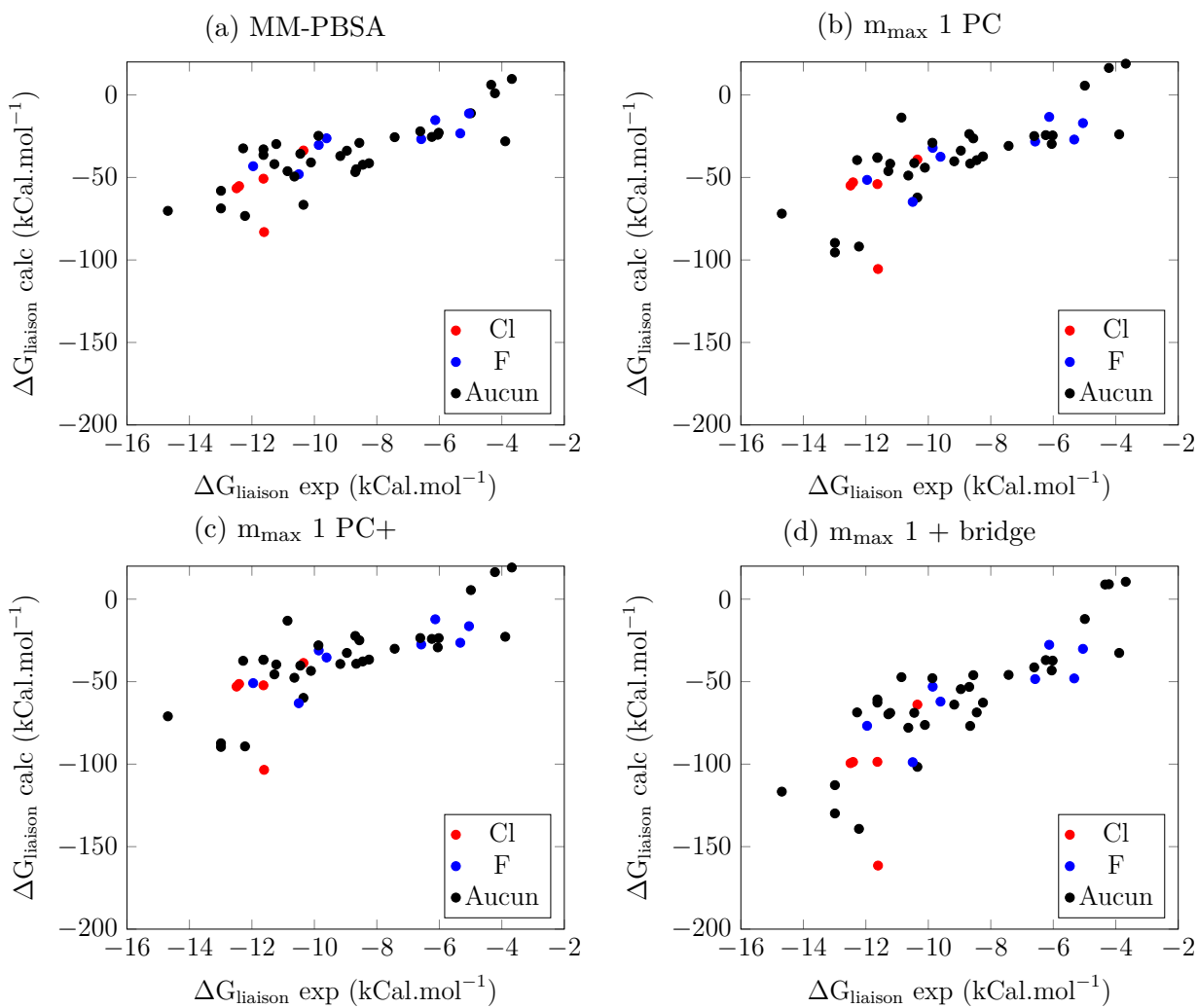
\begin{figure}[ht]
    \begin{subfigure}[b]{0.50\textwidth}
        \centering
        \caption{MM-PBSA}
        \resizebox{\linewidth}{!}{
          \begin{tikzpicture}
            \begin{axis}[
                    xlabel= $\Delta \mathrm{G_{liaison}\ exp}\ (\mathrm{kCal.mol}^{-1}$),
                    ylabel= $\Delta \mathrm{G_{liaison}\ calc}\ (\mathrm{kCal.mol}^{-1}$),
                    legend pos=south east,
                    xmin = -16, xmax = -2, ymin = -200, ymax = 20,
              ]
              \addplot[
        scatter,only marks,scatter src=explicit symbolic,
        scatter/classes={
            Fluor={red},
            Chlore={blue},
            None={black}
}
    ] table[x index=0, y index=1, meta index=10]{chapters/Applications/datas/MMPBSA/results.csv};
            \legend{Cl, F, Aucun,};
            \end{axis}
          \end{tikzpicture}
}
     \end{subfigure}
     \begin{subfigure}[b]{0.50\textwidth}
        \centering
        \caption{$\mathrm{m}_\mathrm{max}$ 1 PC}
        \resizebox{\linewidth}{!}{
          \begin{tikzpicture}
            \begin{axis}[
                    xlabel= $\Delta \mathrm{G_{liaison}\ exp}\ (\mathrm{kCal.mol}^{-1}$),
                    ylabel= $\Delta \mathrm{G_{liaison}\ calc}\ (\mathrm{kCal.mol}^{-1}$),
                    legend pos=south east,
                    xmin = -16, xmax = -2, ymin = -200, ymax = 20,
              ]
              \addplot[
        scatter,only marks,scatter src=explicit symbolic,
        scatter/classes={
            Fluor={red},
            Chlore={blue},
            None={black}
}
    ] table[x index=0, y index=2, meta index=10]{chapters/Applications/datas/MMPBSA/results.csv};
            \legend{Cl, F, Aucun,};
            \end{axis}
          \end{tikzpicture}
}
     \end{subfigure}
     \begin{subfigure}[b]{0.50\textwidth}
        \centering
        \caption{$\mathrm{m}_\mathrm{max}$ 1 PC+}
        \resizebox{\linewidth}{!}{
          \begin{tikzpicture}
            \begin{axis}[
                    xlabel= $\Delta \mathrm{G_{liaison}\ exp}\ (\mathrm{kCal.mol}^{-1}$),
                    ylabel= $\Delta \mathrm{G_{liaison}\ calc}\ (\mathrm{kCal.mol}^{-1}$),
                    legend pos=south east,
                    xmin = -16, xmax = -2, ymin = -200, ymax = 20,
              ]
              \addplot[
        scatter,only marks,scatter src=explicit symbolic,
        scatter/classes={
            Fluor={red},
            Chlore={blue},
            None={black}
}
    ] table[x index=0, y index=3, meta index=10]{chapters/Applications/datas/MMPBSA/results.csv};
            \legend{Cl, F, Aucun,};
            \end{axis}
          \end{tikzpicture}
}
     \end{subfigure}
     \begin{subfigure}[b]{0.50\textwidth}
        \centering
        \caption{$\mathrm{m}_\mathrm{max}$ 1 + bridge}
        \resizebox{\linewidth}{!}{
          \begin{tikzpicture}
            \begin{axis}[
                    xlabel= $\Delta \mathrm{G_{liaison}\ exp}\ (\mathrm{kCal.mol}^{-1}$),
                    ylabel= $\Delta \mathrm{G_{liaison}\ calc}\ (\mathrm{kCal.mol}^{-1}$),
                    legend pos=south east,
                    xmin = -16, xmax = -2, ymin = -200, ymax = 20,
              ]
              \addplot[
        scatter,only marks,scatter src=explicit symbolic,
        scatter/classes={
            Fluor={red},
            Chlore={blue},
            None={black}
}
    ] table[x index=0, y index=4, meta index=10]{chapters/Applications/datas/MMPBSA/results.csv};
            \legend{Cl, F, Aucun,};
            \end{axis}
          \end{tikzpicture}
}
     \end{subfigure}
    \caption[Énergie libre de liaison calculée par MM-PBSA et par MM-MDFT en fonction du type d'halogène.]{Énergie libre de liaison calculée par MM-PBSA et par MM-MDFT en fonction du type d'halogène. Les valeurs calculées sont comparées aux valeurs expérimentales pour différents paramètres de la fonctionnelle et pour chacun des complexes étudiés. Les complexes dont le ligand comporte au moins un Chlore sont représentés en rouge. Ceux qui comportent au moins un Fluor sont en bleu et ceux ne comportant pas d'halogène sont en noir.}
    \label{fig:MMPBSA_MMMDFT_complete_by_halogene}
\end{figure}

Les systèmes étudiés sont composés de nombreux halogènes et ne sont pas représentatifs de l'espace chimique. Afin de nous assurer que MDFT n'était pas biaisé par rapport à certains composés, nous avons dans un premier temps coloré les points en fonction des halogènes qu'il contenaient (voir figure \ref{fig:MMPBSA_MMMDFT_complete_by_halogene}). Dans un second temps nous avons coloré les composés contenant au moins un atome de Souffre. Dans les deux cas, les différentes catégories sont uniformément réparties dans le nuage de point. Le souffre ainsi que les différents halogènes présents sont donc correctement modélisés.

\subsection{Perspectives}
Dans cette étude préliminaire, nous n'avons considéré qu'une seule conformation par complexe, la conformation après minimisation de la structure cristallographique. À cause des conditions expérimentales nécessaires à la résolution de structures de systèmes biologiques en 3 dimensions, les conformations obtenues peuvent être éloignées de celles présentes en solution dans les conditions standards. Afin de minimiser ces différences, il est possible de calculer une trajectoire de dynamique moléculaire ou Monte Carlo du complexe dans le vide et d'en extraire différentes conformations à intervalles réguliers. 
L'énergie libre de solvatation du système correspond, dans ce cas, à la moyenne des énergies libres de solvatation calculées pour chaque conformation. Pour MM-PBSA, l'utilisation de cette méthode, appelée communément "l'approximation de trajectoire unique", permet d'améliorer la corrélation de MM-PBSA à $\mathrm{R}^2$=0,71. Au moment de la rédaction de ce rapport, des calculs similaires étaient en cours pour MM-MDFT.

\begin{table}[ht]
  \begin{center}
    \begin{tabular}{c c}
      \hline & \\[-1em]\hline
       fonctionnelle  & R²  \\
      \hline
       MM-PBSA  & 0,66 \\
       $\mathrm{m}_\mathrm{max}$ 1 PC  & 0,61  \\
       $\mathrm{m}_\mathrm{max}$ 1 PC+  & 0,62  \\
       $\mathrm{m}_\mathrm{max}$ 1 + bridge  & 0,69  \\
      \hline & \\[-1em]\hline%
    \end{tabular}
  \end{center}
  \caption{Coefficient de corrélation entre les valeurs d'énergie libre de liaisons calculées et les valeurs expérimentales.}
  \label{tab:MMPBSA_correlation}  
\end{table}

\clearpage

\begin{figure}[ht]
    \begin{subfigure}[b]{0.50\textwidth}
        \centering
        \caption{MM-PBSA}
        \resizebox{\linewidth}{!}{
          \begin{tikzpicture}
            \begin{axis}[
                    xlabel= $\Delta \mathrm{G_{liaison}\ exp}\ (\mathrm{kCal.mol}^{-1}$),
                    ylabel= $\Delta \mathrm{G_{liaison}\ calc}\ (\mathrm{kCal.mol}^{-1}$),
                    legend pos=south east,
                    xmin = -16, xmax = -2, ymin = -200, ymax = 20,
              ]
              \addplot[
        scatter,only marks,scatter src=explicit symbolic,
        scatter/classes={
            Sulfure={red},
            None={black}
}
    ] table[x index=0, y index=1, meta index=11]{chapters/Applications/datas/MMPBSA/results.csv};
            \legend{Avec S, Sans S};
            \end{axis}
          \end{tikzpicture}
}
     \end{subfigure}
     \begin{subfigure}[b]{0.50\textwidth}
        \centering
        \caption{$\mathrm{m}_\mathrm{max}$ 1 PC}
        \resizebox{\linewidth}{!}{
          \begin{tikzpicture}
            \begin{axis}[
                    xlabel= $\Delta \mathrm{G_{liaison}\ exp}\ (\mathrm{kCal.mol}^{-1}$),
                    ylabel= $\Delta \mathrm{G_{liaison}\ calc}\ (\mathrm{kCal.mol}^{-1}$),
                    legend pos=south east,
                    xmin = -16, xmax = -2, ymin = -200, ymax = 20,
              ]
              \addplot[
        scatter,only marks,scatter src=explicit symbolic,
        scatter/classes={
            Sulfure={red},
            None={black}
}
    ] table[x index=0, y index=2, meta index=11]{chapters/Applications/datas/MMPBSA/results.csv};
            \legend{Avec S, Sans S};
            \end{axis}
          \end{tikzpicture}
}
     \end{subfigure}
     \begin{subfigure}[b]{0.50\textwidth}
        \centering
        \caption{$\mathrm{m}_\mathrm{max}$ 1 PC+}
        \resizebox{\linewidth}{!}{
          \begin{tikzpicture}
            \begin{axis}[
                    xlabel= $\Delta \mathrm{G_{liaison}\ exp}\ (\mathrm{kCal.mol}^{-1}$),
                    ylabel= $\Delta \mathrm{G_{liaison}\ calc}\ (\mathrm{kCal.mol}^{-1}$),
                    legend pos=south east,
                    xmin = -16, xmax = -2, ymin = -200, ymax = 20,
              ]
              \addplot[
        scatter,only marks,scatter src=explicit symbolic,
        scatter/classes={
            Sulfure={red},
            None={black}
}
    ] table[x index=0, y index=3, meta index=11]{chapters/Applications/datas/MMPBSA/results.csv};
            \legend{Avec S, Sans S};
            \end{axis}
          \end{tikzpicture}
}
     \end{subfigure}
     \begin{subfigure}[b]{0.50\textwidth}
        \centering
        \caption{$\mathrm{m}_\mathrm{max}$ 1 + bridge}
        \resizebox{\linewidth}{!}{
          \begin{tikzpicture}
            \begin{axis}[
                    xlabel= $\Delta \mathrm{G_{liaison}\ exp}\ (\mathrm{kCal.mol}^{-1}$),
                    ylabel= $\Delta \mathrm{G_{liaison}\ calc}\ (\mathrm{kCal.mol}^{-1}$),
                    legend pos=south east,
                    xmin = -16, xmax = -2, ymin = -200, ymax = 20,
              ]
              \addplot[
        scatter,only marks,scatter src=explicit symbolic,
        scatter/classes={
            Sulfure={red},
            None={black}
}
    ] table[x index=0, y index=4, meta index=11]{chapters/Applications/datas/MMPBSA/results.csv};
            \legend{Avec S, Sans S};
            \end{axis}
          \end{tikzpicture}
}
     \end{subfigure}
    \caption[Énergie libre de liaison calculée par MM-PBSA et par MM-MDFT en fonction de la présence de Souffre.]{Énergie libre de liaison calculée par MM-PBSA et par MM-MDFT en fonction de la présence de Souffre. Les valeurs calculées sont comparées aux valeurs expérimentales pour différents paramètres de la fonctionnelle pour chacun des complexes étudiés. En rouge les complexes dont le ligand comporte un Souffre, en noir ceux n'en comportant pas.}
    \label{fig:MMPBSA_MMMDFT_complete_by_souffre}
\end{figure}
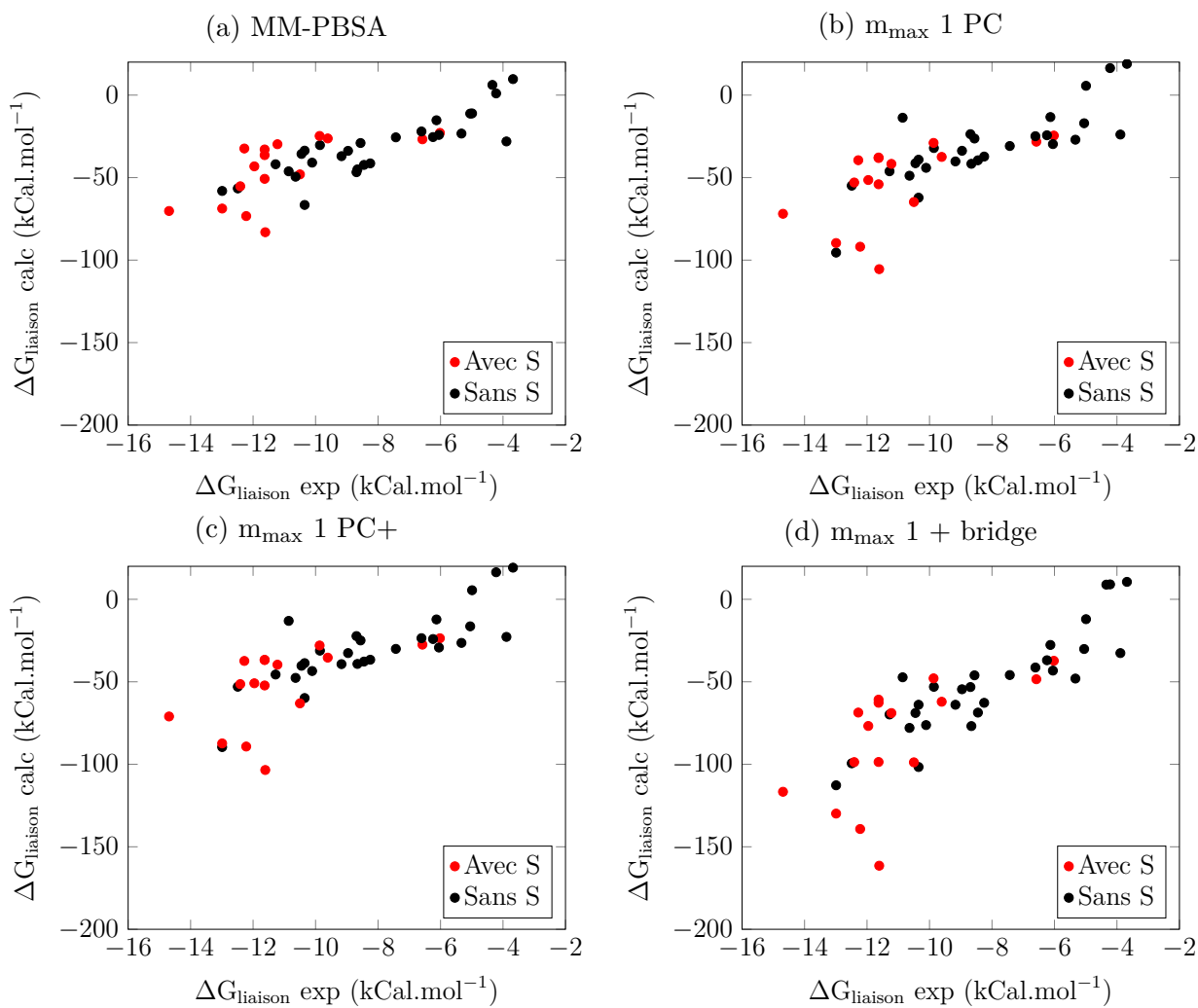

\clearpage
\strut
\vspace{10\baselineskip}

\boitemagique{A Retenir}{
Dans ce chapitre, nous avons montré qu'il est possible avec MDFT (i) d'étudier avec précision des systèmes biologiques, (ii) de retrouver les molécules d'eau expérimentales et les poches à l'intérieur de protéines et (iii) d'améliorer la prédiction d'énergie libre de liaison en solution tout en fournissant la structure de solvatation.
}

\part{Conclusion et perspectives}

\clearemptydoublepage
\chapter{Conclusion}
\label{chap:conclusion}
Le développement d'un nouveau médicament est un processus long et co\^uteux. Entre la détermination d'une cible thérapeutique et la mise sur le marché d'un nouveau médicament, plus de dix ans de recherche sont nécessaires pour un coût supérieur à un milliard d'euros.
L'accélération de ce processus et donc la réduction de son coût reste un enjeu majeur. Pour y parvenir, les simulations numériques, peu co\^uteuses et rapides, sont massivement utilisées. Malgré cela, elles restent limitées, en partie à cause de la quantité très importante de molécules de solvant à considérer.

La théorie de la fonctionnelle de la densité moléculaire permet d'étudier la solvatation de composés de n'importe quelle taille et de n'importe quelle forme. Elle permet en quelques secondes seulement d'obtenir à la fois l'énergie libre de solvatation et une carte détaillée de la densité d'équilibre autour de ce soluté.
Ces grandeurs étant à la base de nombreux autres calculs utilisés par l'industrie pharmaceutique, la MDFT ouvre donc une autre voie d'optimisation de ces process.

Durant ma thèse, mon travail a consisté à effectuer les premiers pas vers des applications biologiques. Cette thèse s'est déroulée en trois grandes étapes. La première consistait à adapter la théorie à des macro-molécules biologiques. Pour cela, nous avons développé une version à symétrie sphérique de la théorie de la fonctionnelle de la densité moléculaire. Cette version, simplifiée et plus rapide, nous a permis de paramétriser un nouveau bridge: le bridge gros grain.
Ce bridge, basé sur une densité gros grain, ajoute de la consistance thermodynamique à nos modèles, en reproduisant une pression et une tension de surface correctes. Il améliore également le calcul de l’énergie libre de solvatation et la prédiction de la structure du solvant, sur les petites molécules mais également sur des molécules plus grosses comme des systèmes biologiques.

La seconde étape consistait à l'adaptation du code. En effet, l'étude de composés de plusieurs milliers d'atomes a mis en évidence différentes limites techniques. Afin de dépasser ces limites, nous avons dû adapter, optimiser et parralléliser le code MDFT.

Enfin, la dernière étape consistait à évaluer nos développements théoriques et numériques sur des systèmes d'interêt biologique. Pour cela, trois études ont été menées. La première: le benchmark de MDFT sur un ensemble de 604 composés de type médicament. Nous avons ainsi mis en évidence que la correction de pression \textit{PC+} n'est plus aujourd'hui adaptée à la théorie au niveau HNC. La meilleure précision dans la prédiction de l'énergie libre de solvation de composés de type médicaments est obtenue à l'aide de la théorie dans l'approximation HNC couplée à la correction de pression \textit{PC} et pour une valeur de $\mathrm{m}_\mathrm{max}$=3.
Les deux autres applications ont permis d'évaluer MDFT sur des systèmes biologiques plus importants. Nous avons ainsi montré qu'il est possible avec MDFT (i) d’étudier avec précision des systèmes biologiques, (ii) de retrouver les molécules d’eau expérimentales et les poches à l’intérieur de protéines et (iii) d’améliorer la prédiction d’énergie libre de liaison en solution tout en fournissant la structure de solvatation. 

\boitesimple{En conclusion, durant cette thèse, nous avons adapté la théorie MFDT et son code associé afin de permettre une étude rapide et précise de systèmes biologiques. L'ensemble de ces travaux constituent un premier pas et ouvre une nouvelle voie d'application pour MDFT: la recherche de médicament.}

\chapter{Perspectives}
\label{chap:perspectives}
Les résultats obtenus durant cette thèse sont très encourageants. Cependant, MDFT reste une voie ouverte de recherche qui offre encore de nombreuses possibilitées et il reste encore beaucoup à faire. Nous proposons ici un ensemble de perspectives non exhaustives qui viennent s'ajouter à celles déjà connues de MDFT sur les petits composés.

\section{MM-MDFT: l'approximation de trajectoire unique}
Les premiers résultats obtenus dans le cadre de la dérivation de MM-PBSA en MM-MDFT sont encourageants. En effet, l'utilisation de MDFT permet actuellement uniquement l'apport d'informations supplémentaires au travers de la structure du solvant. La précision ainsi que le temps de calcul restent les mêmes pour ces deux méthodes. Dans cette étude préliminaire, nous n'avons considéré qu'une seule conformation par complexe: la conformation après minimisation de la structure cristallographique. Pour aller plus loin et ainsi augmenter la précision des résultats obtenus, il est possible de calculer une trajectoire de dynamique moléculaire ou Monte Carlo du complexe dans le vide et d'en extraire différentes conformations à intervalles réguliers. L'énergie libre de solvatation du système correspond, dans ce cas, à la moyenne des énergies libres de solvatation calculées pour chaque conformation. Cette technique, nommée approximation de trajectoire unique est connue pour fortement améliorer les résultats obtenus. Au moment de la rédaction de ce manuscrit, ces calculs sont en cours.

\section{Une étude plus complète des ions}
Le benchmark de MDFT sur des composés de type médicament, et plus particulièrement sur les ions, a permis de mettre en évidence une limite importante de MDFT: les charges partielles. Jusqu'ici nous avons uniquement étudié quelques ions monovalents. Les premiers résultats semblent indiquer que l'erreur pourrait dépendre uniquement de la charge du composé. Afin de confirmer cette tendance et ainsi mieux comprendre et donc corriger ce défaut, la prochaine étape indispensable est l'étude d'une base de données d'ions plus importante non restreinte à des composés monovalents.

\section{Machine learning}
Le développement de l'outil \textit{MDFT Database Tool} permet une obtention rapide, simple et automatique de résultats sur des bases de données de plusieurs milliers de composés. Ce développement a ainsi permis la création de banques de données de référence indispensables à la mise en place d'outils de \textit{machine learning}. Face à l'augmentation exponentielle du nombre de données disponibles, les outils de \textit{machine learning} semblent inévitables. Les premiers résultats obtenus par Sohvi Luukkonen pendant son stage sont d'ailleurs très encourageants.

\section{MDFT pour le \textit{drug design}}
Enfin, cette thèse a ouvert une nouvelle voie pour MDFT: la recherche pharmaceutique. Suite à ces développements, plusieurs applications directes sont actuellement envisageables. Les étapes suivantes consistent donc: (i) à coupler MDFT avec les techniques existantes (docking, virtual screening, ...). La substitution dans ces logiciels d'un solvant implicite par MDFT permettrait l'apport d'informations moléculaires supplémentaires pour des temps de calcul similaires. Et (ii) proposer des alternatives plus précises à certains calculs basés sur l'énergie libre de solvatation comme le calcul du logP ou encore celui du logBBB.

\section{Couplages de MDFT}
En dehors du cadre de cette thèse, la perspective principale de MDFT consiste au remplacement de solvants implicites à différentes échelles. Pour cela deux projets sont actuellement en cours.
\subsection{\`A l'échelle microscopique}
Le premier projet consiste à coupler MDFT à des méthodes quantiques et donc à l'échelle microscopique. En pratique, la méthode quantique, comme la DFT électronique génère une structure électronique pour une petite molécule. À partir de cette structure (qui correspond au potentiel extérieur dans MDFT) il est ensuite possible de prédire la structure du solvant autour de ce composé et ainsi d'affiner la strucutre électronique du composé étudié.

\subsection{\`A l'échelle mésoscopique}
Le second projet consiste à coupler MDFT au logiciel de résolution des équations hydrodynamiques, Laboetie\cite{Levesque_accounting_2013,Vanson_unexpected_2015,Asta_transient_2017}. Laboetie, basé sur la méthode de Lattice-Boltzmann, propage une fonctionnelle (bien plus simple que celle minimisée par MDFT) d'une densité en liquide. Laboetie est unique car il tient compte de la spécificité chimique pour étudier du transport dit "réactif". En d'autres termes, il permet l'étude de la dynamique de particules qui peuvent s'adsorber et se désorber de surfaces.
MDFT permettrait ici une meilleure modélisation de la surface en calculant, à l'échelle moléculaire, les constantes cinétiques d'adsorption et de désorption qui sont aujourd'hui fournies en paramètre d'entrée.


\clearemptydoublepage
\thispagestyle{empty}
This work was supported by the Energy oriented Centre of Excellence (EoCoE), grant agreement number 676629, funded within the Horizon2020 framework of the European Union

\clearemptydoublepage
\renewcommand{\thesubsection}{\Alph{chapter}}
\appendix

\part*{Annexes}
\addcontentsline{toc}{chapter}{Annexes}

\chapter{Calcul du gradient de la fonctionnelle}
Afin d'améliorer la lisibilité, nous nous affranchirons dans ces annexes de $(\boldsymbol{r},\Omega)$. Nous remplacerons donc $\rho(\boldsymbol{r},\Omega)$ par $\rho$, $\phi(\boldsymbol{r},\Omega)$ par $\phi$ et $\gamma(\boldsymbol{r},\Omega)$ par $\gamma$.
\label{chap:annexes:grad}

\clearpage
\section{Fonctionnelle idéale}
\label{sec:annexes:grad:id}
\begin{eqnarray}
\beta \delta \mathcal{F}_\mathrm{id}[\rho] &=& \beta \mathcal{F}_\mathrm{id}[\rho + \delta \rho] -\beta \mathcal{F}_\mathrm{id}[\rho] \\
&=& \int\mathrm{d}\boldsymbol{r}\mathrm{d}\Omega\ [\rho + \delta \rho]\ln(\frac{\rho + \delta \rho}{\rho_0})- [\Delta\rho + \delta \rho] \\
& & - \int\mathrm{d}\boldsymbol{r}\mathrm{d}\Omega\ \rho\ln(\frac{\rho}{\rho_0})-\Delta\rho \nonumber \\
&=& \int\mathrm{d}\boldsymbol{r}\mathrm{d}\Omega\ \rho \ln(\frac{\rho + \delta \rho}{\rho_0}) + \delta \rho\ln(\frac{\rho + \delta \rho}{\rho_0}) - \Delta\rho - \delta\rho \\
& & - \rho\ln(\frac{\rho}{\rho_0}) + \Delta\rho \nonumber \\
&=& \int\mathrm{d}\boldsymbol{r}\mathrm{d}\Omega\ \rho \ln(\frac{\rho + \delta \rho}{\rho_0}) + \delta \rho\ln(\frac{\rho + \delta \rho}{\rho_0}) - \delta\rho - \rho\ln(\frac{\rho}{\rho_0})\nonumber
\end{eqnarray}
Or
\begin{eqnarray}
\ln(\frac{\rho + \delta \rho}{\rho_0}) = \ln(\frac{\rho + \delta \rho}{\rho}\frac{\rho}{\rho_0}) = \ln(\frac{\rho + \delta \rho}{\rho})+\ln(\frac{\rho}{\rho_0}) = \ln(1+\frac{\delta \rho}{\rho})+\ln(\frac{\rho}{\rho_0})
\end{eqnarray}
Et, comme $\frac{\delta \rho}{\rho}$ tend vers 0, il est possible de faire le développement de Taylor de $\ln(1+\frac{\delta \rho}{\rho})$ qui nous donne:
\begin{eqnarray}
\ln(1+\frac{\delta \rho}{\rho}) = \frac{\delta \rho}{\rho} + \mathcal{O}(\delta\rho^{2})
\end{eqnarray}
On injecte ce développement l'équation précédente:
\begin{eqnarray}
\beta \delta \mathcal{F}_\mathrm{id}[\rho] &=& \int\mathrm{d}\boldsymbol{r}\mathrm{d}\Omega\ \rho \ln(\frac{\rho + \delta \rho}{\rho_0})  + \delta \rho\ln(\frac{\rho + \delta \rho}{\rho_0}) - \delta\rho - \rho\ln(\frac{\rho}{\rho_0})\\
& & + \mathcal{O}(\delta\rho^{2}) \nonumber \\
&=& \int\mathrm{d}\boldsymbol{r}\mathrm{d}\Omega\ \rho [\frac{\delta \rho}{\rho} + \ln(\frac{\rho}{\rho_0})] + \delta \rho[\frac{\delta \rho}{\rho} + \ln(\frac{\rho}{\rho_0})] - \delta\rho \\
& & - \rho\ln(\frac{\rho}{\rho_0}) + \mathcal{O}(\delta\rho^{2}) \nonumber \\
&=& \int\mathrm{d}\boldsymbol{r}\mathrm{d}\Omega\ \delta\rho + \rho \ln(\frac{\rho}{\rho_0}) + \frac{\delta \rho^2}{\rho} + \delta \rho \ln(\frac{\rho}{\rho_0}) - \delta\rho \\
& & - \rho\ln(\frac{\rho}{\rho_0}) + \mathcal{O}(\delta\rho^{2}) \nonumber \\
&=& \int\mathrm{d}\boldsymbol{r}\mathrm{d}\Omega\ \delta \rho \ln(\frac{\rho}{\rho_0}) + \mathcal{O}(\delta\rho^{2})
\end{eqnarray}
On obtient finalement:
\boitesimple{
\begin{eqnarray}
\beta \frac{\delta \mathcal{F}_\mathrm{id}[\rho]}{\delta \rho} &=& \ln(\frac{\rho}{\rho_0})
\end{eqnarray}
}

Le gradient de la partie idéale s'annule en $\rho=\rho_0$. Cette partie tend donc vers un système homogène de densité $\rho_0$.

\clearpage
\section{Fonctionnelle extérieure}
\label{sec:annexes:grad:ext}
\begin{eqnarray}
\delta \mathcal{F}_\mathrm{ext}[\rho] &=& \mathcal{F}_\mathrm{ext}[\rho + \delta \rho] -\mathcal{F}_\mathrm{ext}[\rho] \\
&=& \int\mathrm{d}\boldsymbol{r}\mathrm{d}\Omega\ (\rho+\delta\rho)\phi - \int\mathrm{d}\boldsymbol{r}\mathrm{d}\Omega\ \rho\phi  \\
&=& \int\mathrm{d}\boldsymbol{r}\mathrm{d}\Omega\ \delta\rho\phi
\end{eqnarray}
On obtient finalement:
\boitesimple{
\begin{eqnarray}
\beta \frac{\delta \mathcal{F}_\mathrm{ext}[\rho]}{\delta \rho} &=& \phi
\end{eqnarray}
}

On voit ici que la partie idéale favorisera une faible densité pour une valeur élevée de potentiel $\phi$ et des densités fortes pour des valeurs de potentiels faibles.

\clearpage
\section{Fonctionnelle d'excès}
\label{sec:annexes:grad:exc}
\begin{eqnarray}
\beta \delta \mathcal{F}_\mathrm{exc}[\rho] &=& \beta \mathcal{F}_\mathrm{exc}[\rho + \delta \rho, \rho^\prime] -\beta \mathcal{F}_\mathrm{exc}[\rho, \rho^\prime] \\
& & + \beta \mathcal{F}_\mathrm{exc}[\rho, \rho^\prime + \delta \rho^\prime] -\beta \mathcal{F}_\mathrm{exc}[\rho, \rho^\prime] \nonumber\\
&=& 2 \beta \mathcal{F}_\mathrm{exc}[\rho + \delta \rho, \rho^\prime] - 2 \beta \mathcal{F}_\mathrm{exc}[\rho, \rho^\prime] \\
&=& -\int\mathrm{d}\boldsymbol{r}\mathrm{d}\Omega\ (\Delta\rho+\delta\rho) \gamma  + \int\mathrm{d}\boldsymbol{r}\mathrm{d}\Omega\ (\Delta\rho \gamma) \\
&=& -\int\mathrm{d}\boldsymbol{r}\mathrm{d}\Omega\ \delta\rho \gamma
\end{eqnarray}
On obtient finalement:
\boitesimple{
\begin{eqnarray}
\beta\frac{\delta \mathcal{F}_\mathrm{exc}[\rho]}{\delta \rho} = - \gamma
\end{eqnarray}
}

\clearpage
\section{Fonctionnelle de bridge}
\label{sec:annexes:grad:bridge}
\begin{eqnarray}
\hat{\rho}(\vec{k})&=&HT[\rho(\vec{r})]  \\
\bar{\hat{\rho}}(\vec{k})&=&\hat{\rho}(\vec{k})K(\vec{k})   \\
\bar{\rho}(\vec{r})&=&HT^{-1}[\bar{\hat{\rho}}(\vec{k})]  \\
\Delta\bar{\rho}(\vec{r})&=&\bar{\rho}(\vec{r})-\rho_{0}  \\
\bar{F}_{b}[\rho(\vec{r})]&=&A\delta\bar{\rho}(\vec{r})^{3}+B\bar{\rho}(\vec{r})^{2}\delta\bar{\rho}(\vec{r})^{4}  \\
\frac{\delta\bar{F}_{b}[\rho(\vec{r})]}{\delta\rho(\vec{r})}&=&3A\delta\bar{\rho}(\vec{r})^{2}+2B\bar{\rho}(\vec{r})\delta\bar{\rho}(\vec{r})^{4}+4B\bar{\rho}(\vec{r})^{2}\delta\bar{\rho}(\vec{r})^{3}  \\
\frac{\delta\bar{\hat{F}}_{b}[\rho(\vec{r})]}{\delta\rho(\vec{r})}&=&HT[\frac{\delta\bar{F}_{b}[\rho(\vec{r})]}{\delta\rho(\vec{r})}]  \\
\frac{\delta\hat{F}_{b}[\rho(\vec{r})]}{\delta\rho(\vec{r})}&=&\frac{\delta\bar{\hat{F}}_{b}[\rho(\vec{r})]}{\delta\bar{\rho}(\vec{r})}K(\vec{k})  \\
\frac{\delta F_{b}[\rho(\vec{r})]}{\delta\rho(\vec{r})}&=&HT^{-1}\frac{\delta\hat{F}_{b}[\rho(\vec{r})]}{\delta\rho(\vec{r})} 
\end{eqnarray}




\chapter{Mesures statistiques}
\label{chap:annexes:stats}

Cinq mesures statistiques sont disponibles avec \textit{MDFT Database Tool} pour quantifier les performances de la dynamique moléculaire ou de MDFT: l'erreur quadratique moyenne RMSE, le $\mathrm{P}_{\mathrm{bias}}$ ainsi que 3 coefficients de corrélation: Pearson R, Spearman $\rho$ et Kendall $\tau$.

\paragraph{Root-mean-squared error (RMSE)} correspond à la racine de l'erreur quadratique moyenne. Cette métrique permet de mesurer la différence entre les valeurs calculées et les valeurs de référence. Elle est définie comme:

\begin{equation}
    RMSE = \sqrt{MSE} = \sqrt{\frac{\sum_i{(\hat{y}_i-y_i)^2}}{n}}
\end{equation}

\noindent avec $\hat{y}_i$ la valeur calculée, $y_i$ la valeur de référence et n la taille de l'échantillon.

\paragraph{Le coefficient de corrélation de Pearson ($R$)} la corrélation linéaire entre deux variables $X$ et $Y$. Il est définit comme

\begin{equation}
    R = \frac{cov(X,Y)}{\sigma_X \sigma_Y}
\end{equation}

\noindent avec $cov(X,Y)=E[(X-E[X])(Y-E(Y))]$ la covariance entre $X$ et $Y$, $\sigma_X$ et $\sigma_Y$ la déviation standard de $X$ et $Y$. $R$ varie entre -1 and 1. La magnitude de $|R|$ mesure la qualité de la corrélation linéraire et le signe de $R$ correspond au signe de la pente. Le coefficient de détermination $R^2$ généralement utilisé correspond au carré du coefficient $R$ de Perason.

\paragraph{Le coefficient de corrélation de rang de Spearman ($\rho$)}  mesure le dépendance statistique non paramétrique entre deux variables $X$ and $Y$. Il est utilisé lorsque deux variables statistiques semblent corrélées sans que la relation entre les deux variables soit de type affine. Pour y parvenir, il calcule le coefficient de corrélation entre les rangs des différentes valeurs et non entre les valeurs elles mêmes. Il est définit comme :

\begin{equation}
    \rho = \frac{cov(rg_X,rg_Y)}{\sigma_{rg_X} \sigma_{rg_Y}}
\end{equation}

\noindent avec $cov(rg_X,rg_Y)$ la covariance entre les rangs des variables $rg_X$ et $rg_Y$ et $\sigma_{rg_X}$ et $\sigma_{rg_Y}$ la déviation standard des rangs de ces variables. 

\paragraph{Le coefficient de corrélation de Kendall ($\tau$)}  est une autre mesure basée sur le rang des valeurs. Il est définit comme:

\begin{equation}
    \tau = \frac{n_{con}-n_{dis}}{n(n-1)/2}
\end{equation}

\noindent avec $n_{con}$ le nombre de paires concordantes, $n_{dis}$ le nombre de paires discordantes et $n$ le nombre de pairs total. Lorsque de l'étude de chaque paire possible $[(X_i,Y_i)(X_j,Y_j)]$, on compte concordant toute paire respectant $Y_i>Y_j$ si $X_i>X_j$ ou $Y_i<Y_j$ si $X_i<X_j$. Les autres paires sont comptées discordantes.

\paragraph{Le pourcentage de biais ($P_{bias}$)}, exprimé en pourcentage, mesure la tendance moyenne relative des valeurs calculées à être supérieures ou inférieures aux valeurs de référence. Il est définit comme:

\begin{equation}
    P_{bias} = \frac{\sum_i(Y_{i}^{obs}-Y_{i}^{sim})*100}{\sum_i(Y_{i}^{obs})}
\end{equation}

\noindent avec $Y_{i}^{sim}$ les valeurs calculées et $Y_{i}^{obs}$ les valeurs de référence. Une valeur de biais positive indique que le modèle à tendance à surestimer les valeurs calculées alors qu'une valeur négative indique que les valeurs sont sous-estimées.

\clearemptydoublepage
\backmatter
\printbibliography

\tikzexternaldisable

\end{document}